\documentclass[a4paper,11pt]{article}
\pdfoutput=1 

\usepackage{jcappub} 

\usepackage[T1]{fontenc} 

\usepackage{empheq}

\usepackage{listings}

\usepackage{tablefootnote}

\newcommand*\widefbox[1]{\fbox{\hspace{2em}#1\hspace{2em}}}
\newcommand{\Kd}{\delta^{\rm K}}

\title{\boldmath A Model for the Redshift-Space Galaxy 4-Point Correlation Function}

\author[1,3]{William Ortolá Leonard,} 
\author[2,3]{Zachary Slepian,}
\author[2,4]{and Jiamin Hou}

\affiliation[1]{Department of Physics, University of Florida,\\2001 Museum Rd., Gainesville, FL 32611, USA}
\affiliation[2]{Department of Astronomy, University of Florida,\\211 Bryant Space Science Center, Gainesville, FL 32611, USA}
\affiliation[3]{Lawrence Berkeley National Laboratory,\\1 Cyclotron Road, Berkeley, CA 94720, USA}
\affiliation[4]{Max-Planck-Institut für Extraterrestrische Physik,\\
Postfach 1312, Giessenbachstrasse 1, 85748 Garching bei München, Germany}
\emailAdd{wortola@ufl.edu}
\emailAdd{zslepian@ufl.edu}
\emailAdd{jiamin.hou@ufl.edu}

\abstract{
The field of cosmology is entering an epoch of unparalleled wealth of observational data thanks to galaxy surveys such as DESI, Euclid, and Roman. Therefore, it is essential to have a firm theoretical basis that allows the effective analysis of the data. With this purpose, we compute the nonlinear, gravitationally-induced connected galaxy 4-point correlation function (4PCF) at the tree level in Standard Perturbation Theory (SPT), including redshift-space distortions (RSD). We begin from the trispectrum and take its inverse Fourier transform into configuration space, exploiting the isotropic basis functions of \cite{Iso_fun}. We ultimately reduce the configuration-space expression to low-dimensional radial integrals of the power spectrum. This model will enable the use of the BAO feature in the connected 4PCF to sharpen our constraints on the expansion history of the Universe. It will also offer an additional avenue for determining the galaxy bias parameters, and thus tighten our cosmological constraints by breaking degeneracies. Survey geometry can be corrected in the 4PCF, and many systematics are localized, which is an advantage over data analysis with the trispectrum. }

\begin{document}
\maketitle
\flushbottom

\section{Introduction}
\label{sec:intro}
\subsection{Current and Future Data Landscape}

Cosmology is entering a golden age, driven by a wealth of current and upcoming data. These data will be obtained through three different types of experiments to measure the Large-Scale Structure (LSS) of the Universe. We have ground-based photometric experiments such as the Kilo-Degree Survey~\citep[KiDS;][]{KiDS}, the Dark Energy Survey~\citep[DES;][]{DES}, and the Legacy Survey of Space and Time~\citep[LSST;][]{LSST} at Vera Rubin Observatory. Moreover, we also have ground-based spectroscopic experiments such as the Dark Energy Spectroscopic Instrument ~\citep[DESI;][]{DESI}, the Hobby Eberly Telescope Dark Energy Experiment ~\citep[HETDEX;][]{HETDEX}, the Large Sky Area Multi-Object Fiber Spectroscopic Telescope  ~\citep[LAMOST;][]{LAMOST}, the Prime Focus Spectrograph ~\citep[PFS;][]{PFS} and the 4-meter Multi-Object Spectroscopic Telescope ~\citep[4MOST;][]{4MOST}. Finally, we also have space-based experiments, including the Euclid Satellite Mission~\citep[Euclid;][]{Euclid} and the Nancy Grace Roman Space Telescope ~\citep[Roman;][]{Roman}. From this current and upcoming data, we can gain more insight into the origin and the evolution of the Universe, and address fundamental questions such as the nature of dark energy and dark matter.

\subsection{Past, Present, and Future Work on the Trispectrum and 4PCF}
With this wealth of data, it is critical to develop theoretical models required to extract parameter constraints. In the past, data from galaxy surveys such as the Sloan Digital Sky Survey~\citep[SDSS;][]{SDSS} and its cosmological programs, Baryon Oscillation Spectroscopic Survey ~\citep[BOSS;][]{BOSS} and extended BOSS ~\citep[eBOSS;][]{eBOSS} have been analyzed using the 2-point correlation function (2PCF)~\cite{EisenBAO,Hou2018eboss, Hou2021eboss, Bautista2021eboss, Tamone2020eboss,Satpathy, Grieb, Sanchez}, or its Fourier-space counterpart, the power spectrum \cite{Cole05, deMattia2021eboss, Neveux2020eboss, Beutler}. 

However, due to non-linear gravitational evolution, the galaxies' distribution becomes non-Gaussian at late times. The 2PCF does not fully capture the non-Gaussian distribution. Therefore, to access this information, we must employ higher-order correlation functions \cite{Bernardeau_review,Higher_order_info,Ivanov_Bispectrum}. Previous works have modeled the 3PCF~\cite{SlepianBAO} and its Fourier-space analog, the bispectrum~\cite{Scoccimarro1998}. 3-point statistics have shown to be a powerful tool to break the degeneracy of the neutrino mass with galaxy bias \cite{Farshad_neutrino, Aviles}, constrain higher-order galaxy biasing parameters \cite{3PCFbiasConstraint, LadoRSDBispectrum, Eggemeier_Bias, D'Amico_Bispectrum}, test modified gravity theories \cite{AvilesMG, modified_gr2, Hou2023:MGreview}, and study alternative dark energy models \cite{DESTUDY, TsedrikDE}. 

The galaxy 4-Point Correlation Function (4PCF) was first defined and estimated in~\cite{Fry4PCFdef} using the Lick and Zwicky catalogues \cite{LickCatalogue, ZwickyCatalogue}. Later,~\cite{Fry4PCF_BBGKY1, Fry4PCF_BBGKY2} provided a model of the 4PCF for the Born-Bogoliubov-Green-Kirkwood-Yvon (BBGKY) hierarchy. More recently, other works have modeled the Fourier space counterpart of the 4PCF, the trispectrum, using perturbation theory at tree level \cite{Gualdi1, Gualdi2} as well as at one-loop order using effective field theory (EFT) \cite{Bertolini_Trispectrum}. There have also been measurements of the 4PCF in BOSS \cite{Sabiu_4PCF_measurment, Philcox_4PCF_measurment}, as well as the integrated trispectrum \cite{itrispectrum_measurment}; the galaxy 4PCF was first proposed as a probe of parity violation in \cite{cahn_prl} and measured on data in \cite{Parity-odd} and subsequently in \cite{Philcox-Parity} with covariance as in \cite{hou-analytic-covar}; use of the even-parity 4PCF to calibrate the covariance of the odd-parity 4PCF was proposed in \cite{Parity-odd}, and using Baryon Acoustic Oscillations (BAO) in the odd-parity 4PCF in \cite{hou_bao}. 


In this paper we present the first model of the even-parity, redshift-space galaxy 4PCF, sourced by non-linear structure formation under gravity. With this model, we can study the Baryon Acoustic Oscillation (BAO) features, which provide a standard ruler to study the cosmic expansion history \cite{BondBAO1, BondBAO2, PeeblesBAO, SunyaevBAO, EisenBAO,HuBAO}.


These proposed studies rely on accurate knowledge of galaxies' 3D positions. However, when measuring  galaxies' distance from us along the line of sight, it is assumed they are commoving with the expansion of the Universe. In reality, though, they also have peculiar velocities. Peculiar velocities cause an additional Doppler shift. If we do not take this into account we will infer a wrong line of sight distance, causing so-called ``redshift-space distortions'' (RSD) in the recovered map \cite{Hamilton}. Therefore, the 4PCF model of this work includes RSD. Since our model includes RSD (which ultimately stem from growth of structure) and galaxy biasing, we can measure the matter density of the Universe, the logarithmic derivative of the linear growth rate, $f$, and galaxy bias parameters up to third order in Eulerian Standard Perturbation Theory (SPT).

\subsection{Plan of the Paper}
This paper is structured as follows. In $\S$\ref{Sec: SPT}, we present the formalism used to model the density fluctuations in the distribution of matter and of galaxies. We then present the tree-level trispectrum including RSD. In $\S$\ref{Section T3}, we show the redshift-space terms that comprise the tree-level contribution from $T_{3111}(\mathbf{k}_1, \mathbf{k}_2, \mathbf{k}_3, \mathbf{k}_4)$ and we show how to convert all of the terms into configuration space. We follow the same procedure in $\S$\ref{Section T2} for $T_{2211}(\mathbf{k}_1, \mathbf{k}_2, \mathbf{k}_3, \mathbf{k}_4)$. Finally, we complement the analyses with several appendices showing mathematical details needed for the main text’s derivations: conventions of the paper, decoupling denominators, use of the isotropic basis functions, numerical integration for the radial integrals we will find in our results, and finally, analytic integration of these radial integrals. 

\section{Computing the \textit{State of the Art} Redshift-Space Galaxy Tree-Level Trispectrum}\label{Sec: SPT}
This section briefly reviews the aspects of Standard Perturbation Theory (SPT) and galaxy biasing, and computes the state of the art galaxy redshift-space Trispectrum; it also establishes the notation used throughout this work.

We begin with the idea that the LSS we observe in the Universe forms from small primordial fluctuations that evolve in an expanding Universe under the force of gravity. Therefore, we start from the collisonless Boltzmann equation \cite{Bernardeau_review}: 

\begin{align}\label{eq:Boltzmann's}
\frac{df(\mathbf{x,p},\tau)}{dt} = \frac{1}{a(\tau)} \frac{ \partial f}{ \partial \tau} + \frac{p^i}{m \; a(\tau)^{2}} \frac{ \partial f}{ \partial x^i}- m \frac{\partial f}{\partial p^i} \frac{ \partial \Phi }{ \partial x_i }=0,
\end{align}
where the above equation is an approximation since we are neglecting the collision of Baryons, which is a valid assumption for the scales at which we will apply our model. We are using the Einstein summation convention, $f$ is the phase space density of particles, $m$ and $p^i$ are respectively the mass and $i^{\rm th}$ momentum component of the matter particles, $a(\tau)$ is the scale factor, and $\tau$ is the conformal time, which is defined relative to the comoving observer's time as $dt = a(\tau)d\tau$. $\Phi$ is the cosmological gravitational potential:

\begin{align}
\Phi(\mathbf{x},\tau) \equiv \frac{1}{2} \frac{\partial \mathcal{H}}{\partial \tau} x^2 + \phi (\mathbf{x},\tau), \nonumber
\end{align}
with $\phi$ being the Newtonian potential induced by the local mass density $\rho(\mathbf{r})$ and $\mathcal{H}\equiv aH = da/dt$ \cite{Bernardeau_review}.

The standard approach to the Boltzmann equation is taking velocity moments. The zeroth velocity moment corresponds to integrating Eq. (\ref{eq:Boltzmann's}) over the 3D momentum, which results in: 

\begin{align}
\frac{ \partial \rho}{\partial \tau} + \frac{1}{a(\tau)} \frac{\partial (\rho v^i)}{\partial x^i} = 0,
\end{align}
where
\begin{align}
\rho(\mathbf{x},\tau) \equiv m\int d^{3}\mathbf{p} \; f(\mathbf{x,p},\tau), \nonumber
\end{align}
is the  matter density and 

\begin{align}
v^i(\mathbf{x},\tau) = \frac{1}{\rho(\mathbf{x},\tau)}\int d^{3}\mathbf{p} \; p^i f(\mathbf{x,p},\tau), \nonumber    
\end{align}
is the $i^{\rm th}$ component of its peculiar velocity. Using $\rho(\mathbf{x},\tau) = \overline{\rho}(1+\delta(\mathbf{x},\tau))$, with $\overline{\rho}$ being the mean comoving matter density and $\delta(\mathbf{x},\tau)$---known as the density contrast---being the density fluctuation relative to $\overline{\rho}$, we arrive at the continuity equation:

\begin{align}\label{eq:Continuity}
\frac{ \partial \delta(\mathbf{x},\tau)}{\partial \tau}  + \frac{1}{a(\tau)} \frac{\partial \left[\left(1+\delta(\mathbf{x},\tau) \right) v^i\right]}{\partial x^i} = 0. 
\end{align}

Next, we evaluate the first moment of the Boltzmann equation, which amounts to weighting Eq. (\ref{eq:Boltzmann's}) by the momentum vector and performing a 3D integral over momentum. This procedure results in the Euler equation:

\begin{align}\label{eq:Euler}
\frac{\partial  v^i}{\partial \tau} + \mathcal{H} v^i+ v^j \frac{\partial v^i}{\partial x^j} + \frac{\partial \Phi}{\partial x^i} = 0.
\end{align}

We now have a system of two coupled partial differential equations, Eqs. (\ref{eq:Continuity}) and (\ref{eq:Euler}), which are complicated to solve. Therefore, we continue by taking their divergences and Fourier Transforming the results---using the definition of velocity divergence $\theta \equiv \partial_i v^{i}(\mathbf{x},\tau)$---to obtain:

\begin{align}\label{eq:Continuity_Fourier}
\frac{\partial \widetilde{\delta}(\mathbf{k},\tau)}{\partial \tau} + \widetilde{\theta}(\mathbf{k},\tau) = - \int \frac{d^{3}\mathbf{k}_1 d^{3}\mathbf{k}_2}{(2\pi)^{6}} \; \delta_{\rm D}^{\left[3\right]}(\mathbf{k} - \mathbf{k}_1 - \mathbf{k}_2) \alpha(\mathbf{k}_1,\mathbf{k}_2)\widetilde{\theta}(\mathbf{k}_1,\tau)\widetilde{\delta}(\mathbf{k}_2,\tau)
\end{align}
and 

\begin{align}\label{eq:Euler_Fourier}
&\frac{\partial \widetilde{\theta}(\mathbf{k},\tau)}{\partial \tau} +\mathcal{H} \widetilde{\theta}(\mathbf{k},\tau) + \frac{3}{2} \Omega_{\rm m} \mathcal{H}^2(\tau) \widetilde{\delta}(\mathbf{k},\tau) = - \int \frac{d^{3}\mathbf{k}_1 d^{3}\mathbf{k}_2}{(2\pi)^{6}} \; \delta_{\rm D}^{\left[3\right]}(\mathbf{k} - \mathbf{k}_1 - \mathbf{k}_2) \nonumber \\ 
& \qquad \qquad \qquad \qquad \qquad \qquad \qquad \qquad \qquad \qquad \times \beta(\mathbf{k}_1,\mathbf{k}_2)\widetilde{\theta}(\mathbf{k}_1,\tau)\widetilde{\theta}(\mathbf{k}_2,\tau),
\end{align}
with the tilde denoting Fourier transform. We used the Poisson equation: 
\begin{align}
\frac{\partial^2 \Phi}{\partial x^2 } (\mathbf{x},\tau)= \frac{3}{2} \mathcal{H}^2(\tau)\Omega_{\rm m}(\tau) \delta(\mathbf{x},\tau),     
\end{align}
to write Eq. (\ref{eq:Euler_Fourier}) in terms of the density contrast, $\delta_{\rm D}$ to denote the 3D Dirac delta function, and $\alpha$ and $\beta$ are defined via \cite{Bernardeau_review}: 

\begin{align}
\alpha(\mathbf{k}_1,\mathbf{k}_2) \equiv \frac{\mathbf{k}_1\cdot (\mathbf{k}_1 + \mathbf{k}_2)}{k_{1}^2}, \qquad \qquad \beta(\mathbf{k}_1,\mathbf{k}_2) \equiv \frac{(\mathbf{k}_1 + \mathbf{k}_2)^2 (\mathbf{k}_1 \cdot \mathbf{k}_2)}{2 k_1^2 k_2^2},
\end{align}
while 

\begin{align}
\Omega_{\rm m}(\tau) \equiv \frac{\rho_{\rm}(\tau)}{\rho_{\rm crit.}(\tau)} \equiv \frac{8\pi G}{3 H^{2}(\tau)}\rho_{\rm}(\tau). \nonumber
\end{align}

The scale-free nature of collapse in matter domination (Einstein-de Sitter) ensures the factorizability of the space and time dependence of the expansion \cite{ Bernardeau1995, Bernardeau_review, Bouchet}. Therefore, Eqs. (\ref{eq:Continuity_Fourier}) and (\ref{eq:Euler_Fourier}) are solved with the following perturbative expansion:  

\begin{align}
\widetilde{\delta}(\mathbf{k},\tau) = \sum_{n=1}^{\infty} a^n(\tau)\widetilde{\delta}^{(n)}(\mathbf{k}), \qquad \qquad \widetilde{\theta}(\mathbf{k},\tau) = -\mathcal{H}(\tau) \sum_{n=1}^{\infty} a^n(\tau)\widetilde{\theta}^{(n)}(\mathbf{k}). 
\end{align}

We have \cite{Jain}: 

\begin{align}\label{eq:delta_n}
&\widetilde{\delta}^{(n)}(\mathbf{k}) = \int d^{3}\mathbf{q}_1\cdots\int d^{3}\mathbf{q}_{ n} \;\delta_{\rm D}^{\left[3\right]}(\mathbf{k}-\mathbf{q}_1 - \cdots -\mathbf{q}_{ n}) \nonumber \\  
& \qquad \quad \quad \times F_{ \rm s}^{( n)}(\mathbf{q}_1,\cdots,\mathbf{q}_{ n})\widetilde{\delta}^{(1)}(\mathbf{q}_1) \cdots \widetilde{\delta}^{(1)}(\mathbf{q}_{ n})
\end{align}
and 

\begin{align}\label{eq:theta_n}
&\widetilde{\theta}^{( n)}(\mathbf{k}) = \int d^{3}\mathbf{q}_1\cdots\int d^{3}\mathbf{q}_{ n}\; \delta_{\rm D}^{\left[3\right]}(\mathbf{k}-\mathbf{q}_1 - \cdots-\mathbf{q}_{ n})\nonumber \\  
& \qquad \quad \quad \times  G_{\rm s}^{({ n})}(\mathbf{q}_1,\cdots,\mathbf{q}_{ n})\widetilde{\delta}^{(1)}(\mathbf{q}_1)\cdots\widetilde{\delta}^{(1)}(\mathbf{q}_{ n}),
\end{align}
with $\widetilde{\delta}^{(1)}$ representing the linear density field. $F_{\rm s}^{({ n})}$ and $G_{\rm s}^{({ n})}$ are symmetrized kernels that characterize the coupling between different wave vectors and are given by \cite{Jeong}: 

\begin{align}
&F_{\rm s}^{({ n})}(\mathbf{q}_1,\cdots,\mathbf{q}_{ n}) = \frac{1}{n!} \sum_{\sigma} F^{({ n})}(\mathbf{q}_{\sigma_1},\cdots ,\mathbf{q}_{\sigma_n}) \nonumber \\ 
&  \qquad \qquad  \qquad \; \; \; \; \; = \frac{1}{n!} \sum_{\sigma} \sum_{m=1}^{n-1} \frac{G^{ (m)}(\mathbf{q}_1,\cdots,\mathbf{q}_{ m})}{(2n+3)(n-1)}\left[ (2n+1)\alpha(\mathbf{k}_1,\mathbf{k}_2) F^{ (n-m)}(\mathbf{q}_{\rm m+1},\cdots,\mathbf{q}_{ n}) \right. \nonumber \\ 
& \qquad \qquad \qquad \qquad \qquad \qquad  \quad \left. + 2 \beta(\mathbf{k}_1,\mathbf{k}_2) G^{ (n-m)}(\mathbf{q}_{\rm m+1},\cdots,\mathbf{q}_{ n})\right]
\end{align}
and 

\begin{align}
&G_{\rm s}^{({ n})}(\mathbf{q}_1,\cdots ,\mathbf{q}_{ n}) = \frac{1}{n!} \sum_{\sigma} G^{( n)}(\mathbf{q}_{\sigma_1},\cdots,\mathbf{q}_{\sigma_{ n}}) \nonumber \\ 
&  \qquad \qquad  \qquad \; \; \; \; \; = \frac{1}{n!} \sum_{\sigma} \sum_{m=1}^{n-1} \frac{G^{ (m)}(\mathbf{q}_1,\cdots,\mathbf{q}_{ m})}{(2n+3)(n-1)}\left[ 3\alpha(\mathbf{k}_1,\mathbf{k}_2) F^{(n-m)}(\mathbf{q}_{ m+1},\cdots,\mathbf{q}_{ n}) \right. \nonumber \\ 
& \qquad \qquad \qquad \qquad \qquad \qquad \quad \left. + 2 n \beta(\mathbf{k}_1,\mathbf{k}_2) G^{(n-m)}(\mathbf{q}_{ m+1},\cdots,\mathbf{q}_{ n})\right],
\end{align}
with the sum over $\sigma$ being for all the permutations $\sigma \equiv (\sigma_1,\cdots,\sigma_{ n})$ of the set $\left\{ 1,\cdots,n \right\}$. \textit{I.e.}, each $\sigma_{i}$ represents a specific permutation of the set $\left\{ 1,\cdots,n \right\}$ and we need to sum over all permutations of this set. We divide by the number of permutations, $n!$, to obtain the average. Also, $\mathbf{k}_1\equiv \mathbf{q}_1 + \cdots+ \mathbf{q}_{ m}$, $\mathbf{k}_2 \equiv \mathbf{q}_{ m+1} + \cdots+\mathbf{q}_{ n}$, $\mathbf{k} = \mathbf{k}_1 + \mathbf{k}_2$, and $F^{(1)} = G^{(1)} = 1$. For the rest of this paper we will suppress the subscript s and always refer to the symmetrized kernels, except where otherwise noted. Below we show the second- and third-order kernels \cite{Farshad_neutrino}: 

\begin{align}\label{eq:F2_def}
F^{(2)}\left(\mathbf{k}_2,\mathbf{k}_3\right) = \frac{17}{21} \mathcal{L}_{0}(\mathbf{\widehat{k}}_2 \cdot \mathbf{\widehat{k}}_3) + \frac{1}{2} \left(\frac{k_3}{k_2}+\frac{k_2}{k_3}\right)\mathcal{L}_{1}(\mathbf{\widehat{k}}_2 \cdot \mathbf{\widehat{k}}_3) + \frac{4}{21}\mathcal{L}_{2}(\mathbf{\widehat{k}}_2 \cdot \mathbf{\widehat{k}}_3),
\end{align}

\begin{align}\label{eq:G2_def}
G^{(2)}\left(\mathbf{k}_2,\mathbf{k}_3\right) = \frac{13}{21} \mathcal{L}_{0}(\mathbf{\widehat{k}}_2 \cdot \mathbf{\widehat{k}}_3) + \frac{1}{2} \left(\frac{k_3}{k_2}+\frac{k_2}{k_3}\right)\mathcal{L}_{1}(\mathbf{\widehat{k}}_2 \cdot \mathbf{\widehat{k}}_3) + \frac{8}{21}\mathcal{L}_{2}(\mathbf{\widehat{k}}_2 \cdot \mathbf{\widehat{k}}_3),
\end{align}

\begin{align}
 & F^{(3)}\left( \mathbf{k}_1,\mathbf{k}_2,\mathbf{k}_3\right) = \frac{2k_{123}^{2}}{54} \left[ \frac{\mathcal{L}_{1}(\mathbf{\widehat{k}}_1\cdot \mathbf{\widehat{k}}_{23})}{k_1 k_{23}}G^{(2)}\left( \mathbf{k}_2,\mathbf{k}_3\right) + 2 \;{\rm perm.} \right]  \nonumber \\ 
& \qquad   + \frac{7}{54} \left[ \frac{1}{k_{23}} \left\{ k_1 \mathcal{L}_1 (\mathbf{\widehat{k}}_1\cdot \mathbf{\widehat{k}}_{23}) + k_2 \mathcal{L}_1 (\mathbf{\widehat{k}}_2\cdot \mathbf{\widehat{k}}_{23}) + k_3 \mathcal{L}_1 (\mathbf{\widehat{k}}_3\cdot \mathbf{\widehat{k}}_{23}) \right\} G^{(2)}\left( \mathbf{k}_2,\mathbf{k}_3\right) + 2 \; {\rm perm.}\right]  \nonumber \\ 
& \qquad+ \frac{7}{54} \left[ \frac{1}{k_{1}} \left\{ k_1  + k_2 \mathcal{L}_1 (\mathbf{\widehat{k}}_2\cdot \mathbf{\widehat{k}}_1) + k_3 \mathcal{L}_1 (\mathbf{\widehat{k}}_3\cdot \mathbf{\widehat{k}}_1) \right\} F^{(2)}\left( \mathbf{k}_2,\mathbf{k}_3\right) + 2 \; {\rm perm.}\right],
\end{align}

\begin{align}
 & G^{(3)}\left( \mathbf{k}_1,\mathbf{k}_2,\mathbf{k}_3\right) = \frac{k_{123}^{2}}{9} \left[ \frac{\mathcal{L}_{1}(\mathbf{\widehat{k}}_1\cdot \mathbf{\widehat{k}}_{23})}{k_1 k_{23}}G^{(2)}\left( \mathbf{k}_2,\mathbf{k}_3\right) + 2 \;{\rm perm.} \right]  \nonumber \\ 
& \qquad   + \frac{1}{18} \left[ \frac{1}{k_{23}} \left\{ k_1 \mathcal{L}_1 (\mathbf{\widehat{k}}_1\cdot \mathbf{\widehat{k}}_{23}) + k_2 \mathcal{L}_1 (\mathbf{\widehat{k}}_2\cdot \mathbf{\widehat{k}}_{23}) + k_3 \mathcal{L}_1 (\mathbf{\widehat{k}}_3\cdot \mathbf{\widehat{k}}_{23}) \right\} G^{(2)}\left( \mathbf{k}_2,\mathbf{k}_3\right) + 2 \; {\rm perm.}\right]  \nonumber \\ 
& \qquad+ \frac{1}{18} \left[ \frac{1}{k_{1}} \left\{ k_1  + k_2 \mathcal{L}_1 (\mathbf{\widehat{k}}_2\cdot \mathbf{\widehat{k}}_1) + k_3 \mathcal{L}_1 (\mathbf{\widehat{k}}_3\cdot \mathbf{\widehat{k}}_1) \right\} F^{(2)}\left( \mathbf{k}_2,\mathbf{k}_3\right) + 2 \; {\rm perm.}\right],
\end{align}
with $\mathbf{k}_{ij}\equiv \mathbf{k}_i + \mathbf{k}_j$ and $\mathbf{\widehat{k}}_{ij} \equiv \widehat{\mathbf{k}_i + \mathbf{k}_j}$. As shown in the third-order kernels above, in order to symmetrize them we need to permute the term shown over all the other combinations of the wave-vectors. We explicitly show the term with interchange symmetry between the wave-vectors $\mathbf{k}_2$ and $\mathbf{k}_3$; the summation over the two permutations makes reference to the terms symmetrized over $\mathbf{k}_1$ and $\mathbf{k}_2$ and $\mathbf{k}_1$ and $\mathbf{k}_3$. 

Our work throughout this paper will involve analysing and taking the inverse Fourier transform of the above kernels. Therefore we define general forms for the second and third-order kernels as:

\begin{align}\label{eq:W2_definition}
&W^{(2)}\left(\mathbf{k}_2, \mathbf{k}_3\right) = \overline{c}_{0,0}^{(2)}\mathcal{L}_{0}(\mathbf{\widehat{k}}_2 \cdot \mathbf{\widehat{k}}_3) + \overline{c}_{1,1}^{(2)} \mathcal{L}_{1}(\mathbf{\widehat{k}}_2 \cdot \mathbf{\widehat{k}}_3) \left( \frac{k_3}{k_2} +  \frac{k_2}{k_3}\right) + \overline{c}_{2,0}^{(2)} \mathcal{L}_{2}(\mathbf{\widehat{k}}_2 \cdot \mathbf{\widehat{k}}_3) \\ \nonumber
& \qquad \qquad \qquad  = \sum_{j=0}^{2}\sum_{n=-1}^{1} \overline{c}_{j,n}^{(W)}\mathcal{L}_{j}(\mathbf{\widehat{k}}_2 \cdot \mathbf{\widehat{k}}_3)\,k_{2}^{n}\,k_{3}^{-n}, 
\end{align}

\begin{align}\label{eq:W3_definition}
 & W^{(3)}\left( \mathbf{k}_1,\mathbf{k}_2,\mathbf{k}_3\right) = D\;k_{123}^{2} \left[ \frac{\mathcal{L}_{1}(\mathbf{\widehat{k}}_1\cdot \mathbf{\widehat{k}}_{23})}{k_1 k_{23}}G^{(2)}\left( \mathbf{k}_2,\mathbf{k}_3\right) + 2 \;{\rm perm.} \right]  \nonumber \\ 
& \qquad   + D_1 \left[ \frac{1}{k_{23}} \left\{ k_1 \mathcal{L}_1 (\mathbf{\widehat{k}}_1\cdot \mathbf{\widehat{k}}_{23}) + k_2 \mathcal{L}_1 (\mathbf{\widehat{k}}_2\cdot \mathbf{\widehat{k}}_{23}) + k_3 \mathcal{L}_1 (\mathbf{\widehat{k}}_3\cdot \mathbf{\widehat{k}}_{23}) \right\} G^{(2)}\left( \mathbf{k}_2,\mathbf{k}_3\right) + 2 \; {\rm perm.}\right]  \nonumber \\ 
& \qquad+ D_1\left[ \frac{1}{k_{1}} \left\{ k_1  + k_2 \mathcal{L}_1 (\mathbf{\widehat{k}}_2\cdot \mathbf{\widehat{k}}_1) + k_3 \mathcal{L}_1 (\mathbf{\widehat{k}}_3\cdot \mathbf{\widehat{k}}_1) \right\} F^{(2)}\left( \mathbf{k}_2,\mathbf{k}_3\right) + 2 \; {\rm perm.}\right],
\end{align}
with  $\overline{c}_{1,1} = \overline{c}_{1,-1}$, and  $\overline{c}_{0,1} = \overline{c}_{0,-1} = \overline{c}_{1,0} = \overline{c}_{2,1} = \overline{c}_{2,-1} = 0$ in order to match Eq. (\ref{eq:W2_definition}) to Eq. (\ref{eq:F2_def}) and Eq. (\ref{eq:G2_def}). $\overline{c}_{j,n}$ represents the numerical factors in front of the Legendre polynomials in the terms $F^{(2)}\left(\mathbf{k}_2,\mathbf{k}_3\right)$ and $G^{(2)}\left(\mathbf{k}_2,\mathbf{k}_3\right)$ \cite{Farshad_neutrino}. The subscripts $j$ and $n$ are used with the purpose of getting the correct order of Legendre polynomial, obtaining the correct power in the $k_i$, and the correct numerical fraction simultaneously. In the third-order kernel, the $D_{i}$ are constants and should not be confused with the linear growth rate $D$.

With this in hand, we can compute the galaxy trispectrum after we account for the peculiar velocities in the linear density field evolution and introduce RSD. The redshift-space galaxy density contrast is expressed in Eq. (5)\footnote{In going from Eq. (4) to (5) in \cite{Scoccimarro1998}, the authors expand the exponential with the velocity term in a power series. Then, the products in the expansion and the terms in between parentheses are Fourier transformed which allows them to introduce a Dirac delta function as a Fourier transform of unity, via an integral over their $\mathbf{x}$. These steps result in their Eq. (5).} of \cite{Scoccimarro1998} as: 

\begin{align}\label{eq:delta_sg_1}
&\widetilde{\delta}_{\rm g,s}(\mathbf{k}) = \sum_{n=1}^{\infty}\int d^{3}\mathbf{k}_1 \cdots d^{3}\mathbf{k}_n \;\delta_{\rm D}^{(3)}(\mathbf{k} - \mathbf{k}_1 - \cdots - \mathbf{k}_n) \left[\widetilde{\delta}_{\rm g} (\mathbf{k}_1) + f \mu_{1}^{2} \widetilde{\theta}(\mathbf{k}_1)\right]\nonumber \\
&\qquad\qquad \quad \times \frac{(f\mu k)^{n-1}}{(n-1)!}\frac{\mu_2}{k_2}\widetilde{\theta}(\mathbf{k}_2)\cdots\frac{\mu_n}{k_n}\widetilde{\theta}(\mathbf{k}_n),
\end{align}
with $\widetilde{\delta}_{\rm g}$ representing the galaxy density contrast and $\mu_{i}\equiv \widehat{\bf k}_{i}\cdot \widehat{\bf z}$, where $\widehat{\bf z}$ is the line of sight. We will assume the galaxy density contrast can be expanded in a Eulerian bias model \cite{Bias1, Bias2, Bias3} in terms of the matter density field:

\begin{align}\label{eq:bias}
\widetilde{\delta}_{\rm g} = b_1 \widetilde{\delta}_{\rm m} + \frac{b_2}{2} \widetilde{\delta}_{\rm m}^2 + b_{\rm s} S^{(2)} + \frac{b_3}{6}\widetilde{\delta}_{\rm m}^{3} + b_{\mathcal{G}_3} \mathcal{G}^{(3)} + b_{\delta\mathcal{G}_2} \widetilde{\delta}\mathcal{G}^{(2)} + b_{\Gamma_3} \Gamma^{(3)},
\end{align}
with $b_1$ being the \textit{linear bias} parameter, which describes how the galaxy density traces the matter's density fluctuation linearly, $b_2$ the \textit{quadratic bias} parameter, which describes how the galaxy density traces the matter's density fluctuation square, \textit{etc}. $b_0$ ensures $\left<\widetilde{\delta}_{\rm g}\right> = 0$, but is omitted above since it does not enter connected correlation functions \cite{Scoccimarro1998}; the work presented in this paper is the connected 4PCF. The $S^{(2)},\;\mathcal{G}^{(2)},\;\mathcal{G}^{(3)}$ and $\Gamma^{(3)}$ kernels have been defined in Eqs. (\ref{eq:S2_Equation})-(\ref{eq:Cappital_Letter_gamma}). Therefore, we can re-write Eq. (\ref{eq:delta_sg_1}) as:  

\begin{align}
&\widetilde{\delta}_{\rm g,s}(\mathbf{k})= \sum_{n=1}^{\infty}\int d^{3}\mathbf{k}_1 \cdots d^{3}\mathbf{k}_n\;\delta_{\rm D}^{(3)}(\mathbf{k} - \mathbf{k}_1 - \cdots - \mathbf{k}_n) \left[\sum_{m} \frac{b_{m}}{m!}\widetilde{\delta}_{\rm m}^{m} (\mathbf{k}_1) + f \mu_{1}^{2} \widetilde{\theta}(\mathbf{k}_1)\right]\nonumber \\
&\qquad\qquad \quad \times \frac{(f\mu k)^{n-1}}{(n-1)!}\frac{\mu_2}{k_2}\widetilde{\theta}(\mathbf{k}_2)\cdots\frac{\mu_n}{k_n}\widetilde{\theta}(\mathbf{k}_n). 
\end{align}

Next, we expand $\widetilde{\delta}_{\rm m}$ and $\widetilde{\theta}$ in perturbation series to arrive at: 

\begin{align}\label{eq:delta_sg_expansion}
&\widetilde{\delta}_{\rm g,s}(\mathbf{k})= \sum_{n=1}^{\infty}\int d^{3}\mathbf{k}_1 \cdots d^{3}\mathbf{k}_n\; \delta_{\rm D}^{(3)}(\mathbf{k} - \mathbf{k}_1 - \cdots - \mathbf{k}_n) \left[\sum_{m} \frac{b_{m}}{m!}\left(\sum_{n_1=0}^{\infty}\widetilde{\delta}^{(n_1)}(\mathbf{k}_1) \right)^{m} \right. \nonumber \\
&\qquad\qquad \quad \left. + f \mu_{1}^{2} \sum_{n'_1=0}^{\infty}\widetilde{\theta}^{(n'_1)}(\mathbf{k}_1)\right] \frac{(f\mu k)^{n-1}}{(n-1)!}\frac{\mu_2}{k_2}\sum_{n'_2=0}\widetilde{\theta}^{(n'_2)}(\mathbf{k}_2)\cdots\frac{\mu_n}{k_n}\sum_{n'_n=0}\widetilde{\theta}^{(n'_n)}(\mathbf{k}_n). 
\end{align}

In the interest of simplifying the notation we write the above equation as:

\begin{align}
\widetilde{\delta}_{\rm g,s}(\mathbf{k})  = \sum_{n=1}^{\infty} \widetilde{\delta}_{\rm g,s}^{(n)}(\mathbf{k})   \nonumber
\end{align}
with $\widetilde{\delta}_{\rm g,s}^{(n)}$ given in the same format as Eq. (\ref{eq:delta_n}):

\begin{align}\label{eq:delta_sg_z_kernel}
&\widetilde{\delta}_{\rm g,s}^{(n)} (\mathbf{k}) =  \int d^{3}\mathbf{k}_1 \cdots d^{3}\mathbf{k}_n \;\delta_{\rm D}^{(3)}(\mathbf{k} - \mathbf{k}_1 - \cdots - \mathbf{k}_n) Z^{(n)}(\mathbf{k}_1,\cdots,\mathbf{k}_n) \nonumber \\
& \qquad\qquad \quad \times \widetilde{\delta}^{(1)}(\mathbf{k}_1)\cdots\widetilde{\delta}^{(1)}(\mathbf{k}_n),
\end{align}
where the redshift-space kernels $Z^{(n)}$ can be obtained by inserting Eqs. (\ref{eq:delta_n}) and (\ref{eq:theta_n}) into Eq. (\ref{eq:delta_sg_expansion}) and comparing with Eq. (\ref{eq:delta_sg_z_kernel}); $Z^{({ n})}$ up to third is given by Eqs.~(\ref{eq:Z1}--\ref{eq:Z3}). Therefore, the redshift-space galaxy trispectrum is given by:

\begin{align}\label{eq:Trispectrum}
&(2\pi)^{3}\delta_{ \rm D}^{\left[3\right]}(\mathbf{k}_{1234}) T\left(\mathbf{k}_1, \mathbf{k}_2, \mathbf{k}_3, \mathbf{k}_4\right) =\left< \widetilde{\delta}_{\rm g,s}(\mathbf{k}_1) \widetilde{\delta}_{\rm g,s}(\mathbf{k}_2) \widetilde{\delta}_{\rm g,s}(\mathbf{k}_3) \widetilde{\delta}_{\rm g,s}(\mathbf{k}_4) \right> \nonumber \\  
&= \sum_{n_1,n_2,n_3,n_4}\left< \widetilde{\delta}_{\rm g,s}^{(n_1)}(\mathbf{k}_1) \widetilde{\delta}_{\rm g,s}^{(n_2)}(\mathbf{k}_2) \widetilde{\delta}_{\rm g,s}^{(n_3)}(\mathbf{k}_3) \widetilde{\delta}_{\rm g,s}^{(n_4)}(\mathbf{k}_4) \right> \nonumber \\ 
& = \quad \;\; \left< \widetilde{\delta}_{\rm g,s}^{(1)}(\mathbf{k}_1) \widetilde{\delta}_{\rm g,s}^{(1)}(\mathbf{k}_2) \widetilde{\delta}_{\rm g,s}^{(1)}(\mathbf{k}_3) \widetilde{\delta}_{\rm g,s}^{(1)}(\mathbf{k}_4) \right> \nonumber \\
&\quad+ \left(\left< \widetilde{\delta}_{\rm g,s}^{(3)}(\mathbf{k}_1) \widetilde{\delta}_{\rm g,s}^{(1)}(\mathbf{k}_2) \widetilde{\delta}_{\rm g,s}^{(1)}(\mathbf{k}_3) \widetilde{\delta}_{\rm g,s}^{(1)}(\mathbf{k}_4) \right> + 3\;{\rm perm.}\right) \nonumber \\
&  \quad + \left(\left< \widetilde{\delta}_{\rm g,s}^{(2)}(\mathbf{k}_1) \widetilde{\delta}_{\rm g,s}^{(2)}(\mathbf{k}_2) \widetilde{\delta}_{\rm g,s}^{(1)}(\mathbf{k}_3) \widetilde{\delta}_{\rm g,s}^{(1)}(\mathbf{k}_4) \right> + 5\;{\rm perm.} \right) + \mathcal{O}((\widetilde{\delta}^{(1)})^{8}),
\end{align}
with $\mathbf{k}_{1234} \equiv\mathbf{k}_1+\mathbf{k}_2+\mathbf{k}_3+\mathbf{k}_4 $. The first term in the third equality is the disconnected piece of the trispectrum, which means all of the density contrast are linear and we can apply Wick's Theorem to reduce it to a product of two power spectrum. The next two terms form the connected trispectrum, also known as the tree-level trispectrum---$T_{3111}\left(\mathbf{k}_1, \mathbf{k}_2, \mathbf{k}_3, \mathbf{k}_4\right)$ and $T_{2211}\left(\mathbf{k}_1, \mathbf{k}_2, \mathbf{k}_3, \mathbf{k}_4\right)$. In this paper we will focus on these two terms. Therefore, using the above equation and the derivation in Appendix \ref{sec:Tree-Level_Trispectrum}, we find \cite{Gualdi1, Gualdi2}:

\begin{align} \label{eq:T3}
&T_{3111}\left(\mathbf{k}_1, \mathbf{k}_2, \mathbf{k}_3, \mathbf{k}_4\right) = \; 2Z_{\rm ns}^{(3)}\left(\mathbf{k}_1; \mathbf{k}_2, \mathbf{k}_3\right)Z^{(1)}\left(\mathbf{k}_1\right) Z^{(1)}\left(\mathbf{k}_2\right) Z^{(1)}\left(\mathbf{k}_3\right) P(k_1) P(k_2) P(k_3) \nonumber \\ 
 & \qquad \qquad \qquad \qquad \qquad  + 11\;{\rm perm.} 
 \end{align}
 and

\begin{align} \label{eq:T2}
&T_{2211}\left(\mathbf{k}_1, \mathbf{k}_2, \mathbf{k}_3, \mathbf{k}_4\right) = 4 Z^{(1)}\left(\mathbf{k}_1\right) Z^{(1)}\left(\mathbf{k}_2\right) P(k_1)P(k_2) \left\{Z^{(2)}\left(\mathbf{k}_1, -\mathbf{k}_{13}\right)Z^{(2)}\left(\mathbf{k}_2, \mathbf{k}_{13}\right) P(k_{13}) \right. \nonumber \\ 
&\qquad \qquad \qquad  \qquad \qquad   \left.  + Z^{(2)}\left(\mathbf{k}_1, -\mathbf{k}_{14}\right) Z^{(2)}\left(\mathbf{k}_2, \mathbf{k}_{14}\right) P(k_{14})\right\}+ 5\;{\rm perm.}  
\end{align}
with $\mathbf{k}_{ij} \equiv \mathbf{k}_i + \mathbf{k}_j$, and with $P(k_{i})$ the linear matter power spectrum with $k_i \equiv \left| \mathbf{k}_i\right|$. We use the subscript ``ns'' in the term $Z^{(3)}$ and the semi-colon in front of the wave-vector $\mathbf{k}_1$ to emphasize this redshift-space kernel is not symmetrized and $\mathbf{k}_1$ will play a "special" role in our analyses before we account for the remaining 11 permutations\footnote{We have 11 permutations since there is interchange symmetry between the wave-vectors appearing as the second and third arguments of the third-order kernel.}. 

We note that in \cite{Gualdi1, Gualdi2}, the authors obtain a factor of 6 in front of the term $Z^{(3)}$ and therefore have 3 permutations instead of 11 permutations as we do. The reason for this discrepancy is the authors in \cite{Gualdi1, Gualdi2} assume a symmetry between the position of the arguments of the wave vectors in the third-order kernel when using Wick's Theorem to expand the density contrasts in products of power spectrum. We do not assume the symmetry in the positional arguments since $Z^{(3)}$ is a non-symmetrized kernel. We show in Appendix \ref{sec:Tree-Level_Trispectrum} how to compute $T_{3111}$ correctly. The redshift-space kernels are \cite{Gualdi1, Gualdi2}:

\begin{align}\label{eq:Z1}
&Z^{(1)}\left(\mathbf{k}_i\right) = b_1 + f \mu_i^{2}, \\ \label{eq:Z2}
&Z^{(2)}\left(\mathbf{k}_1, \mathbf{k}_2\right) = b_1 F^{(2)}\left(\mathbf{k}_1, \mathbf{k}_2\right) + f \mu_{12}^{2} G^{(2)}\left(\mathbf{k}_1, \mathbf{k}_2\right) + \frac{b_2}{2} + \frac{b_s}{2} S^{(2)}\left(\mathbf{k}_1,\mathbf{k}_2\right) \nonumber \\
& \qquad + \frac{b_1f}{2} \left[ \mu_{1}^{2} + \mu_{2}^{2} + \mu_{1}\mu_{2} \left(\frac{k_1}{k_2} + \frac{k_2}{k_1}\right)\right] + f^2 \left[\mu_{1}^{2}\mu_{2}^{2} + \frac{\mu_{1}\mu_{2}}{2}\left(\mu_{1}^{2}\frac{k_1}{k_2} + \mu_{2}^{2}\frac{k_2}{k_1}\right) \right], \\ \label{eq:Z3}
&Z_{\rm ns}^{(3)}\left(\mathbf{k}_1; \mathbf{k}_2, \mathbf{k}_3\right) = b_1 F^{(3)}\left(\mathbf{k}_1, \mathbf{k}_2, \mathbf{k}_3\right) + f \mu_{123}^{2}G^{(3)}\left(\mathbf{k}_1, \mathbf{k}_2, \mathbf{k}_3\right)  \nonumber \\ 
&\qquad + b_1 f \left(F^{(2)}\left(\mathbf{k}_2, \mathbf{k}_3\right) \left[ \mu_{1}^{2} + \mu_{1}\mu_{23} \frac{k_{23}}{k_1}\right] + G^{(2)}\left(\mathbf{k}_2, \mathbf{k}_3\right) \left[ \mu_{23}^{2} + \mu_{1}\mu_{23} \frac{k_{1}}{k_{23}}\right]\right)  \nonumber \\ 
&\qquad + f^2 G^{(2)}\left(\mathbf{k}_2, \mathbf{k}_3\right) \left( 2\mu_1^2 \mu_{23}^{2} + \mu_{1}\mu_{23}^{3} \frac{k_{23}}{k_{1}}+\mu_{23}\mu_{1}^{3} \frac{k_{1}}{k_{23}}\right) \nonumber  \\
&\qquad + F^{(2)}\left(\mathbf{k}_2, \mathbf{k}_3\right) \left(2b_2 + 4 b_s S^{(2)}\left(\mathbf{k}_2,\mathbf{k}_3\right)\right)  + \left(\frac{b_2 f}{2} + b_s f S^{(2)}\left(\mathbf{k}_2,\mathbf{k}_3\right) \right)\left(\mu_{1}^{2} + \mu_{1}\mu_{23}\frac{k_{23}}{k_1}\right) \nonumber \\
& \qquad +b_1 f \left[ \mu_{2}^{2}\mu_{3}^{2} + \mu_{1}\mu_{2}^{2}\mu_{3}\frac{k_3}{k_1} + \mu_{1}\mu_{3}^{2}\mu_{2}\frac{k_2}{k_1} +\frac{1}{2} \left( \mu_1\mu_{2}^{3}\frac{k_2}{k_1} + \mu_1\mu_{3}^{3}\frac{k_3}{k_1} + \mu_{1}^{2}\mu_{2}\mu_{3}\frac{k_{1}^{2}}{k_2 k_3} \right) \right] \nonumber \\ 
& \qquad +f^3 \left[ \mu_{1}^{2}\mu_{2}^{2}\mu_{3}^{2} + \frac{3}{2}\left( \mu_{2}^{2}\mu_{3}^{3}\mu_{1}\frac{k_3}{k_1}+\mu_{2}^{3}\mu_{3}^{2}\mu_{1}\frac{k_2}{k_1} + \frac{1}{2} \mu_{1}^{4}\mu_{2}\mu_{3}\frac{k_{1}^{2}}{k_2 k_3}\right)\right] + b_3\nonumber \\ 
& \qquad + b_{\delta \mathcal{G}^{(2)}} \mathcal{G}^{(2)}\left(\mathbf{k}_2, \mathbf{k}_3\right) + b_{\mathcal{G}^{(3)}} \mathcal{G}^{(3)}\left(\mathbf{k}_1 \mathbf{k}_2, \mathbf{k}_3\right) + b_{\Gamma^{(3)}} \Gamma^{(3)}\left(\mathbf{k}_1 \mathbf{k}_2, \mathbf{k}_3\right),
\end{align}
with $\mu_{ij} \equiv \mathbf{\widehat{k}}_{ij} \cdot \mathbf{\widehat{z}} = \left(k_i \mu_i + k_j \mu_j \right)/k_{ij}$ and $k_{ij} = \left| \mathbf{k}_{ij}\right|$, and $f$ the logarithmic derivative of the linear growth rate. The $S^{(2)},\;\mathcal{G}^{(2)},\;\mathcal{G}^{(3)}$, and $\Gamma^{(3)}$ kernels are given by \cite{Gualdi1, Gualdi2, Kobayashi}:

\begin{align} \label{eq:S2_Equation}
S^{(2)}\left(\mathbf{k}_{1}, \mathbf{k}_{2}\right) = \frac{(\mathbf{k}_1 \cdot \mathbf{k}_2 )^2}{k_{1}^{2}k_{2}^{2}} - \frac{1}{3},
\end{align}

\begin{align}\label{eq:FancyG2_def}
\mathcal{G}^{(2)}\left(\mathbf{k}_{1}, \mathbf{k}_{2}\right) = \frac{(\mathbf{k}_1 \cdot \mathbf{k}_2)^2 }{k_{1}^{2}k_{2}^{2}} - 1,
\end{align}

\begin{align}\label{eq:FancyG3}
\mathcal{G}^{(3)}(\mathbf{k}_{1}, \mathbf{k}_{2}, \mathbf{k}_{3}) = 2 (\mathbf{\widehat{k}}_1 \cdot \mathbf{\widehat{k}}_2)(\mathbf{\widehat{k}}_3 \cdot \mathbf{\widehat{k}}_2)(\mathbf{\widehat{k}}_1 \cdot \mathbf{\widehat{k}}_3) - \left [ (\mathbf{\widehat{k}}_1 \cdot \mathbf{\widehat{k}}_2)^2 +  (\mathbf{\widehat{k}}_3 \cdot \mathbf{\widehat{k}}_2)^2  + (\mathbf{\widehat{k}}_1 \cdot \mathbf{\widehat{k}}_3)^2\right],
\end{align}
and 
\begin{align}\label{eq:Cappital_Letter_gamma}
\Gamma^{(3)}(\mathbf{k}_1, \mathbf{k}_2, \mathbf{k}_3) = \mathcal{G}^{(2)} (\mathbf{k}_1, \mathbf{k}_2+\mathbf{k}_3) \left[ F^{(2)}(\mathbf{k}_2,\mathbf{k}_3) - G^{(2)}(\mathbf{k}_2,\mathbf{k}_3) \right].
\end{align}

We also introduce a new kernel---for simplicity of our analysis in $\S$\ref{Section T2}---defined with the fifth and sixth term in $Z^{(2)}(\mathbf{k}_1,\mathbf{k}_2)$: 

\begin{align}\label{eq:gamma}
&\gamma^{(2)}(\mathbf{k}_{1}, \mathbf{k}_2) \equiv \frac{b_1 f}{2}\left[ \mu_1^2 + \mu_2^2 + \mu_1 \mu_2 \left(\frac{k_1}{k_2}+ \frac{k_2}{k_1}\right) \right]  \nonumber \\ 
& \qquad\qquad\qquad + f^2 \left[ \mu_1^2\mu_2^2 + \frac{\mu_1 \mu_2}{2}\left(\mu_1^2\frac{k_1}{k_2}+\mu_2^2 \frac{k_2}{k_1}\right)\right],
\end{align}
which we term the second-order gamma kernel.

\begin{figure}[h]
\centering
\includegraphics[scale=0.22]{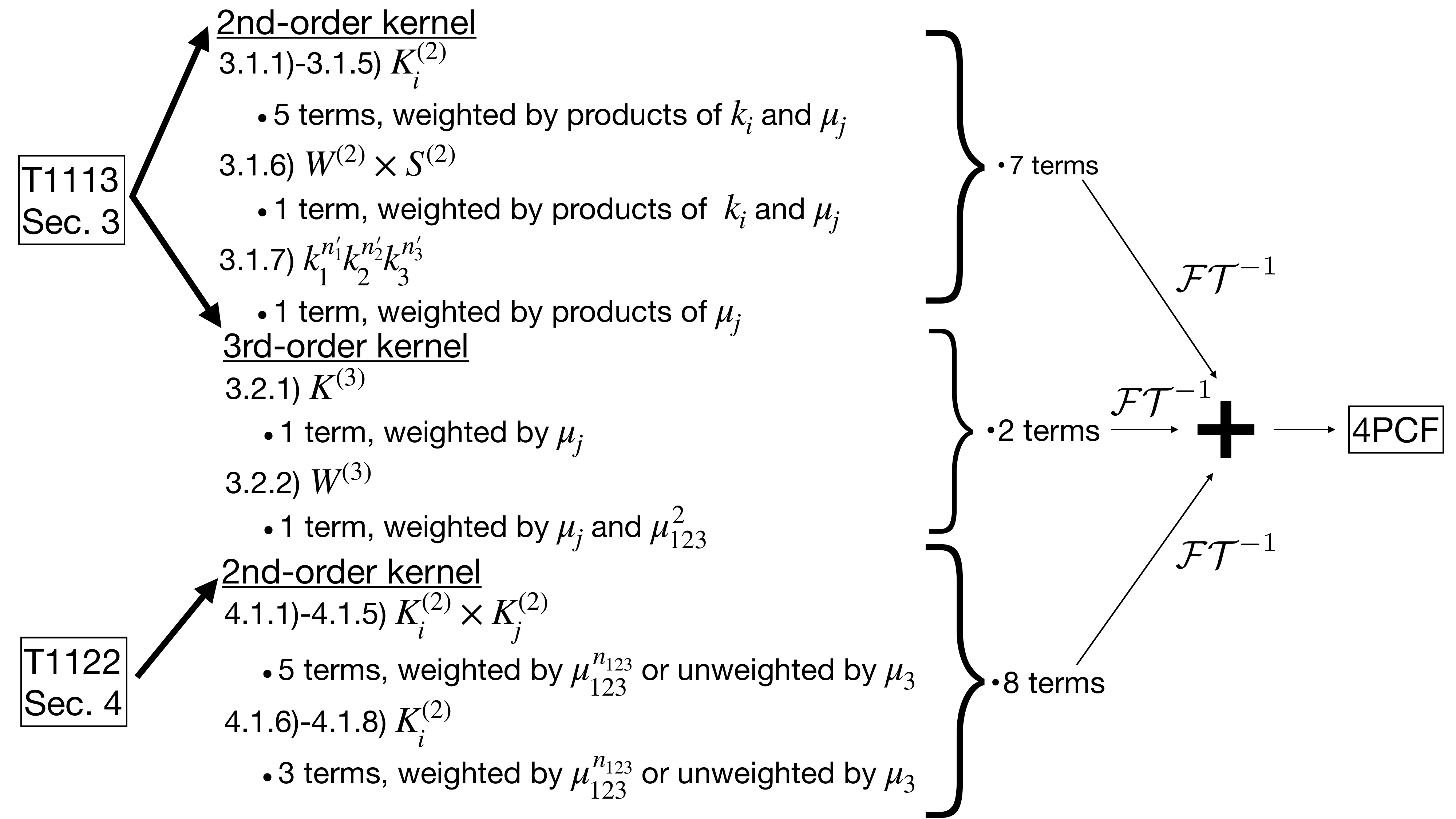}
\caption{Flowchart depicting the structure of this paper. We read this flow chart from left to right and from top to bottom. Our starting point is the tree-level trispectrum $T_{3111}$; we analyze it and take the inverse Fourier transform of its nine representative terms in $\S$\ref{Section T3}. We denote some of the second-order and third-order kernels of this section by $K_{i}^{(2)}$ and $K^{(3)}$, respectively. We then move on to $T_{2211}$; we analyze it and take the inverse Fourier transform of its eight representative terms in $\S$\ref{Section T2}; we denote all the second-order kernels of this section by $K_{i}^{(2)}$. Finally, we add up the resulting 17 terms to find the complete configuration space tree-level 4PCF as shown in $\S$\ref{sec:4PCF_Main_Result}.}
\label{fig:Flowchart_4PCF}
\end{figure}
\newpage
 
\section{Analysis $\&$ Inverse Fourier Transform of $T_{3111}$} \label{Section T3}
We begin the analysis by excluding the bias parameters and logarithmic growth rate $f$ from our considerations, since they are constants and can simply be accounted for after we have finished. We expand the full $T_{3111}$ and search for patterns. This leads us to nine terms, which reproduce it fully. We denote the resulting equations $R_{i,(3)}$, with the subscript 3 indicating that the term is found from the term $T_{3111}$ and $i$ indicating which term was evaluated in the list below:  

\begin{enumerate}
    \item  $K^{(2)}\left(\mathbf{k}_2, \mathbf{k}_3\right) \mu_{1}^{n_1}\mu_{2}^{n_2}\mu_{3}^{n_3}$ \qquad \qquad \quad \; \(\rightarrow\) $\S$\ref{sec:T3.1}
    \item $K^{(2)}\left(\mathbf{k}_2, \mathbf{k}_3\right) k_1^{-1} k_{23} \;  \mu_{23}\; \mu_{1}^{n_1}\mu_{2}^{n_2}\mu_{3}^{n_3}$ \; \(\rightarrow\)\;$\S$\ref{s:methods.1.1}
    \item $W^{(2)}\left(\mathbf{k}_2, \mathbf{k}_3\right) \mu_{23}^{2}\mu_{1}^{n_1}\mu_{2}^{n_2}\mu_{3}^{n_3}$ \quad \quad \qquad  \(\rightarrow\)\;$\S$\ref{sec:T3.1.3}
    \item $W^{(2)}\left(\mathbf{k}_2, \mathbf{k}_3\right) k_{1}k_{23}^{-1}\;  \mu_{23}\;\mu_{1}^{n_1}\mu_{2}^{n_2}\mu_{3}^{n_3}$ \;\;  \(\rightarrow\)\;$\S$\ref{sec:T3.1.4}
    \item $W^{(2)}\left(\mathbf{k}_2, \mathbf{k}_3\right) k_1^{-1}\mu_{23}^{3}\; k_{23}\;\mu_{1}^{n_1}\mu_{2}^{n_2}\mu_{3}^{n_3}$ \; \(\rightarrow\)\;$\S$\ref{sec:T3.1.5}
    \item  $W^{(2)}\left(\mathbf{k}_2, \mathbf{k}_3\right) S^{(2)}\left(\mathbf{k}_2,\mathbf{k}_3\right)\mu_{1}^{n_1}\mu_{2}^{n_2}\mu_{3}^{n_3}$ \(\rightarrow\)\;\;$\S$\ref{sec:T3.1.6}
    \item $k_{1}^{n'_1}k_{2}^{n'_2}k_{3}^{n'_3}\mu_{1}^{n_1}\mu_{2}^{n_2}\mu_{3}^{n_3}$ \qquad \qquad \qquad \quad \(\rightarrow\) $\S$\ref{sec:T3.1.7}
    \item $K^{(3)}\left(\mathbf{k}_1,\mathbf{k}_2, \mathbf{k}_3\right) \mu_{1}^{n_1}\mu_{2}^{n_2}\mu_{3}^{n_3} $ \quad \quad \;\; \; \; \(\rightarrow\) $\S$\ref{sec:T3.1.8}
    \item $W^{(3)}\left(\mathbf{k}_1, \mathbf{k}_2, \mathbf{k}_3\right) \mu_{123}^{2} \mu_{1}^{n_1}\mu_{2}^{n_2}\mu_{3}^{n_3}$  \quad \;\;\; \(\rightarrow\)\;$\S$\ref{sec:T3.1.9}
\end{enumerate}

We have defined $K^{(2)} \equiv \left\{W^{(2)},\mathcal{G}^{(2)}, S^{(2)} \right\}$ in term 1, $K^{(2)} \equiv \left\{W^{(2)}, S^{(2)}\right\}$ in term 2 and $K^{(3)} = \{ W^{(3)}, \mathcal{G}^{(3)}, \Gamma^{(3)}\}$ in term 8. We remind the reader that the $W$ kernels have been defined in Eqs. (\ref{eq:W2_definition}) and (\ref{eq:W3_definition}), for  second and third order, respectively. We also provide Table \ref{table:1} with all the coefficients needed to understand the analysis of this section and reference the equation that defines each coefficient.   

\begin{table} [h!]
\centering
\begin{tabular}{ |p{3cm}|p{3cm}| p{7cm}|}
 \hline
 \multicolumn{3}{|c|}{Table of Coefficients} \\
 \hline
 Coefficient & Equation & Definition \\
 \hline
 &&\\
 $c_{j,n}^{(W)}$  &  \ref{eq:W2} & $W^{(2)}$ kernel coefficients.   \\
 &&\\
 $C_{L_1,L_2,L_3}$ & \ref{eq:Constant_from_exp} & Coefficient of plane-wave expansion when projected onto the isotropic basis functions. \\
  &&\\
 $c_{g}^{\mathcal{G}^{(2)}}$ & \ref{eq:mathcal_G2} & $\mathcal{G}^{(2)}$ kernel coefficient. \\
  &&\\
 $b_{n''}$ & \ref{eq:b_n''_def} & Binomial coefficient. \\ 
  &&\\
  $g_{n_{12},n_{13},n_{23}}$ & \ref{eq:constant_g} & $\mathcal{G}^{(3)}$ kernel coefficient. \\
  &&\\
  $\gamma^{n_{12},n_{13}}$ & \ref{eq:gamma_const_def} & $\Gamma^{(3)}$ kernel coefficient. \\
  &&\\
 $t_{a_1,a_2,a_3}^{(2)}$ & \ref{eq:mu_123_squared} & Trinomial coefficients.\\
  &&\\
 $s_{\ell_1}^{(\rm I)}$ & \ref{eq:k23} &  Isotropic basis function coefficient.\\ 
  &&\\
$c_{j}^{(n)}$ & \ref{eq:dot_to_iso_eq} & Coefficient for the expansion of a dot product into the isotropic basis function. \\
  &&\\
 $\mathcal{C}_{j_1,j_2,j_3}^{n_1,n_2,n_3}$ & \ref{eq:C_cons} & Coefficient from averaging over the line of sight.\\
  &&\\
 $\Upsilon$ & \ref{eq:Upsilon_splitting} & Coefficient from the splitting of position-space vector and wave-vector isotropic basis function. \\ 
  &&\\
 $ \mathcal{G}_{\ell_1,\ell_2,\ell_3} $ &   \ref{eq:Product_of_2_iso}, \ref{eq:Product_of_3_iso} and \ref{eq:Product_of_n_iso} & Modified Gaunt integral from reduction of products of isotropic basis functions. \\ 
 \hline
\end{tabular} 
\caption{Table with the coefficients relevant for the inverse Fourier transform of  $T_{3111}$.}
\label{table:1}
\end{table}

\subsection{Second-Order Kernel}
\subsubsection{Second-Order Kernel Unweighted}\label{sec:T3.1}
\paragraph{$K^{(2)} = W^{(2)}$} \mbox{}\\ 
We begin with the analysis of term 1 in the list at the beginning of this section with $K^{(2)} = W^{(2)}$, and simplify our future calculations in this paper by using Eq. (\ref{eq:dot_to_iso_eq}) and writing $W^{(2)}\left(\mathbf{k}_2, \mathbf{k}_3\right)$ (Eq. (\ref{eq:W2_definition})) in terms of the two-argument isotropic basis functions, $\mathcal{P}_{j}(\mathbf{\widehat{k}}_2,\mathbf{\widehat{k}}_3)$, as:

\begin{align}\label{eq:W2}
&W^{(2)}\left(\mathbf{k}_2, \mathbf{k}_3\right) = 4\pi \sum_{j=0}^{2}\sum_{n=-1}^{1} \overline{c}_{j,n}^{(W)}\sqrt{2j+1}(-1)^{j} \mathcal{P}_{j}(\mathbf{\widehat{k}}_2,\mathbf{\widehat{k}}_3)\,k_{2}^{n}\,k_{3}^{-n} \nonumber \\  
& \qquad \qquad \quad \;\;\; \equiv 4 \pi \sum_{j,n}  c_{j,n}^{(W)} \mathcal{P}_{j} (\mathbf{\widehat{k}}_2,\mathbf{\widehat{k}}_3)  \,k_{2}^{n}\,k_{3}^{-n},  
\end{align}
with $c_{j,n}^{(W)}$ being defined by the last equality and the isotropic basis functions using Eq. (\ref{eq:iso_gen}). 

We proceed to average over the line of sight given the statistical isotropy of the Universe and write the result of doing this shown in Eq. (\ref{eq:avg_z}) as:

\begin{align}\label{eq:W2_with_ave_mu}
&W^{(2)}\left(\mathbf{k}_2, \mathbf{k}_3\right) \left< \mu_{1}^{n_1}\mu_{2}^{n_2}\mu_{3}^{n_3} \right>_{\rm l.o.s}  \nonumber \\ 
&= (4\pi)^{4} \sum_{j,n} \sum_{j_1,j_2,j_3 } c_{j,n}^{(W)} \mathcal{C}_{j_1,j_2,j_3}^{n_1,n_2,n_3} \mathcal{P}_{j}(\mathbf{\widehat{k}}_2,\mathbf{\widehat{k}}_3) \mathcal{P}_{j_1,j_2,j_3}(\mathbf{\widehat{k}}_1,\mathbf{\widehat{k}}_2,\mathbf{\widehat{k}}_3) k_{2}^{n} k_{3}^{-n}  \nonumber \\ 
& = (4\pi)^{9/2} \sum_{j,n} \sum_{j_1,j_2,j_3 } \sum_{J_2=0}^{j_2+j}\sum_{J_3=0}^{j_3+j} c_{j,n}^{(W)} \mathcal{C}_{j_1,j_2,j_3}^{n_1,n_2,n_3} \mathcal{G}_{j_1,J_2,J_3} \mathcal{P}_{j_1,J_2,J_3} (\mathbf{\widehat{k}}_1,\mathbf{\widehat{k}}_2,\mathbf{\widehat{k}}_3) k_{2}^{n} k_{3}^{-n}. 
\end{align}

The sum over each $j_i$ runs from 0 to $n_i$ as shown in Eq. (\ref{eq:avg_z}). Given $\mathcal{P}_{j}(\mathbf{\widehat{k}}_{2}, \mathbf{\widehat{k}}_{3})= \sqrt{4\pi}\mathcal{P}_{0,j,j}(\mathbf{\widehat{k}}_1,\mathbf{\widehat{k}}_{2}, \mathbf{\widehat{k}}_{3})$, we were able to use the result in Eq. (\ref{eq:Product_of_2_iso}) to go from the first to the second equality. Taking the inverse Fourier transform of the above expression including the power spectrum, $P(k_1)P(k_2)P(k_3)$, results in the integrals:

\begin{align}\label{eq:R_1}
&R_{1,(3)}^{\left[n_1\right], \left[n_2\right], \left[n_3\right]}(\mathbf{r}_1, \mathbf{r}_2, \mathbf{r}_3) =  {\rm FT}^{-1} \left\{W^{(2)}\left(\mathbf{k}_2, \mathbf{k}_3\right) \left< \mu_{1}^{n_1}\mu_{2}^{n_2}\mu_{3}^{n_3} \right>_{\rm l.o.s}P(k_1)P(k_2)P(k_3) \right\} (\mathbf{r}_1, \mathbf{r}_2, \mathbf{r}_3)  \nonumber \\ 
& \qquad \qquad \qquad = (4\pi)^{9/2} \sum_{j,n} \sum_{j_1,j_2,j_3 } \sum_{J_2}\sum_{J_3} c_{j,n}^{(W)} \mathcal{C}_{j_1,j_2,j_3}^{n_1,n_2,n_3}\mathcal{G}_{j_1,J_2,J_3} \nonumber \\ 
& \qquad \qquad \qquad \qquad \times \int\frac{d^{3}\mathbf{k}_1\,d^{3}\mathbf{k}_2\,d^{3}\mathbf{k}_3}{(2 \pi)^{9}} \exp \left( -i\sum_{i=1}^{3} \mathbf{k}_{i} \cdot r_{i}\right) \mathcal{P}_{j_1,J_2,J_3} (\mathbf{\widehat{k}}_1,\mathbf{\widehat{k}}_2,\mathbf{\widehat{k}}_3) \nonumber \\ 
& \qquad \qquad \qquad \qquad \times k_{2}^{n} k_{3}^{-n} P(k_1)P(k_2)P(k_3).  
\end{align}

We evaluate this integral by first expanding the exponential into the isotropic basis using the plane-wave expansion:

\begin{align}\label{eq:exp_expansion_iso}
&\exp \left\{-i \sum_{j=1}^{3} \mathbf{k}_j  \cdot \mathbf{ r}_j \right\} = (4\pi)^3 \sum_{\ell'_1,\ell'_2,\ell'_3} C_{\ell'_1,\ell'_2,\ell'_3} \mathcal{P}_{\ell'_1}(\mathbf{\widehat{k}}_1,\mathbf{\widehat{r}}_1) \mathcal{P}_{\ell'_2}(\mathbf{\widehat{k}}_2,\mathbf{\widehat{r}}_2) \mathcal{P}_{\ell'_3}(\mathbf{\widehat{k}}_3,\mathbf{\widehat{r}}_3) \nonumber \\ 
& \qquad \qquad \qquad \qquad \qquad \qquad \times j_{\ell'_1}(k_1 r_1)j_{\ell'_2}(k_2 r_2)j_{\ell'_3}(k_3 r_3) \nonumber \\
&\qquad \qquad \qquad \qquad\quad\; = (4\pi)^{3}\sum_{L_1,L_2,L_3} C_{L_1,L_2,L_3} \Upsilon_{L_1,L_2,L_3} \mathcal{P}_{L_1,L_2,L_3}(\mathbf{\widehat{r}}_1,\mathbf{\widehat{r}}_2,\mathbf{\widehat{r}}_3)\nonumber \\ 
& \qquad \qquad \qquad \qquad \qquad \qquad \times  \mathcal{P}_{L_1,L_2,L_3}(\mathbf{\widehat{k}}_1,\mathbf{\widehat{k}}_2,\mathbf{\widehat{k}}_3) j_{L_1}(k_1 r_1)j_{L_2}(k_2 r_2)j_{L_3}(k_3 r_3).
\end{align}

In the first equality we have used Eq. (\ref{eq:dot_to_iso_eq}) to rewrite the dot products in terms of the isotropic basis and in the last equality we have used Eq. (\ref{eq:product_iso_kr_into_k_r}) to separate the unit vector and unit wave-vector angular dependence into two distinct 3-argument isotropic basis function. We have defined:  
\begin{align} \label{eq:Constant_from_exp}
C_{\ell'_1,\ell'_2,\ell'_3} \equiv i^{\ell'_1+\ell'_2+\ell'_3} \; (-1)^{\ell'_1+\ell'_2+\ell'_3}\sqrt{(2\ell'_1+1)(2\ell'_2+1)(2\ell'_3+1)}. 
 \end{align}

\begin{figure}[h]
\centering
\includegraphics[scale=0.4]{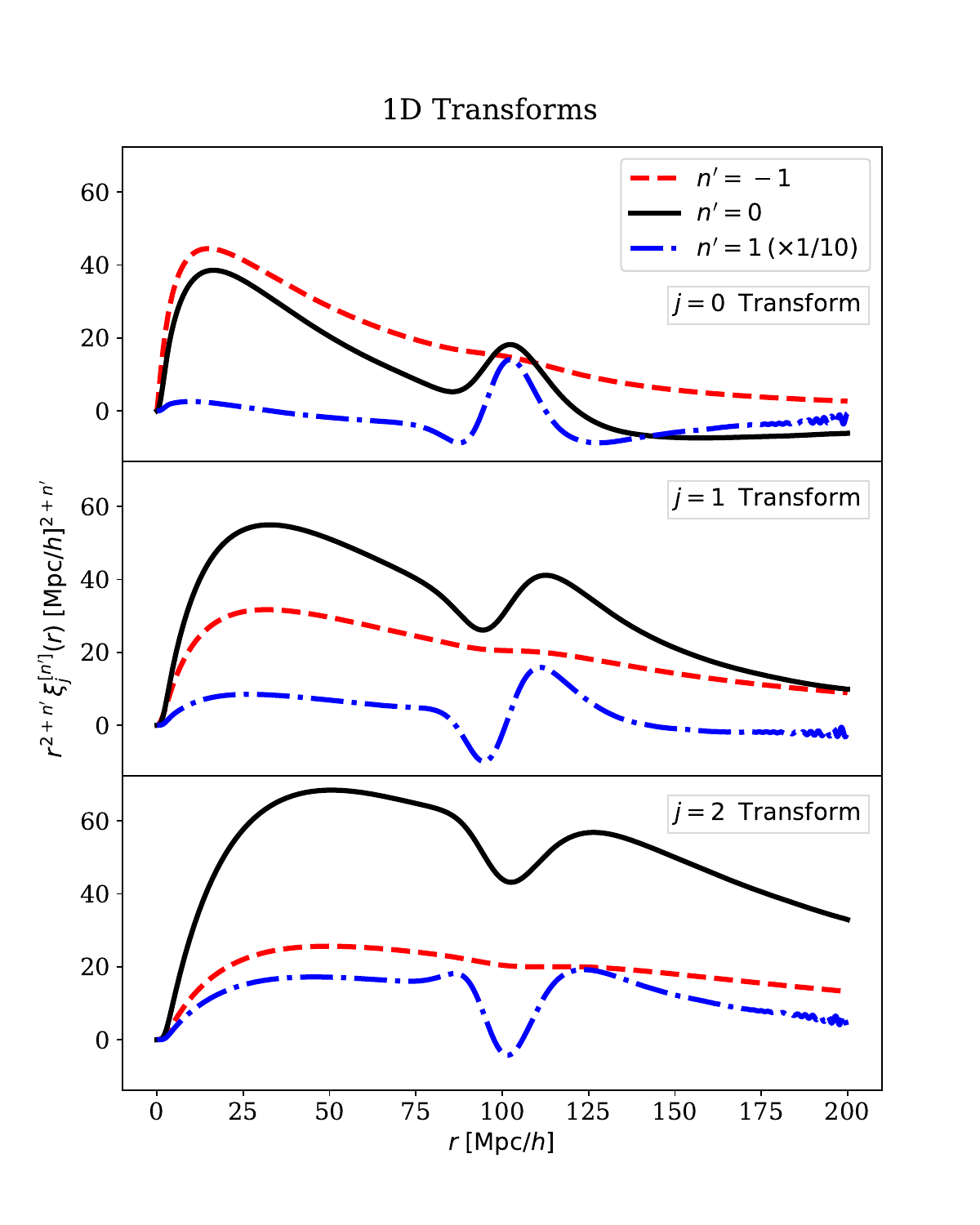}
\caption{1D plot of the integral Eq. (\ref{eq:1D-radial}) evaluated at different values of the power $n'$ and spherical Bessel function order $j$. Each panel shows the integral for a given $j$, and within each panel we show different powers $n'$. This integral falls off as $1/r^{2+n'}$, as we show in Appendix \ref{Sec: Radial Integrals}; therefore we have weighted the integral by $r^{2+n'}$ to take out its fall-off. In the \textit{top panel}, the black solid line for $n'=0$, $j=0$ integral reproduces the 2-Point Correlation Function (2PCF) as expected. In the \textit{lower panel}, the black solid line for the $n'=0\;j=2$ reproduces the quadrupole of the 2PCF but with a sign flip. The sign flip occurs because, when taking the inverse Fourier transform of the power spectrum to find the 2PCF's  quadrupole, the integration picks up a $(-i)^{\ell}$ from the plane-wave expansion, which is later evaluated at $\ell=2$. We also find that the curves with $n'=0,1$ in the \textit{middle panel} correspond to the derivatives of the curves with $n'=-1,0$ in the \textit{top panel}, as we show in Appendix \ref{Sec: Radial Integrals}. It is also notable that the $n'=-1$ curves are smoother than the curves with $n'=0,1$. The reason for this is that $n'=-1$ brings a factor of $1/k$ in the integrand, which serves as a broad low-pass filter, smoothing the curves. Meanwhile, the $n'=0,1$ curves have factors of $k$ that serve as high-pass filters, bringing out more sharp features in the larger scales. The blue dotted line has been re-scaled by a factor of $1/10$ in all three panels.}
\label{fig:1-d_int}
\end{figure}

Finally, we insert Eq.~(\ref{eq:exp_expansion_iso}) into Eq.~(\ref{eq:R_1}) and perform the angular integrals using orthogonality of the isotropic basis to find:

\begin{empheq}[box=\widefbox]{align}\label{eq:R1,3}
&R_{1.1,(3)}^{\left[n_1\right], \left[n_2\right], \left[n_3\right]}(\mathbf{r}_1; \mathbf{r}_2, \mathbf{r}_3) = (4\pi)^{9/2} \sum_{j,n} \sum_{j_1,j_2,j_3 } \sum_{L_2,L_3=0}^{j+j_2,j+j_3} c_{j,n}^{(W)} \mathcal{C}_{j_1,j_2,j_3}^{n_1,n_2,n_3} \nonumber \\ 
& \qquad  \qquad \qquad \qquad \quad \times \mathcal{G}_{j_1,L_2,L_3} \Upsilon_{j_1,L_2,L_3}C_{j_1,L_2,L_3} \nonumber \\ 
& \qquad  \qquad \qquad \qquad \quad \times \mathcal{P}_{j_1,L_2,L_3} (\mathbf{\widehat{r}}_1,\mathbf{\widehat{r}}_2,\mathbf{\widehat{r}}_3) \xi_{j_1}^{\left[n'_1 = 0 \right]}(r_{1})\xi_{L_2}^{\left[n \right]}(r_{2}) \xi_{L_3}^{\left[-n \right]}(r_{3}),
\end{empheq}
with coefficients given in Table \ref{table:1}. Orthogonality of the isotropic basis gave $j_1=L_1$, $J_2 = L_2$ and $J_3=L_3$. We have also introduced the radial integral: 

\begin{align}\label{eq:1D-radial}
&\xi_{j}^{\left[n'\right]}(r_{i}) \equiv \int_{0}^{\infty}\frac{dk_i}{2\pi^{2}}\; k_{i}^{n'+ 2}j_{j}(k_ir_i)P(k_i). 
\end{align}
If we choose a power-law power spectrum, $P(k_i) \sim 1/k_i$ we find analytic solutions for several values of $n'$ and $j$ for Eq. (\ref{eq:1D-radial}) in \cite{Indefinite_Integrals_SBF, Sph_Bessel_Integral_kiersten, Sph_Bessel_int_Rami}. Therefore, in Appendix \ref{Sec: Radial Integrals}, we have analyzed several special cases to explain the behavior shown in Figure \ref{fig:1-d_int}, where we display Eq. (\ref{eq:1D-radial}) with the true power spectrum.

\paragraph{$K^{(2)} = \mathcal{G}^{(2)}$} \mbox{}\\
We proceed by first expressing $\mathcal{G}^{(2)}$ in terms of the Legendre polynomials as we did with $W^{(2)}$:

\begin{align}
&\mathcal{G}^{(2)}\left(\mathbf{k}_{2}, \mathbf{k}_{3}\right) = \frac{(\mathbf{k}_2 \cdot \mathbf{k}_3)^2 }{k_{1}^{2}k_{2}^{2}} - 1  =  \frac{2}{3} \left( \mathcal{L}_{2}(\mathbf{\widehat{k}}_2 \cdot \mathbf{\widehat{k}}_3 ) - \mathcal{L}_{0}(\mathbf{\widehat{k}}_2 \cdot \mathbf{\widehat{k}}_3 ) \right)\nonumber \\ 
& \qquad \qquad  \quad \;= \sum_{g=0}^{2} \overline{c}_{g}^{(\mathcal{G}^{(2)})} \mathcal{L}_{g}(\mathbf{\widehat{k}}_2 \cdot \mathbf{\widehat{k}}_3),
\end{align}
which in turn allows us to express it in terms of the isotropic basis functions using Eq.(\ref{eq:dot_to_iso_eq}): 

\begin{align}\label{eq:mathcal_G2}
&\mathcal{G}^{(2)}\left(\mathbf{k}_{2}, \mathbf{k}_{3}\right) =\sum_{g} \overline{c}_{g}^{(\mathcal{G}^{(2)})} \mathcal{L}_{g}(\mathbf{\widehat{k}}_2 \cdot \mathbf{\widehat{k}}_3)  = 4\pi\sum_{g} \overline{c}_{g}^{(\mathcal{G}^{(2)})}(-1)^g  \sqrt{2g+1} \; \mathcal{P}_{g}(\mathbf{\widehat{k}}_2, \mathbf{\widehat{k}}_3) \nonumber \\
& \qquad \qquad \quad \; \equiv  4\pi \sum_{g} c_{g}^{(\mathcal{G}^{(2)})} \mathcal{P}_{g}(\mathbf{\widehat{k}}_2, \mathbf{\widehat{k}}_3),
\end{align}
where $c_{g}^{(\mathcal{G}^{(2)})}$ has been defined in the last equality. We can immediately see that $\mathcal{G}^{(2)}$ has the same structure as $W^{(2)}$. Therefore, including all the $\mu$ and power spectra terms, and taking its inverse Fourier transform results in: 

\begin{empheq}[box=\widefbox]{align}\label{eq:R1.2,3}
&R_{1.2,(3)}^{\left[n_1\right], \left[n_2\right], \left[n_3\right]}(\mathbf{r}_1; \mathbf{r}_2, \mathbf{r}_3) = (4\pi)^{9/2} \sum_{g} \sum_{j_1,j_2,j_3 } \sum_{L_2,L_3=0}^{g+j_2,g+j_3} c_{g}^{(\mathcal{G}^{(2)})}  \mathcal{C}_{j_1,j_2,j_3}^{n_1,n_2,n_3} \nonumber \\ 
& \qquad  \qquad \qquad \qquad \quad \times \mathcal{G}_{j_1,L_2,L_3} \Upsilon_{j_1,L_2,L_3}C_{j_1,L_2,L_3} \nonumber \\ 
& \qquad  \qquad \qquad \qquad \quad \times \mathcal{P}_{j_1,L_2,L_3} (\mathbf{\widehat{r}}_1,\mathbf{\widehat{r}}_2,\mathbf{\widehat{r}}_3) \xi_{j_1}^{\left[ 0 \right]}(r_{1})\xi_{L_2}^{\left[0 \right]}(r_{2}) \xi_{L_3}^{\left[0 \right]}(r_{3}),
\end{empheq}
with coefficients given in Table \ref{table:1}. The sum over each $j_i$ runs from 0 to $n_i$ and radial integrals are once again defined by Eq. (\ref{eq:1D-radial}). 

\paragraph{$K^{(2)} = S^{(2)}$} \mbox{}\\
Analyzing $S^{(2)}$, we find:

\begin{align}\label{eq:S2_isotropic}
&S^{(2)}\left(\mathbf{k}_{2}, \mathbf{k}_{3}\right) = \frac{2 \sqrt{5}}{3} \mathcal{P}_{2}(\mathbf{\widehat{k}}_2, \mathbf{\widehat{k}}_3), 
\end{align}
which also has the same structure as $W^{(2)}$. Therefore, including all the $\mu$ and power spectra terms, and taking its inverse Fourier transform results in: 

\begin{empheq}[box=\widefbox]{align}\label{eq:R1.3,3}
&R_{1.3,(3)}^{\left[n_1\right], \left[n_2\right], \left[n_3\right]}(\mathbf{r}_1; \mathbf{r}_2, \mathbf{r}_3) = \frac{2 \sqrt{5}\; (4\pi)^{7/2}}{3} \sum_{j_1,j_2,j_3 } \sum_{L_2,L_3=0}^{2+j_2,2+j_3}   \nonumber \\ 
& \qquad  \qquad \qquad \qquad \quad \times \mathcal{C}_{j_1,j_2,j_3}^{n_1,n_2,n_3} \mathcal{G}_{j_1,L_2,L_3} \Upsilon_{j_1,L_2,L_3}C_{j_1,L_2,L_3} \nonumber \\ 
& \qquad  \qquad \qquad \qquad \quad \times \mathcal{P}_{j_1,L_2,L_3} (\mathbf{\widehat{r}}_1,\mathbf{\widehat{r}}_2,\mathbf{\widehat{r}}_3) \xi_{j_1}^{\left[ 0 \right]}(r_{1})\xi_{L_2}^{\left[0 \right]}(r_{2}) \xi_{L_3}^{\left[0 \right]}(r_{3}),
\end{empheq}
with coefficients given in Table \ref{table:1}. The sum over each $j_i$ runs from 0 to $n_i$ and radial integrals are once again defined by Eq. (\ref{eq:1D-radial}). 

\subsubsection{Second-Order Kernel Weighted by $k_{1}^{-1}k_{23}$} \label{s:methods.1.1}
\paragraph{$K^{(2)} = W^{(2)}$} \mbox{}\\
\qquad For term 2 in the list at the beginning of this section with $K^{(2)} = W^{(2)}$, we begin by analysing $\mu_{23}$:

\begin{equation} \label{eq:mu_23}
\mu_{23} = \mathbf{\widehat{k}}_{23}\cdot \widehat{\bf z} = \frac{\mathbf{k}_2 \cdot \mathbf{\widehat{z}} + \mathbf{k}_3 \cdot \mathbf{\widehat{z}}}{k_{23}} = \frac{k_2 \mu_2 + k_3 \mu_3}{k_{23}} 
\end{equation}
which implies that
\begin{equation}
\mu_{23}k_{23} = k_2 \mu_2 + k_3 \mu_3 = \sum_{n'=0}^{1} \frac{k_{3}^{n'}\mu_{3}^{n'}}{k_{2}^{n'-1}\mu_{2}^{n'-1}}.
\end{equation}

Therefore, we can now compute term 2  with $K^{(2)} = W^{(2)}$ as:

\begin{align}
&W^{(2)}\left(\mathbf{k}_2, \mathbf{k}_3\right) \frac{\mu_{23}k_{23}}{k_1}\mu_{1}^{n_1}\mu_{2}^{n_2}\mu_{3}^{n_3}  = \sum_{n'} W^{(2)}\left(\mathbf{k}_2, \mathbf{k}_3\right) \frac{k_{2}^{1-n'} k_{3}^{n'}}{k_{1}}\mu_{1}^{n_1}\mu_{2}^{n_2 - n' +1}\mu_{3}^{n_3+n'}.
\end{align}
Finally, adding the power spectrum for $k_1, k_2$ and $k_3$, and comparing with term 1 gives the result of taking term 2's inverse Fourier transform: 

\begin{empheq}[box=\widefbox]{align}\label{eq:R2,3}
&R_{2.1,(3)}^{\left[n_1\right], \left[n_2\right], \left[n_3\right]}(\mathbf{r}_1; \mathbf{r}_2, \mathbf{r}_3) =  (4\pi)^{9/2} \sum_{j,n,n'} \sum_{j_1,j_2,j_3 } \sum_{L_2,L_3=0}^{j+j_2,j+j_3} c_{j,n}^{(W)}  \nonumber \\ 
& \qquad  \qquad \qquad \qquad   \times \mathcal{C}_{j_1,j_2,j_3}^{n_1,n_2-n'+1,n_3+n'}\mathcal{G}_{j_1,L_2,L_3} \Upsilon_{j_1,L_2,L_3}C_{L_1,L_2,L_3} \nonumber \\ 
& \qquad  \qquad \qquad \qquad   \times  \mathcal{P}_{j_1,L_2,L_3} (\mathbf{\widehat{r}}_1,\mathbf{\widehat{r}}_2,\mathbf{\widehat{r}}_3) \xi_{j_1}^{\left[-1 \right]}(r_{1})\xi_{L_2}^{\left[n-n'+1 \right]}(r_{2}) \xi_{L_3}^{\left[n'-n \right]}(r_{3}).
\end{empheq}
with coefficients defined in Table \ref{table:1}. The sum over $j$ and $n$ is finite and the range of each is shown in Eq. (\ref{eq:W2}), the sum over $j_1$ runs from 0 to $n_1$, while the sum for $j_2$ runs from 0 to $n_2-n'+1$, and $j_3$ from 0 to $n_3 + n'$. We show the definition for the radial integrals in Eq. (\ref{eq:1D-radial}).

\paragraph{$K^{(2)} = S^{(2)}$} \mbox{}\\
Given the structure of $S^{(2)}$ in Eq. (\ref{eq:S2_isotropic}), we can directly compute the result of Fourier transforming term 2 with $K^{(2)} = S^{(2)}$:

\begin{empheq}[box=\widefbox]{align}\label{eq:R2.2,3}
&R_{2.2,(3)}^{\left[n_1\right], \left[n_2\right], \left[n_3\right]}(\mathbf{r}_1; \mathbf{r}_2, \mathbf{r}_3) =  \frac{2 \sqrt{5}\; (4\pi)^{7/2}}{3} \sum_{n'} \sum_{j_1,j_2,j_3 } \sum_{L_2,L_3=0}^{2+j_2,2+j_3} \nonumber \\ 
& \qquad  \qquad \qquad \qquad   \times \mathcal{C}_{j_1,j_2,j_3}^{n_1,n_2-n'+1,n_3+n'}\mathcal{G}_{j_1,L_2,L_3} \Upsilon_{j_1,L_2,L_3}C_{L_1,L_2,L_3} \nonumber \\ 
& \qquad  \qquad \qquad \qquad   \times  \mathcal{P}_{j_1,L_2,L_3} (\mathbf{\widehat{r}}_1,\mathbf{\widehat{r}}_2,\mathbf{\widehat{r}}_3) \xi_{j_1}^{\left[-1 \right]}(r_{1})\xi_{L_2}^{\left[1-n' \right]}(r_{2}) \xi_{L_3}^{\left[n' \right]}(r_{3}),
\end{empheq}
with coefficients defined in Table \ref{table:1}. The sum over $j$ and $n$ is finite and the range of each is shown in Eq. (\ref{eq:W2}), the sum over $j_1$ runs from 0 to $n_1$, while the sum for $j_2$ runs from 0 to $n_2-n'+1$, and $j_3$ from 0 to $n_3 + n'$. We show the definition for the radial integrals in Eq. (\ref{eq:1D-radial}).

\subsubsection{Second-Order Kernel Weighted by $\mu_{23}^{2}$} \label{sec:T3.1.3}
\qquad With term 3 in the list at the beginning of this section, our calculations become more complicated. This term has $\mu_{23}^{2}$, in Eq. (\ref{eq:mu_23}) we found an expression for $\mu_{23}$, squaring that expression and using the binomial theorem leads us to write term 3 as: 

\begin{align}\label{eq:cn'}
&W^{(2)}\left(\mathbf{k}_2, \mathbf{k}_3\right) \mu_{23}^{2}\mu_{1}^{n_1}\mu_{2}^{n_2}\mu_{3}^{n_3} = W^{(2)}\left(\mathbf{k}_2, \mathbf{k}_3\right)\sum_{n''} b_{n''} \frac{k_{2}^{2-n''} k_{3}^{n''}}{k_{23}^{2}}\mu_{1}^{n_1}\mu_{2}^{n_2-n''+2}\mu_{3}^{n_3+n''},
\end{align}
with $b_{n''}$ being the binomial coefficient:

\begin{align}\label{eq:b_n''_def}
b_{n''} \equiv \binom{2}{n''}. 
\end{align}

Using Eq. (\ref{eq:k23}) to decouple the denominator $k_{23}^{2}$, averaging over the line of sight, and using Eq. (\ref{eq:W2}) to expand $W^{(2)}\left(\mathbf{k}_2, \mathbf{k}_3\right)$, we find:

\begin{align}
&W^{(2)}\left(\mathbf{k}_2, \mathbf{k}_3\right)\left<\mu_{23}^{2}\mu_{1}^{n_1}\mu_{2}^{n_2}\mu_{3}^{n_3}\right>_{\rm l.o.s}  \nonumber \\ 
&=  (4\pi)^{4} \sum_{j,n,n''} \sum_{j_1,j_2,j_3} \sum_{\ell_2} b_{n''}\;c_{j,n}^{(W)}\; \mathcal{C}_{j_1,j_2,j_3}^{n_1,n_2-n''+2,n_3+n''}\; s_{\ell_2}^{(\rm I)} \mathcal{P}_{j}(\mathbf{\widehat{k}}_2,\mathbf{\widehat{k}}_3) \mathcal{P}_{j_1,j_2,j_3}(\mathbf{\widehat{k}}_1,\mathbf{\widehat{k}}_2,\mathbf{\widehat{k}}_3)\mathcal{P}_{\ell_2}(\mathbf{\widehat{k}}_2,\mathbf{\widehat{k}}_3) \nonumber \\
& \qquad \times k_{2}^{-n''+n+2} k_{3}^{n''-n} \int_{0}^{\infty} dr \; r \; j_{\ell_2}(k_2r)j_{\ell_2}(k_3r).
\end{align}

Since $\mathcal{P}_{j}(\mathbf{\widehat{k}}_2,\mathbf{\widehat{k}}_3) = \sqrt{4\pi}\;\mathcal{P}_{0,j,j}(\mathbf{\widehat{k}}_1, \mathbf{\widehat{k}}_2,\mathbf{\widehat{k}}_3)$,\footnote{The same applies to $\mathcal{P}_{\ell_2}(\mathbf{\widehat{k}}_2,\mathbf{\widehat{k}}_3)$.} we use the result in Eq. (\ref{eq:Product_of_3_iso}) to obtain:
\begin{align}
&W^{(2)}\left(\mathbf{k}_2, \mathbf{k}_3\right)\left<\mu_{23}^{2}\mu_{1}^{n_1}\mu_{2}^{n_2}\mu_{3}^{n_3}\right>_{\rm l.o.s}  \nonumber \\ 
& = (4\pi)^{5} \sum_{j,n,n''} \sum_{j_1,j_2,j_3} \sum_{\ell_2} \sum_{L''_2,L''_3}  b_{n''}\; c_{j,n}^{(W)}\; \mathcal{C}_{j_1,j_2,j_3}^{n_1,n_2-n''+2,n_3+n''}\; s_{\ell_2}^{(\rm I)}\; \mathcal{G}_{j_1,L''_2,L''_3} \mathcal{P}_{j_1,L''_2,L''_3}(\mathbf{\widehat{k}}_1,\mathbf{\widehat{k}}_2,\mathbf{\widehat{k}}_3) \nonumber \\
& \qquad \times k_{2}^{n-n''+2} k_{3}^{n''-n}  \int_{0}^{\infty}dr \; r\; j_{\ell_2}(k_2r)j_{\ell_2}(k_3r).
\end{align}
Then, we perform the inverse Fourier transform integrals using Eq. (\ref{eq:exp_expansion_iso}) to obtain:

\begin{empheq}[box=\widefbox]{align}\label{eq:R5,3}
&R_{3,(3)}^{\left[n_1\right], \left[n_2\right], \left[n_3\right]}(\mathbf{r}_1; \mathbf{r}_2, \mathbf{r}_3) = (4\pi)^{5}\sum_{j,n,n''} \sum_{j_1,j_2,j_3} \sum_{\ell_2} \sum_{L_2,L_3} b_{n''} c_{j,n}^{(W)} s_{\ell_2}^{(\rm I)} \nonumber \\ 
& \qquad \qquad \qquad \qquad \times \mathcal{C}_{j_1,j_2,j_3}^{n_1,n_2-n''+2,n_3+n''}  C_{j_1,L_2,L_3} \mathcal{G}_{j_1,L_2,L_3} \Upsilon_{j_1,L_2,L_3}\nonumber \\ 
& \qquad \qquad \qquad \qquad \times \mathcal{P}_{j_1,L_2,L_3} (\mathbf{\widehat{r}}_1,\mathbf{\widehat{r}}_2,\mathbf{\widehat{r}}_3) \xi_{j_1}^{\left[0 \right]}(r_{1}) F_{(\ell_2),L_2,L_3}^{\left[n-n''+2\right],\left[ n''-n\right]}(r_2,r_3),
\end{empheq}
with coefficients defined in Table \ref{table:1}. The sums over $j$ and $n$ are finite and the range of each is shown in Eq. (\ref{eq:W2}). The sum over $j_1$ runs from 0 to $n_1$, while the sum for $j_2$ runs from 0 to $n_2-n''+2$ and $j_3$ from 0 to $n_3 + n''$. The sum over $\ell_2$ runs to infinity and since $L_2$ and $L_3$ run from 0 to $j+j_2+\ell_2$ and $j+j_3+\ell_2$, respectively, they run to infinity, as well.  The radial integral for $r_1$ is defined by Eq. (\ref{eq:1D-radial}), while the radial integrals for $r_2$ and $r_3$ are defined as:   
\begin{align} \label{eq:2D-radial}
F_{(\ell_2),L_2,L_3}^{\left[n'_2\right],\left[ n'_3\right]}(r_2,r_3) \equiv \int_{0}^{\infty} dr \;r\; f_{\ell_2,L_2}^{\left[n'_2\right]}(r,r_2) f_{\ell_2,L_3}^{\left[n'_3\right]}(r,r_3),
\end{align} 
where we have used parentheses around the order $ \ell_2$ of the spherical Bessel to indicate that it is being integrated out and does not directly affect the resulting variables $r_2$ and $r_3$. The $f$ integral is defined as:

\begin{align}\label{eq:2_Sph.Bess_int}
f_{\ell,L}^{\left[n'\right]}(r,r_i) \equiv \int_{0}^{\infty}\frac{dk_i}{2 \pi^2} \; k_{i}^{n'+2} j_{\ell}(k_ir)j_{L}(k_i r_i)P(k_i).
\end{align}

\begin{figure}[h]
\centering
\includegraphics[scale=0.7]{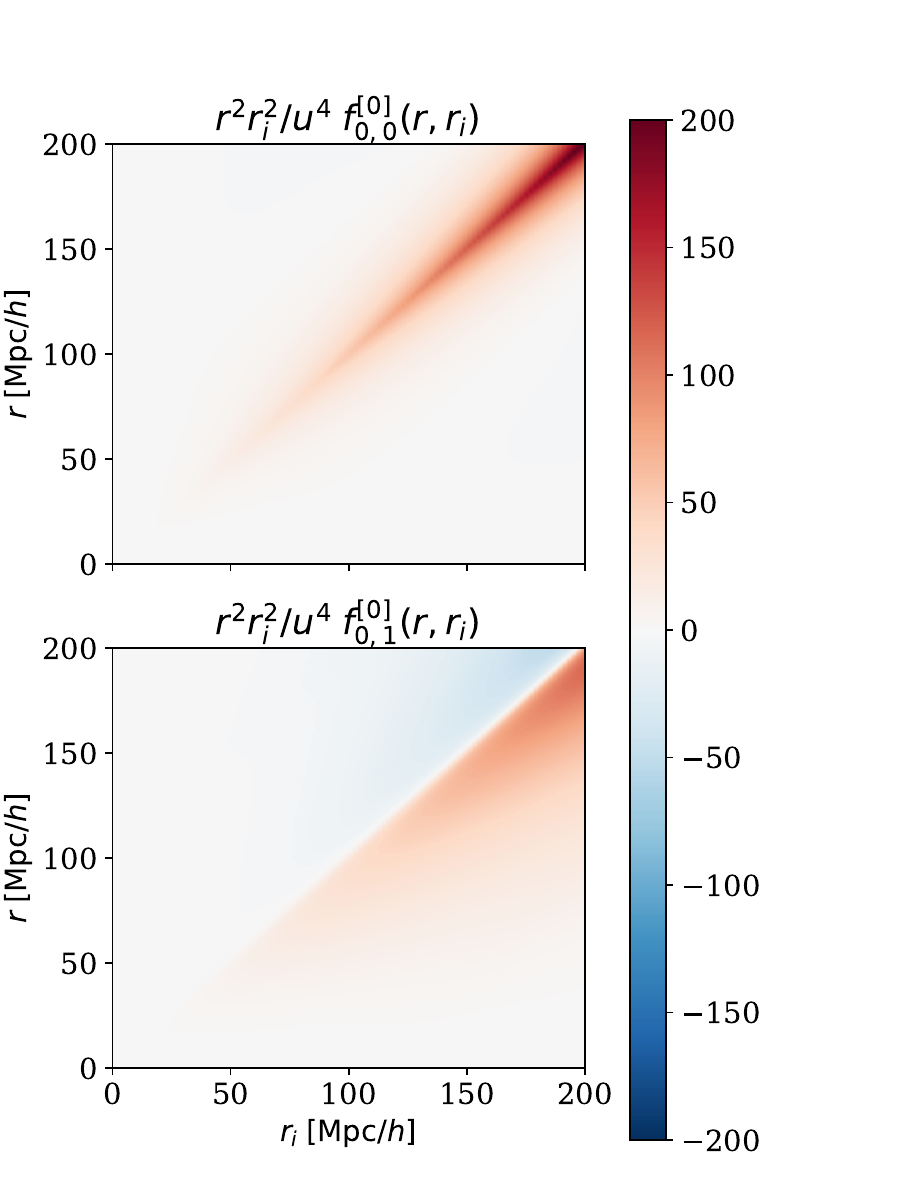}
\caption{Here, we show Eq. (\ref{eq:2_Sph.Bess_int}) for $n'=0$, $\ell = 0$ and $L = \left\{0,1\right\}$. The \textit{upper panel} shows the integral for $L=0$, while the \textit{lower panel} shows the integral for $L=1$. Since the 4PCF can be approximated as the square of the 2PCF on large scales, $(\xi_{0})^{2}(r) \sim (1/r^2)^2$, we have weighted the integral by $r^2r_i^2/u^{4}$, with $u \equiv 10 \;\left[{\rm Mpc}/h\right]$, to take out its fall-off. The \textit{upper panel} shows the integral is largest along the diagonal; this is because the integral $f_{0,0}^{\left[0\right]}$ can be approximated as a Dirac delta function, as is shown in Appendix \ref{Sec: Radial Integrals}. The \textit{lower panel}, for $L=1$, in contrast, shows the integral is largest in the off-diagonal elements; we demonstrate why in Appendix \ref{Sec: Radial Integrals}. The \textit{lower panel} is also not symmetric under exchange of axes (\textit{i.e.}, $r_i \leftrightarrow r$); this asymmetry stems from choosing $\ell = 0$ for the sBF order corresponding to the variable $r$ but $L=1$ for the sBF order corresponding to the variable $r_i$. }
\label{fig:fint2}
\end{figure}

\begin{figure}[h]
\centering
\includegraphics[scale=0.7]{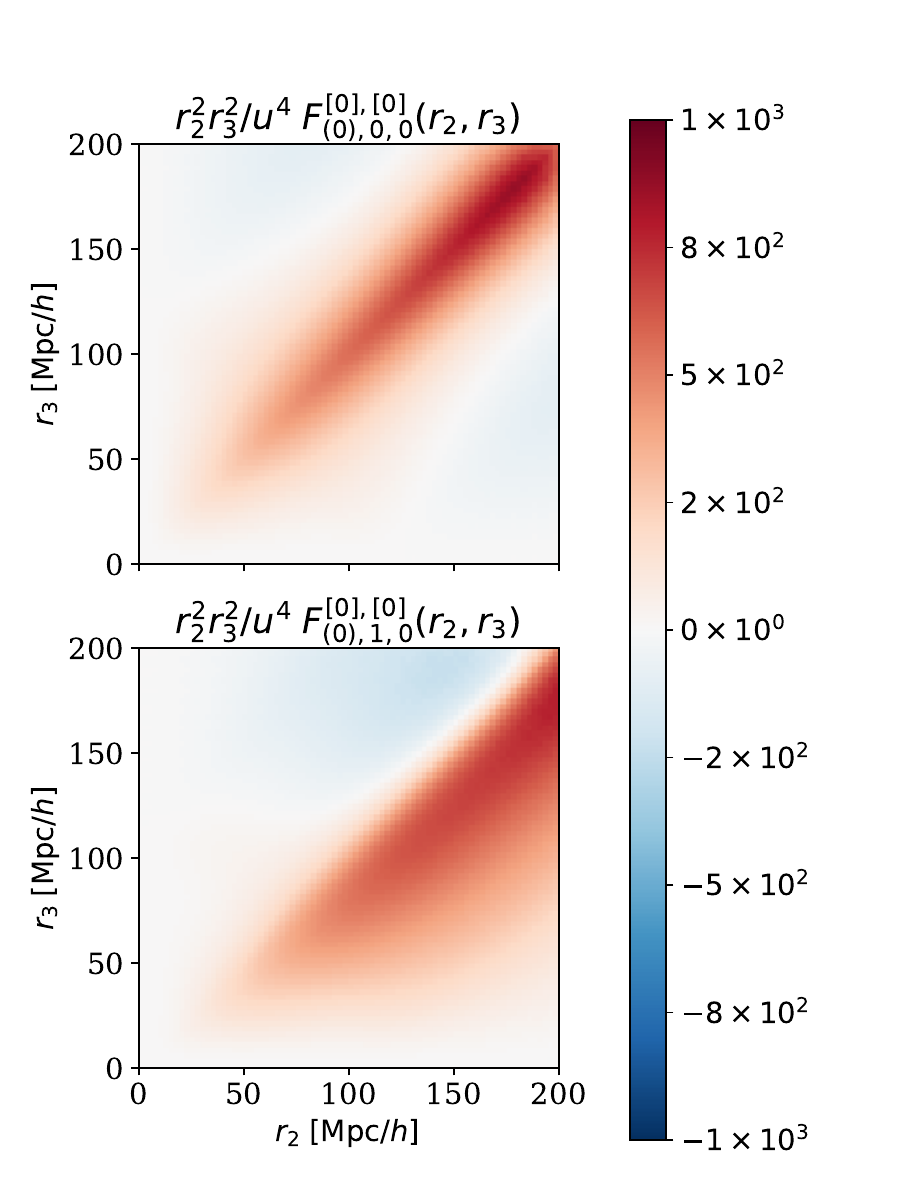}
\caption{2D plot of integral Eq. (\ref{eq:2D-radial}) for $n'_2=0$, $n'_3=0$, $\ell_2 = 0$, $L_2=\left\{0,1\right\}$ and $L_3 = 0$.  The \textit{upper panel} shows the integral for $L_2=0$, while the \textit{lower panel} shows the integral for $L_2=1$. We place $\ell_2$ in parentheses to identify it as the sBF order that does not contribute directly to the sBFs whose arguments contain $r_2$ and $r_3$. Since the 4PCF can be approximated as the square of the 2PCF on large scales, $(\xi_{0})^{2}(r) \sim (1/r^2)^2$, we have weighted the integral by $r^2r_i^2/u^{4}$, with $u \equiv 10 \;\left[{\rm Mpc}/h\right]$, to take out its fall-off. The \textit{upper panel} shows the integral is largest along the diagonal. Since $F$ is the integral of the product of two $f$ integrals (Eq. (\ref{eq:2_Sph.Bess_int})) we can approximate it as a Dirac delta function, as shown in Appendix \ref{Sec: Radial Integrals}. The \textit{lower panel}, for $L_2=1$, shows the integral is largest in the off-diagonal elements. Once again, the fact that $F$ is the integral of the product of two $f$ integrals explains why we only see non-diagonal behavior; we demonstrate this explicitly in Appendix \ref{Sec: Radial Integrals}. The \textit{lower panel} is also not symmetric under exchange of axes (\textit{i.e.}, $r_2 \leftrightarrow r_3$); this asymmetry stems from choosing $L_3 = 0$ for the sBF order corresponding to the variable $r_3$ and $L=1$ for the sBF order corresponding to the variable $r_2$. }
\label{fig:2D_int}
\end{figure}

If we choose a power-law power spectrum, $P(k_i) \sim 1/k_i$ we find analytic solutions, for some values of $n'$, $n'_2$, $n'_3$, $\ell$, $L$, $\ell_2$, $L_2$ and $L_3$ for Eq. (\ref{eq:2D-radial}) and Eq. (\ref{eq:2_Sph.Bess_int}) in \cite{ Indefinite_Integrals_SBF, Sph_Bessel_Integral_kiersten, Sph_Bessel_int_Rami}. Therefore, in Appendix \ref{Sec: Radial Integrals}, we have analyzed several special cases to explain the behavior shown in Figure \ref{fig:fint2} and Figure \ref{fig:2D_int}, where we display Eq. (\ref{eq:2_Sph.Bess_int}) and Eq. (\ref{eq:2D-radial}) with the true power spectrum.  

\subsubsection{Second-Order Kernel Weighted by $k_{1}k_{23}^{-1}\mu_{23}$}\label{sec:T3.1.4}
\qquad For term 4 in the list at the beginning of this section, we start by using Eq. (\ref{eq:mu_23}) which allows us to generalize the term as:

\begin{equation}
    W^{(2)}\left(\mathbf{k}_2, \mathbf{k}_3\right) \frac{\mu_{23}k_1}{k_{23}}\mu_{1}^{n_1}\mu_{2}^{n_2}\mu_{3}^{n_3} = \sum_{n'=0}^{1}  W^{(2)}\left(\mathbf{k}_2, \mathbf{k}_3\right) \frac{k_1\; k_{2}^{1-n'} \;k_{3}^{n'}}{k_{23}^{2}}\mu_{1}^{n_1}\mu_{2}^{n_2-n'+1}\mu_{3}^{n_3+n'}.
\end{equation}
This has the same structure as term 3, but with an additional factor of $k_1$ and no $b_{n''}$ coefficient. Therefore, we can easily read off the result for term 4 as: 

\begin{empheq}[box=\widefbox]{align}\label{eq:R6,3}
&R_{4,(3)}^{\left[n_1\right], \left[n_2\right], \left[n_3\right]}(\mathbf{r}_1; \mathbf{r}_2, \mathbf{r}_3) =(4\pi)^{5} \sum_{j,n,n'} \sum_{j_1,j_2,j_3} \sum_{\ell_2} \sum_{L_2,L_3} c_{j,n}^{(W)} s_{\ell_2}^{(\rm I)} \nonumber \\ 
& \qquad \qquad \qquad \qquad \times \mathcal{C}_{j_1,j_2,j_3}^{n_1,n_2,n_3}  C_{j_1,L_2,L_3} \mathcal{G}_{j_1,L_2,L_3} \Upsilon_{j_1,L_2,L_3} \nonumber \\ 
& \qquad \qquad \qquad \qquad \times \mathcal{P}_{j_1,L_2,L_3} (\mathbf{\widehat{r}}_1,\mathbf{\widehat{r}}_2,\mathbf{\widehat{r}}_3) \xi_{j_1}^{\left[1 \right]}(r_{1}) F_{(\ell_2),L_2,L_3}^{\left[n-n'+1\right],\left[ n'-n\right]}(r_2,r_3),
\end{empheq}
with the only difference that we are substituting the sum over $n''$ with the sum over $n'$, which implies the sum over $j_2$ will run from 0 to $n_2-n'+1$ and $j_3$ from 0 to $n_3+n'$, instead. The rest of sums follow the same logic as explained for term 5 and all coefficients are given in Table \ref{table:1}. The radial integrals are defined in Eq. (\ref{eq:1D-radial}) and Eq. (\ref{eq:2D-radial}).

\subsubsection{Second-Order Kernel Weighted by $k_1^{-1}k_{23}\mu_{23}^{3}$}\label{sec:T3.1.5}
\qquad For term 5 in the list at the beginning of this section, we use Eq. (\ref{eq:mu_23}) and the binomial theorem again to simplify the expression of $\mu_{23}^{3}$ and write the term 5 as:

\begin{equation}\label{eq:W2_mu^3}
    W^{(2)}\left(\mathbf{k}_2, \mathbf{k}_3\right) \frac{\mu_{23}^{3}k_{23}}{k_{1}}\mu_{1}^{n_1}\mu_{2}^{n_2}\mu_{3}^{n_3} \equiv \sum_{n''=0}^{3} W^{(2)}\left(\mathbf{k}_2, \mathbf{k}_3\right) b_{n''} \frac{k_{2}^{3 -n''} k_{3}^{n''}}{k_1 k_{23}^{2}}\mu_{1}^{n_1}\mu_{2}^{n_2-n''+3}\mu_{3}^{n_3+n''}.
\end{equation}
In the above expression, we find the same structure as term 3, but with an additional factor $1/k_1$. Therefore, comparing Eq. (\ref{eq:W2_mu^3}) with term 3 we obtain:

\begin{empheq}[box=\widefbox]{align}\label{eq:R7,3}
&R_{5,(3)}^{\left[n_1\right], \left[n_2\right], \left[n_3\right]}(\mathbf{r}_1; \mathbf{r}_2, \mathbf{r}_3) = (4\pi)^{5}\sum_{j,n,n''} \sum_{j_1,j_2,j_3}\sum_{\ell_2} \sum_{L_2,L_3} b_{n''} c_{j,n}^{(W)} s_{\ell_2}^{(\rm I)} \nonumber \\
& \qquad  \qquad \qquad \qquad \quad  \times \mathcal{C}_{j_1,j_2,j_3}^{n_1,n_2,n_3}  C_{j_1,L_2,L_3} \mathcal{G}_{j_1,L_2,L_3} \Upsilon_{j_1,L_2,L_3} \nonumber \\
& \qquad  \qquad \qquad \qquad \quad  \times \mathcal{P}_{j_1,L_2,L_3} (\mathbf{\widehat{r}}_1,\mathbf{\widehat{r}}_2,\mathbf{\widehat{r}}_3)\xi_{j_1}^{\left[-1 \right]}(r_{1}) F_{(\ell_2),L_2,L_3}^{\left[n-n''+3\right],\left[ n''-n\right]}(r_2,r_3),
\end{empheq}
with coefficients given in Table \ref{table:1}. The sums follow the same logic as in term 3 and radial integrals are defined in Eq. (\ref{eq:1D-radial}) and Eq. (\ref{eq:2D-radial}).

\subsubsection{Product of Second-Order Kernel with Tidal Tensor Kernel}\label{sec:T3.1.6}
\qquad We evaluate term 6 in the list at the beginning of this section where the structure for the tidal tensor kernel, $S^{(2)}$ is given by Eq. (\ref{eq:S2_isotropic}). Taking the product of $S^{(2)}\left(\mathbf{k}_2,\mathbf{k}_3\right)$ with $W^{(2)}\left(\mathbf{k}_2,\mathbf{k}_3\right)$ and all the $\mu_{i}$ factors, averaging over the line of sight we find:

\begin{align}\label{eq:W2S2_term4}
&W^{(2)}\left(\mathbf{k}_2,\mathbf{k}_3\right)S^{(2)}\left(\mathbf{k}_2,\mathbf{k}_3\right)\left<\mu_{1}^{n_1}\mu_{2}^{n_2}\mu_{3}^{n_3}\right>_{\rm l.o.s} \nonumber \\ 
& = (4\pi)^{2} \frac{2\;\sqrt{5}}{3} \sum_{j,n} \sum_{j_1,j_2,j_3} c_{j,n}^{(W)}  \mathcal{C}_{j_1,j_2,j_3}^{n_1,n_2,n_3} \mathcal{P}_{j}(\mathbf{\widehat{k}}_2,\mathbf{\widehat{k}}_3) \mathcal{P}_{2}(\mathbf{\widehat{k}}_2,\mathbf{\widehat{k}}_3) \mathcal{P}_{j_1,j_2,j_3} (\mathbf{\widehat{r}}_1,\mathbf{\widehat{r}}_2,\mathbf{\widehat{r}}_3) k_{2}^{n} k_{3}^{-n}  \nonumber \\ 
& = (4\pi)^{3} \frac{2\;\sqrt{5}}{3} \sum_{j,n} \sum_{j_1,j_2,j_3} \sum_{J_2=0}^{j_2+2+j} \sum_{J_3=0}^{j_3+2+j} c_{j,n}^{(W)} c_{j_s}^{(S)} \mathcal{C}_{j_1,j_2,j_3}^{n_1,n_2,n_3} \mathcal{G}_{j_1,J_2,J_3} \nonumber \\ 
& \qquad \qquad \qquad \times \mathcal{P}_{j_1,J_2,J_3} (\mathbf{\widehat{r}}_1,\mathbf{\widehat{r}}_2,\mathbf{\widehat{r}}_3) k_{2}^{n} k_{3}^{-n}.
\end{align}

To obtain last equality we used Eq. (\ref{eq:Product_of_3_iso}) with $ \mathcal{P}_{j}(\mathbf{\widehat{k}}_2,\mathbf{\widehat{k}}_3) = \sqrt{4\pi}\;\mathcal{P}_{0,j,j}(\mathbf{\widehat{k}}_1, \mathbf{\widehat{k}}_2,\mathbf{\widehat{k}}_3)$ and $\mathcal{P}_{j_s}(\mathbf{\widehat{k}}_2,\mathbf{\widehat{k}}_3)= \sqrt{4\pi}\;\mathcal{P}_{0,j_s,j_s}(\mathbf{\widehat{k}}_1, \mathbf{\widehat{k}}_2,\mathbf{\widehat{k}}_3)$. Comparing Eq. (\ref{eq:W2S2_term4}) with term 1 we find:

\begin{empheq}[box=\widefbox]{align}\label{eq:R4,3}
&R_{6,(3)}^{\left[n_1\right], \left[n_2\right], \left[n_3\right]}(\mathbf{r}_1; \mathbf{r}_2, \mathbf{r}_3) = (4\pi)^{3} \frac{2\;\sqrt{5}}{3} \sum_{j,n} \sum_{j_1,j_2,j_3} \sum_{L_2,L_3} c_{j,n}^{(W)}  \mathcal{C}_{j_1,j_2,j_3}^{n_1,n_2,n_3}    \nonumber \\ 
& \qquad  \qquad \quad  \qquad \qquad  \qquad \times C_{j_1,L_2,L_3} \mathcal{G}_{j_1,L_2,L_3}\Upsilon_{j_1,L_2,L_3}   \nonumber \\ 
& \qquad  \qquad \quad   \qquad \qquad  \qquad \times\mathcal{P}_{j_1,L_2,L_3} (\mathbf{\widehat{r}}_1,\mathbf{\widehat{r}}_2,\mathbf{\widehat{r}}_3) \xi_{j_1}^{\left[0 \right]}(r_{1})\xi_{L_2}^{\left[n \right]}(r_{2}) \xi_{L_3}^{\left[-n \right]}(r_{3}),
\end{empheq}
with coefficients defined in Table \ref{table:1}. Rhe sums over $j$ and $n$ are finite and the range of each is shown in Eq. (\ref{eq:W2}). The sum over each $j_i$ runs from 0 to $n_i$. The orthogonality of the isotropic basis functions resulted in $L_2 = J_2$ and $L_3 =  J_3$, which means  the range of each is the same as the one specified in Eq. (\ref{eq:W2S2_term4}). The radial integrals have been defined in Eq. (\ref{eq:1D-radial}).

\subsubsection{Products of $k_i$}\label{sec:T3.1.7}
\qquad For term 7 in the list at the beginning of this section we have:
\begin{align}
k_{1}^{n'_1}k_{2}^{n'_2}k_{3}^{n'_3}\left<\mu_{1}^{n_1}\mu_{2}^{n_2}\mu_{3}^{n_3}\right>_{\rm l.o.s} = (4 \pi)^{3} k_{1}^{n'_1}k_{2}^{n'_2}k_{3}^{n'_3} \sum_{j_1,j_2,j_3 }\mathcal{C}_{j_1,j_2,j_3}^{n_1,n_2,n_3} \mathcal{P}_{j_1,j_2,j_3}(\mathbf{\widehat{k}}_1,\mathbf{\widehat{k}}_2,\mathbf{\widehat{k}}_3). 
\end{align}

Performing the inverse Fourier transform we obtain: 

\begin{empheq}[box=\widefbox]{align}\label{eq:R3_T3111}
&R_{7,(3)}^{\left[n'_1\right], \left[n'_2\right], \left[n'_3\right], \left[n_1\right], \left[n_2\right], \left[n_3\right]}(\mathbf{r}_1; \mathbf{r}_2, \mathbf{r}_3) \nonumber  \\ 
& \qquad \qquad \quad \;\;\;= {\rm FT}^{-1}\left\{k_{1}^{n'_1}k_{2}^{n'_2}k_{3}^{n'_3}\left<\mu_{1}^{n_1}\mu_{2}^{n_2}\mu_{3}^{n_3}\right>_{\rm l.o.s}P(k_1)P(k_2)P(k_3)\right\} \nonumber  \\ 
& \qquad \qquad \quad \;\;\; = (4 \pi)^{3} \sum_{j_1,j_2,j_3 }\mathcal{C}_{j_1,j_2,j_3}^{n_1,n_2,n_3}  \Upsilon_{j_1,L_2,L_3}C_{j_1,j_2,j_3} \nonumber \\
& \qquad \qquad \qquad \qquad  \times \mathcal{P}_{j_1,j_2,j_3}(\mathbf{\widehat{r}}_1,\mathbf{\widehat{r}}_2,\mathbf{\widehat{r}}_3) \xi_{j_1}^{\left[n'_1 \right]}(r_{1})\xi_{j_2}^{\left[n'_2 \right]}(r_{2}) \xi_{j_3}^{\left[n'_3 \right]}(r_{3}),
\end{empheq}
with coefficients defined in Table \ref{table:1}. The sum over $j$ and $n$ is finite and the range of each is shown in Eq. (\ref{eq:W2}). The sum over each $j_i$ runs from 0 to $n_i$. We show the definition for the radial integrals in Eq. (\ref{eq:1D-radial}).

\subsection{Third-Order Kernel}
\subsubsection{Third-Order Kernel Unweighted}\label{sec:T3.1.8}
\paragraph{$K^{(3)} = W^{(3)}$ with Inverse $k_{23}^{n'_{23}}$; $n'_{23}=2$} \mbox{}\\
\qquad
For term 8 in the list at the beginning of this section we start by analyzing the third order kernel $K^{(3)} = W^{(3)}$, defined in Eq. (\ref{eq:W3_definition}). This kernel is composed of dot products of the form, $\mathbf{\widehat{k}}_{i} \cdot \mathbf{\widehat{k}}_{23} = \left(k_2 (\mathbf{\widehat{k}}_2 \cdot \mathbf{\widehat{k}}_i) + k_3 (\mathbf{\widehat{k}}_3 \cdot \mathbf{\widehat{k}}_i )\right)/k_{23}$, allowing us to conclude every term in $W^{(3)}\left( \mathbf{k}_1,\mathbf{k}_2,\mathbf{k}_3\right)$ can be reproduced by the general expression:

\begin{equation}\label{eq:W3_expression}
W^{(3)}\left( \mathbf{k}_1,\mathbf{k}_2,\mathbf{k}_3\right) \rightarrow \frac{k_{1}^{n'_1}k_{2}^{n'_2}k_{3}^{n'_3}}{k_{23}^{n'_{23}}} (\mathbf{\widehat{k}}_1 \cdot \mathbf{\widehat{k}}_2)^{n_{12}}(\mathbf{\widehat{k}}_1 \cdot \mathbf{\widehat{k}}_3)^{n_{13}} (\mathbf{\widehat{k}}_2 \cdot \mathbf{\widehat{k}}_3)^{n_{23}} W^{(2)}\left(\mathbf{k}_2,\mathbf{k}_3\right), 
\end{equation}
for $n'_{23} = \left\{0,2\right\}$. 

We start with $n'_{23} = 2$; we will indicate this with $W^{(3)\left[n'_{23} = 2\right]}\left(\mathbf{k}_1,\mathbf{k}_2,\mathbf{k}_3\right)$. Using Eq. (\ref{eq:dot_to_iso_eq}) to expand the dot products into the isotropic basis, we find:

\begin{align}\label{eq:W_3}
&W^{(3)\left[2\right]}\left( \mathbf{k}_1,\mathbf{k}_2,\mathbf{k}_3\right) \left<\mu_{1}^{n_1}\mu_{2}^{n_2}\mu_{3}^{n_3}\right>_{\rm l.o.s} \nonumber \\ 
& \rightarrow \frac{k_{1}^{n'_1}k_{2}^{n'_2}k_{3}^{n'_3}}{k_{23}^{2}} (\mathbf{\widehat{k}}_1 \cdot \mathbf{\widehat{k}}_2)^{n_{12}}(\mathbf{\widehat{k}}_1 \cdot \mathbf{\widehat{k}}_3)^{n_{13}} (\mathbf{\widehat{k}}_2 \cdot \mathbf{\widehat{k}}_3)^{n_{23}} W_{2}\left( \mathbf{k}_2,\mathbf{k}_3\right)  \left<\mu_{1}^{n_1}\mu_{2}^{n_2}\mu_{3}^{n_3}\right>_{\rm l.o.s} \nonumber \\
&=(4\pi)^{7}\sum_{j_1,j_2,j_3}\sum_{j_{12},j_{13},j_{23}}\sum_{j,n} \mathcal{C}_{j_1,j_2,j_3}^{n_1,n_2,n_3} c_{j_{12}}^{(n_{12})}c_{j_{13}}^{(n_{13})}c_{j_{23}}^{(n_{23})} c_{j,n}^{(W)} \mathcal{P}_{j_1,j_2,j_3}(\mathbf{\widehat{k}}_1,\mathbf{\widehat{k}}_2,\mathbf{\widehat{k}}_3) \nonumber \\ 
& \qquad \times \mathcal{P}_{j}(\mathbf{\widehat{k}}_2,\mathbf{\widehat{k}}_3) \mathcal{P}_{j_{12}}(\mathbf{\widehat{k}}_1,\mathbf{\widehat{k}}_2) \mathcal{P}_{j_{13}}(\mathbf{\widehat{k}}_1,\mathbf{\widehat{k}}_3)\mathcal{P}_{j_{23}}(\mathbf{\widehat{k}}_2,\mathbf{\widehat{k}}_3) \frac{k_{1}^{n'_1}k_{2}^{n'_2+n}k_{3}^{n'_3-n}}{k_{23}^{2}},
\end{align}
with the $j_{12}$, $j_{13}$ and $j_{23}$ running from 0 to $n_{12}$, $n_{13}$ and $n_{23}$, respectively. We use Eq. (\ref{eq:k23}) to decouple the denominator in the last line and find:

\begin{align} \label{eq:W3_initial}
&W^{(3)\left[2\right]}\left( \mathbf{k}_1,\mathbf{k}_2,\mathbf{k}_3\right) \left<\mu_{1}^{n_1}\mu_{2}^{n_2}\mu_{3}^{n_3}\right>_{\rm l.o.s} \nonumber \\ 
& \rightarrow (4\pi)^{8}\sum_{j_1,j_2,j_3}\sum_{j_{12},j_{13},j_{23}}\sum_{j,n} \sum_{\ell_2} \mathcal{C}_{j_1,j_2,j_3}^{n_1,n_2,n_3} c_{j_{12}}^{(n_{12})}c_{j_{13}}^{(n_{13})}c_{j_{23}}^{(n_{23})} c_{j,n}^{(W)} s_{\ell_2}^{(\rm I)}\mathcal{P}_{j_1,j_2,j_3}(\mathbf{\widehat{k}}_1,\mathbf{\widehat{k}}_2,\mathbf{\widehat{k}}_3) \nonumber \\ 
& \qquad \times \mathcal{P}_{j}(\mathbf{\widehat{k}}_2,\mathbf{\widehat{k}}_3)\mathcal{P}_{j_{12}}(\mathbf{\widehat{k}}_1,\mathbf{\widehat{k}}_2) \mathcal{P}_{j_{13}}(\mathbf{\widehat{k}}_1,\mathbf{\widehat{k}}_3)\mathcal{P}_{j_{23}}(\mathbf{\widehat{k}}_2,\mathbf{\widehat{k}}_3) \mathcal{P}_{l_2}(\mathbf{\widehat{k}}_2,\mathbf{\widehat{k}}_3) \nonumber \\ 
& \qquad \times k_{1}^{n'_1}k_{2}^{n'_2+n}k_{3}^{n'_3-n} \int_{0}^{\infty} dr\; r\; j_{\ell_2}(k_2r)j_{\ell_2}(k_3r).
\end{align}

We now can combine all the isotropic basis functions into a single 3-argument isotropic basis function using Eq. (\ref{eq:Product_of_n_iso}) with $N=6$ to obtain:

\begin{align}\label{eq:dec_W3}
&W^{(3)\left[2\right]}\left( \mathbf{k}_1,\mathbf{k}_2,\mathbf{k}_3\right) \left<\mu_{1}^{n_1}\mu_{2}^{n_2}\mu_{3}^{n_3}\right>_{\rm l.o.s} \nonumber \\ 
& \rightarrow (4\pi)^{21/2}\sum_{j_1,j_2,j_3}\sum_{j_{12},j_{13},j_{23}}\sum_{j,n} \sum_{J'_1,J'_2,J'_3} \sum_{\ell_2} \mathcal{C}_{j_1,j_2,j_3}^{n_1,n_2,n_3} c_{j_{12}}^{(n_{12})}c_{j_{13}}^{(n_{13})}c_{j_{23}}^{(n_{23})} s_{\ell_2}^{(\rm I)} c_{j,n}^{(W)} \mathcal{G}_{J'_1,J'_2,J'_3} \nonumber \\ 
& \qquad \times \mathcal{P}_{J'_1,J'_2,J'_3} (\mathbf{\widehat{k}}_1,\mathbf{\widehat{k}}_2,\mathbf{\widehat{k}}_3) k_{1}^{n'_1}k_{2}^{n'_2+n}k_{3}^{n'_3-n} \int_{0}^{\infty}j_{\ell_2}(k_2r)j_{\ell_2}(k_3r) r dr.
\end{align}

Finally, we take the inverse Fourier transform of Eq. (\ref{eq:dec_W3}) using the same approach as for term 3 and find: 

\begin{empheq}[box=\widefbox]{align}\label{eq:R8,3,1}
&R_{8,(3);\left[2\right],\left[n_{12}\right], \left[n_{13}\right], \left[n_{23}\right]}^{ \left[n_1\right], \left[n_2\right], \left[n_3\right]}(\mathbf{r}_1; \mathbf{r}_2, \mathbf{r}_3) \rightarrow (4\pi)^{21/2}\sum_{j_1,j_2,j_3}\sum_{j_{12},j_{13},j_{23}}\sum_{j,n} \sum_{\ell_2} \sum_{L_1,L_2,L_3}  \nonumber \\ 
& \qquad \qquad \qquad \qquad\times c_{j,n}^{(W)} s_{\ell_2}^{(\rm I)} \mathcal{C}_{j_1,j_2,j_3}^{n_1,n_2,n_3} c_{j_{12}}^{(n_{12})}c_{j_{13}}^{(n_{13})}c_{j_{23}}^{(n_{23})} C_{L_1,L_2,L_3} \nonumber \\ 
& \qquad \qquad \qquad \qquad\times \mathcal{G}_{L_1,L_2,L_3} \Upsilon_{L_1,L_2,L_3} \nonumber \\ 
& \qquad \qquad \qquad \qquad\times \mathcal{P}_{L_1,L_2,L_3} (\mathbf{\widehat{r}}_1,\mathbf{\widehat{r}}_2,\mathbf{\widehat{r}}_3) \xi_{L_1}^{\left[n'_1 \right]}(r_{1}) F_{(\ell_2),L_2,L_3}^{\left[n+n'_2\right],\left[ n'_3-n\right]}(r_2,r_3),
\end{empheq}
with coefficients given in Table \ref{table:1}. The sum over $j$ and $n$ is finite and the range of each is shown in Eq. (\ref{eq:W2}). The sum over each $j_i$ runs from 0 to $n_i$, while the sums for $\ell_2$, $L_2$, and $L_3$ run from 0 to infinity; the sum for $L_1$ runs from 0 to $j_1 + j_{12} + j_{13}+j$. Also, radial integrals are once again defined by Eq. (\ref{eq:1D-radial}) and Eq. (\ref{eq:2D-radial}). The subscripts $\left[n'_{23}=2\right],\left[n_{12}\right], \left[n_{13}\right], \left[n_{23}\right]$ in the left-hand side of Eq. (\ref{eq:R8,3,1}) are internal dependencies that arises from the third-order kernel; the other superscripts are the external dependencies from the term we are analyzing. We use the same notation throughout this paper for other terms that exhibit a dependency coming from a kernel. 

\paragraph{$K^{(3)} = W^{(3)}$ with Inverse $k_{23}^{n'_{23}}$; $n'_{23}=0$}\mbox{}\\

We continue with $n'_{23}=0$ using Eq. (\ref{eq:Product_of_n_iso}) with $N=5$ to combine the isotropic basis functions, and then using Eq. (\ref{eq:W3_initial}) we find:

\begin{align}
&W^{(3)\left[0\right]}\left( \mathbf{k}_1,\mathbf{k}_2,\mathbf{k}_3\right) \left<\mu_{1}^{n_1}\mu_{2}^{n_2}\mu_{3}^{n_3}\right>_{\rm l.o.s} \nonumber \\ 
&\rightarrow (4\pi)^{9}\sum_{j_1,j_2,j_3}\sum_{j_{12},j_{13},j_{23}}\sum_{j,n} \sum_{J'_1,J'_2,J'_3} \mathcal{C}_{j_1,j_2,j_3}^{n_1,n_2,n_3} c_{j_{12}}^{(n_{12})}c_{j_{13}}^{(n_{13})}c_{j_{23}}^{(n_{23})} c_{j,n}^{(W)} \mathcal{G}_{J'_1,J'_2,J'_3} \nonumber \\ 
& \qquad \qquad \qquad \qquad \qquad \qquad \quad   \times \mathcal{P}_{J'_1,J'_2,J'_3} (\mathbf{\widehat{k}}_1,\mathbf{\widehat{k}}_2,\mathbf{\widehat{k}}_3) k_{1}^{n'_1}k_{2}^{n'_2+n}k_{3}^{n'_3-n}. 
\end{align}

Proceeding to take the inverse Fourier transform of the above expression and using the orthogonality of the isotropic basis functions we find:

\begin{empheq}[box=\widefbox]{align}\label{eq:R8,3,2}
&R_{8.1,(3);\left[0\right],\left[n_{12}\right], \left[n_{13}\right], \left[n_{23}\right]}^{ \left[n_1\right], \left[n_2\right], \left[n_3\right]}(\mathbf{r}_1; \mathbf{r}_2, \mathbf{r}_3) = (4\pi)^{9}\sum_{j_1,j_2,j_3}\sum_{j_{12},j_{13},j_{23}}\sum_{j,n} \sum_{L_1,L_2,L_3}  \nonumber\\
& \qquad \qquad \qquad \qquad \quad  \times c_{j,n}^{(W)} C_{L_1,L_2,L_3}\mathcal{C}_{j_1,j_2,j_3}^{n_1,n_2,n_3} c_{j_{12}}^{(n_{12})}c_{j_{13}}^{(n_{13})}c_{j_{23}}^{(n_{23})}\nonumber\\
& \qquad \qquad \qquad \qquad \quad \times\mathcal{G}_{L_1,L_2,L_3}  \Upsilon_{L_1,L_2,L_3}\nonumber\\
& \qquad \qquad \qquad \qquad \quad  \times \mathcal{P}_{L_1,L_2,L_3} (\mathbf{\widehat{r}}_1,\mathbf{\widehat{r}}_2,\mathbf{\widehat{r}}_3)\xi_{L_1}^{\left[n'_1 \right]}(r_{1})\xi_{L_2}^{\left[n+n'_2 \right]}(r_{2}) \xi_{L_3}^{\left[n'_3-n \right]}(r_{3}),
\end{empheq}
with coefficients given in Table \ref{table:1}. All the sums are the same as in Eq. (\ref{eq:R8,3,1}), except for $L_1$ running from 0 to $j_1 + j_{12}+j_{13}$, $L_2$ from 0 to $j_2 + j_{12}+j_{23}+j$ and $L_3$ running from 0 to $j_3 + j_{13}+j_{23}+j$. The radial integrals are given in Eq. (\ref{eq:1D-radial}). 
 
\paragraph{$K^{(3)} = \mathcal{G}^{(3)}$ }\mbox{}\\
We evaluate now the third-order kernel with $K^{(3)} = \mathcal{G}^{(3)}$, where $\mathcal{G}^{(3)}$ is defined in Eq. (\ref{eq:FancyG3}) and can be written as:

\begin{align}\label{eq:constant_g}
\mathcal{G}^{(3)} (\mathbf{k}_1, \mathbf{k}_2, \mathbf{k}_3) = \sum_{n_{12},n_{13},n_{23} = 0}^{2} g_{n_{12},n_{13},n_{23}}(\mathbf{\widehat{k}}_1 \cdot \mathbf{\widehat{k}}_2)^{n_{12}}(\mathbf{\widehat{k}}_1 \cdot \mathbf{\widehat{k}}_3)^{n_{13}} (\mathbf{\widehat{k}}_2 \cdot \mathbf{\widehat{k}}_3)^{n_{23}}, 
\end{align}
where the constant $g$ is non zero for $g_{1,1,1} = 2$, and $g_{2,0,0} = g_{0,2,0} = g_{0,0,2} = -1$; all other combinations are zero. Using Eq. (\ref{eq:dot_to_iso_eq}) to expand the dot products into isotropic basis functions we find:

\begin{align}
&\mathcal{G}^{(3)} (\mathbf{k}_1, \mathbf{k}_2, \mathbf{k}_3) =  (4\pi)^3 \sum_{n_{12},n_{13},n_{23}} \sum_{j_{12},j_{13},j_{23}=0}^{n_{12},n_{13},n_{23}} g_{n_{12},n_{13},n_{23}} c_{j_{12}}^{(n_{12})} c_{j_{13}}^{(n_{13})} c_{j_{23}}^{(n_{23})} \nonumber \\
& \qquad \qquad \qquad \qquad \times \mathcal{P}_{j_{12}}(\mathbf{\widehat{k}}_1, \mathbf{\widehat{k}}_2) \mathcal{P}_{j_{13}}(\mathbf{\widehat{k}}_1, \mathbf{\widehat{k}}_3) \mathcal{P}_{j_{23}}(\mathbf{\widehat{k}}_2, \mathbf{\widehat{k}}_3).
\end{align}
Including all the $\mu$ terms, and combining all isotropic basis functions into one using Eq. (\ref{eq:Product_of_n_iso}) for N=4, we obtain:

\begin{align}
&\mathcal{G}^{(3)} (\mathbf{k}_1, \mathbf{k}_2, \mathbf{k}_3) \left<\mu_1^{n_1}\mu_2^{n_2}\mu_3^{n_3}\right>_{\rm l.o.s} = (4\pi)^{15/2} \sum_{n_{12},n_{13},n_{23}} \sum_{j_{12},j_{13},j_{23}}\sum_{j_1,j_2,j_3} \sum_{J_1,J_2,J_3} g_{n_{12},n_{13},n_{23}}\nonumber \\
& \qquad \qquad \qquad \qquad \qquad \qquad \times  c_{j_{12}}^{(n_{12})} c_{j_{13}}^{(n_{13})} c_{j_{23}}^{(n_{23})} \mathcal{C}_{j_1,j_2,j_3}^{n_1,n_2,n_3} \mathcal{G}_{J_1,J_2,J_3} \mathcal{P}_{J_1,J_2,J_3}(\mathbf{\widehat{k}}_1, \mathbf{\widehat{k}}_2, \mathbf{\widehat{k}}_3). 
\end{align}
Finally, including the power spectra and taking the inverse Fourier transform, we find:

\begin{empheq}[box=\widefbox]{align}\label{eq:R8,3.2}
&R_{8.2,(3)}^{ \left[n_1\right], \left[n_2\right], \left[n_3\right]}(\mathbf{r}_1; \mathbf{r}_2, \mathbf{r}_3) = (4\pi)^{15/2} \sum_{n_{12},n_{13},n_{23}} \sum_{j_{12},j_{13},j_{23}}\sum_{j_1,j_2,j_3} \sum_{L_1,L_2,L_3} \nonumber \\
& \qquad \qquad \qquad \qquad \qquad \qquad \times g_{n_{12},n_{13},n_{23}} c_{j_{12}}^{(n_{12})} c_{j_{13}}^{(n_{13})} c_{j_{23}}^{(n_{23})} \mathcal{C}_{j_1,j_2,j_3}^{n_1,n_2,n_3} \nonumber \\
& \qquad \qquad \qquad \qquad \qquad \qquad \times C_{L_1,L_2,L_3}\mathcal{G}_{L_1,L_2,L_3} \Upsilon_{L_1,L_2,L_3} \nonumber \\
& \qquad \qquad \qquad \qquad \qquad \qquad \times \mathcal{P}_{L_1,L_2,L_3}(\mathbf{\widehat{r}}_1, \mathbf{\widehat{r}}_2, \mathbf{\widehat{r}}_3) \xi_{L_1}^{[0]}(r_1)\xi_{L_2}^{[0]}(r_2)\xi_{L_3}^{[0]}(r_3). 
\end{empheq}
with coefficients given in Table \ref{table:1}. The sum over each $j_i$ runs from 0 to $n_i$, while the sums for $L_i$ runs from 0 to $j_i + j_{ij} + j_{ik}$; $i \neq j \neq k$. Also, the radial integrals are once again defined by Eq. (\ref{eq:1D-radial}). 

\paragraph{$K^{(3)} = \Gamma^{(3)}$ }\mbox{}\\
Lastly, we have to evaluate term 8 for $K^{(3)} = \Gamma^{(3)}$, where $\Gamma^{(3)}$ is given by Eq. (\ref{eq:Cappital_Letter_gamma}). The two second-order kernels, $F^{(2)}$ and $G^{(2)}$, in Eq. (\ref{eq:Cappital_Letter_gamma}) can be expressed in general with the $W^{(2)}$ kernel. Therefore, we will only evaluate $W^{(2)}$ once. With this in mind, we expand the $\mathcal{G}^{(2)}$ kernel to find: 
\begin{align}
\Gamma^{(3)} (\mathbf{k}_1, \mathbf{k}_2, \mathbf{k}_3) \rightarrow \left[ \frac{(\mathbf{\widehat{k}}_1\cdot\mathbf{k}_2)^2 + (\mathbf{\widehat{k}}_1\cdot\mathbf{k}_3)^2 + 2(\mathbf{\widehat{k}}_1\cdot\mathbf{k}_2)(\mathbf{\widehat{k}}_1\cdot\mathbf{k}_3) }{k_{23}^{2}} -1 \right]W^{(2)}(\mathbf{k}_2, \mathbf{k}_3).
\end{align}
The second term in parenthesis is the same as term 1 in the list at the beginning of this section (evaluated at $K=W^{(2)}$), which means its result is given by Eq. (\ref{eq:R1,3}). We will not include this term in what follows of derivation. Simplifying the first expression in parenthesis, we obtain: 
\begin{align}\label{eq:gamma_const_def}
\Gamma^{(3)} (\mathbf{k}_1, \mathbf{k}_2, \mathbf{k}_3) \rightarrow  \sum_{n_{12},n_{13}=0}^{2}\frac{\gamma^{n_{12},n_{13}}(\mathbf{\widehat{k}}_1\cdot\mathbf{\widehat{k}}_2)^{n_{12}} (\mathbf{\widehat{k}}_1\cdot\mathbf{\widehat{k}}_3)^{n_{13}} k_2^{n_{12}}k_3^{n_{13}}}{k_{23}^2} W^{(2)}(\mathbf{k}_2, \mathbf{k}_3),
\end{align}
where $\gamma$ is a constant that ensures only the correct terms are non zero; only for specific combinations of $n_{12} = n_{13} = \left\{0,1,2\right\}$. Using Eq. (\ref{eq:dot_to_iso_eq}) to expand the dot products in terms of isotropic functions, we find:
\begin{align}
&\Gamma^{(3)} (\mathbf{k}_1, \mathbf{k}_2, \mathbf{k}_3) \rightarrow (4\pi)^4  \sum_{n_{12},n_{13}} \sum_{j_{12},j_{13}} \sum_{n,j}\sum_{\ell_2}   \gamma^{n_{12},n_{13}} s_{\ell_2}^{(\rm I)} c_{j,n}^{(W)}  c_{j_{12}}^{(n_{12})}c_{j_{13}}^{(n_{13})} \nonumber \\
& \qquad \qquad \qquad \qquad \times \mathcal{P}_{j}(\mathbf{k}_2, \mathbf{k}_3)\mathcal{P}_{\ell_2}(\mathbf{k}_2, \mathbf{k}_3) \mathcal{P}_{j_{12}}(\mathbf{\widehat{k}}_1, \mathbf{\widehat{k}}_2) \mathcal{P}_{j_{13}}(\mathbf{\widehat{k}}_1, \mathbf{\widehat{k}}_3) \nonumber \\
& \qquad \qquad \qquad \qquad \times k_2^{n_{12}}k_3^{n_{13}}\int_{0}^{\infty} dr\; r\; j_{\ell_2}(rk_2) j_{\ell_2}(rk_3 ).  
\end{align}

Including the averaged $\mu$ terms and using Eq. (\ref{eq:Product_of_n_iso}), with $N=5$, to express all the isotropic basis functions into a single one, we find:
\begin{align}
&\Gamma^{(3)} (\mathbf{k}_1, \mathbf{k}_2, \mathbf{k}_3) \left<\mu_1^{n_1}\mu_2^{n_2}\mu_3^{n_3}\right> \rightarrow  (4\pi)^9   \sum_{n_{12},n_{13}} \sum_{j_{12},j_{13}}\sum_{n,j}\sum_{\ell_2} \sum_{j_1,j_2,j_3}\sum_{J_1,J_2,J_3}  \gamma^{n_{12},n_{13}} s_{\ell_2}^{(\rm I)} \nonumber \\
& \qquad \qquad \qquad\qquad \qquad \qquad\qquad\times  c_{j,n}^{(W)}  c_{j_{12}}^{(n_{12})}c_{j_{13}}^{(n_{13})}\mathcal{C}_{j_1,j_2,j_3}^{n_1,n_2,n_3}\mathcal{G}_{J_1,J_2,J_3} \mathcal{P}_{J_1,J_2,J_3}(\mathbf{\widehat{k}}_1,\mathbf{\widehat{k}}_2,\mathbf{\widehat{k}}_3)   \nonumber \\
& \qquad \qquad\qquad\qquad\qquad \qquad\qquad\times k_2^{n_{12}}k_3^{n_{13}}\int_{0}^{\infty} dr\; r\; j_{\ell_2}(rk_2) j_{\ell_2}(rk_3 ).  
\end{align}

Finally, taking the inverse Fourier transform of the above equation and including the power spectra, we obtain:
\begin{empheq}[box=\widefbox]{align}\label{eq:R8,3.3}
&R_{8.3,(3)}^{ \left[n_1\right], \left[n_2\right], \left[n_3\right]}(\mathbf{r}_1; \mathbf{r}_2, \mathbf{r}_3) \rightarrow (4\pi)^{9} \sum_{n_{12},n_{13}}\sum_{j_{12},j_{13}} \sum_{n,j}\sum_{\ell_2} \sum_{j_1,j_2,j_3}\sum_{L_1,L_2,L_3} \nonumber \\
& \qquad \qquad\qquad  \qquad\qquad\times \gamma^{n_{12},n_{13}} c_{j,n}^{(W)} s_{\ell_2}^{(\rm I)} c_{j_{12}}^{(n_{12})}c_{j_{13}}^{(n_{13})} \mathcal{C}_{j_1,j_2,j_3}^{n_1,n_2,n_3}\mathcal{G}_{L_1,L_2,L_3}\nonumber \\
& \qquad \qquad\qquad  \qquad\qquad\times  C_{L_1,L_2,L_3} \Upsilon_{L_1,L_2,L_3}\mathcal{P}_{L_1,L_2,L_3}(\mathbf{\widehat{r}}_1,\mathbf{\widehat{r}}_2, \mathbf{\widehat{r}}_3) \nonumber \\
& \qquad \qquad\qquad  \qquad\qquad \times \xi_{L_1}^{[0]}(r_1) F_{(\ell_2),L_2,L_3}^{[n-n_{12}],[n_{13}-n]}(r_2,r_3),
\end{empheq}
with coefficients given in Table \ref{table:1}. The sum over $j$ and $n$ is finite and the range of each is shown in Eq. (\ref{eq:W2}). The sum over each $j_i$ runs from 0 to $n_i$, while the sums for $\ell_2$, $L_2$, and $L_3$ run from 0 to infinity; the sum for $L_1$ runs from 0 to $j_1 + j_{12} + j_{13}$. Also, radial integrals are once again defined by Eq. (\ref{eq:1D-radial}) and Eq. (\ref{eq:2D-radial}). 

\subsubsection{Third-Order Kernel Weighted by $\mu_{123}^{2}$}\label{sec:T3.1.9}
\qquad Term 9 in the list at the beginning of this section is:

\begin{align}
W^{(3)}(\mathbf{k}_2,\mathbf{k}_3)\mu_{123}^{2}\mu_{1}^{n_1}\mu_{2}^{n_2}\mu_{3}^{n_3},\nonumber
\end{align}

so we begin with the analysis by looking at the structure of $\mu_{123}$:

\begin{equation}\label{eq:mu_123}
    \mu_{123} = \mathbf{\widehat{k}}_{123}\cdot\mathbf{\widehat{z}} = \frac{(\mathbf{k}_1 + \mathbf{k}_2 + \mathbf{k}_3)\cdot\mathbf{\widehat{z}}}{k_{123}} = \frac{k_1 \mu_{1} + k_2 \mu_{2}+k_3 \mu_{3}}{k_{123}}.
\end{equation}

Using the trinomial expansion on the above expression implies $\mu_{123}^{2}$ is: 
 
\begin{equation}\label{eq:mu_123_squared}
\mu_{123}^{2} \equiv \sum_{a_1,a_2,a_3=0}^{2} t_{a_1,a_2,a_3}^{(2)} \frac{(k_1\mu_1)^{a_1}(k_2\mu_2)^{a_2}(k_3\mu_3)^{a_3}}{k_{123}^{2}},
\end{equation}
with $t_{a_1,a_2,a_3}^{(2)} \equiv 2/(a_1! a_2!a_3!)$ and the dummy indices satisfying $a_1+a_2+a_3 =2$. 
\paragraph{Inverse $k_{23}^{n'_{23}}$; $n'_{23}=2$}\mbox{}\\

\begin{figure}[h]
\centering
\includegraphics[scale=0.6]{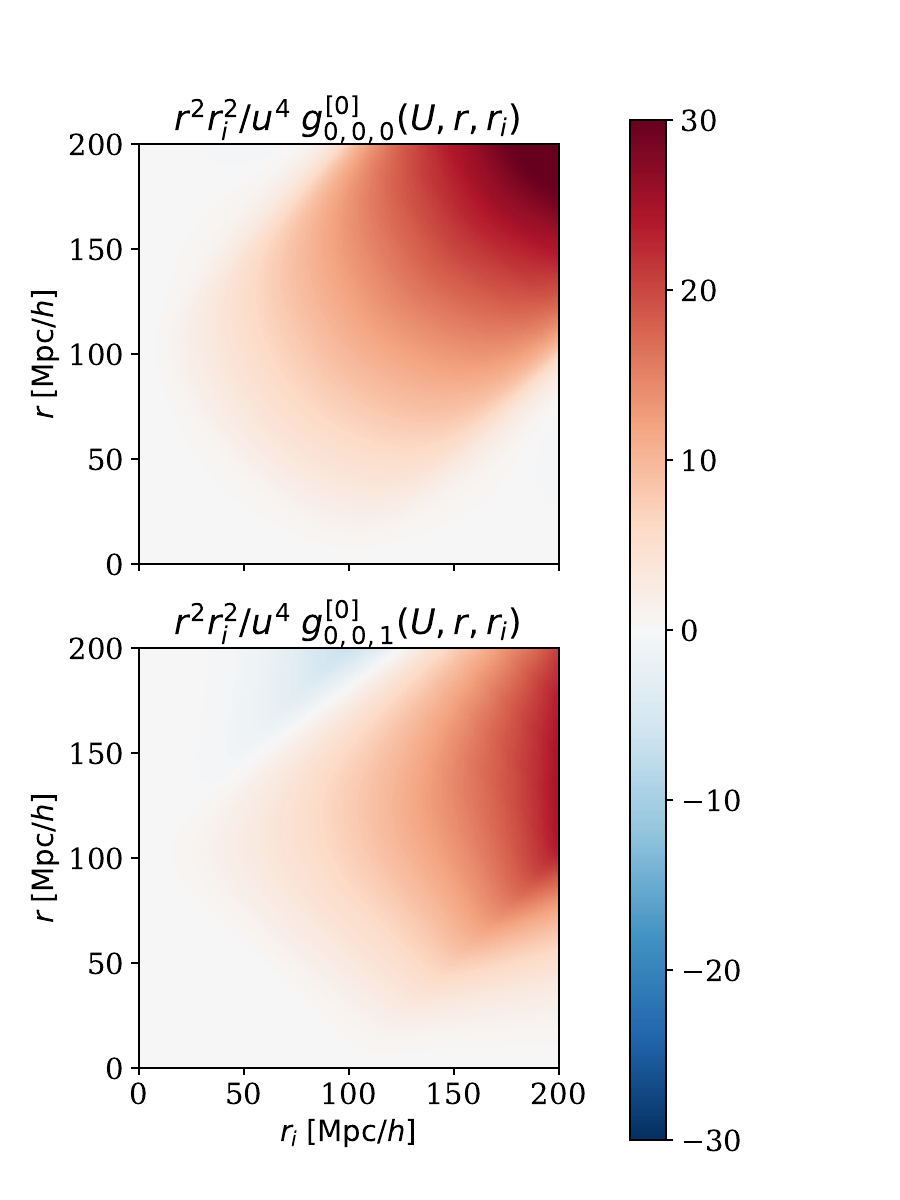}
\caption{Here, we show Eq. (\ref{eq:3_Sph.Bess_int}) for fixed $r^{'} = U \equiv 100 \;\left[{\rm Mpc}/h\right]$, $n' = 0$ $\ell'=0$,  $\ell = 0$ and $L = \left\{0,1\right\}$.The \textit{upper panel} shows the integral for $L=0$, while the \textit{lower panel} shows the integral for $L=1$. Since the 4PCF can be approximated as the square of the 2PCF on large scales, $(\xi_{0})^{2}(r) \sim (1/r^2)^2$, we have weighted the integral by $r^2r_i^2/u^{4}$, with $u \equiv 10 \;\left[{\rm Mpc}/h\right]$ to take out its fall-off. Both, the \textit{upper} and \textit{lower} panel show the behavior of the integral creates a rectangular boundary. We demonstrate analytically in Appendix \ref{Sec: Radial Integrals} that the rectangular behavior arises from the product of two "top-hat" functions for the \textit{upper panel} and a product of a "top hat" with a Heaviside function for the \textit{lower panel}. Evaluation using \texttt{Mathematica} from our analytic results is in Figure \ref{fig:g000 analytical}  and Figure \ref{fig:g001 analytical}. The \textit{lower panel} is also not symmetric under exchange of axes (\textit{i.e.}, $r_i \leftrightarrow r$), which results from choosing $\ell = 0$ for the sBF order corresponding to the variable $r$ and $L=1$ for the sBF order corresponding to the variable $r_i$.}
\label{fig:gint2}
\end{figure}

\begin{figure}[h]
\centering
\includegraphics[scale=0.7]{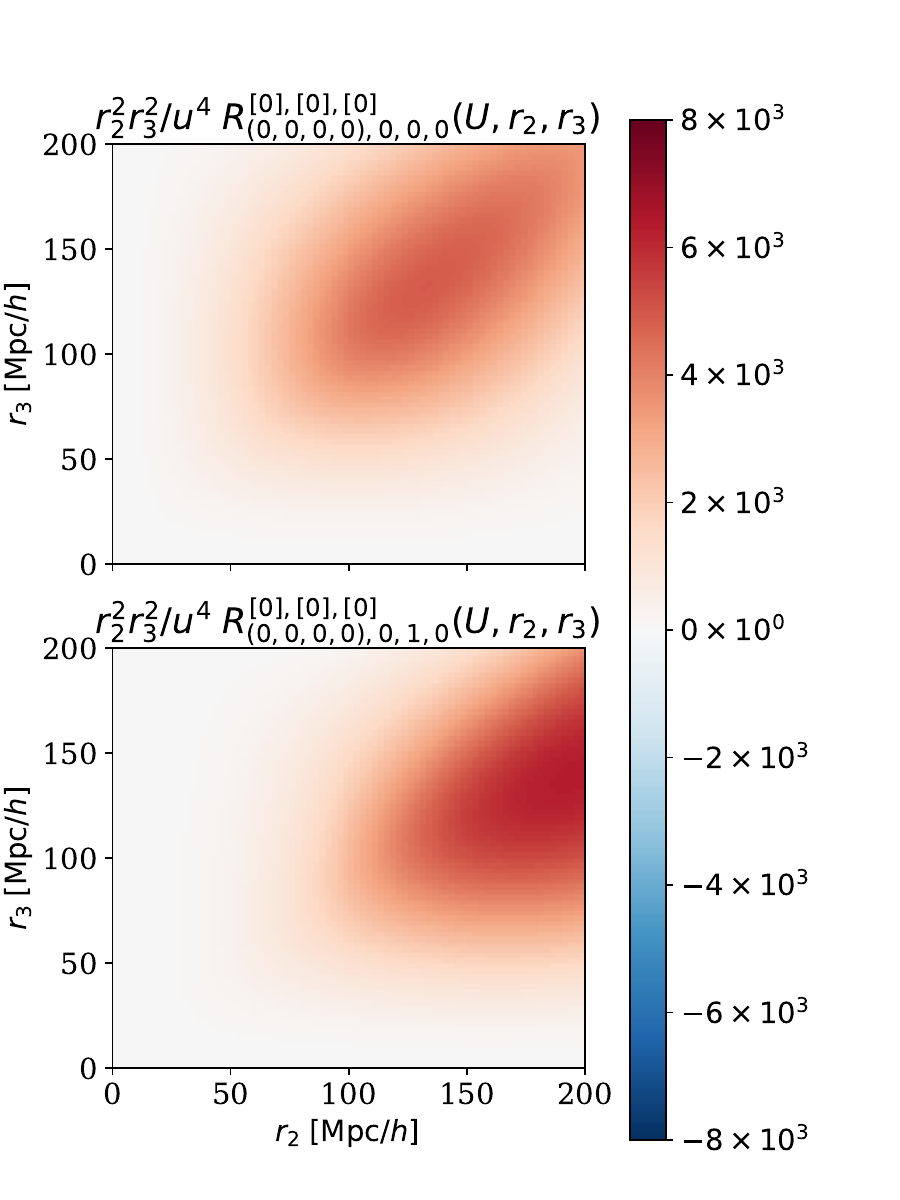}
\caption{2D plot of integral Eq. (\ref{eq:R_int_T3111}) for fixed $r_1= U \equiv 100 \;\left[{\rm Mpc}/h\right]$, $n'_1=n'_2=n'_3 =0$, $\ell_2 =\ell'_1 =\ell'_2 =\ell'_3 =0 $, $L_1 = 0$, $L_2 = \left\{ 0,1\right\}$ and $L_3=0$. We place $\ell_2$, $\ell'_1 $, $\ell'_2 $ and $\ell'_3 $ in parentheses to identify them as the sBF orders that do not contribute directly to the $r_1,\;r_2$ and $r_3$ variables. The \textit{upper panel} shows the integral for $L_2=0$, while the \textit{lower panel} shows the integral for $L_2=1$. Since the 4PCF can be approximated as the square of the 2PCF on large scales, $(\xi_{0})^{2}(r) \sim (1/r^2)^2$, we have weighted the integral by $r^2r_i^2/u^{4}$, with $u \equiv 10 \;\left[{\rm Mpc}/h\right]$ to take out its fall-off. Both the \textit{upper} and \textit{lower panel} show a blob that appears at about $r_2 = r_3 \simeq U $ and decreases in magnitude slightly as one moves away from its center. We demonstrate analytically in Appendix \ref{Sec: Radial Integrals} that the behavior on both of these panels arises from the product of two "top-hat" functions times $1/r$ for the \textit{upper panel} and a product of a "top hat" with a Heaviside times a function defined in Eq. (\ref{eq:I_function_from_g001}) for the \textit{lower panel}. Evaluation using \texttt{Mathematica} from our analytic results is in Figures \ref{fig:R0_analytical}  and \ref{fig:R1_analytical}. The \textit{lower panel} is also not symmetric under exchange of axes (\textit{i.e.}, $r_2 \leftrightarrow r_3$), which results from choosing $L_3 = 0$ for the sBF order corresponding to the variable $r_3$ and $L_2=1$ for the sBF order corresponding to the variable $r_2$. }
\label{fig:R_int}
\end{figure}

The $W^{(3)}$ kernel has a $1/k_{23}^{n'_{23}}$ factor---with $n'_{23}=\left\{0,2\right\}$---in its structure as shown in Eq. (\ref{eq:W3_expression}), therefore we follow the analysis of term 9 by choosing $n'_{23}=2$. We obtain: 

\begin{align}
&W^{(3)\left[2\right]}\left(\mathbf{k}_1,\mathbf{k}_2,\mathbf{k}_3\right) \left<\mu_{123}^{2}\mu_{1}^{n_1}\mu_{2}^{n_2}\mu_{3}^{n_3}\right>_{\rm l.o.s} \nonumber \\ 
&\rightarrow (4\pi)^{25/2}\sum_{a_1,a_2,a_3}\sum_{j_1,j_2,j_3}\sum_{j_{12},j_{13},j_{23}}\sum_{j,n} \sum_{\ell_2} \sum_{\ell'_1,\ell'_2,\ell'_3}\sum_{J'_1,J'_2,J'_3}  \mathcal{C}_{j_1,j_2,j_3}^{n_1,n_2,n_3} c_{j_{12}}^{(n_{12})}c_{j_{13}}^{(n_{13})}c_{j_{23}}^{(n_{23})} c_{j,n}^{(W)} s_{\ell_2}^{(\rm I)}  \nonumber \\ 
& \qquad \qquad  \times \mathcal{C}_{\ell'_1,\ell'_2,\ell'_3}t_{a_1,a_2,a_3}^{(2)}\Upsilon_{J'_1,J'_2,J'_3} \mathcal{P}_{J'_1,J'_2,J'_3} (\mathbf{\widehat{k}}_1,\mathbf{\widehat{k}}_2,\mathbf{\widehat{k}}_3) k_{1}^{n'_1+a_1}k_{2}^{n'_2+n+a_2}k_{3}^{n'_3-n+a_3} \nonumber \\ 
& \qquad \qquad  \times \int_{0}^{\infty} dr\; r\; j_{\ell_2}(k_2r)j_{\ell_2}(k_3r)   \int_{0}^{\infty} dr' r' j_{\ell'_1}(k_1r')j_{\ell'_2}(k_2r')j_{\ell'_3}(k_3r'),
\end{align}
after we have decoupled $1/k_{23}^2$ and $1/k_{123}^2 $ using equations (\ref{eq:k23}) and (\ref{eq:k123}), respectively. We have also combined the seven isotropic basis functions into a single one using Eq. (\ref{eq:Product_of_n_iso}) for $N=7$. The sum over $j_1$ runs from 0 to $n_1+a_1$, over $j_2$ runs from 0 to $n_2+a_2$ and over $j_3$ runs from 0 to $n_3+a_3$. The sum over $J'_1$ runs from 0 to $j_1+j_{12}+j_{13}+\ell'_1$, over $J'_2$ runs from 0 to $j_2+j_{12}+j_{23}+j+\ell_2+\ell'_2$ and over $J'_3$ runs from 0 to $j_3+j_{13}+j_{23}+j+\ell_2+\ell'_3$. We continue by including the power spectrum and writing the integrals for the inverse Fourier transforms. After expanding the complex exponentials into the isotropic basis functions as we did for Eq. (\ref{eq:exp_expansion_iso}), we perform the angular integral to find:

\begin{empheq}[box=\widefbox]{align}\label{eq:R_9T3111}
&R_{9,(3);\left[2\right],\left[n_{12}\right], \left[n_{13}\right], \left[n_{23}\right]}^{ \left[n_1\right], \left[n_2\right], \left[n_3\right]}(\mathbf{r}_1; \mathbf{r}_2, \mathbf{r}_3) \rightarrow (4\pi)^{25/2}\sum_{a_1,a_2,a_3} \sum_{j_1,j_2,j_3}\sum_{j_{12},j_{13},j_{23}}  \nonumber \\ 
& \qquad  \qquad \qquad  \times \sum_{j,n} \sum_{l_2} \sum_{\ell'_1,\ell'_2,\ell'_3} \sum_{L_1,L_2,L_3} s_{\ell_2}^{(\rm I)} c_{j,n}^{(W)} c_{j_{12}}^{(n_{12})}c_{j_{13}}^{(n_{13})}c_{j_{23}}^{(n_{23})} \mathcal{C}_{j_1,j_2,j_3}^{n_1,n_2,n_3}\nonumber \\ 
& \qquad  \qquad \qquad \times \mathcal{C}_{\ell'_1,\ell'_2,\ell'_3}  t_{a_1,a_2,a_3}^{(2)}  C_{L_1,L_2,L_3} \mathcal{G}_{L_1,L_2,L_3}\Upsilon_{L_1,L_2,L_3} \nonumber \\ 
& \qquad   \qquad \qquad \times \mathcal{P}_{L_1,L_2,L_3} (\mathbf{\widehat{r}}_1,\mathbf{\widehat{r}}_2,\mathbf{\widehat{r}}_3) R_{(\ell_2,\ell'_1,\ell'_2,\ell'_3),L_1,L_2,L_3}^{\left[n'_1+a_1\right],\left[n'_2+n+a_2\right],\left[n'_3-n+a_3\right]}(r_1,r_2,r_3),
\end{empheq}
where we have used the orthogonality of the isotropic basis functions to obtain $J'_1=L_1$, $J'_2=L_2$, and $J'_3=L_3$. The subscripts $\left[n’_{23} = 2\right],\left[n_{12}\right], \left[n_{13}\right], \left[n_{23}\right]$ in the left-hand side of Eq. (\ref{eq:R_9T3111}) are internal dependencies that arises from the third-order kernel; the other superscripts are the external dependencies from the term we are analyzing. The coefficients are given in Table \ref{table:1} and the radial integral is:

\begin{align}\label{eq:R_int_T3111}
&R_{(\ell_2,\ell'_1,\ell'_2,\ell'_3),L_1,L_2,L_3}^{\left[n'_1\right],\left[n'_2\right],\left[n'_3\right]}(r_1,r_2,r_3) \equiv \int_{0}^{\infty} dr' \; r' \;f_{\ell'_1,L_1}^{\left[n'_1\right]}(r',r_1) \int_{0}^{\infty} dr \;r \nonumber \\ 
& \qquad \qquad \qquad \qquad \qquad \qquad \qquad \qquad \times g_{\ell'_2,\ell_2,L_2}^{\left[n'_2\right]}(r',r,r_2) g_{\ell'_3,\ell_2,L_3}^{\left[n'_3\right]}(r',r,r_3). 
\end{align}
We have used parentheses in the radial integral around some of the orders of the spherical Bessel to indicate that these are being integrated out and do not directly affect the resulting variables $r_1$, $r_2$ and $r_3$. The definition of $f_{\ell,L}$ is given by Eq. (\ref{eq:2_Sph.Bess_int}) and the definition of $g_{\ell',\ell,L}$ is:

\begin{align}\label{eq:3_Sph.Bess_int}
g_{\ell',\ell,L}^{\left[n'\right]}(r',r,r_i) \equiv \int_{0}^{\infty} \frac{dk_i}{2\pi^2} \; k_i^{n'+2} j_{\ell'}(k_ir')j_{\ell}(k_ir)j_{L}(k_ir_i) P(k_i).
\end{align}
If we choose a power-law power spectrum, $P(k_i) \sim 1/k_i$ we find analytic solutions, for some values of $n'$, $n'_2$, $n'_3$, $\ell$, $\ell'$, $L$, $\ell_2$, $\ell'_1$, $\ell'_2$, $\ell'_3$, $L_1$, $L_2$ and $L_3$ for Eq. (\ref{eq:R_int_T3111}) and Eq. (\ref{eq:3_Sph.Bess_int}) in \cite{DeLaBella3sbf, J.Chellino, Indefinite_Integrals_SBF, Sph_Bessel_Integral_kiersten, Sph_Bessel_int_Rami, Fabrikant}. Therefore, in Appendix \ref{Sec: Radial Integrals}, we have analyzed several special cases to explain the behavior shown in Figure \ref{fig:gint2} and Figure \ref{fig:R_int}, where we display Eq. (\ref{eq:3_Sph.Bess_int}) and Eq. (\ref{eq:R_int_T3111}) with the true power spectrum. 

\clearpage

\paragraph{Inverse $k_{23}^{n'_{23}}$; $n'_{23}=0$}\mbox{}\\

Continuing with $n'_{23}=0$, we obtain: 
\begin{align}
&W_{3}^{\left[0\right]}\left(\mathbf{k}_1,\mathbf{k}_2,\mathbf{k}_3\right) \left<\mu_{123}^{2}\mu_{1}^{n_1}\mu_{2}^{n_2}\mu_{3}^{n_3}\right>_{\rm l.o.s} \nonumber \\ 
&\rightarrow (4\pi)^{11} \sum_{a_1,a_2,a_3} \sum_{j_1,j_2,j_3}\sum_{j_{12},j_{13},j_{23}}\sum_{j,n} \sum_{\ell'_1,\ell'_2,\ell'_3} \sum_{J'_1,J'_2,J'_3} \mathcal{C}_{j_1,j_2,j_3}^{n_1,n_2,n_3} c_{j_{12}}^{(n_{12})}c_{j_{13}}^{(n_{13})}c_{j_{23}}^{(n_{23})} c_{j,n}^{(W)} t_{a_1,a_2,a_3}^{(2)}  \nonumber \\ 
& \qquad \qquad \times \mathcal{C}_{\ell'_1,\ell'_2,\ell'_3} \mathcal{G}_{J'_1,J'_2,J'_3} \mathcal{P}_{J'_1,J'_2,J'_3} (\mathbf{\widehat{k}}_1,\mathbf{\widehat{k}}_2,\mathbf{\widehat{k}}_3)
 k_{1}^{n'_1+a_1}k_{2}^{n'_2+n+a_2}k_{3}^{n'_3-n+a_3} \nonumber \\ 
& \qquad \qquad \times \int_{0}^{\infty} dr'\; r' \;j_{\ell'_1}(k_1r')j_{\ell'_2}(k_2r')j_{\ell'_3}(k_3r') .
\end{align}

\begin{figure}[h]
\centering
\includegraphics[scale=0.7]{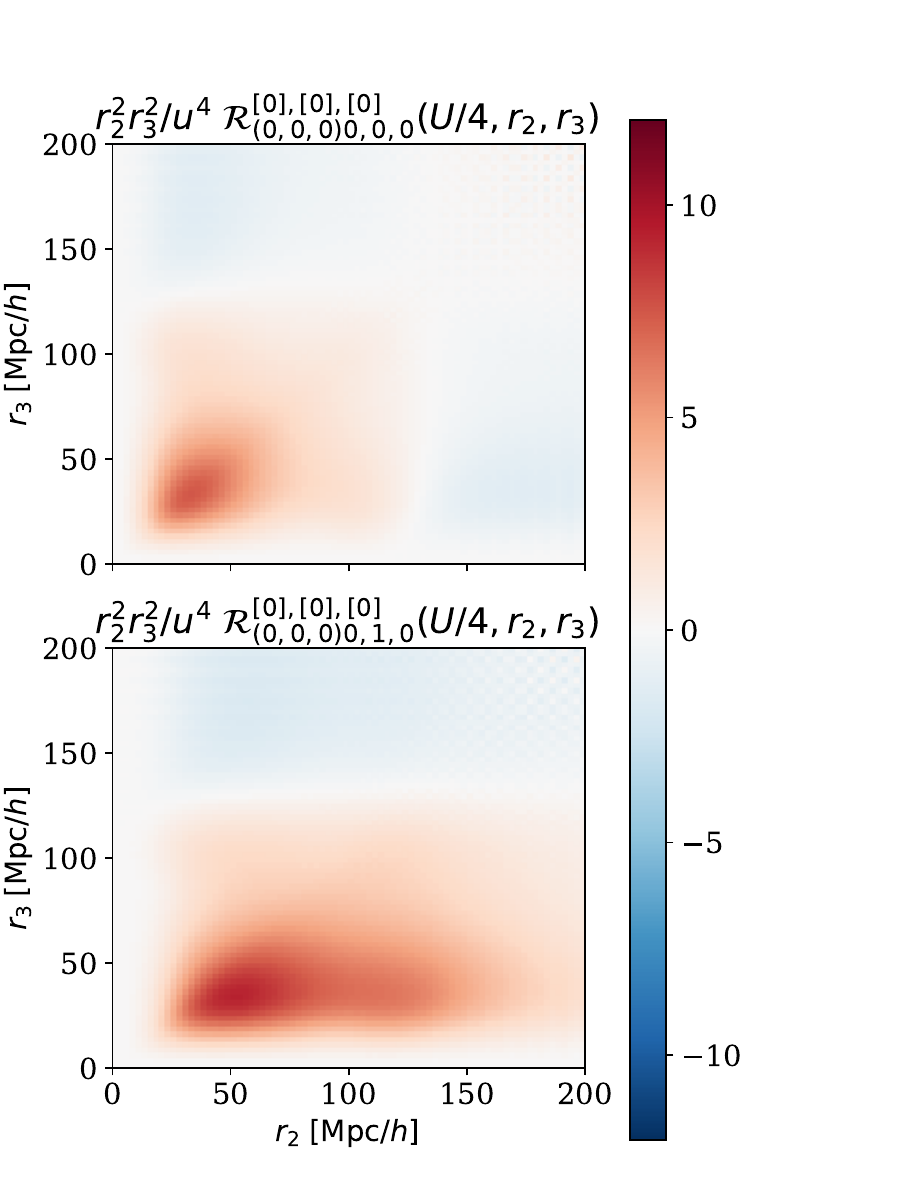}
\caption{Here we show Eq. (\ref{eq:fancy_R_int}) for fixed $r_1= U/4 \equiv 25 \;\left[{\rm Mpc}/h\right]$, $n'_1=n'_2=n'_3 =0$, $\ell'_1 =\ell'_2 =\ell'_3 =0 $, $L_1 = 0$, $L_2 = \left\{ 0,1\right\}$ and $L_3=0$. We place $\ell'_1 $, $\ell'_2 $ and $\ell'_3 $ in parentheses to identify them as the sBF orders that do not pertain directly to the $r_1,\;r_2$ and $r_3$ variables. The \textit{upper panel} shows the integral for $L_2=0$, while the \textit{lower panel} shows the integral for $L_2=1$. Since the 4PCF can be approximated as the square of the 2PCF on large scales, $(\xi_{0})^{2}(r) \sim (1/r^2)^2$, we have weighted the integral by $r^2r_i^2/u^{4}$, with $u \equiv 10 \;\left[{\rm Mpc}/h\right]$ to take out its fall-off. Both the \textit{upper} and \textit{lower panel} show a blob that appears at about $r_2 = r_3 \simeq U/4 $. We demonstrate analytically in Appendix \ref{Sec: Radial Integrals} that the behavior on the \textit{upper panel} arises because we can approximate $\mathcal{R}_{(0,0,0),0,0,0}^{\left[0\right],\left[0\right],\left[0\right]}$ as a product of two Dirac delta functions dependent on $r_1,r_2$ and $r_3$, while the behavior in the \textit{lower panel} is described by approximating $\mathcal{R}_{(0,0,0),0,1,0}^{\left[0\right],\left[0\right],\left[0\right]}$ as one Dirac delta function in the region $r_2>r_1$. The \textit{lower panel} is also not symmetric under exchange of axes (\textit{i.e.}, $r_2 \leftrightarrow r_3$), which results from choosing $L_3 = 0$ for the sBF order corresponding to the variable $r_3$ and $L_2=1$ for the sBF order corresponding to the variable $r_2$. This Figure is the first of three; we show the three figures to show that the choice of $r_1$ moves the location of the blob, as we expect from our analytic results in Appendix \ref{Sec: Radial Integrals}.}
\label{fig:fancy_R_int_down}
\end{figure}

\begin{figure}[h]
\centering
\includegraphics[scale=0.7]{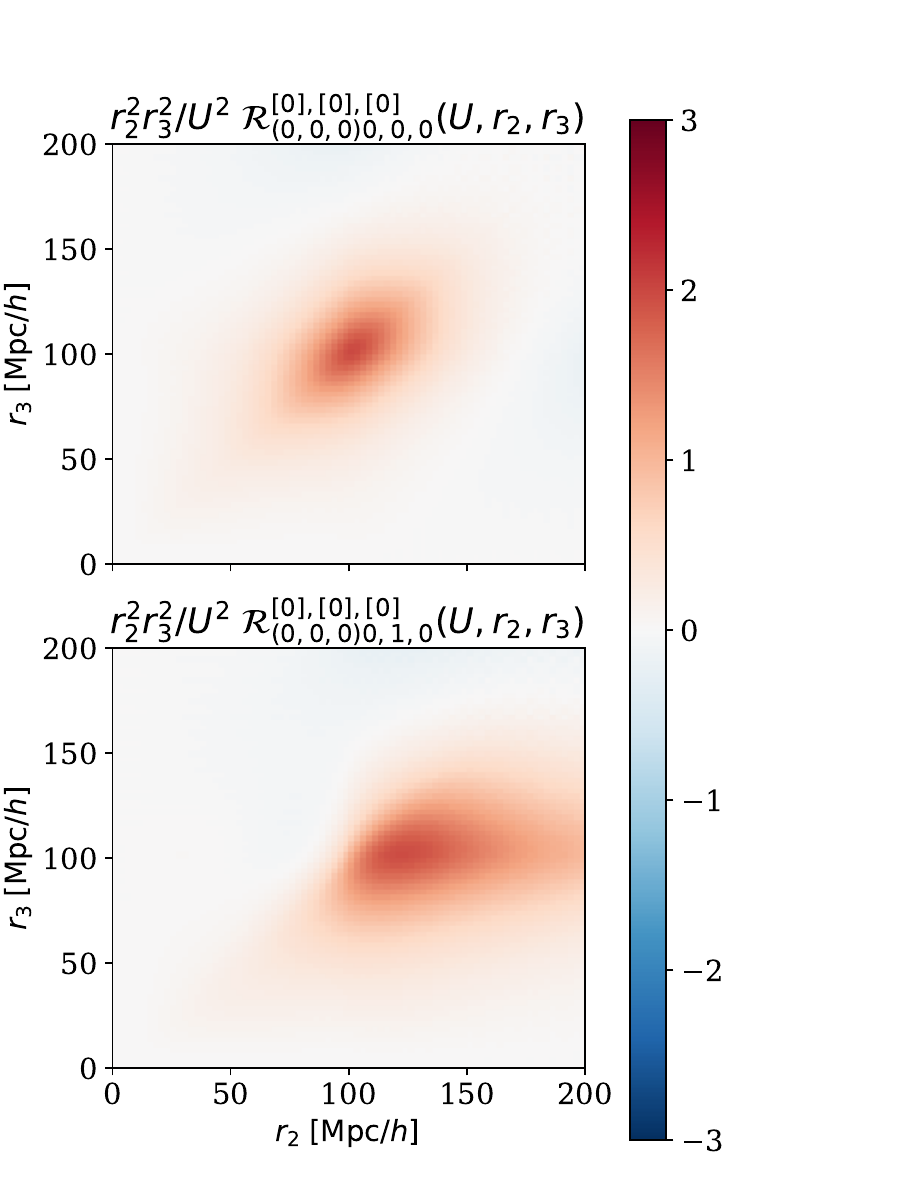}
\caption{Here, we show Eq. (\ref{eq:fancy_R_int}) for fixed $r_1= U \equiv 100 \;\left[{\rm Mpc}/h\right]$, $n'_1=n'_2=n'_3 =0$, $\ell'_1 =\ell'_2 =\ell'_3 =0 $, $L_1 = 0$, $L_2 = \left\{ 0,1\right\}$ and $L_3=0$. We place $\ell'_1 $, $\ell'_2 $ and $\ell'_3 $ in parentheses to identify them as the sBF orders that do not contribute directly to the $r_1,\;r_2$ and $r_3$ variables. The \textit{upper panel} shows the integral for $L_2=0$, while the \textit{lower panel} shows the integral for $L_2=1$. Since the 4PCF can be approximated as the square of the 2PCF on large scales, $(\xi_{0})^{2}(r) \sim (1/r^2)^2$, we have weighted the integral by $r^2r_i^2/u^{4}$, with $u \equiv 10 \;\left[{\rm Mpc}/h\right]$ to take out its fall-off. Both the \textit{upper} and \textit{lower panel} show a blob that appears at about $r_2 = r_3 \simeq U $. We demonstrate analytically in Appendix \ref{Sec: Radial Integrals} that the behavior on the \textit{upper panel} arises because we can approximate $\mathcal{R}_{(0,0,0),0,0,0}^{\left[0\right],\left[0\right],\left[0\right]}$ as a product of two Dirac delta functions dependent on $r_1,r_2$ and $r_3$, while the behavior of the \textit{lower panel} is described by approximating $\mathcal{R}_{(0,0,0),0,1,0}^{\left[0\right],\left[0\right],\left[0\right]}$ as one Dirac delta function but only in the region $r_2>r_1$. The \textit{lower panel} is also not symmetric under exchange of axes (\textit{i.e.}, $r_2 \leftrightarrow r_3$), which results from choosing $L_3 = 0$ for the sBF order corresponding to the variable $r_3$ and $L_2=1$ for the sBF order corresponding to the variable $r_2$. This Figure is the second of three; we show the three figures to show that the choice of $r_1$ moves the location of the blob, as we expect from our analytic results in Appendix \ref{Sec: Radial Integrals}.}
\label{fig:fancy_R_int}
\end{figure}

\begin{figure}[h]
\centering
\includegraphics[scale=0.7]{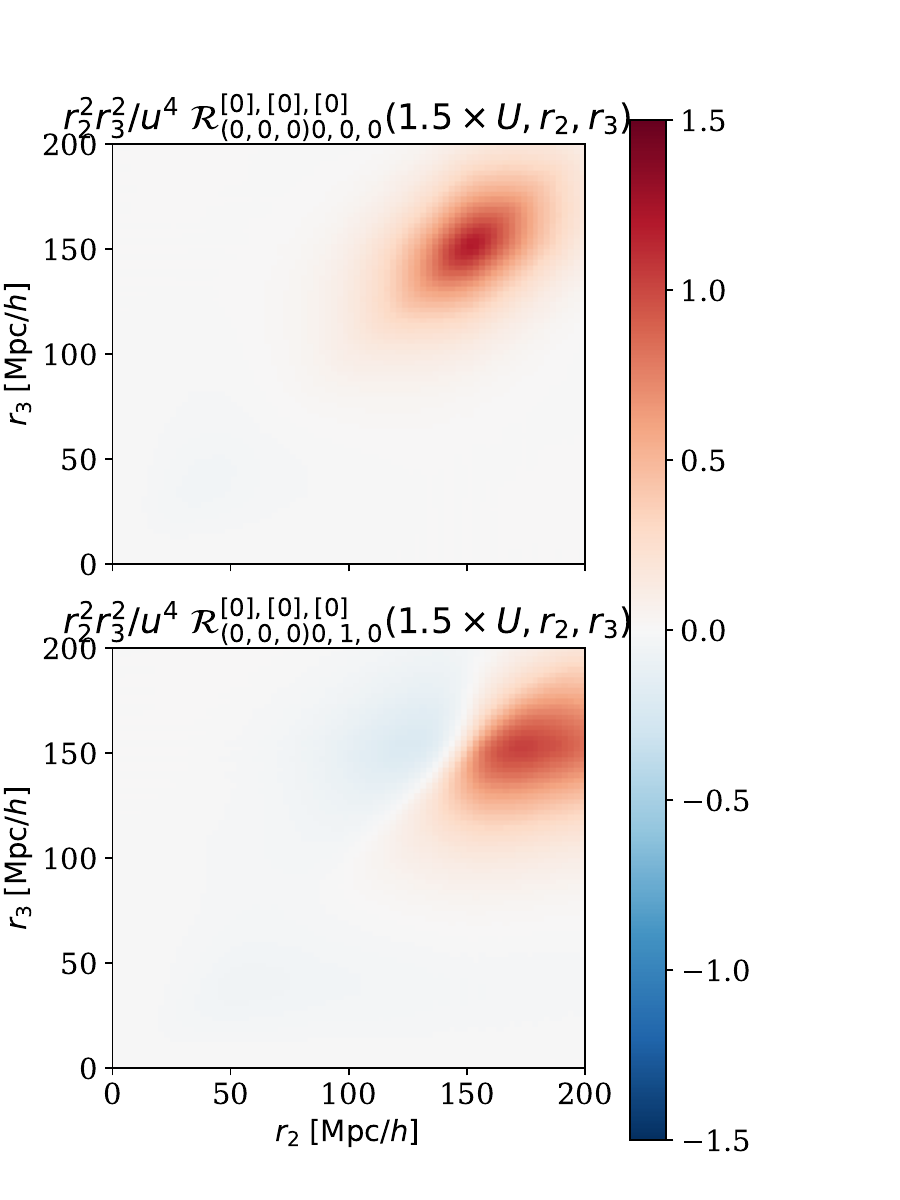}
\caption{Here, we show Eq. (\ref{eq:fancy_R_int}) for fixed $r_1= 1.5 \times U \equiv 150 \;\left[{\rm Mpc}/h\right]$, $n'_1=n'_2=n'_3 =0$, $\ell'_1 =\ell'_2 =\ell'_3 =0 $, $L_1 = 0$, $L_2 = \left\{ 0,1\right\}$ and $L_3=0$. We place $\ell'_1 $, $\ell'_2 $ and $\ell'_3 $ in parentheses to identify them as the sBF orders that do not contribute directly to the $r_1,\;r_2$ and $r_3$ variables. The \textit{upper panel} shows the integral for $L_2=0$, while the \textit{lower panel} shows the integral for $L_2=1$. Since the 4PCF can be approximated as the square of the 2PCF on large scales, $(\xi_{0})^{2}(r) \sim (1/r^2)^2$, we have weighted the integral by $r^2r_i^2/u^{4}$, with $u \equiv 10 \;\left[{\rm Mpc}/h\right]$ to take out its fall-off. Both the \textit{upper} and \textit{lower panel} show a blob that appears at about $r_2 = r_3 \simeq 1.5 \times U $. We demonstrate analytically in Appendix \ref{Sec: Radial Integrals} that the behavior on the \textit{upper panel} arises because we can approximate $\mathcal{R}_{(0,0,0),0,0,0}^{\left[0\right],\left[0\right],\left[0\right]}$ as a product of two Dirac delta functions dependent on $r_1,r_2$ and $r_3$, while the behavior of the \textit{lower panel} is described by approximating $\mathcal{R}_{(0,0,0),0,1,0}^{\left[0\right],\left[0\right],\left[0\right]}$ as one Dirac delta function but only in the region $r_2>r_1$. The \textit{lower panel} is also not symmetric under the exchange of axes (\textit{i.e.}, $r_2 \leftrightarrow r_3$), which results from choosing $L_3 = 0$ for the sBF order corresponding to the variable $r_3$ and $L_2=1$ for the sBF order corresponding to the variable $r_2$. This Figure is the third of three; we show the three figures to show that the choice of $r_1$ moves the location of the blob, as we expect from our analytic results in Appendix \ref{Sec: Radial Integrals}.}
\label{fig:fancy_R_int_up}
\end{figure}

Then, taking the inverse Fourier transforms and including the power spectrum, we find: 

\begin{empheq}[box=\widefbox]{align}\label{eq:R_9,0,T3}
&R_{9,(3);\left[0\right],\left[n_{12}\right], \left[n_{13}\right], \left[n_{23}\right] }^{ \left[n_1\right], \left[n_2\right], \left[n_3\right]}(\mathbf{r}_1; \mathbf{r}_2, \mathbf{r}_3) \rightarrow (4\pi)^{11}\sum_{a_1,a_2,a_3} \sum_{j_1,j_2,j_3}\sum_{j_{12},j_{13},j_{23}}\sum_{j,n}   \nonumber \\
& \qquad \qquad \qquad \times \sum_{\ell'_1,\ell'_2,\ell'_3} \sum_{L_1,L_2,L_3} c_{j,n}^{(W)} c_{j_{12}}^{(n_{12})}c_{j_{13}}^{(n_{13})}c_{j_{23}}^{(n_{23})} 
 t_{a_1,a_2,a_3}^{(2)} \mathcal{C}_{j_1,j_2,j_3}^{n_1,n_2,n_3}  \nonumber \\
& \qquad \qquad \qquad \times 
\mathcal{C}_{\ell'_1,\ell'_2,\ell'_3} C_{L_1,L_2,L_3} \mathcal{G}_{L_1,L_2,L_3}\Upsilon_{L_1,L_2,L_3} \nonumber \\
& \qquad \qquad  \qquad \times \mathcal{P}_{L_1,L_2,L_3} (\mathbf{\widehat{r}}_1,\mathbf{\widehat{r}}_2,\mathbf{\widehat{r}}_3) \mathcal{R}_{(\ell'_1,\ell'_2,\ell'_3),L_1,L_2,L_3}^{\left[n'_1+a_1\right],\left[n'_2+n+a_2\right],\left[n'_3-n+a_3\right]}(r_1,r_2,r_3),
\end{empheq}
with the sum over $L_1$ running from 0 to $j_1+j_{12}+j_{13}+\ell'_1$, over $L_2$ running from 0 to $j_2+j_{12}+j_{23}+j+ \ell'_2$ and over $L_3$ running from 0 to $j_3+j_{13}+j_{23}+j+ \ell'_3$. The coefficients are given in Table \ref{table:1} and the radial integral defined as: 

\begin{align}\label{eq:fancy_R_int}
&\mathcal{R}_{(\ell'_1,\ell'_2,\ell'_3),L_1,L_2,L_3}^{\left[n'_1\right],\left[n'_2\right],\left[n'_3\right]}(r_1,r_2,r_3) \equiv \int_{0}^{\infty} dr'\; r' \;f_{\ell'_1,L_1}^{\left[n'_1\right]}(r',r_1)\nonumber \\ 
& \qquad \qquad \qquad \qquad \qquad \qquad \qquad \qquad \qquad \times f_{\ell'_2,L_2}^{\left[n'_2\right]}(r',r_2) \; f_{\ell'_3,L_3}^{\left[n'_3\right]}(r',r_3),
\end{align}
for which the $f_{\ell,L}$'s are defined in Eq. (\ref{eq:2_Sph.Bess_int}). Parentheses have been used around some of the orders of the spherical Bessel to indicate that these are being integrated out and do not directly affect the resulting variables $r_1$, $r_2$ and $r_3$. If we choose a power-law power spectrum, $P(k_i) \sim 1/k_i$, we find analytic solutions for some values of $n'_1$, $n'_2$, $n'_3$, $\ell'_1$, $\ell'_2$, $\ell'_3$, $L_1$, $L_2$ and $L_3$ for Eq. (\ref{eq:fancy_R_int}) in \cite{Indefinite_Integrals_SBF, Sph_Bessel_Integral_kiersten, Sph_Bessel_int_Rami}. Therefore, in Appendix \ref{Sec: Radial Integrals}, we have analyzed several special cases to explain the behavior shown in Figures \ref{fig:fancy_R_int_down}-\ref{fig:fancy_R_int_up}, where we display Eq. (\ref{eq:fancy_R_int}) with the true power spectrum. As can be understood from these three figures, and Eq. (\ref{eq:Fancy_R0_analytical}) and Eq. (\ref{eq:Fancy_R1_analytical}), the location of the blob on the figures depends on  our choice of $r_1$.  

\clearpage

\section{Analysis $\&$ Inverse Fourier Transform of $T_{2211}$} \label{Section T2}
\qquad We now continue our analysis with the trispectrum term $T_{2211}$, given in Eq. (\ref{eq:T2}). We see that in between parentheses we have two terms with the same structure but different arguments. We will only analyze the first term, since the second one will then be given by replacing the arguments. This leads us to split $T_{2211}$ into eight different terms which capture it fully.  We denote the resulting equations with $R_{i,(2)}$, with the 2 indicating the term is found from the $T_{2211}$ term and $i$ indicating which term in the below list was evaluated: 

\begin{enumerate}
    \item $W^{(2)}\left(\mathbf{k}_1, -\mathbf{k}_{13}\right) W^{(2)}\left(\mathbf{k}_2, \mathbf{k}_{13}\right)\mu_{1}^{n_1}\mu_{2}^{n_2}\mu_{3}^{n_3}\mu_{123}^{n_{123}} $  \;\;\; \(\rightarrow\) \; $\S$\ref{sec:T4.1.1}
    \item $S^{(2)}\left(\mathbf{k}_{i}, \pm \mathbf{k}_{13}\right) W^{(2)}\left(\mathbf{k}_{j}, \pm \mathbf{k}_{13}\right) \mu_{1}^{n_1}\mu_{2}^{n_2}\mu_{3}^{n_3}\mu_{123}^{n_{123}}$ \quad \(\rightarrow\) \; $\S$\ref{sec:T4.1.2}
    \item $W^{(2)}\left(\mathbf{k}_{i}, \pm \mathbf{k}_{13}\right) \gamma^{(2)}\left(\mathbf{k}_{j}, \pm \mathbf{k}_{13}\right)\mu_{1}^{n_1}\mu_{2}^{n_2}\mu_{3}^{n_3}\mu_{123}^{n_{123}} $ \quad \(\rightarrow\) \; $\S$\ref{sec:T4.1.3}
    \item $S^{(2)}\left(\mathbf{k}_1, -\mathbf{k}_{13}\right) S^{(2)}\left(\mathbf{k}_2, \mathbf{k}_{13}\right)\mu_{1}^{n_1}\mu_{2}^{n_2} $ \qquad \quad \qquad \;\;\(\rightarrow\) \; $\S$\ref{sec:T4.1.4}
    \item $S^{(2)}\left(\mathbf{k}_{i}, \pm \mathbf{k}_{13}\right) \gamma^{(2)}\left(\mathbf{k}_{j}, \pm \mathbf{k}_{13}\right)\mu_{1}^{n_1}\mu_{2}^{n_2}$ \qquad \qquad \quad \(\rightarrow\) \; $\S$\ref{sec:T4.1.5}
    \item $\gamma^{(2)}\left(\mathbf{k}_1, -\mathbf{k}_{13}\right) \gamma^{(2)}\left(\mathbf{k}_2, \mathbf{k}_{13}\right) \mu_{1}^{n_1}\mu_{2}^{n_2}$. \quad \;\;\;\;\;\;\; \qquad \(\rightarrow\) \; $\S$\ref{sec:T4.1.6}
    \item $W^{(2)}\left(\mathbf{k}_{i}, \pm \mathbf{k}_{13}\right) \mu_{1}^{n_1}\mu_{2}^{n_2}\mu_{3}^{n_3} \mu_{123}^{n_{123}}$ \qquad \qquad \qquad \quad \;\(\rightarrow\) \; $\S$\ref{sec:T4.1.7}
    \item $S^{(2)}\left(\mathbf{k}_{i}, \pm \mathbf{k}_{13}\right) \mu_{1}^{n_1}\mu_{2}^{n_2} $ \qquad \qquad \qquad \qquad \qquad \;\;\;\;\;\(\rightarrow\) \; $\S$\ref{sec:T4.1.8}
\end{enumerate}
Some of these terms have the argument $(\mathbf{k}_i,\pm \mathbf{k}_{13})$, for $i = \left \{ 1,2\right\}$ such that when $i=1$ we have $(\mathbf{k}_1,-\mathbf{k}_{13})$ and when $i=2$ we have $(\mathbf{k}_2, \mathbf{k}_{13})$; $\mathbf{k}_{13} \equiv \mathbf{k}_1 + \mathbf{k}_3$.

\begin{table} [h!]
\centering
\begin{tabular}{ |p{3cm}|p{3cm}| p{7cm}  |}
 \hline
 \multicolumn{3}{|c|}{Table of Coefficients} \\
 \hline
 Coefficient & Equation & Definition \\
 \hline
 $C_{L_1,L_2,L_3}$ & \ref{eq:Constant_from_exp} & Plane-wave expansion coefficient. \\
 & & \\
 $w$  &  \ref{eq:W2W2}  & Coefficient indicating argument of $W^{(2)}$. \\
 & & \\
 $C_{s}$ & \ref{eq:gen_S2} & Coefficient indicating argument of $S^{(2)}$. \\
  & &\\
 $g$ & \ref{eq:gamma_gen} & Coefficient indicating argument of $\gamma^{(2)}$. \\
 & & \\
 $s_{\ell}^{(\rm I)}$ & \ref{eq:k23} & Isotropic basis function coefficient. \\
 & & \\
 $\mathcal{C}_{\ell'_1,\ell'_2,\ell'_3}$ & \ref{eq:k123} & Decoupling $\mathbf{k}_{123}$ coefficient.\\
 &&\\
 $c_{j}^{(n)}$ & \ref{eq:dot_to_iso_eq} & Coefficient for the expansion of a dot product into the isotropic basis function. \\
  & &\\
 $\Upsilon$ & \ref{eq:Upsilon_splitting} & Coefficient from the splitting of vector and
wave-vector isotropic basis function. \\ 
  & &\\
  $\mathcal{C}_{j_1,j_2,j_3}^{n_1,n_2,n_3}$ & \ref{eq:C_cons} & Averaging over line of sight coefficient. \\
  & &\\
 $ \mathcal{G}_{\ell_1,\ell_2,\ell_3} $ &   \ref{eq:Product_of_2_iso}, \ref{eq:Product_of_3_iso} and \ref{eq:Product_of_n_iso} & Modified Gaunt integral from reduction of
products of isotropic basis functions. \\ 
 \hline
\end{tabular} 
\caption{Table with the coefficients resulting from the calculations of the inverse Fourier transform of $T_{2211}$ (first column), the equations that define these coefficients (second column), and the origin/meaning (third column).}
\label{table:2}
\end{table}

\subsection{Second-Order Kernel}
\subsubsection{Product of Two Second-Order Kernels Weighted by $\mu_{123}^{n_{123}}$}\label{sec:T4.1.1}
\qquad Given that the argument of $W^{(2)}$ in term 1 contains a sum of two wave vectors we must analyze its structure again. From Eq. (\ref{eq:W2_definition}) we have:

\begin{equation}
\begin{split}
W^{(2)}\left(\mathbf{k}_1, -\mathbf{k}_{13}\right) = \overline{c}_{0,0} \mathcal{L}_0(\mathbf{\widehat{k}}_1 \cdot \mathbf{\widehat{k}}_{13}) - \overline{c}_{1,1} \mathcal{L}_1(\mathbf{\widehat{k}}_1 \cdot \mathbf{\widehat{k}}_{13}) \left(\frac{k_1}{k_{13}} + \frac{k_{13}}{k_1}\right) + \overline{c}_{2,0} \mathcal{L}_2(\mathbf{\widehat{k}}_1 \cdot \mathbf{\widehat{k}}_{13}),
\end{split}
\end{equation}
and since: 

\begin{equation}
\begin{split}
\mathbf{\widehat{k}}_1 \cdot \mathbf{\widehat{k}}_{13} = \frac{\mathbf{\widehat{k}}_1 \cdot (\mathbf{k}_1+\mathbf{k}_3)}{k_{13}} = \frac{k_1}{k_{13}} + \frac{k_3 (\mathbf{\widehat{k}}_1 \cdot \mathbf{\widehat{k}}_3)}{k_{13}},
\end{split}
\end{equation}
we can express $W^{(2)}(\mathbf{k}_1,-\mathbf{k}_{13})$ in general as: 

\begin{equation}
\begin{split}
W^{(2)}\left(\mathbf{k}_1, -\mathbf{k}_{13}\right) \rightarrow \frac{ k_{1}^{n'_1}k_{3}^{n'_3}}{k_{13}^{n'_{13}}}(\mathbf{\widehat{k}}_1 \cdot \mathbf{\widehat{k}}_{3})^{n_{13}},
\end{split}
\end{equation}
with $n'_{13} = \left\{ 0,2 \right\}$. Following the same logic for the argument of $W^{(2)}\left(\mathbf{k}_2, \mathbf{k}_{13}\right)$, we find: 

\begin{equation}
\begin{split}
W^{(2)}\left(\mathbf{k}_2, \mathbf{k}_{13}\right) \rightarrow \frac{k_{1}^{n''_1}k_{2}^{n''_2}k_{3}^{n''_3}}{k_{13}^{n''_{13}}}(\mathbf{\widehat{k}}_1 \cdot \mathbf{\widehat{k}}_{2})^{n'''_{12}}(\mathbf{\widehat{k}}_2 \cdot \mathbf{\widehat{k}}_{3})^{n'''_{23}}(\mathbf{\widehat{k}}_1 \cdot \mathbf{\widehat{k}}_{3})^{n'''_{13}}. 
\end{split}
\end{equation}

This implies that for term 1, by multiplying both $W^{(2)}$ kernels, we obtain: 

\begin{align}\label{eq:W2W2}
&W^{(2)}\left(\mathbf{k}_1, -\mathbf{k}_{13}\right)W^{(2)}\left(\mathbf{k}_2, \mathbf{k}_{13}\right) \nonumber \\
&\rightarrow \frac{w^{n'_1,n'_2,n'_3,n'_{13}}_{n_{12},n_{13},n_{23}} k_{1}^{n'_1}k_{2}^{n'_2}k_{3}^{n'_3}}{k_{13}^{n'_{13}}}(\mathbf{\widehat{k}}_1 \cdot \mathbf{\widehat{k}}_{2})^{n_{12}}(\mathbf{\widehat{k}}_2 \cdot \mathbf{\widehat{k}}_{3})^{n_{23}}(\mathbf{\widehat{k}}_1 \cdot \mathbf{\widehat{k}}_{3})^{n_{13}},
\end{align}
where $w$ is a constant to indicate which term from $W^{(2)}$ we are referring to; we will suppress the superscripts of $w$ for the rest of this work. Also, we have combined all powers into a single one, and  $n'_{13} =\left\{ 0, 2, 4 \right\}$. Therefore for term 1 once we expand $\mu_{123}^{n_{123}}$ as in Eq. (\ref{eq:mu_123}):

\begin{align} \label{eq:W2W2_mu's}
&W^{(2)}\left(\mathbf{k}_1, -\mathbf{k}_{13}\right) W^{(2)}\left(\mathbf{k}_2, \mathbf{k}_{13}\right)\left<\mu_{1}^{n_1}\mu_{2}^{n_2}\mu_{3}^{n_3}\mu_{123}^{n_{123}}\right>_{\rm l.o.s} \nonumber \\ 
& \rightarrow (4\pi)^6 \sum_{j_1,j_2,j_3}\sum_{j_{12},j_{13},j_{23}} w\;\mathcal{C}_{j_1,j_2,j_3}^{n_1,n_2,n_3} c_{j_{12}}^{(n_{12})}c_{j_{13}}^{(n_{13})}c_{j_{23}}^{(n_{23})} \mathcal{P}_{j_1,j_2,j_3}(\mathbf{\widehat{k}}_1,\mathbf{\widehat{k}}_2,\mathbf{\widehat{k}}_3) \mathcal{P}_{j_{12}}(\mathbf{\widehat{k}}_1,\mathbf{\widehat{k}}_2) \nonumber \\ 
&\qquad\qquad \qquad \qquad \qquad \times \mathcal{P}_{j_{13}}(\mathbf{\widehat{k}}_1,\mathbf{\widehat{k}}_3)\mathcal{P}_{j_{23}}(\mathbf{\widehat{k}}_2,\mathbf{\widehat{k}}_3) \frac{k_{1}^{n'_1}k_{2}^{n'_2}k_{3}^{n'_3}}{k_{13}^{n'_{13}}k_{123}^{n_{123}}}, 
\end{align}
where we have expanded all the dot products into the isotropic basis functions following Eq. (\ref{eq:dot_to_iso_eq}).

\paragraph{Inverse $k_{13}^{n'_{13}}$ and $k_{123}^{n_{123}}$; $n'_{13}\neq0$ and $n_{123}\neq0$} \mbox{}\\

\begin{figure}[h]
\centering
\includegraphics[scale=0.6]{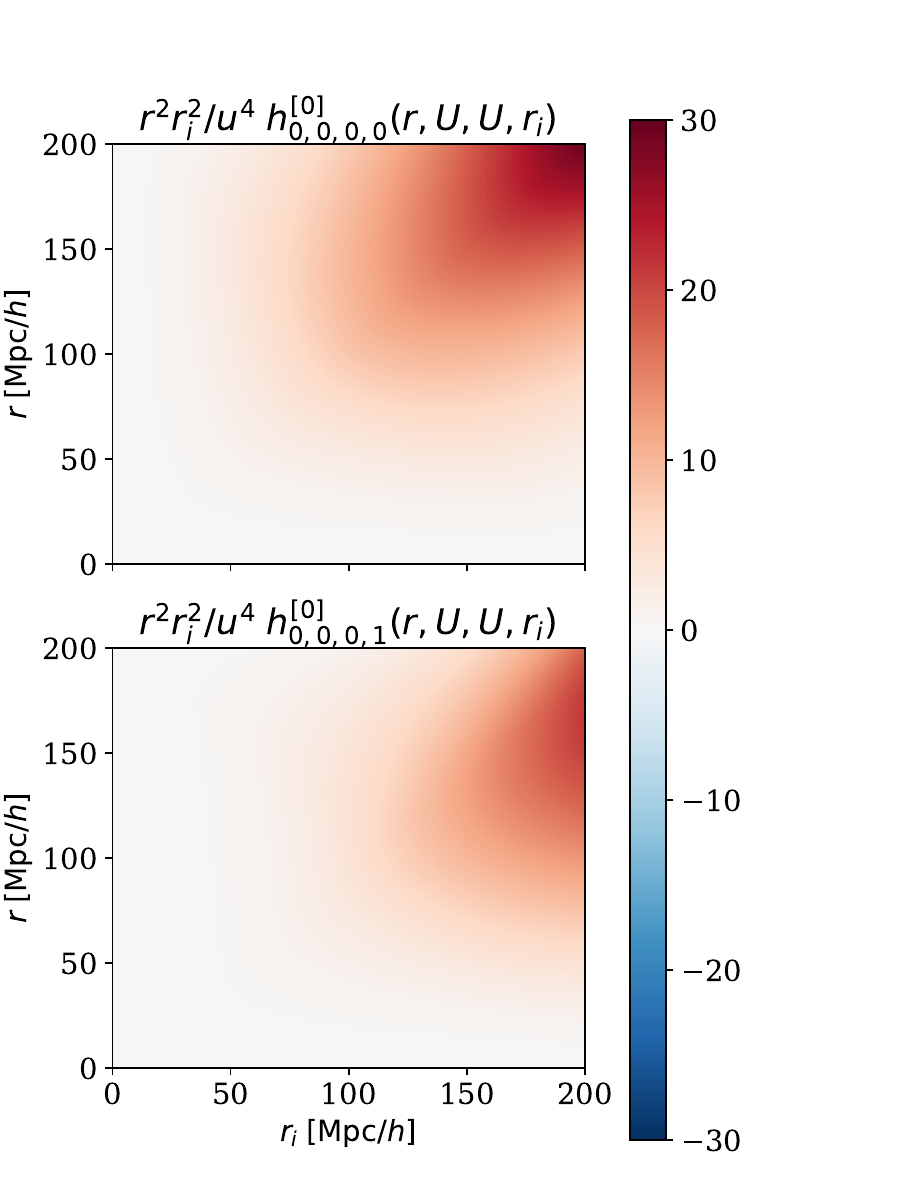}
\caption{Here, we show Eq. (\ref{eq:4_Sph.Bessel_int}) for fixed $r' = s = U \equiv 100 \;\left[{\rm Mpc}/h\right]$, $n'=0$, $\ell =\ell' =\ell'' =0 $ and $L = \left\{ 0,1\right\}$. The \textit{upper panel} shows the integral for $L=0$, while the \textit{lower panel} shows the integral for $L=1$. Since the 4PCF can be approximated as the square of the 2PCF on large scales, $(\xi_{0})^{2}(r) \sim (1/r^2)^2$, we have weighted the integral by $r^2r_i^2/u^{4}$ to take out its fall-off. Both the \textit{upper} and \textit{lower panel} show a blob that appears at about $r_2 = r_3 \simeq U $ and increases monotonically. We demonstrate analytically in Appendix \ref{Sec: Radial Integrals} that the behavior on the \textit{upper panel} arises from the product of two "top-hat" functions. The behavior of the \textit{lower panel} arises from the product of a "top hat" with a Heaviside times a function defined in Eq. (\ref{eq:I_function_from_g001}). The \textit{lower panel} is also not symmetric under the exchange of axes (\textit{i.e.}, $r \leftrightarrow r_i$), which results from choosing $\ell = 0$ for the sBF order corresponding to the variable $r$ and $L=1$ for the sBF order corresponding to the variable $r_2$.}
\label{fig:h_int}
\end{figure}

\begin{figure}[h]
\centering
\includegraphics[scale=0.7]{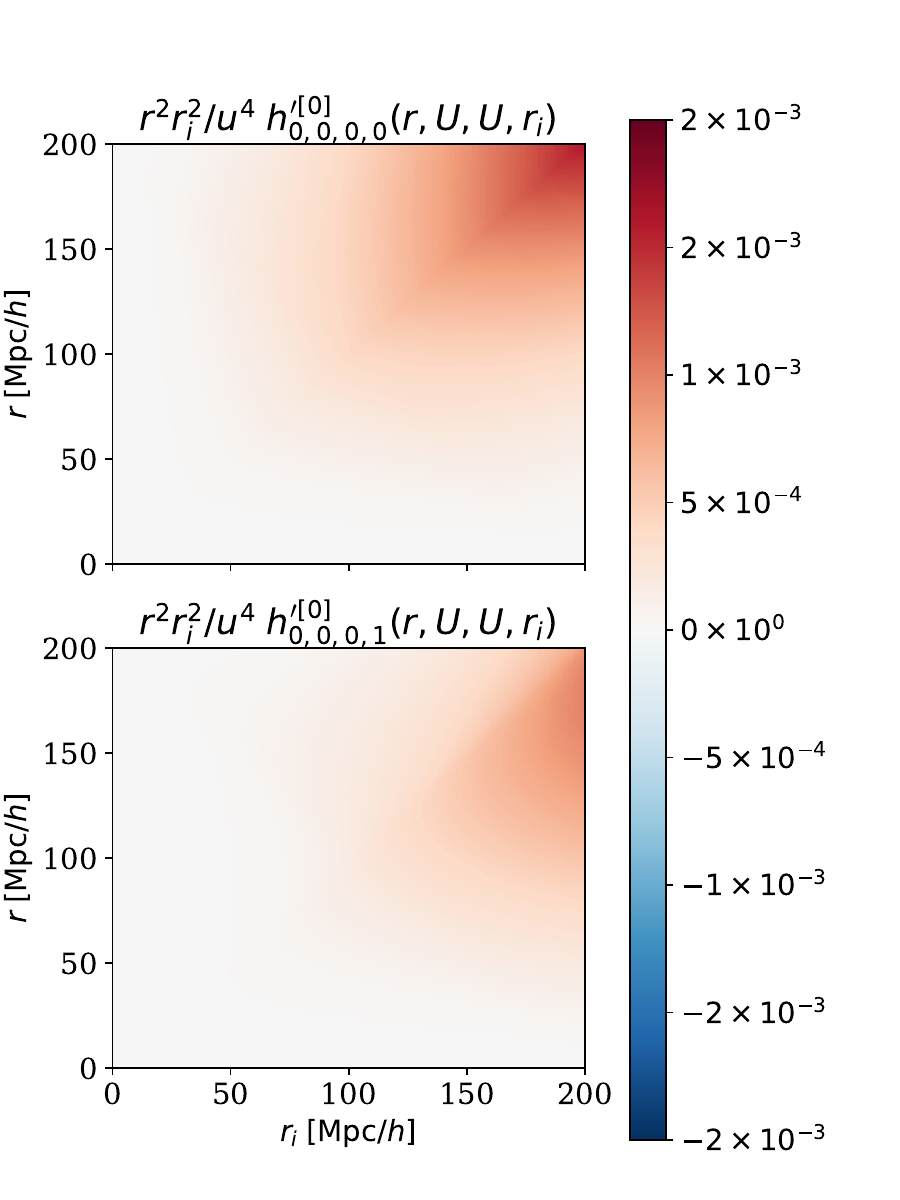}
\caption{Here, we show Eq. (\ref{eq:prime_4_sph.Bessel_int})  for fixed $r' = s = U \equiv 100 \;\left[{\rm Mpc}/h\right]$, $n'=0$, $\ell =\ell' =\ell'' =0 $ and $L = \left\{ 0,1\right\}$. The \textit{upper panel} shows the integral for $L=0$, while the \textit{lower panel} shows the integral for $L=1$. Since the 4PCF can be approximated as the square of the 2PCF on large scales, $(\xi_{0})^{2}(r) \sim (1/r^2)^2$, we have weighted the integral by $r^2r_i^2/u^{4}$ to take out its fall-off. Both the \textit{upper} and \textit{lower panel} show a blob that appears at about $r_2 = r_3 \simeq U $ and increases monotonically. We demonstrate analytically in Appendix \ref{Sec: Radial Integrals} that the behavior on the \textit{upper panel} arises from the product of two "top-hat" functions. The behavior of the \textit{lower panel} arises from a product of a "top hat" with a Heaviside times a function defined in Eq. (\ref{eq:I_function_from_g001}). Evaluation using \texttt{Mathematica} from our analytic results is in Figures \ref{fig:h0000 analytical} and \ref{fig:h0001_analytical}. The \textit{lower panel} is also not symmetric under the exchange of axes (\textit{i.e.}, $r \leftrightarrow r_i$), which results from choosing $\ell = 0$ for the sBF order corresponding to the variable $r$ and $L=1$ for the sBF order corresponding to the variable $r_2$. This figure is almost identical to Figure \ref{fig:h_int} as a result of both radial integrals' ($h$ and $h^{'}$) being the same, except that Eq. (\ref{eq:prime_4_sph.Bessel_int}) does not have a power spectrum in its integrand but Eq. (\ref{eq:4_Sph.Bessel_int}) does.}
\label{fig:hprime_int}
\end{figure}

\begin{figure}[h]
\centering
\includegraphics[scale=0.7]{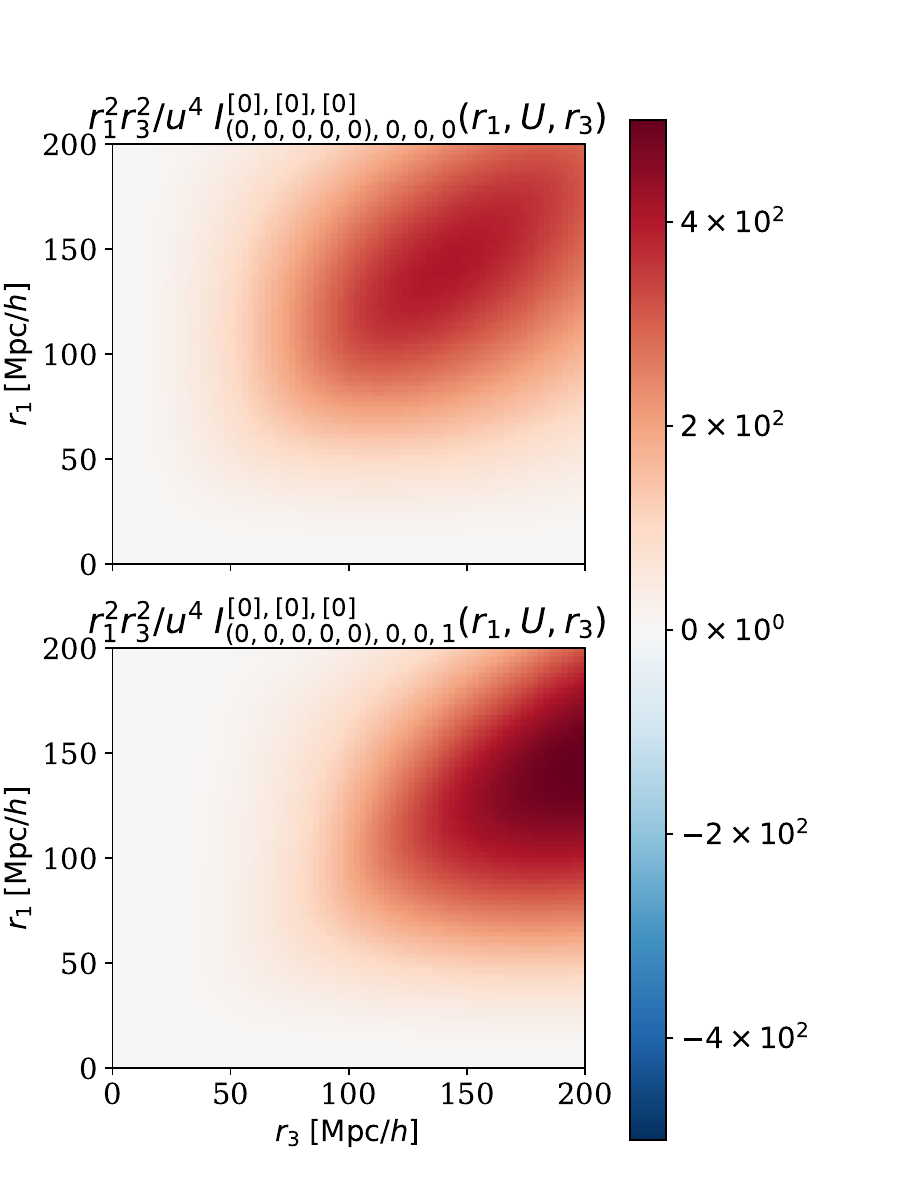}
\caption{2D plot of integral (\ref{eq:R1_radial_n13_n_123}) for fixed $r_2= U \equiv 100 \;\left[{\rm Mpc}/h\right]$, $n'_1=n'_2=n'_3 =0$, $\ell_1 =\ell'_1 =\ell'_2 = \ell'_3 = \ell''_1 = 0 $, $L_1 =0$, $L_2 = 0$ and $L_3= \left\{ 0,1\right\}$. We place $\ell_1$, $\ell'_1 $, $\ell'_2 $ and $\ell'_3 $  and $\ell''_1$ in parentheses to identify them as the sBF orders that do not contribute directly to the $r_1,\;r_2$ and $r_3$ variables. The \textit{upper panel} shows the integral for $L_3=0$, while the \textit{lower panel} shows the integral for $L_3=1$. Since the 4PCF can be approximated as the square of the 2PCF on large scales, $(\xi_{0})^{2}(r) \sim (1/r^2)^2$, we have weighted the integral by $r^2r_i^2/u^{4}$ to take out its fall-off. Both the \textit{upper} and \textit{lower panel} show a blob that appears at about $r_1 = r_3 \simeq U $ and decreases its magnitude slightly as one moves away from its center. We demonstrate analytically in Appendix \ref{Sec: Radial Integrals} that the behavior on both of these panels arises from the product of two "top-hat" functions times $1/r$ for the \textit{upper panel} and a product of a "top hat" with a Heaviside times a function defined in Eq. (\ref{eq:I_function_from_g001}) for the \textit{lower panel}. Evaluation using \texttt{Mathematica} of the \textit{upper panel} from our analytic result is in Figure \ref{fig:R0_analytical}, since we demonstrate in Eq. (\ref{eq:I0_sim_R0}) that $R_{(0,0,0,0),0,0,0}^{\left[0\right], \left[0\right], \left[0\right]} \sim I_{(0,0,0,0,0),0,0,0}^{\left[0\right], \left[0\right], \left[0\right]}$. For the \textit{lower panel} our analytic result analog is in Figure \ref{fig:I001_analytical}. The \textit{lower panel} is also not symmetric under the exchange of axes (\textit{i.e.}, $r_1 \leftrightarrow r_3$), which results from choosing $L_1 = 0$ for the sBF order corresponding to the variable $r_1$ and $L_3=1$ for the sBF order corresponding to the variable $r_3$. }
\label{fig:I_int}
\end{figure}

We evaluate Eq. (\ref{eq:W2W2_mu's}) for $n_{13}\neq0$ and $n_{123}\neq0$, so we continue by decoupling and expanding the terms in the denominator using the results in equations (\ref{eq:k23}) and (\ref{eq:k123}):

\begin{align}
&W^{(2)}\left(\mathbf{k}_1, -\mathbf{k}_{13}\right) W^{(2)}\left(\mathbf{k}_2, \mathbf{k}_{13}\right)\left<\mu_{1}^{n_1}\mu_{2}^{n_2}\mu_{3}^{n_3}\mu_{123}^{n_{123}}\right>_{\rm l.o.s} \nonumber \\ 
& \rightarrow (4\pi)^6 \sum_{j_1,j_2,j_3}\sum_{j_{12},j_{13},j_{23}} \sum_{l_1} \sum_{\ell'_1,\ell'_2,\ell'_3}  \frac{(256 \pi^{3} \; i^{n'_{13}+n_{123}}) \;w \;s_{\ell_1}^{(\rm I)} \mathcal{C}_{j_1,j_2,j_3}^{n_1,n_2,n_3} c_{j_{12}}^{(n_{12})}c_{j_{13}}^{(n_{13})}c_{j_{23}}^{(n_{23})} \mathcal{C}_{\ell'_1,\ell'_2,\ell'_3}}{\Gamma(n'_{13}-1)\Gamma(n_{123}-1)\left[ (-1)^{n'_{13} -1} + 1 \right]\left[ (-1)^{n_{123} -1} + 1 \right]} \nonumber \\ 
& \qquad \qquad  \times \mathcal{P}_{j_1,j_2,j_3}(\mathbf{\widehat{k}}_1,\mathbf{\widehat{k}}_2,\mathbf{\widehat{k}}_3) \mathcal{P}_{j_{12}}(\mathbf{\widehat{k}}_1,\mathbf{\widehat{k}}_2) \mathcal{P}_{j_{13}}(\mathbf{\widehat{k}}_1,\mathbf{\widehat{k}}_3) \nonumber \\ 
& \qquad \qquad  \times \mathcal{P}_{j_{23}}(\mathbf{\widehat{k}}_2,\mathbf{\widehat{k}}_3) \mathcal{P}_{\ell_2}(\mathbf{\widehat{k}}_2,\mathbf{\widehat{k}}_3) \mathcal{P}_{\ell'_1,\ell'_2,\ell'_3}(\mathbf{\widehat{k}}_1,\mathbf{\widehat{k}}_2,\mathbf{\widehat{k}}_3)  \nonumber \\
&  \qquad \qquad \times k_{1}^{n'_1}k_{2}^{n'_2}k_{3}^{n'_3}  \int_{0}^{\infty} dr' \;r'^{\;n_{123}-1} \; \int_{0}^{\infty}dr\; r^{n'_{13}-1} \;j_{\ell_1}(k_1r) \nonumber \\
&  \qquad \qquad \times j_{\ell_1}(k_3r) j_{\ell'_1}(k_1r')j_{\ell'_2}(k_2r')j_{\ell'_3}(k_3r'). 
\end{align}
Next, we proceed to write the inverse Fourier transform integrals: 

\begin{align}\label{eq:R1_FT_inv_inital}
& R_{1,(2);\left[n’_{1}\right], \left[n’_{2}\right], \left[n’_{3}\right],\left[n'_{13}\right],\left[n_{12}\right], \left[n_{13}\right], \left[n_{23}\right] }^{\left[n_{123}\right], \left[n_{1}\right], \left[n_{2}\right], \left[n_{3}\right]}(\mathbf{r}_1, \mathbf{r}_2, \mathbf{r}_3) \nonumber \\ 
& = {\rm FT}^{-1}\left\{W^{(2)}\left(\mathbf{k}_{1}, - \mathbf{k}_{13}\right) W^{(2)}\left(\mathbf{k}_{2}, \mathbf{k}_{13}\right) \left<\mu_{1}^{n_1}\mu_{2}^{n_2} \mu_{3}^{n_3}\mu_{123}^{n_{123}}\right>_{\rm l.o.s} P(k_1)P(k_2)P(k_{13}) \right\} \nonumber \\ 
&\qquad   \rightarrow \sum_{j_1,j_2,j_3}\sum_{j_{12},j_{23},j_{13}}\sum_{\ell'_1,\ell'_2,\ell'_3}\sum_{\ell_1,\ell''_1}\sum_{J_1,J_2,J_3} \sum_{L_1,L_2,L_3} (4\pi)^{8} (256 \pi^{3})i^{n'_{13}+n_{123}}\nonumber \\ 
& \qquad   \times  \frac{s_{\ell''_1}^{(\rm I)} \;w\; s_{\ell_1}^{(\rm I)}\; \mathcal{C}_{j_1,j_2,j_3}^{n_1,n_2,n_3} c_{j_{12}}^{(n_{12})}c_{j_{13}}^{(n_{13})}c_{j_{23}}^{(n_{23})} \mathcal{C}_{\ell'_1,\ell'_2,\ell'_3} C_{L_1,L_2,L_3} \mathcal{G}_{J_1,J_2,J_3}\Upsilon_{L_1,L_2,L_3}}{\Gamma(n'_{13}-1)\Gamma(n_{123}-1)\left[ (-1)^{n'_{13} -1} + 1 \right]\left[ (-1)^{n_{123} -1} + 1 \right]}  \nonumber \\ 
& \qquad   \times \mathcal{P}_{L_1,L_2,L_3}(\mathbf{\widehat{r}}_1,\mathbf{\widehat{r}}_2,\mathbf{\widehat{r}}_3) \int_{0}^{\infty} dr' \; r'^{n_{123}-1}\int_{0}^{\infty}dr \;r^{n'_{13}-1}\int_{0}^{\infty}ds \; s^2\; \xi_0(s)  \nonumber \\ 
& \qquad   \times \int \frac{d^{3}\mathbf{k}_1\,d^{3}\mathbf{k}_2\,d^{3}\mathbf{k}_3}{(2\pi)^{9}}k_{1}^{n'_1}k_{2}^{n'_2}k_{3}^{n'_3} j_{L_1}(k_1r_1) j_{L_2}(k_2r_2) j_{L_3}(k_3r_3)  \nonumber \\ 
& \qquad    \times j_{\ell_1}(k_1r) j_{\ell_1}(k_3r) j_{\ell'_1}(k_1r')j_{\ell'_2}(k_2r')j_{\ell'_3}(k_3r') j_{\ell''_1}(k_1s)j_{\ell''_1}(k_3s) \nonumber \\ 
& \qquad   \times P(k_1)P(k_2) \mathcal{P}_{J_1,J_2,J_3}(\mathbf{\widehat{k}}_1,\mathbf{\widehat{k}}_2,\mathbf{\widehat{k}}_3) \mathcal{P}_{L_1,L_2,L_3}(\mathbf{\widehat{k}}_1,\mathbf{\widehat{k}}_2,\mathbf{\widehat{k}}_3), 
\end{align}
for which we have decoupled the power spectrum $P(k_{13})$ using the result in Eq. (\ref{eq:Decoupled_P(k13)}) and have combined all seven isotropic basis---excluding the isotropic basis from the exponent, $\mathcal{P}_{L_1,L_2,L_3}(\mathbf{\widehat{k}}_1,\mathbf{\widehat{k}}_2,\mathbf{\widehat{k}}_3)$---into a single one, as in Eq. (\ref{eq:Product_of_n_iso}) for $n=7$. The subscripts $\left[n’_{1}\right], \left[n’_{2}\right], \left[n’_{3}\right], \left[n’_{13} \right],\left[n_{12}\right], \left[n_{13}\right], \left[n_{23}\right]$ in the left-hand side of Eq. (\ref{eq:R1_FT_inv_inital}) are internal dependencies that arises from the structure of the kernels we are evaluating; the other superscripts are the external dependencies from the term we are analyzing. Therefore, using the orthogonality of the isotropic basis we find:

\begin{empheq}[box=\widefbox]{align}\label{eq: R1_T2}
&R_{1,(2);\left[n’_{1}\right], \left[n’_{2}\right], \left[n’_{3}\right], \left[n'_{13}\right],\left[n_{12}\right], \left[n_{13}\right], \left[n_{23}\right] }^{\left[n_{123}\right], \left[n_1\right], \left[n_2\right], \left[n_3\right]}(\mathbf{r}_1, \mathbf{r}_2, \mathbf{r}_3) \nonumber\\ 
& \rightarrow \sum_{j_1,j_2,j_3}\sum_{j_{12},j_{23},j_{13}}\sum_{\ell'_1,\ell'_2,\ell'_3}\sum_{\ell_1,\ell''_1} \sum_{L_1,L_2,L_3} (4\pi)^{8} \nonumber \\ 
&  \qquad \qquad \times (256 \pi^{3})i^{n'_{13}+n_{123}} \;s_{\ell''_1}^{(\rm I)} \;w\; s_{\ell_1}^{(\rm I)}\;\mathcal{C}_{j_1,j_2,j_3}^{n_1,n_2,n_3}\nonumber\\ 
& \qquad \qquad  \times c_{j_{12}}^{(n_{12})}c_{j_{13}}^{(n_{13})}c_{j_{23}}^{(n_{23})} \mathcal{C}_{\ell'_1,\ell'_2,\ell'_3} C_{L_1,L_2,L_3} \mathcal{G}_{L_1,L_2,L_3}\Upsilon_{L_1,L_2,L_3}\nonumber \\ 
& \qquad \qquad  \times  \left\{ \Gamma(n'_{13}-1)\Gamma(n_{123}-1)\right\}^{-1}\nonumber \\ 
& \qquad \qquad  \times  \left\{ \left[ (-1)^{n'_{13} -1} - 1 \right]\left[ (-1)^{n_{123} -1} - 1 \right]\right\}^{-1} \nonumber \\ 
& \qquad \qquad  \times \mathcal{P}_{L_1,L_2,L_3}(\mathbf{\widehat{r}}_1,\mathbf{\widehat{r}}_2,\mathbf{\widehat{r}}_3) I_{(\ell_1,\ell'_1,\ell'_2,\ell'_3,\ell''_1),L_1,L_2,L_3}^{\left[n'_1\right],\left[n'_2\right],\left[n'_3\right]}(r_1,r_2,r_3),
\end{empheq}
with coefficients given in Table \ref{table:2} and the radial integral defined as 

\begin{align}\label{eq:R1_radial_n13_n_123}
&I_{(\ell_1,\ell'_1,\ell'_2,\ell'_3,\ell''_1),L_1,L_2,L_3}^{\left[n'_1\right],\left[n'_2\right],\left[n'_3\right]}(r_1,r_2,r_3) \equiv \int_{0}^{\infty} dr' \;r'^{n_{123}-1} f_{\ell'_2,L_2}^{\left[n'_2\right]}(r',r_2) \int_{0}^{\infty}dr \;r^{n_{13}-1}\int_{0}^{\infty}ds\; s^2 \;\xi_0(s) \nonumber \\ 
& \quad \qquad \qquad \qquad\qquad \qquad \qquad \qquad   \times  h_{\ell_1,\ell'_1,\ell''_1,L_1}^{\left[n'_1\right]} (r,r',s,r_1) h_{\ell_1,\ell'_3,\ell''_1,L_3}^{\prime \left[n'_3\right]}(r,r',s,r_3), 
\end{align}
for which the definition of $f_{\ell,L}$ is given by Eq. (\ref{eq:2_Sph.Bess_int}) and we have defined $h$ as:

\begin{align}\label{eq:4_Sph.Bessel_int}
h_{\ell,\ell',\ell'',L}^{\left[n'\right]} (r,r',s,r_i) = \int_{0}^{\infty}\frac{dk_i}{2\pi^2} \; k_i^{n'+2} j_{\ell}(k_i r) j_{\ell'}(k_i r') j_{\ell''}(k_i s) j_{L}(k_i r_i) P(k_i), 
\end{align}
and $h^{\prime}$ as
\begin{align}\label{eq:prime_4_sph.Bessel_int}
&h_{\ell,\ell',\ell'',L}^{\prime \left[n'\right]}(r,r',s,r_i) = \int_{0}^{\infty}\frac{dk_i}{2\pi^2}\; k_i^{n'+2} j_{\ell}(k_i r) j_{\ell'}(k_i r') j_{\ell''}(k_i s) j_{L}(k_i r_i). 
\end{align}
analytic solutions to equations (\ref{eq:4_Sph.Bessel_int})---with $P(k_i) = 1/k_i$---and (\ref{eq:prime_4_sph.Bessel_int}) are in \cite{Sph_Bessel_int_Rami, Fabrikant} for special cases of $n'$ and $\ell,\ell'\ell'',L$.  In Appendix \ref{Sec: Radial Integrals}, we have made an analysis for a subset of special cases to try to explain the behavior observed in Figures \ref{fig:h_int} -- \ref{fig:I_int} when plotting the above expression. We have used parentheses in the radial integral around some of the orders of the spherical Bessel to indicate that these are being integrated out and do not directly affect the resulting variables $r_1$, $r_2$ and $r_3$.

\clearpage

\paragraph{Inverse $k_{13}^{n'_{13}}$ and $k_{123}^{n_{123}}$; $n'_{13} = 0$ and $n_{123}\neq 0$}\mbox{}\\

To evaluate term 1 in the list at the beginning of this section with $n'_{13} = 0$ and $n_{123}\neq 0$, we insert these values in equation (\ref{eq:R1_FT_inv_inital}), yielding:

\begin{align}
&R_{1,(2);\left[n’_{1}\right], \left[n’_{2}\right], \left[n’_{3}\right], \left[0\right],\left[n_{12}\right], \left[n_{13}\right], \left[n_{23}\right] }^{\left[n_{123}\right], \left[n_1\right], \left[n_2\right], \left[n_3\right]}(\mathbf{r}_1, \mathbf{r}_2, \mathbf{r}_3)\nonumber \\
&= {\rm FT}^{-1}\left\{W^{(2)}\left(\mathbf{k}_{1}, - \mathbf{k}_{13}\right) W^{(2)}\left(\mathbf{k}_{2}, \mathbf{k}_{13}\right) \left<\mu_{1}^{n_1}\mu_{2}^{n_2} \mu_{3}^{n_3}\right>_{\rm l.o.s} P(k_1)P(k_2)P(k_{13}) \right\} \nonumber \\ 
&\qquad \qquad  \rightarrow \sum_{j_1,j_2,j_3}\sum_{j_{12},j_{23},j_{13}}\sum_{\ell'_1,\ell'_2,\ell'_3}\sum_{\ell''_1}\sum_{J_1,J_2,J_3} \sum_{L_1,L_2,L_3} (4\pi)^{15/2} (32^2 \pi) \; i^{n_{123}} \nonumber \\ 
& \qquad \qquad  \times  \frac{s_{\ell''_1}^{(\rm I)} \;w\; \mathcal{C}_{j_1,j_2,j_3}^{n_1,n_2,n_3} c_{j_{12}}^{(n_{12})}c_{j_{13}}^{(n_{13})}c_{j_{23}}^{(n_{23})} \mathcal{C}_{\ell'_1,\ell'_2,\ell'_3} C_{L_1,L_2,L_3} \mathcal{G}_{J_1,J_2,J_3}\Upsilon_{L_1,L_2,L_3}}{\Gamma(n_{123}-1)\left[ (-1)^{n_{123} -1} + 1 \right]}  \nonumber \\ 
& \qquad \qquad  \times \mathcal{P}_{L_1,L_2,L_3}(\mathbf{\widehat{r}}_1,\mathbf{\widehat{r}}_2,\mathbf{\widehat{r}}_3) \int_{0}^{\infty} dr'\; r'^{n_{123}-1}\;\int_{0}^{\infty}ds \;s^2 \;\xi_0(s)  \nonumber \\ 
& \qquad \qquad  \times \int \frac{d^{3}\mathbf{k}_1\,d^{3}\mathbf{k}_2\,d^{3}\mathbf{k}_3}{(2\pi)^{9}}\;k_{1}^{n'_1}k_{2}^{n'_2}k_{3}^{n'_3} j_{L_1}(k_1r_1) j_{L_2}(k_2r_2) j_{L_3}(k_3r_3)  \nonumber \\ 
& \qquad  \qquad  \times j_{\ell'_1}(k_1r')j_{\ell'_2}(k_2r')j_{\ell'_3}(k_3r') j_{\ell''_1}(k_1s)j_{\ell''_1}(k_3s) \nonumber \\ 
& \qquad \qquad  \times P(k_1)P(k_2) \mathcal{P}_{J_1,J_2,J_3}(\mathbf{\widehat{k}}_1,\mathbf{\widehat{k}}_2,\mathbf{\widehat{k}}_3) \mathcal{P}_{L_1,L_2,L_3}(\mathbf{\widehat{k}}_1,\mathbf{\widehat{k}}_2,\mathbf{\widehat{k}}_3). 
\end{align}

\begin{figure}[h]
\centering
\includegraphics[scale=0.65]{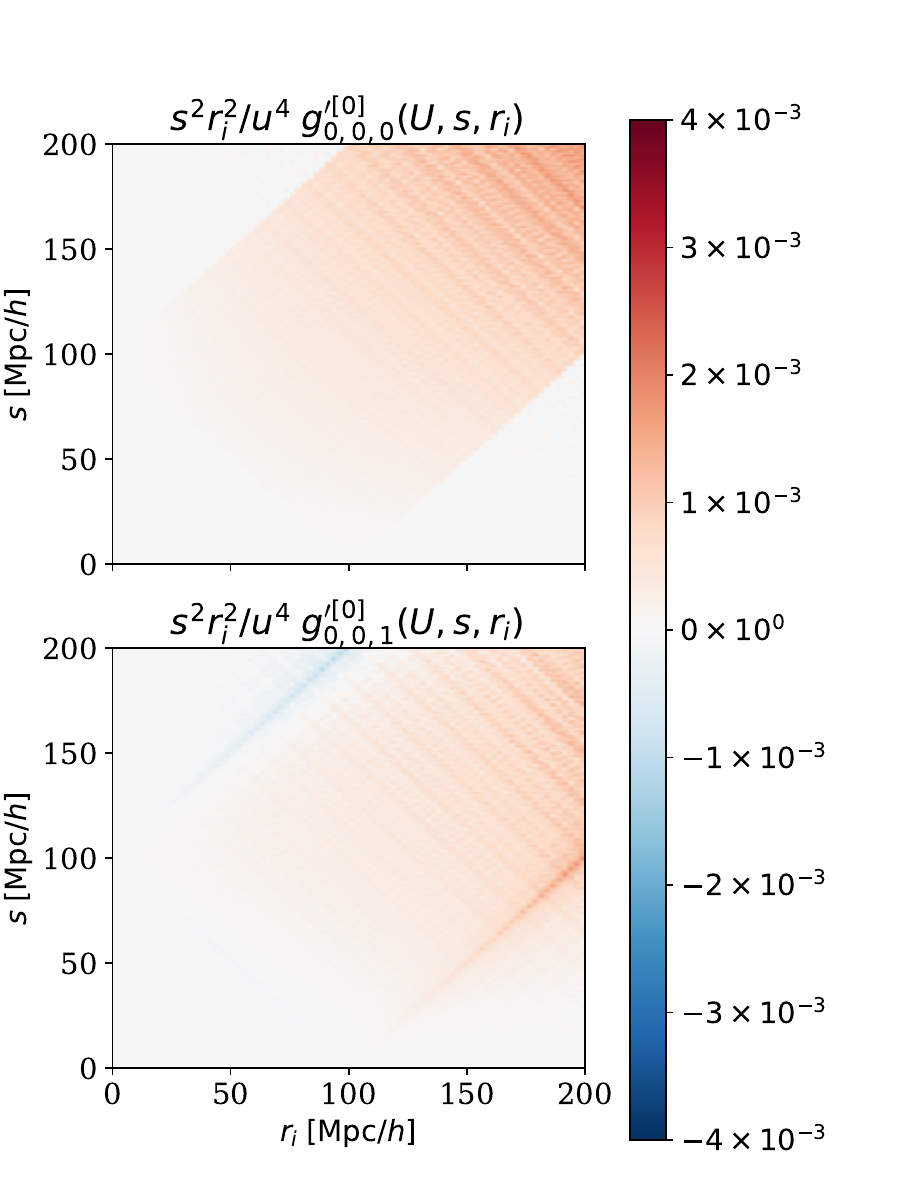}
\caption{Here, we show Eq. (\ref{eq:3_prime_Sph.Bess_int}) for fixed $r'= U \equiv 100 \left[{\rm Mpc}/h\right]$, $n'=0$, $\ell'_3 = \ell''_1 = 0 $ and $L = \left\{ 0,1\right\}$. The \textit{upper panel} shows the integral for $L=0$, while the \textit{lower panel} shows the integral for $L=1$. Since the 4PCF can be approximated as the square of the 2PCF on large scales, $(\xi_{0})^{2}(r) \sim (1/r^2)^2$, we have weighted the integral by $r^2r_i^2/u^{4}$ to take out its fall-off. As with Figure \ref{fig:gint2} we have a rectangular behavior in the \textit{upper} and \textit{lower} panel. We demonstrate analytically in Appendix \ref{Sec: Radial Integrals} that the rectangular behavior on both of these panels arises from the product of two ``top-hat'' functions for the \textit{upper panel} and a product of a ``top hat'' with a Heaviside function for the \textit{lower panel}. Evaluation using  \texttt{Mathematica} from our analytic results is in Figures \ref{fig:g000 analytical}  and \ref{fig:g001 analytical}. The \textit{lower panel} is also not symmetric under exchange of axes (\textit{i.e.}, $r_i \leftrightarrow r$), which results from choosing $\ell = 0$ for the sBF order corresponding to the variable $r$ and $L=1$ for the sBF order corresponding to the variable $r_i$. This plot is almost identical to the plot in Figure \ref{fig:gint2} since the radial integrals Eq. (\ref{eq:3_Sph.Bess_int}) and Eq. (\ref{eq:3_prime_Sph.Bess_int}) are the same except that Eq. (\ref{eq:3_prime_Sph.Bess_int}) does not have a power spectrum in its integrand.}
\label{fig:gprime_int}
\end{figure}

\begin{figure}[h]
\centering
\includegraphics[scale=0.7]{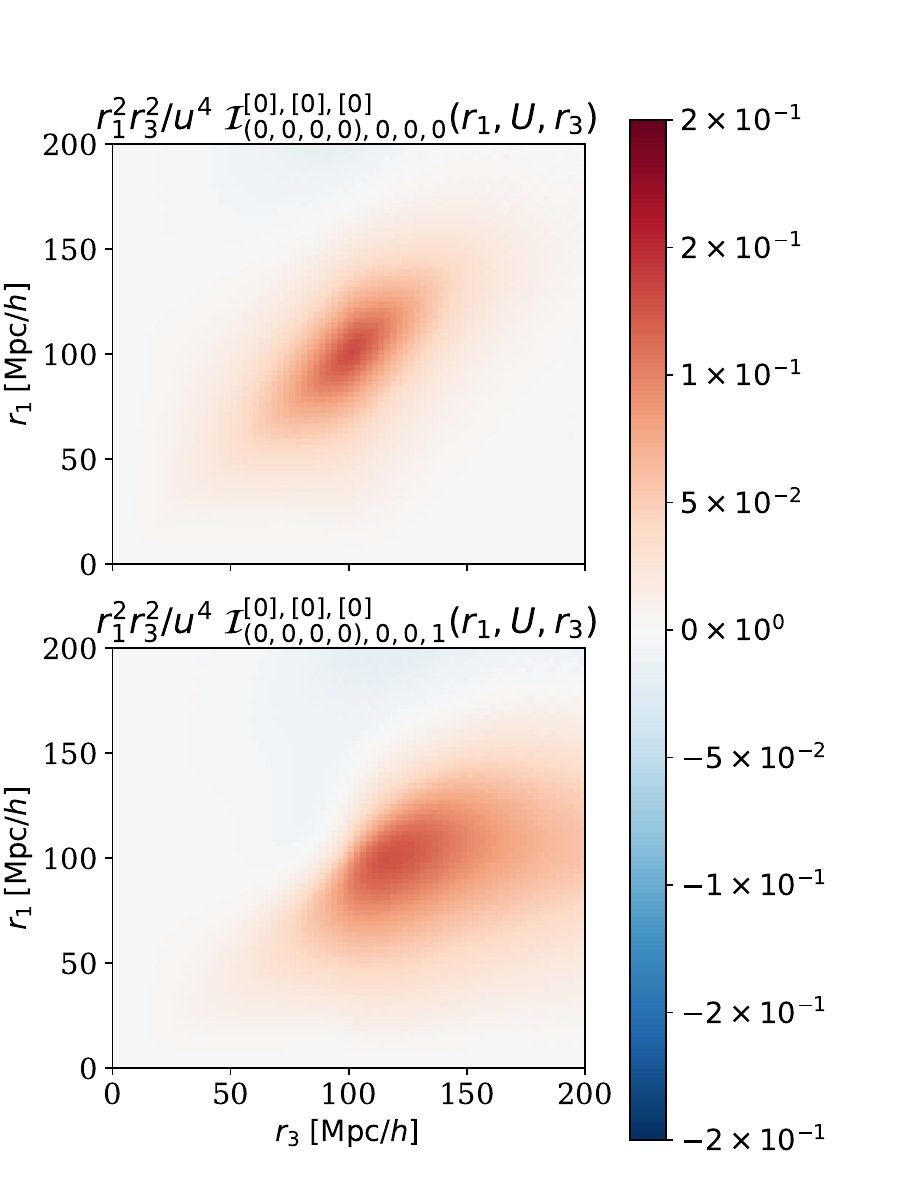}
\caption{Here, we show Eq. (\ref{eq:radial_n123_R1_T2}) for fixed $r_2= U \equiv 100 \;{\rm Mpc}/h$, $n'_1=n'_2=n'_3 =0$, $ \ell'_1 =\ell'_2 = \ell'_3 = \ell''_1 = 0 $, $L_1 = 0$, $L_2 = 0$ and $L_3=\left\{ 0,1\right\}$. We place $\ell'_1 $, $\ell'_2 $, $ \ell'_3 $ and $\ell''_1$ in parentheses to identify them as the sBF orders that do not contribute directly to the $r_1,\;r_2$ and $r_3$ variables. The \textit{upper panel} shows the integral for $L_3=0$, while the \textit{lower panel} shows the integral for $L_3=1$. Since the 4PCF can be approximated as the square of the 2PCF on large scales, $(\xi_{0})^{2}(r) \sim (1/r^2)^2$, we have weighted the integral by $r^2r_i^2/u^{4}$ to take out its fall-off. Both the \textit{upper} and \textit{lower panel} show a blob that appears at about $r_2 = r_3 \simeq U $. We demonstrate analytically in Appendix \ref{Sec: Radial Integrals} that the behavior on the \textit{upper panel} arises because we can approximate $\mathcal{I}_{(0,0,0,0),0,0,0}^{\left[0\right],\left[0\right],\left[0\right]}$ as a product of two Dirac delta functions dependent on $r_1,r_2$ and $r_3$, which is the same approximation we find with $\mathcal{R}_{(0,0,0),0,0,0}$. We find the \textit{lower panel}, $\mathcal{I}_{(0,0,0,0),0,0,1}^{\left[0\right],\left[0\right],\left[0\right]}$, to be described by the same expression as for  $\mathcal{R}_{(0,0,0),0,1,0}$. Our results in Appendix \ref{Sec: Radial Integrals} clearly show why Figure \ref{fig:R_int} and this figure are nearly identical. The \textit{lower panel} is also not symmetric under exchange of axes (\textit{i.e.}, $r_1 \leftrightarrow r_3$), which results from choosing $L_3 = 0$ for the sBF order corresponding to the variable $r_3$ and $L_2=1$ for the sBF order corresponding to the variable $r_2$. In the same manner as in Figures \ref{fig:fancy_R_int_down}-\ref{fig:fancy_R_int_up}, if we change the set value of $r_2$, the blob also moves to reflect this change.}
\label{fig:fancy_I_int}
\end{figure}

Performing the angular integrals we arrive at: 

\begin{empheq}[box=\widefbox]{align}\label{eq: R1_n13=0_T2}
&R_{1,(2);\left[n’_{1}\right], \left[n’_{2}\right], \left[n’_{3}\right], \left[0\right],\left[n_{12}\right], \left[n_{13}\right], \left[n_{23}\right] }^{\left[n_{123}\right], \left[n_1\right], \left[n_2\right], \left[n_3\right]}(\mathbf{r}_1, \mathbf{r}_2, \mathbf{r}_3) \nonumber \\ 
&  \rightarrow \sum_{j_1,j_2,j_3}\sum_{j_{12},j_{23},j_{13}}\sum_{\ell'_1,\ell'_2,\ell'_3} \sum_{\ell''_1} \sum_{L_1,L_2,L_3} (4\pi)^{15/2} (32\pi^2) \; i^{n_{123}}\nonumber \\ 
&  \qquad   \times \frac{s_{\ell''_1}^{(\rm I)} \;w\; \;\mathcal{C}_{j_1,j_2,j_3}^{n_1,n_2,n_3}c_{j_{12}}^{(n_{12})}c_{j_{13}}^{(n_{13})}c_{j_{23}}^{(n_{23})} \mathcal{C}_{\ell'_1,\ell'_2,\ell'_3}}{\Gamma(n_{123}-1)\left[ (-1)^{n_{123} -1} - 1 \right]}\nonumber \\ 
& \qquad   \times  C_{L_1,L_2,L_3} \mathcal{G}_{L_1,L_2,L_3}\Upsilon_{L_1,L_2,L_3}\nonumber \\ 
& \qquad   \times  \mathcal{P}_{L_1,L_2,L_3}(\mathbf{\widehat{r}}_1,\mathbf{\widehat{r}}_2,\mathbf{\widehat{r}}_3) \mathcal{I}_{(\ell'_1,\ell'_2,\ell'_3,\ell''_1),L_1,L_2,L_3}^{\left[n'_1\right],\left[n'_2\right],\left[n'_3\right]}(r_1,r_2,r_3)
\end{empheq}
with coefficients given in Table \ref{table:2} and the radial integral defined as: 

\begin{align}\label{eq:radial_n123_R1_T2}
&\mathcal{I}_{(\ell'_1,\ell'_2,\ell'_3,\ell''_1),L_1,L_2,L_3}^{\left[n'_1\right],\left[n'_2\right],\left[n'_3\right]}(r_1,r_2,r_3) \equiv \int_{0}^{\infty} dr'\; r'^{n_{123}-1} \;f_{\ell'_2,L_2}^{\left[n'_2\right]}(r',r_2) \int_{0}^{\infty}ds\; s^2 \; \xi_0(s)   \nonumber \\
&\quad \qquad \qquad \qquad \qquad \qquad \qquad \quad \times g_{\ell'_1,\ell''_1,L_1}^{\left[n'_1\right]} (r',s,r_1) g_{\ell'_3,\ell''_1,L_3}^{\prime \left[n'_3\right]}(r',s,r_3),
\end{align}
with $f_{\ell,L}$ and $g_{\ell',\ell'',L}$ defined in equations (\ref{eq:2_Sph.Bess_int}) and (\ref{eq:3_Sph.Bess_int}), respectively. In addition, we have defined:

\begin{align}\label{eq:3_prime_Sph.Bess_int}
g_{\ell'_3,\ell''_1,L_3}^{\prime \left[n'_3\right]}(r',s,r_3) \equiv \int_{0}^{\infty} \frac{dk_3}{2\pi^2} \; k_3^{n'_3+2} j_{\ell'_3}(k_3r')j_{\ell''_1}(k_3r)j_{L_3}(k_3r_3) .
\end{align}
analytic solutions to Eq. (\ref{eq:3_prime_Sph.Bess_int}) are in  \cite{DeLaBella3sbf, J.Chellino, Sph_Bessel_Integral_kiersten, Sph_Bessel_int_Rami, Fabrikant}. In Appendix~\ref{Sec: Radial Integrals}, we have made an analysis for a subset of special cases to try to explain the behavior observed in Figure \ref{fig:gprime_int} and Figure \ref{fig:fancy_I_int} when plotting the above expression. Parentheses have been used around some of the orders of the spherical Bessel to indicate that these are being integrated out and do not directly affect the resulting variables $r_1$, $r_2$ and $r_3$.

\clearpage

\paragraph{Inverse $k_{13}^{n'_{13}}$ and $k_{123}^{n_{123}}$; $n'_{13} \neq 0$ and $n_{123} = 0$}\mbox{}\\

To evaluate term 1 in the list at the beginning of this section with $n'_{13} \neq 0$ and $n_{123} = 0$, we insert these values in equation (\ref{eq:R1_FT_inv_inital}), yielding:

\begin{align}
&R_{1,(2);\left[n’_{1}\right], \left[n’_{2}\right], \left[n’_{3}\right], \left[n'_{13}\right],\left[n_{12}\right], \left[n_{13}\right], \left[n_{23}\right] }^{\left[0\right],  \left[n_1\right], \left[n_2\right], \left[n_3\right]}(\mathbf{r}_1, \mathbf{r}_2, \mathbf{r}_3)\nonumber \\
&= {\rm FT}^{-1}\left\{W^{(2)}\left(\mathbf{k}_{1}, - \mathbf{k}_{13}\right) W^{(2)}\left(\mathbf{k}_{2}, \mathbf{k}_{13}\right) \left<\mu_{1}^{n_1}\mu_{2}^{n_2} \mu_{3}^{n_3}\right>_{\rm l.o.s} P(k_1)P(k_2)P(k_{13}) \right\} \nonumber \\ 
&\qquad \qquad  \rightarrow \sum_{j_1,j_2,j_3}\sum_{j_{12},j_{23},j_{13}}\sum_{\ell_1,\ell''_1}\sum_{J_1,J_2,J_3} \sum_{L_1,L_2,L_3} (4\pi)^{8} (8 \pi) \; i^{n'_{13}} \nonumber \\ 
& \qquad \qquad  \times  \frac{s_{\ell''_1}^{(\rm I)} \;cs_{\ell_1}^{(\rm I)}\;w\; \mathcal{C}_{j_1,j_2,j_3}^{n_1,n_2,n_3} c_{j_{12}}^{(n_{12})}c_{j_{13}}^{(n_{13})}c_{j_{23}}^{(n_{23})} C_{L_1,L_2,L_3} \mathcal{G}_{J_1,J_2,J_3}\Upsilon_{L_1,L_2,L_3}}{\Gamma(n'_{13}-1)\left[ (-1)^{n'_{13} -1} + 1 \right]}  \nonumber \\ 
& \qquad \qquad  \times \mathcal{P}_{L_1,L_2,L_3}(\mathbf{\widehat{r}}_1,\mathbf{\widehat{r}}_2,\mathbf{\widehat{r}}_3) \int_{0}^{\infty} dr \; r^{n'_{13}-1}\;\int_{0}^{\infty}ds\;s^2 \; \xi_0(s)  \nonumber \\ 
& \qquad \qquad  \times \int \frac{d^{3}\mathbf{k}_1\,d^{3}\mathbf{k}_2\,d^{3}\mathbf{k}_3}{(2\pi)^{9}}\;k_{1}^{n'_1}k_{2}^{n'_2}k_{3}^{n'_3} j_{L_1}(k_1r_1) j_{L_2}(k_2r_2) j_{L_3}(k_3r_3)  \nonumber \\ 
& \qquad  \qquad  \times j_{\ell_1}(k_1r)j_{\ell_1}(k_3r) j_{\ell''_1}(k_1s)j_{\ell''_1}(k_3s) \nonumber \\ 
& \qquad \qquad  \times P(k_1)P(k_2) \mathcal{P}_{J_1,J_2,J_3}(\mathbf{\widehat{k}}_1,\mathbf{\widehat{k}}_2,\mathbf{\widehat{k}}_3) \mathcal{P}_{L_1,L_2,L_3}(\mathbf{\widehat{k}}_1,\mathbf{\widehat{k}}_2,\mathbf{\widehat{k}}_3). 
\end{align}

\begin{figure}[h]
\centering
\includegraphics[scale=0.7]{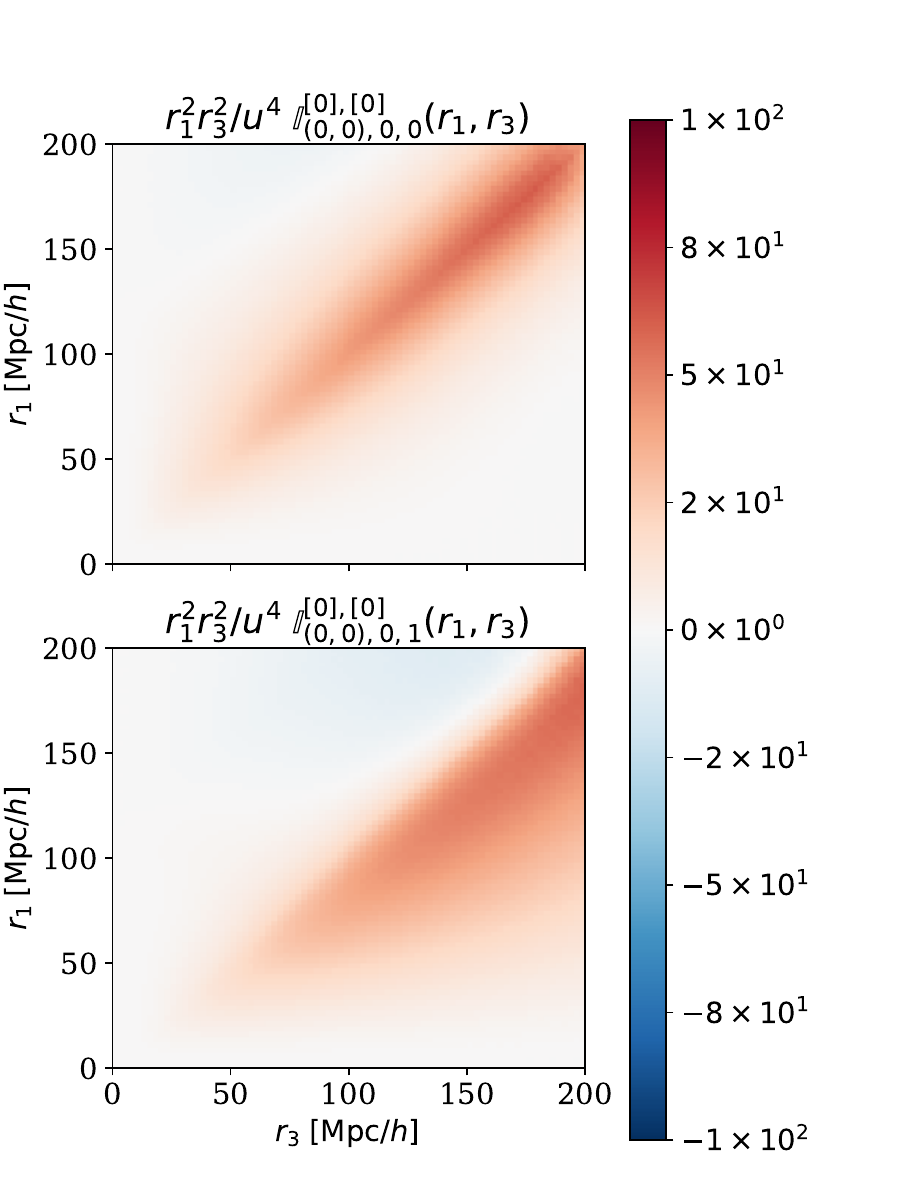}
\caption{2D plot of integral Eq. (\ref{eq:radial_n13_R1_T2}) for $n'_1=n'_3 =0$, $ \ell_1 = \ell''_1 = 0 $, $L_1 = \left\{ 0,1\right\}$ and $L_3=0$. We place $\ell_1 $ and $\ell''_1$ in parentheses to identify them as the sBF orders that do not contribute directly to the $r_1,\;r_2$ and $r_3$ variables. The \textit{upper panel} shows the integral for $L_3=0$, while the \textit{lower panel} shows the integral for $L_3=1$. Since the 4PCF can be approximated as the square of the 2PCF on large scales, $(\xi_{0})^{2}(r) \sim (1/r^2)^2$, we have weighted the integral by $r^2r_i^2/u^{4}$ to take out its fall-off. In the \textit{upper panel} the integral is largest along the diagonal, while in the \textit{lower panel} the integral is largest in the off-diagonal elements. We demonstrate analytically in Appendix \ref{Sec: Radial Integrals} that the behavior on the \textit{upper panel} arises because we can approximate $\mathbb{I}_{(0,0),0,0}^{\left[0\right],\left[0\right]}$ as a Dirac delta function dependent on $r_1$ and $r_3$, making the behavior appear only when $r_1=r_3$. For the \textit{lower panel}, $\mathbb{I}_{(0,0),0,1}^{\left[0\right],\left[0\right]}$, we simplify the integral until we find it to be the same expression as those resulting from $f_{1,0}^{\left[0\right]}$, $f_{1,0}^{'\left[0\right]}$ and $F_{(0),1,0}^{\left[0\right],\left[0\right]}$ . Our results in Appendix \ref{Sec: Radial Integrals} clearly show why Figure \ref{fig:fint2}, Figure \ref{fig:2D_int}, Figure \ref{fig:fprime_int}, and this figure are nearly identical. The \textit{lower panel} is also not symmetric under exchange of axes (\textit{i.e.}, $r_1 \leftrightarrow r_3$), which results from choosing $L_1 = 0$ for the sBF order corresponding to the variable $r_1$ and $L_3=1$ for the sBF order corresponding to the variable $r_3$.}
\label{fig:II_int}
\end{figure}

Performing the angular integrals we arrive at: 

\begin{empheq}[box=\widefbox]{align}\label{eq: R1_n123=0_T2}
&R_{1,(2);\left[n’_{1}\right], \left[n’_{2}\right], \left[n’_{3}\right], \left[n'_{13}\right],\left[n_{12}\right], \left[n_{13}\right], \left[n_{23}\right] }^{\left[0\right], \left[n_1\right], \left[n_2\right], \left[n_3\right]}(\mathbf{r}_1, \mathbf{r}_2, \mathbf{r}_3) \nonumber \\ 
& \rightarrow \sum_{j_1,j_2,j_3}\sum_{j_{12},j_{23},j_{13}}\sum_{\ell_1,\ell''_1} \sum_{L_1,L_2,L_3} (4\pi)^{8} (8 \pi) \; i^{n_{123}} \nonumber \\ 
&  \qquad  \times \frac{s_{\ell''_1}^{(\rm I)}s_{\ell_1}^{(\rm I)} \;w\; \;\mathcal{C}_{j_1,j_2,j_3}^{n_1,n_2,n_3}c_{j_{12}}^{(n_{12})}c_{j_{13}}^{(n_{13})}c_{j_{23}}^{(n_{23})}}{\Gamma(n_{13}-1)\left[ (-1)^{n_{13} -1} - 1 \right]}\nonumber \\ 
& \qquad \times  C_{L_1,L_2,L_3} \mathcal{G}_{L_1,L_2,L_3}\Upsilon_{L_1,L_2,L_3}\nonumber \\ 
& \qquad  \times\mathcal{P}_{L_1,L_2,L_3}(\mathbf{\widehat{r}}_1,\mathbf{\widehat{r}}_2,\mathbf{\widehat{r}}_3) \xi_{L_2}^{\left[n'_2\right]}(r_2) \mathbb{I}_{(\ell_1,\ell''_1),L_1,L_3}^{\left[n'_1\right],\left[n'_3\right]}(r_1,r_3),
\end{empheq}
with coefficients given in Table \ref{table:2}, $\xi_L$ given by Eq. (\ref{eq:1D-radial}) and the $\mathbb{I}$ radial integral defined as 

\begin{align}\label{eq:radial_n13_R1_T2}
&\mathbb{I}_{(\ell_1,\ell''_1),L_1,L_3}^{\left[n'_1\right],\left[n'_3\right]}(r_1,r_3) \equiv \int_{0}^{\infty} dr \; r^{n'_{13}-1} \;\int_{0}^{\infty}ds\; s^2 \; \xi_0(s) g_{l_1,l''_1,L_1}^{\left[n'_1\right]} (r,s,r_1) g_{l_1,l''_1,L_3}^{\prime \left[n'_3\right]}(r,s,r_3), 
\end{align}
with $g_{\ell',\ell'',L}$ defined in Eq. (\ref{eq:3_Sph.Bess_int}) and $g_{\ell',\ell'',L}^{'}$ defined in Eq. (\ref{eq:3_prime_Sph.Bess_int}). In appendix $\ref{Sec: Radial Integrals}$, we have made an analysis for a subset of special cases to try to explain the behavior observed in Figure \ref{fig:II_int}  when plotting the above expression. We have used parentheses around some of the orders of the spherical Bessel to indicate that these are being integrated out and do not directly affect the resulting variables $r_1$ and $r_3$. 

\clearpage

\paragraph{Inverse $k_{13}^{n'_{13}}$ and $k_{123}^{n_{123}}$; $n'_{13} = n_{123} = 0$} \mbox{}\\

\begin{figure}[h]
\centering
\includegraphics[scale=0.7]{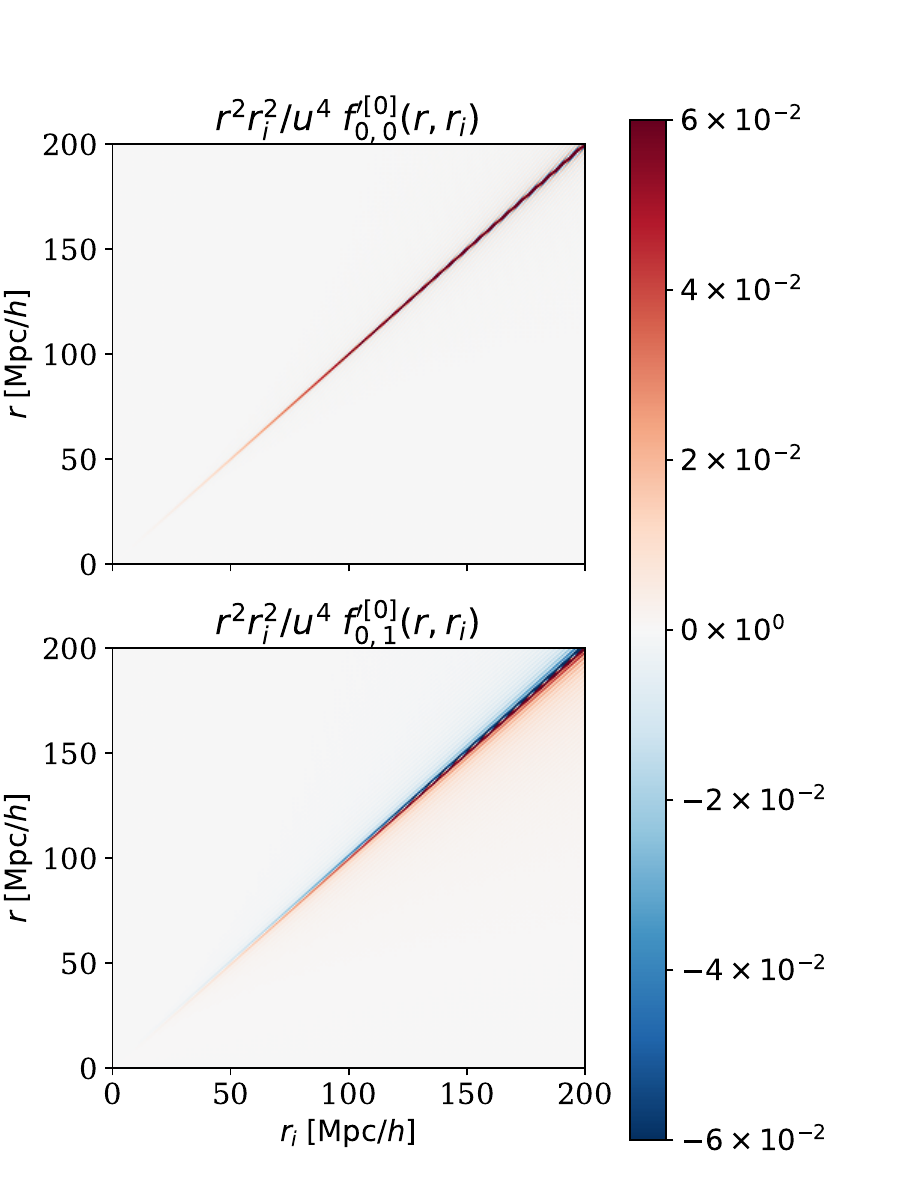}
\caption{Here, we show (\ref{eq:2_Sph.Bess_int_prime}) for $n'=0$, $L_1 = \left\{ 0,1\right\}$ and $L_3=0$. The \textit{upper panel} shows the integral for $L=0$, while the \textit{lower panel} shows the integral for $L=1$. Since the 4PCF can be approximated as the square of the 2PCF on large scales, $(\xi_{0})^{2}(r) \sim (1/r^2)^2$, we have weighted the integral by $r^2r_i^2/u^{4}$, with $u \equiv 10 \;\left[{\rm Mpc}/h\right]$, to take out its fall-off. The \textit{upper panel} shows the integral is largest along the diagonal; this is because the integral $f_{0,0}^{'\left[0\right]}$ can be approximated by a Dirac delta function, as is shown in Appendix \ref{Sec: Radial Integrals}. The \textit{lower panel}, for $L=1$, in contrast shows that the integral is largest in the off-diagonal elements; we demonstrate why in Appendix \ref{Sec: Radial Integrals}. The \textit{lower panel} is also not symmetric under exchange of axes (\textit{i.e.}, $r_i \leftrightarrow r$), which results from choosing $\ell = 0$ for the sBF order corresponding to the variable $r$ and $L=1$ for the sBF order corresponding to the variable $r_i$.}
\label{fig:fprime_int}
\end{figure}

\begin{figure}[h]
\centering
\includegraphics[scale=0.7]{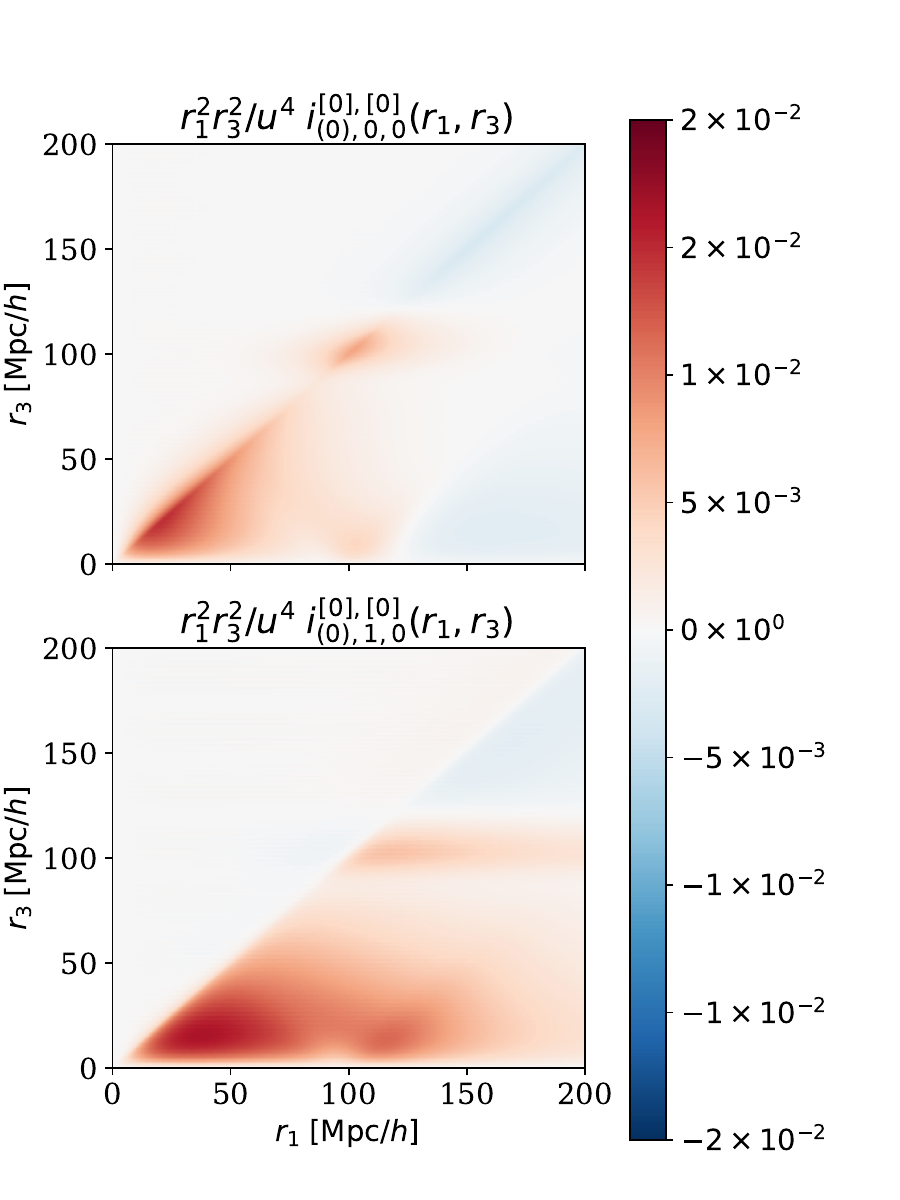}
\caption{Here, we show Eq. (\ref{eq:radial_0_R1_T2}) for $n'_1=n'_3 = 0$, $\ell'' =0$, $L_1 = \left\{ 0,1\right\}$ and $L_3=0$. We place $\ell''_1$ in parentheses to identify it as the sBF order which does not contribute directly to the $r_1$ and $r_3$ variables. The \textit{upper panel} shows the integral for $L_1=0$, while the \textit{lower panel} shows the integral for $L_1=1$. Since the 4PCF can be approximated as the square of the 2PCF on large scales, $(\xi_{0})^{2}(r) \sim (1/r^2)^2$, we have weighted the integral by $r^2r_i^2/u^{4}$ to take out its fall-off. The \textit{upper panel} shows the integral is largest along the diagonal, with most of its highest values in the region between $0$ and $100 \; \left[{\rm Mpc}/h\right]$. At $r_1=r_3 = 100 \; \left[{\rm Mpc}/h\right]$ we see a bump, which, as we prove analytically in Appendix \ref{Sec: Radial Integrals}, is a result of this integral's being expressed in terms of the 2PCF. The \textit{lower panel} shows that the upper-diagonal elements ($r_3>r_1$) are zero, with all the non-zero values of the integral concentrated in the lower-diagonal elements ($r_1>r_3$). Once again most of the integral's highest values lie in the region between $0$ and $100 \; \left[{\rm Mpc}/h\right]$ and the behavior is nearly identical to that of the \textit{upper panel} except there is more off-diagonal power. In Appendix \ref{Sec: Radial Integrals} we prove analytically that the integral on the region $r_3>r_1$ must vanish, while the integral on the region $r_1>r_3$ must behave in a similar fashion to the \textit{upper panel}. The \textit{upper panel}, although this might be unexpected, is not symmetric under the exchange of axes (\textit{i.e.}, $r_1 \leftrightarrow r_3$). We demonstrate in Appendix \ref{Sec: Radial Integrals} that this behavior arises as a result of the $f'$ integral's being independent of the power spectrum while in contrast the $f$ integral contains it. The \textit{lower panel} is also not symmetric under exchange of axes (\textit{i.e.}, $r_1 \leftrightarrow r_3$), which results from choosing $L_3 = 0$ for the sBF order corresponding to the variable $r_3$ and $L_1=1$ for the sBF order corresponding to the variable $r_1$.}
\label{fig:i_int}
\end{figure}

Finally, we evaluate term 1 in the list at the beginning of this section for $n_{13} = n_{123} = 0$; we need not decouple any denominator. However we still have the power spectrum $P(k_{13})$, so we find:

\begin{align}
&R_{1,(2);\left[n’_{1}\right], \left[n’_{2}\right], \left[n’_{3}\right], \left[0\right],\left[n_{12}\right], \left[n_{13}\right], \left[n_{23}\right] }^{\left[0\right], \left[n_1\right], \left[n_2\right], \left[n_3\right]}(\mathbf{r}_1, \mathbf{r}_2, \mathbf{r}_3) \rightarrow \sum_{j_1,j_2,j_3}\sum_{j_{12},j_{23},j_{13}} \sum_{\ell''_1} \sum_{J_1,J_2,J_3} \sum_{L_1,L_2,L_3}  (4\pi)^{15/2} \nonumber \\ 
& \qquad \qquad \qquad \qquad \times s_{\ell''_1}^{(\rm I)}\; w \;\mathcal{C}_{j_1,j_2,j_3}^{n_1,n_2,n_3} c_{j_{12}}^{(n_{12})}c_{j_{13}}^{(n_{13})}c_{j_{23}}^{(n_{23})}C_{L_1,L_2,L_3}\mathcal{G}_{J_1,J_2,J_3}\Upsilon_{L_1,L_2,L_3}  \nonumber \\ 
& \qquad \qquad \qquad \qquad \times \mathcal{P}_{L_1,L_2,L_3}(\mathbf{\widehat{r}}_1,\mathbf{\widehat{r}}_2,\mathbf{\widehat{r}}_3)  \int_{0}^{\infty}ds \; s^2 \; \xi_0(s) \int \frac{d^{3}\mathbf{k}_1\,d^{3}\mathbf{k}_2\,d^{3}\mathbf{k}_3}{(2\pi)^{9}}\;k_{1}^{n'_1}k_{2}^{n'_2}k_{3}^{n'_3} \nonumber \\ 
& \qquad \qquad \qquad \qquad \times j_{L_1}(k_1r_1) j_{L_2}(k_2r_2) j_{L_3}(k_3r_3) j_{\ell''_1}(k_1s)j_{\ell''_1}(k_3s) \nonumber \\ 
& \qquad \qquad \qquad \qquad  \times P(k_1)P(k_2) \mathcal{P}_{J_1,J_2,J_3}(\mathbf{\widehat{k}}_1,\mathbf{\widehat{k}}_2,\mathbf{\widehat{k}}_3) \mathcal{P}_{L_1,L_2,L_3}(\mathbf{\widehat{k}}_1,\mathbf{\widehat{k}}_2,\mathbf{\widehat{k}}_3),
\end{align}
for which we have combined all isotropic basis functions as above but this time following the result in Eq. (\ref{eq:Product_of_n_iso}) with $N=5$, which results in: 

\begin{empheq}[box=\widefbox]{align}\label{eq:R1_0_T2}
&R_{1,(2);\left[n’_{1}\right], \left[n’_{2}\right], \left[n’_{3}\right], \left[0\right],\left[n_{12}\right], \left[n_{13}\right], \left[n_{23}\right] }^{\left[0\right], \left[n_1\right], \left[n_2\right], \left[n_3\right]}(\mathbf{r}_1, \mathbf{r}_2, \mathbf{r}_3)  \nonumber \\ 
& \rightarrow \sum_{j_1,j_2,j_3}\sum_{j_{12},j_{23},j_{13}}\sum_{J_1,J_2,J_3} \sum_{l''_1} \sum_{L_1,L_2,L_3}   (4\pi)^{15/2} \nonumber \\
& \qquad \times s_{\ell''_1}^{(\rm I)} \;w\;
\mathcal{C}_{j_1,j_2,j_3}^{n_1,n_2,n_3} c_{j_{12}}^{(n_{12})}c_{j_{13}}^{(n_{13})}c_{j_{23}}^{(n_{23})} C_{L_1,L_2,L_3} \mathcal{G}_{L_1,L_2,L_3} \Upsilon_{L_1,L_2,L_3}   \nonumber \\ 
& \qquad \times \mathcal{P}_{L_1,L_2,L_3}(\mathbf{\widehat{r}}_1,\mathbf{\widehat{r}}_2,\mathbf{\widehat{r}}_3)   \xi_{L_2}^{\left[n'_2\right]}(r_2) i_{(\ell''_1),L_1,L_3}^{\left[n'_1\right],\left[n'_3\right]}(r_1,r_3),
\end{empheq}
with coefficients given in Table \ref{table:2} and the radial integral defined as:

\begin{align}\label{eq:radial_0_R1_T2}
&i_{(\ell''_1),L_1,L_3}^{\left[n'_1\right]\left[n'_3\right]}(r_1,r_3) \equiv \int_{0}^{\infty}ds\; s^2 \;\xi_0(s)  f_{\ell''_1,L_1}^{\left[n'_1\right]}(s,r_1) f_{\ell''_1,L_3}^{'\left[n'_3\right]}(s,r_3), 
\end{align}
with 

\begin{align}\label{eq:2_Sph.Bess_int_prime}
f_{\ell,L}^{'\left[n'\right]}(r,r_i) \equiv \int_{0}^{\infty}\frac{dk_i}{2 \pi^2} \;k_{i}^{n'+2} j_{\ell}(k_ir)j_{L}(k_i r_i).
\end{align}
$\xi_L$ and $f_{\ell,L}$ are given by Eq. (\ref{eq:1D-radial}) and Eq. (\ref{eq:2_Sph.Bess_int}), respectively. analytic solutions to Eq. (\ref{eq:2_Sph.Bess_int_prime}) are in \cite{Sph_Bessel_Integral_kiersten, Sph_Bessel_int_Rami}. In Appendix \ref{Sec: Radial Integrals}, we have made an analysis for a subset of special cases to try to explain the behavior observed in Figure \ref{fig:fprime_int} and Figure \ref{fig:i_int} when plotting the above expression. The parentheses around $\ell''_1$ indicates it does not contribute to the arguments $r_1$ and $r_3$ directly, only through the variable $s$. We have used parentheses on the radial integral around $\ell''_1$ to indicate that this spherical Bessel order is being integrated out and does not directly affect the resulting variables $r_1$, and $r_3$.

\clearpage

\subsubsection{Product of Second-Order Kernel and Tidal Tensor Kernel Weighted by $\mu_{123}^{n_{123}}$} \label{sec:T4.1.2}
For term 2 in the list at the beginning of this section, we have the SPT kernel, $S^{(2)}\left(\mathbf{k}_{i}, \pm \mathbf{k}_{13}\right)$, with the same logic as in term 1 for its arguments. The structure of this kernel with the new arguments is given by: 

\begin{align}
&S^{(2)}\left(\mathbf{k}_{1}, - \mathbf{k}_{13}\right) = \frac{-\mathbf{k}_1 \cdot(\mathbf{k}_1 + \mathbf{k}_3)}{k_{1}^{2}k_{13}^{2}} - \frac{1}{3} \nonumber \\ 
&\qquad\qquad\qquad= \frac{-1}{k_{13}^{2}} - \frac{k_3}{k_1 k_{13}^{2}}(\mathbf{\widehat{k}}_1 \cdot \mathbf{\widehat{k}}_3) - \frac{1}{3},
\end{align}
likewise

\begin{equation}
\begin{split}
S^{(2)}\left(\mathbf{k}_{2}, \mathbf{k}_{13}\right) = \frac{\mathbf{\widehat{k}}_1 \cdot \mathbf{\widehat{k}}_2}{k_{2}k_{13}^{2}} + \frac{\mathbf{\widehat{k}}_2 \cdot \mathbf{\widehat{k}}_3}{k_{2}k_{13}^{2}} - \frac{1}{3}. 
\end{split}
\end{equation}

We can see that the structure for $S^{(2)}\left(\mathbf{k}_{i}, \pm \mathbf{k}_{13}\right)$ given any set of arguments is: 

\begin{equation}\label{eq:gen_S2}
\begin{split}
S^{(2)}\left(\mathbf{k}_{i}, \pm \mathbf{k}_{13}\right) \rightarrow \frac{C_s^{n'_1,n'_2,n'_3,n'_{13}n_{12},n_{13},n_{23}} k_{1}^{n'_1}k_{2}^{n'_2}k_{3}^{n'_3}}{k_{13}^{n'_{13}}}(\mathbf{\widehat{k}}_1 \cdot \mathbf{\widehat{k}}_{2})^{n_{12}}(\mathbf{\widehat{k}}_2 \cdot \mathbf{\widehat{k}}_{3})^{n_{23}}(\mathbf{\widehat{k}}_1 \cdot \mathbf{\widehat{k}}_{3})^{n_{13}},
\end{split}
\end{equation}
where $C_s$ is a constant to indicate which term from $S^{(2)}$ we are referring to; we will suppress the superscripts of $C_s$ for the rest of this work. We have $n'_{13} = \left\{0,2\right\}$ once more. So, if we now multiply this $S^{(2)}\left(\mathbf{k}_{i}, \pm \mathbf{k}_{13}\right)$ kernel with the $W^{(2)}\left(\mathbf{k}_{j}, \pm \mathbf{k}_{13}\right)$ kernel we obtain the same structure as in Eq. (\ref{eq:W2W2}):

\begin{equation}\label{eq:S2W2}
\begin{split}
S^{(2)}\left(\mathbf{k}_{i}, \pm \mathbf{k}_{13}\right)W^{(2)}\left(\mathbf{k}_{j}, \pm \mathbf{k}_{13}\right) \rightarrow \frac{C_{s} \; w \;k_{1}^{n'_1}k_{2}^{n'_2}k_{3}^{n'_3}}{k_{13}^{n'_{13}}}(\mathbf{\widehat{k}}_1 \cdot \mathbf{\widehat{k}}_{2})^{n_{12}}(\mathbf{\widehat{k}}_2 \cdot \mathbf{\widehat{k}}_{3})^{n_{23}}(\mathbf{\widehat{k}}_1 \cdot \mathbf{\widehat{k}}_{3})^{n_{13}},
\end{split}
\end{equation}
which implies that taking the inverse Fourier transform of 2. will result in Eqs. (\ref{eq: R1_T2}), (\ref{eq: R1_n13=0_T2}), (\ref{eq: R1_n123=0_T2}) and (\ref{eq:R1_0_T2}) multiplied by $C_{s}$: 

\begin{empheq}[box=\widefbox]{align}\label{eq: R2_T2}
&R_{2,(2);\left[n’_{1}\right], \left[n’_{2}\right], \left[n’_{3}\right], \left[n'_{13}\right],\left[n_{12}\right], \left[n_{13}\right], \left[n_{23}\right]}^{\left[n_{123}\right], \left[n_1\right], \left[n_2\right], \left[n_3\right]}(\mathbf{r}_1, \mathbf{r}_2, \mathbf{r}_3)  \nonumber \\ 
&  \rightarrow \sum_{j_1,j_2,j_3}\sum_{j_{12},j_{23},j_{13}}\sum_{\ell'_1,\ell'_2,\ell'_3}\sum_{\ell_1,\ell''_1} \sum_{L_1,L_2,L_3} (4\pi)^{8} \nonumber \\ 
&  \qquad \times (256 \pi^{3})i^{n'_{13}+n_{123}} \;s_{\ell''_1}^{(\rm I)} \;C_s\;w\; s_{\ell_1}^{(\rm I)}\;\mathcal{C}_{j_1,j_2,j_3}^{n_1,n_2,n_3}\nonumber\\ 
& \qquad  \times c_{j_{12}}^{(n_{12})}c_{j_{13}}^{(n_{13})}c_{j_{23}}^{(n_{23})} \mathcal{C}_{\ell'_1,\ell'_2,\ell'_3} C_{L_1,L_2,L_3} \mathcal{G}_{L_1,L_2,L_3}\Upsilon_{L_1,L_2,L_3}\nonumber \\ 
& \qquad  \times  \left\{ \Gamma(n'_{13}-1)\Gamma(n_{123}-1)\right\}^{-1}\nonumber \\ 
& \qquad  \times  \left\{ \left[ (-1)^{n'_{13} -1} - 1 \right]\left[ (-1)^{n_{123} -1} - 1 \right]\right\}^{-1} \nonumber \\ 
& \qquad  \times \mathcal{P}_{L_1,L_2,L_3}(\mathbf{\widehat{r}}_1,\mathbf{\widehat{r}}_2,\mathbf{\widehat{r}}_3) I_{(\ell_1,\ell'_1,\ell'_2,\ell'_3,\ell''_1),L_1,L_2,L_3}^{\left[n'_1\right],\left[n'_2\right],\left[n'_3\right]}(r_1,r_2,r_3).
\end{empheq}

\begin{empheq}[box=\widefbox]{align}\label{eq: R2_n13=0_T2}
&R_{2,(2);\left[n’_{1}\right], \left[n’_{2}\right], \left[n’_{3}\right], \left[0\right],\left[n_{12}\right], \left[n_{13}\right], \left[n_{23}\right]}^{\left[n_{123}\right], \left[n_1\right], \left[n_2\right], \left[n_3\right]}(\mathbf{r}_1, \mathbf{r}_2, \mathbf{r}_3) \nonumber \\ 
& \rightarrow \sum_{j_1,j_2,j_3}\sum_{j_{12},j_{23},j_{13}}\sum_{\ell'_1,\ell'_2,\ell'_3}\sum_{\ell''_1} \sum_{L_1,L_2,L_3} (4\pi)^{15/2} (32\pi^2) \; i^{n_{123}}\nonumber \\ 
&  \qquad  \times \frac{s_{\ell''_1}^{(\rm I)} \;w\; \;\mathcal{C}_{j_1,j_2,j_3}^{n_1,n_2,n_3}c_{j_{12}}^{(n_{12})}c_{j_{13}}^{(n_{13})}c_{j_{23}}^{(n_{23})} \mathcal{C}_{\ell'_1,\ell'_2,\ell'_3}}{\Gamma(n_{123}-1)\left[ (-1)^{n_{123} -1} - 1 \right]}\nonumber \\ 
& \qquad  \times  C_{L_1,L_2,L_3} \mathcal{G}_{L_1,L_2,L_3}\Upsilon_{L_1,L_2,L_3}\nonumber \\ 
& \qquad \times  \mathcal{P}_{L_1,L_2,L_3}(\mathbf{\widehat{r}}_1,\mathbf{\widehat{r}}_2,\mathbf{\widehat{r}}_3) \mathcal{I}_{(\ell'_1,\ell'_2,\ell'_3,\ell''_1),L_1,L_2,L_3}^{\left[n'_1\right],\left[n'_2\right],\left[n'_3\right]}(r_1,r_2,r_3).
\end{empheq}

\begin{empheq}[box=\widefbox]{align}\label{eq: R2_n123=0_T2}
&R_{2,(2);\left[n’_{1}\right], \left[n’_{2}\right], \left[n’_{3}\right], \left[n'_{13}\right],\left[n_{12}\right], \left[n_{13}\right], \left[n_{23}\right]}^{\left[0\right], \left[n_1\right], \left[n_2\right], \left[n_3\right]}(\mathbf{r}_1, \mathbf{r}_2, \mathbf{r}_3) \nonumber \\ 
&  \rightarrow \sum_{j_1,j_2,j_3}\sum_{j_{12},j_{23},j_{13}}\sum_{\ell_1,\ell''_1} \sum_{L_1,L_2,L_3} (4\pi)^{8} (8 \pi) \; i^{n_{123}}\nonumber \\ 
&  \qquad  \times \frac{s_{\ell''_1}^{(\rm I)}s_{\ell_1}^{(\rm I)} \;w\; \;\mathcal{C}_{j_1,j_2,j_3}^{n_1,n_2,n_3}c_{j_{12}}^{(n_{12})}c_{j_{13}}^{(n_{13})}c_{j_{23}}^{(n_{23})}}{\Gamma(n_{13}-1)\left[ (-1)^{n_{13} -1} - 1 \right]}\nonumber \\ 
& \qquad  \times  C_{L_1,L_2,L_3} \mathcal{G}_{L_1,L_2,L_3}\Upsilon_{L_1,L_2,L_3}\nonumber \\ 
& \qquad  \times\mathcal{P}_{L_1,L_2,L_3}(\mathbf{\widehat{r}}_1,\mathbf{\widehat{r}}_2,\mathbf{\widehat{r}}_3) \xi_{L_2}^{\left[n'_2\right]}(r_2) \mathbb{I}_{(\ell_1,\ell''_1),L_1,L_3}^{\left[n'_1\right],\left[n'_3\right]}(r_1,r_3).
\end{empheq}

\begin{empheq}[box=\widefbox]{align}\label{eq:R2_0_T2}
&R_{2,(2);\left[n’_{1}\right], \left[n’_{2}\right], \left[n’_{3}\right], \left[0\right],\left[n_{12}\right], \left[n_{13}\right], \left[n_{23}\right]}^{\left[0\right], \left[n_1\right], \left[n_2\right], \left[n_3\right]}(\mathbf{r}_1, \mathbf{r}_2, \mathbf{r}_3)  \nonumber \\ 
&  \rightarrow \sum_{j_1,j_2,j_3}\sum_{j_{12},j_{23},j_{13}}\sum_{J_1,J_2,J_3} \sum_{l''_1} \sum_{L_1,L_2,L_3}   (4\pi)^{15/2} \nonumber \\ 
& \qquad  \times s_{\ell''_1}^{(\rm I)} \;w\;
\mathcal{C}_{j_1,j_2,j_3}^{n_1,n_2,n_3} c_{j_{12}}^{(n_{12})}c_{j_{13}}^{(n_{13})}c_{j_{23}}^{(n_{23})} C_{L_1,L_2,L_3} \mathcal{G}_{L_1,L_2,L_3} \Upsilon_{L_1,L_2,L_3}   \nonumber \\ 
& \qquad  \times \mathcal{P}_{L_1,L_2,L_3}(\mathbf{\widehat{r}}_1,\mathbf{\widehat{r}}_2,\mathbf{\widehat{r}}_3)   \xi_{L_2}^{\left[n'_2\right]}(r_2) i_{(\ell''_1),L_1,L_3}^{\left[n'_1\right],\left[n'_3\right]}(r_1,r_3).
\end{empheq}

\subsubsection{Product of Second-Order Kernel and Gamma Kernel Weighted by $\mu_{123}^{n_{123}}$}\label{sec:T4.1.3}
\qquad Next, we find the real space version of $W^{(2)}\left(\mathbf{k}_{i}, \pm \mathbf{k}_{13}\right) \gamma^{(2)}\left(\mathbf{k}_{j}, 
\pm \mathbf{k}_{13}\right)\mu_{1}^{n_1}\mu_{2}^{n_2}\mu_{3}^{n_3}\mu_{123}^{n_{123}}$. We start by looking closely at the structure of the term $\gamma^{(2)}\left(\mathbf{k}_1, - \mathbf{k}_{13}\right)$:

\begin{align}
&\gamma^{(2)}\left(\mathbf{k}_1, - \mathbf{k}_{13}\right) = \frac{1}{2}\left[ \mu_1^2 + \mu_{13}^2 - \mu_1 \mu_{13} \left(\frac{k_1}{k_{13}}+ \frac{k_{13}}{k_1}\right) \right]\nonumber \\ 
& \qquad \qquad \qquad \qquad \qquad + \left[ \mu_1^2\mu_{13}^2 - \frac{\mu_1 \mu_{13}}{2}\left(\mu_1^2\frac{k_1}{k_{13}}+\mu_{13}^2 \frac{k_{13}}{k_1}\right)\right],
\end{align}
which applying Eq. (\ref{eq:mu_23}) we obtain: 

\begin{align}\label{eq:cgamma}
\gamma^{(2)}\left(\mathbf{k}_1, - \mathbf{k}_{13}\right) \propto \frac{\mu_1^{n_1}\mu_{3}^{n_3}k_1^{n'_1}k_3^{n'_3}}{k_{13}^{n'_{13}}},
\end{align}
In the same manner, we find: 

\begin{align}
\gamma^{(2)}\left(\mathbf{k}_2, \mathbf{k}_{13}\right) \propto \frac{\mu_1^{n_1}\mu_2^{n_2}\mu_{3}^{n_3}k_1^{n'_1}k_2^{n'_2}k_3^{n'_3}}{k_{13}^{n'_{13}}}.
\end{align}
Therefore, we can take the most general expression for the gamma kernel to be: 

\begin{align}\label{eq:gamma_gen}
\gamma^{(2)}\left(\mathbf{k}_{j}, \pm \mathbf{k}_{13}\right) \rightarrow \frac{g_{n_1,n_2,n_3}^{n'_1,n'_2,n'_3,n'_{13}}\mu_1^{n_1}\mu_2^{n_2}\mu_{3}^{n_3}k_1^{n'_1}k_2^{n'_2}k_3^{n'_3}}{k_{13}^{n'_{13}}},
\end{align}
where $g$ is a constant to indicate which term from $\gamma^{(2)}$ we are referring to; we will suppress the subscripts and superscripts of $g$ for the rest of this work. We have $n'_{13} = \left\{ 0,2 \right\}$, again. Using the structure of $W^{(2)}\left(\mathbf{k}_{i}, \pm \mathbf{k}_{13}\right)$ as given in Eq. (\ref{eq:W2W2}) we find:

\begin{align}
&W^{(2)}\left(\mathbf{k}_{i}, \pm \mathbf{k}_{13}\right) \left<\gamma^{(2)}\left(\mathbf{k}_{j}, 
\pm \mathbf{k}_{13}\right) \mu_{1}^{n_1}\mu_{2}^{n_2}\mu_{3}^{n_3}\mu_{123}^{n_{123}}\right>_{\rm l.o.s} \nonumber \\
&\rightarrow (4\pi)^6 \sum_{j_1,j_2,j_3}\sum_{j_{12},j_{13},j_{23}}\;g\; w \;\mathcal{C}_{j_1,j_2,j_3}^{n_1,n_2,n_3} \mathcal{C}_{j_{1
2},j_{13},j_{23}}^{n_{12},n_{13},n_{23}}\mathcal{P}_{j_1,j_2,j_3}(\mathbf{\widehat{k}}_1,\mathbf{\widehat{k}}_2,\mathbf{\widehat{k}}_3) \mathcal{P}_{j_{12}}(\mathbf{\widehat{k}}_1,\mathbf{\widehat{k}}_2) \nonumber \\ 
&\qquad\qquad \qquad \qquad \qquad \times \mathcal{P}_{j_{13}}(\mathbf{\widehat{k}}_1,\mathbf{\widehat{k}}_3)\mathcal{P}_{j_{23}}(\mathbf{\widehat{k}}_2,\mathbf{\widehat{k}}_3) \frac{k_{1}^{n'_1}k_{2}^{n'_2}k_{3}^{n'_3}}{k_{13}^{n'_{13}}k_{123}^{n_{123}}}, 
\end{align}
which is exactly the same as Eq. (\ref{eq:W2W2_mu's}) multiplied by $g$, implying that the result of term 3 is exactly the same as those given by equations (\ref{eq: R1_T2}), (\ref{eq: R1_n13=0_T2}), (\ref{eq: R1_n123=0_T2}) and (\ref{eq:R1_0_T2}) times the constant $g$: 

\begin{empheq}[box=\widefbox]{align}\label{eq: R3_T2}
&R_{3,(2);\left[n’_{1}\right], \left[n’_{2}\right], \left[n’_{3}\right], \left[n'_{13}\right],\left[n_{12}\right], \left[n_{13}\right], \left[n_{23}\right]}^{\left[n_{123}\right], \left[n_1\right], \left[n_2\right], \left[n_3\right]}(\mathbf{r}_1, \mathbf{r}_2, \mathbf{r}_3) \nonumber \\ 
& \rightarrow \sum_{j_1,j_2,j_3}\sum_{j_{12},j_{23},j_{13}}\sum_{\ell'_1,\ell'_2,\ell'_3}\sum_{\ell_1,\ell''_1} \sum_{L_1,L_2,L_3} (4\pi)^{8} \nonumber \\ 
&  \qquad  \times (256 \pi^{3})i^{n'_{13}+n_{123}} \;s_{\ell''_1}^{(\rm I)} \;g\;w\; s_{\ell_1}^{(\rm I)}\;\mathcal{C}_{j_1,j_2,j_3}^{n_1,n_2,n_3}\nonumber\\ 
& \qquad  \times c_{j_{12}}^{(n_{12})}c_{j_{13}}^{(n_{13})}c_{j_{23}}^{(n_{23})} \mathcal{C}_{\ell'_1,\ell'_2,\ell'_3} C_{L_1,L_2,L_3} \mathcal{G}_{L_1,L_2,L_3}\Upsilon_{L_1,L_2,L_3}\nonumber \\ 
& \qquad  \times  \left\{ \Gamma(n'_{13}-1)\Gamma(n_{123}-1)\right\}^{-1}\nonumber \\ 
& \qquad  \times  \left\{ \left[ (-1)^{n'_{13} -1} - 1 \right]\left[ (-1)^{n_{123} -1} - 1 \right]\right\}^{-1} \nonumber \\ 
& \qquad \times \mathcal{P}_{L_1,L_2,L_3}(\mathbf{\widehat{r}}_1,\mathbf{\widehat{r}}_2,\mathbf{\widehat{r}}_3) I_{(\ell_1,\ell'_1,\ell'_2,\ell'_3,\ell''_1),L_1,L_2,L_3}^{\left[n'_1\right],\left[n'_2\right],\left[n'_3\right]}(r_1,r_2,r_3).
\end{empheq}

\begin{empheq}[box=\widefbox]{align}\label{eq: R3_n13=0_T2}
&R_{3,(2);\left[n’_{1}\right], \left[n’_{2}\right], \left[n’_{3}\right], \left[0\right],\left[n_{12}\right], \left[n_{13}\right], \left[n_{23}\right]}^{\left[n_{123}\right], \left[n_1\right], \left[n_2\right], \left[n_3\right]}(\mathbf{r}_1, \mathbf{r}_2, \mathbf{r}_3) \nonumber \\ 
&  \rightarrow \sum_{j_1,j_2,j_3}\sum_{j_{12},j_{23},j_{13}}\sum_{\ell'_1,\ell'_2,\ell'_3}\sum_{\ell''_1} \sum_{L_1,L_2,L_3} (4\pi)^{15/2} (32\pi^2) \; i^{n_{123}} \nonumber \\ 
&  \qquad \times \frac{s_{\ell''_1}^{(\rm I)} \;w\; \;\mathcal{C}_{j_1,j_2,j_3}^{n_1,n_2,n_3}c_{j_{12}}^{(n_{12})}c_{j_{13}}^{(n_{13})}c_{j_{23}}^{(n_{23})} \mathcal{C}_{\ell'_1,\ell'_2,\ell'_3}}{\Gamma(n_{123}-1)\left[ (-1)^{n_{123} -1} - 1 \right]}\nonumber \\ 
& \qquad  \times  C_{L_1,L_2,L_3} \mathcal{G}_{L_1,L_2,L_3}\Upsilon_{L_1,L_2,L_3}\nonumber \\ 
& \qquad  \times  \mathcal{P}_{L_1,L_2,L_3}(\mathbf{\widehat{r}}_1,\mathbf{\widehat{r}}_2,\mathbf{\widehat{r}}_3) \mathcal{I}_{(\ell'_1,\ell'_2,\ell'_3,\ell''_1),L_1,L_2,L_3}^{\left[n'_1\right],\left[n'_2\right],\left[n'_3\right]}(r_1,r_2,r_3).
\end{empheq}

\begin{empheq}[box=\widefbox]{align}\label{eq: R3_n123=0_T2}
&R_{3,(2);\left[n’_{1}\right], \left[n’_{2}\right], \left[n’_{3}\right], \left[n'_{13}\right],\left[n_{12}\right], \left[n_{13}\right], \left[n_{23}\right]}^{\left[0\right], \left[n_1\right], \left[n_2\right], \left[n_3\right]}(\mathbf{r}_1, \mathbf{r}_2, \mathbf{r}_3) \nonumber \\ 
& \rightarrow \sum_{j_1,j_2,j_3}\sum_{j_{12},j_{23},j_{13}}\sum_{\ell_1,\ell''_1} \sum_{L_1,L_2,L_3} (4\pi)^{8} (8 \pi) \; i^{n_{123}} \nonumber \\ 
&  \qquad \times \frac{s_{\ell''_1}^{(\rm I)}s_{\ell_1}^{(\rm I)} \;w\; \;\mathcal{C}_{j_1,j_2,j_3}^{n_1,n_2,n_3}c_{j_{12}}^{(n_{12})}c_{j_{13}}^{(n_{13})}c_{j_{23}}^{(n_{23})}}{\Gamma(n_{13}-1)\left[ (-1)^{n_{13} -1} - 1 \right]}\nonumber \\ 
& \qquad  \times  C_{L_1,L_2,L_3} \mathcal{G}_{L_1,L_2,L_3}\Upsilon_{L_1,L_2,L_3}\nonumber \\ 
& \qquad \times\mathcal{P}_{L_1,L_2,L_3}(\mathbf{\widehat{r}}_1,\mathbf{\widehat{r}}_2,\mathbf{\widehat{r}}_3) \xi_{L_2}^{\left[n'_2\right]}(r_2) \mathbb{I}_{(\ell_1,\ell''_1),L_1,L_3}^{\left[n'_1\right],\left[n'_3\right]}(r_1,r_3).
\end{empheq}

\begin{empheq}[box=\widefbox]{align}\label{eq:R3_0_T2}
&R_{3,(2);\left[n’_{1}\right], \left[n’_{2}\right], \left[n’_{3}\right], \left[0\right],\left[n_{12}\right], \left[n_{13}\right], \left[n_{23}\right]}^{\left[0\right], \left[n_1\right], \left[n_2\right], \left[n_3\right]}(\mathbf{r}_1, \mathbf{r}_2, \mathbf{r}_3)  \nonumber \\ 
&  \rightarrow \sum_{j_1,j_2,j_3}\sum_{j_{12},j_{23},j_{13}}\sum_{J_1,J_2,J_3} \sum_{l''_1} \sum_{L_1,L_2,L_3}   (4\pi)^{15/2} \nonumber \\ 
& \qquad \times s_{\ell''_1}^{(\rm I)} \;w\;
\mathcal{C}_{j_1,j_2,j_3}^{n_1,n_2,n_3} c_{j_{12}}^{(n_{12})}c_{j_{13}}^{(n_{13})}c_{j_{23}}^{(n_{23})} C_{L_1,L_2,L_3} \mathcal{G}_{L_1,L_2,L_3} \Upsilon_{L_1,L_2,L_3}   \nonumber \\ 
& \qquad  \times \mathcal{P}_{L_1,L_2,L_3}(\mathbf{\widehat{r}}_1,\mathbf{\widehat{r}}_2,\mathbf{\widehat{r}}_3)   \xi_{L_2}^{\left[n'_2\right]}(r_2) i_{(\ell''_1),L_1,L_3}^{\left[n'_1\right],\left[n'_3\right]}(r_1,r_3).
\end{empheq}

\subsubsection{Product of Two Tidal Tensor Kernels without $\mu_{3}$ Weight} \label{sec:T4.1.4}
\qquad We proceed to find the real space version of $S^{(2)}\left(\mathbf{k}_1, - \mathbf{k}_{13}\right) S^{(2)}\left(\mathbf{k}_2, \mathbf{k}_{13}\right)\mu_{1}^{n_1}\mu_{2}^{n_2}$, term 4 in the list provided at the beginning of this section. The structure of $S^{(2)}(\mathbf{k}_i,\pm \mathbf{k}_{13})$ is given in equation Eq. (\ref{eq:gen_S2}) and gives:  

\begin{align}\label{eq:2_S_Product}
&S^{(2)}\left(\mathbf{k}_1, - \mathbf{k}_{13}\right) S^{(2)}\left(\mathbf{k}_2, \mathbf{k}_{13}\right)\left<\mu_{1}^{n_1}\mu_{2}^{n_2}\right>_{\rm l.o.s}  \nonumber \\
&\rightarrow (4\pi)^6 \sum_{j_1}\sum_{j_{12},j_{13},j_{23}}\; C_s \; \mathcal{C}_{j_1,j_1,0}^{n_1,n_1,0} c_{j_{12}}^{(n_{12})}c_{j_{13}}^{(n_{13})}c_{j_{23}}^{(n_{23})}\mathcal{P}_{j_1,j_1,0}(\mathbf{\widehat{k}}_1,\mathbf{\widehat{k}}_2,\mathbf{\widehat{k}}_3) \mathcal{P}_{j_{12}}(\mathbf{\widehat{k}}_1,\mathbf{\widehat{k}}_2) \nonumber \\ 
&\qquad\qquad \qquad \qquad \qquad \times \mathcal{P}_{j_{13}}(\mathbf{\widehat{k}}_1,\mathbf{\widehat{k}}_3)\mathcal{P}_{j_{23}}(\mathbf{\widehat{k}}_2,\mathbf{\widehat{k}}_3) \frac{k_{1}^{n'_1}k_{2}^{n'_2}k_{3}^{n'_3}}{k_{13}^{n'_{13}}}, 
\end{align}
where we have used the result of Eq. (\ref{eq:avg_z}) with $n_3=0$; implying $n_1 = n_2$ and $j_1=j_2$. 

\paragraph{Inverse $k_{13}^{n'_{13}}$; $n'_{13}\neq0$} \mbox{}\\

We can read off the result of performing the inverse Fourier transform of Eq. (\ref{eq:2_S_Product}) for $n'_{13} \neq 0$ from Eq. (\ref{eq: R1_n123=0_T2}):

\begin{empheq}[box=\widefbox]{align}\label{eq:R5,2}
&R_{4,(2);\left[n’_{1}\right], \left[n’_{2}\right], \left[n’_{3}\right], \left[n'_{13}\right],\left[n_{12}\right], \left[n_{13}\right], \left[n_{23}\right] }^{\left[n_1\right], \left[n_2\right]}(\mathbf{r}_1, \mathbf{r}_2, \mathbf{r}_3) \nonumber \\ 
&   \rightarrow \sum_{j_1}\sum_{j_{12},j_{23},j_{13}}\sum_{\ell_1,\ell''_1} \sum_{L_1,L_2,L_3} (4\pi)^{8} (8 \pi)i^{n'_{13}}\nonumber \\ 
&  \qquad \times \frac{ \; s_{\ell_1}^{(\rm I)}\;s_{\ell''_1}^{(\rm I)}\; \mathcal{C}_{j_1,j_1,0}^{n_1,n_2,0} C_s\;}{\Gamma(n'_{13}-1)\left[ (-1)^{n'_{13} -1} - 1 \right]}\nonumber \\ 
& \qquad \times  c_{j_{12}}^{(n_{12})}c_{j_{13}}^{(n_{13})}c_{j_{23}}^{(n_{23})} C_{L_1,L_2,L_3} \mathcal{G}_{L_1,L_2,L_{3}}\Upsilon_{L_1,L_2,L_3}\nonumber \\ 
& \qquad  \times \mathcal{P}_{L_1,L_2,L_3}(\mathbf{\widehat{r}}_1,\mathbf{\widehat{r}}_2,\mathbf{\widehat{r}}_3) \xi_{L_2}^{\left[n'_2\right]}(r_2) \mathbb{I}_{(\ell_1,\ell''_1),L_1,L_3}^{\left[n'_1\right],\left[n'_3\right]}(r_1,r_3).
\end{empheq}

The radial integral $\xi_L$ is defined by Eq. (\ref{eq:1D-radial}), while the radial integral $\mathbb{I}$ is defined by Eq. (\ref{eq:radial_n13_R1_T2}) and with coefficients given in Table \ref{table:2}. In this result we substitute $\mathcal{C}_{j_1,j_2,j_3}^{n_1,n_2,n_3}$ for $c_{j_1}^{(n_1),(n_2)}$. $\mathcal{G}_{L_1,L_2,L_3}$ comes in the above equation from combining all isotropic basis functions into a single one as in Eq. (\ref{eq:Product_of_n_iso}) for $n=6$---excluding the isotropic basis functions that come from the plane-wave expansion and later using the orthogonality of the isotropic basis. 

\paragraph{Inverse $k_{13}^{n'_{13}}$; $n'_{13}=0$}\mbox{}\\

Continuing the analysis of Eq. (\ref{eq:2_S_Product}) with $n'_{13}=0$  and comparing with Eq. (\ref{eq:radial_0_R1_T2}) we realize we have the same structure as long as we set $j_3 =0 $ and make the appropriate constant replacements. Therefore, we find:

\begin{empheq}[box=\widefbox]{align}\label{eq:R5_0,2}
&R_{4,(2);\left[n’_{1}\right], \left[n’_{2}\right], \left[n’_{3}\right], \left[0\right],\left[n_{12}\right], \left[n_{13}\right], \left[n_{23}\right] }^{ \left[n_1\right], \left[n_2\right]}(\mathbf{r}_1, \mathbf{r}_2, \mathbf{r}_3) \nonumber \\ 
& \rightarrow \sum_{j_1,j_2}\sum_{j_{12},j_{23},j_{13}} \sum_{\ell''_1} \sum_{L_1,L_2,L_3}   (4\pi)^{15/2} s_{\ell''_1}^{(\rm I)}\;C_s \nonumber \\ 
& \qquad \times  \mathcal{C}_{j_1,j_1,0}^{n_1,n_1,0} c_{j_{12}}^{(n_{12})}c_{j_{13}}^{(n_{13})}c_{j_{23}}^{(n_{23})} C_{L_1,L_2,L_3} \mathcal{G}_{L_1,L_2,L_{3}}\Upsilon_{L_1,L_2,L_{3}}  \nonumber \\ 
& \qquad \times \mathcal{P}_{L_1,L_2,L_3}(\mathbf{\widehat{r}}_1,\mathbf{\widehat{r}}_2,\mathbf{\widehat{r}}_3) \xi_{L_2}^{\left[n'_2\right]}(r_2) i_{(\ell''_1),L_1,L_3}^{\left[n'_1\right],\left[n'_3\right]}(r_1,r_3),
\end{empheq}
with coefficients given in Table \ref{table:2}.

\subsubsection{Product of Tidal Tensor Kernel and Gamma Kernel without $\mu_3$ Weight} \label{sec:T4.1.5}
\qquad Next, we proceed with the analysis of term 5 in the list at the beginning of this section. The structure of the tidal tensor and gamma kernels for term 5 are given by Eqs. (\ref{eq:gen_S2}) and (\ref{eq:gamma_gen}). Hence, we have:

\begin{align}
&S^{(2)}\left(\mathbf{k}_{i}, \pm \mathbf{k}_{13}\right) \gamma^{(2)}\left(\mathbf{k}_{j}, 
\pm \mathbf{k}_{13}\right) \mu_{1}^{n_1}\mu_{2}^{n_2} \nonumber \\
&\rightarrow \frac{C_{s}\;g\; k_{1}^{n'_1}k_{2}^{n'_2}k_{3}^{n'_3}}{k_{13}^{n'_{13}}}(\mathbf{\widehat{k}}_1 \cdot \mathbf{\widehat{k}}_{2})^{n_{12}}(\mathbf{\widehat{k}}_2 \cdot \mathbf{\widehat{k}}_{3})^{n_{23}}(\mathbf{\widehat{k}}_1 \cdot \mathbf{\widehat{k}}_{3})^{n_{13}}\mu_{1}^{n_1}\mu_{2}^{n_2}\mu_{3}^{n_3}.
\end{align}
 
 Taking a closer look, we see that the above expression has the same structure as term 1, but with $n_{123}=0$. Therefore averaging over the line of sight and including the power spectrum to take the inverse Fourier transform, we find the result for this term is given by Eq.  (\ref{eq: R1_n123=0_T2}) with $n'_{13}\neq0$ and by Eq. (\ref{eq:R1_0_T2}) for $n_{13}=0$, if we just replace $w \rightarrow C_{s}\;g$:

\begin{empheq}[box=\widefbox]{align}\label{eq: R5_n123=0_T2}
&R_{5,(2);\left[n’_{1}\right], \left[n’_{2}\right], \left[n’_{3}\right], \left[n'_{13}\right],\left[n_{12}\right], \left[n_{13}\right], \left[n_{23}\right] }^{ \left[n_1\right], \left[n_2\right], \left[n_3\right]}(\mathbf{r}_1, \mathbf{r}_2, \mathbf{r}_3) \nonumber \\ 
& \rightarrow \sum_{j_1,j_2,j_3}\sum_{j_{12},j_{23},j_{13}}\sum_{\ell_1,\ell''_1} \sum_{L_1,L_2,L_3} (4\pi)^{8} (8 \pi) \; i^{n_{123}} \nonumber \\ 
&  \qquad \times \frac{s_{\ell''_1}^{(\rm I)}s_{\ell_1}^{(\rm I)} \;C_{s}\;g \;\mathcal{C}_{j_1,j_2,j_3}^{n_1,n_2,n_3}c_{j_{12}}^{(n_{12})}c_{j_{13}}^{(n_{13})}c_{j_{23}}^{(n_{23})}}{\Gamma(n_{13}-1)\left[ (-1)^{n_{13} -1} - 1 \right]}\nonumber \\ 
& \qquad  \times  C_{L_1,L_2,L_3} \mathcal{G}_{L_1,L_2,L_3}\Upsilon_{L_1,L_2,L_3}\nonumber \\ 
& \qquad  \times\mathcal{P}_{L_1,L_2,L_3}(\mathbf{\widehat{r}}_1,\mathbf{\widehat{r}}_2,\mathbf{\widehat{r}}_3) \xi_{L_2}^{\left[n'_2\right]}(r_2) \mathbb{I}_{(\ell_1,\ell''_1),L_1,L_3}^{\left[n'_1\right],\left[n'_3\right]}(r_1,r_3).
\end{empheq}

\begin{empheq}[box=\widefbox]{align}\label{eq:R5_0_T2}
&R_{5,(2);\left[n’_{1}\right], \left[n’_{2}\right], \left[n’_{3}\right], \left[0\right],\left[n_{12}\right], \left[n_{13}\right], \left[n_{23}\right]}^{ \left[n_1\right], \left[n_2\right], \left[n_3\right]}(\mathbf{r}_1, \mathbf{r}_2, \mathbf{r}_3)  \nonumber \\ 
& \rightarrow \sum_{j_1,j_2,j_3}\sum_{j_{12},j_{23},j_{13}}\sum_{J_1,J_2,J_3} \sum_{l''_1} \sum_{L_1,L_2,L_3}   (4\pi)^{15/2} \nonumber \\ 
& \qquad \times s_{\ell''_1}^{(\rm I)} \;C_s\;g\;
\mathcal{C}_{j_1,j_2,j_3}^{n_1,n_2,n_3} c_{j_{12}}^{(n_{12})}c_{j_{13}}^{(n_{13})}c_{j_{23}}^{(n_{23})} C_{L_1,L_2,L_3} \mathcal{G}_{L_1,L_2,L_3} \Upsilon_{L_1,L_2,L_3}   \nonumber \\ 
& \qquad  \times \mathcal{P}_{L_1,L_2,L_3}(\mathbf{\widehat{r}}_1,\mathbf{\widehat{r}}_2,\mathbf{\widehat{r}}_3)   \xi_{L_2}^{\left[n'_2\right]}(r_2) i_{(\ell''_1),L_1,L_3}^{\left[n'_1\right],\left[n'_3\right]}(r_1,r_3).
\end{empheq}

\subsubsection{Product of Two Gamma Kernels without $\mu_3$ Weight} \label{sec:T4.1.6}
\qquad We evaluate term 6 in the list at the beginning of this section. The structure for the gamma kernel is given by (\ref{eq:gamma_gen}), so we obtain:

\begin{align}\label{eq:gammagamma6}
\gamma^{(2)}\left(\mathbf{k}_1, -\mathbf{k}_{13}\right) \gamma^{(2)}\left(\mathbf{k}_2, \mathbf{k}_{13}\right) \mu_{1}^{n_1}\mu_{2}^{n_2} \rightarrow \frac{g\;\mu_1^{n_1}\mu_2^{n_2}\mu_{3}^{n_3}k_1^{n'_1}k_2^{n'_2}k_3^{n'_3}}{k_{13}^{n'_{13}}}.
\end{align}

\paragraph{Inverse $k_{13}^{n'_{13}}$; $n'_{13}\neq0$} \mbox{}\\

Analyzing the above equation carefully and comparing with (\ref{eq:W2W2}), we can find our result by setting $j_{12}=j_{13}=j_{23}=n_{123}=0$ and $n'_{13} \neq 0 $, combining all the isotropic basis as in Eq. (\ref{eq:Product_of_3_iso}) and then taking the angular integral to obtain: 

\begin{empheq}[box=\widefbox]{align}\label{eq:R6,2}
&R_{6,(2);\left[n’_{1}\right], \left[n’_{2}\right], \left[n’_{3}\right], \left[n'_{13}\right],\left[n_{12}\right], \left[n_{13}\right], \left[n_{23}\right]}^{ \left[n_1\right], \left[n_2\right], \left[n_3\right]}(\mathbf{r}_1, \mathbf{r}_2, \mathbf{r}_3)  \nonumber \\ 
&  \rightarrow \sum_{j_1,j_2,j_3}\sum_{\ell_1,\ell''_1} \sum_{L_1,L_2,L_3} (4\pi)^{7/2}(8 \pi \;i^{n'_{13}}) \nonumber \\ 
&  \qquad  \times \frac{s_{\ell''_1}^{(\rm I)}s_{\ell_1}^{(\rm I)}\;g\;\mathcal{C}_{j_{1},j_{2},j_{3}}^{n_{1},n_{2},n_{3}}C_{L_1,L_2,L_3}\mathcal{G}_{L_1,L_2,L_{3}}\Upsilon_{L_1,L_2,L_{3}}}{\Gamma(n'_{13}-1)\left[ (-1)^{n'_{13} -1} - 1 \right]}\nonumber \\ 
& \qquad  \times  \mathcal{P}_{L_1,L_2,L_3}(\mathbf{\widehat{r}}_1,\mathbf{\widehat{r}}_2,\mathbf{\widehat{r}}_3) \xi_{L_2}^{\left[n'_2\right]}(r_2) \mathbb{I}_{(\ell_1,\ell''_1),L_1,L_3}^{\left[n'_1\right],\left[n'_3\right]}(r_1,r_3). 
\end{empheq}
The definition of $\xi_L$ is given by Eq. (\ref{eq:1D-radial}) and  $\mathbb{I}_{(\ell_1,\ell''_1),L_1,L_3}$ is given by Eq. (\ref{eq:radial_n13_R1_T2}), with coefficients given in Table \ref{table:2}.

\paragraph{Inverse $k_{13}^{n'_{13}}$; $n'_{13}=0$}\mbox{}\\

Given the structure of Eq. (\ref{eq:gammagamma6}) with $n'_{13}=0$, we can immediately conclude taking its inverse Fourier transform results in Eq. (\ref{eq:R1_0_T2}) but with different constants:

\begin{empheq}[box=\widefbox]{align}\label{eq:R6_0,2}
&R_{6,(2);\left[n’_{1}\right], \left[n’_{2}\right], \left[n’_{3}\right], \left[0\right],\left[n_{12}\right], \left[n_{13}\right], \left[n_{23}\right]}^{\left[n_1\right], \left[n_2\right], \left[n_3\right]}(\mathbf{r}_1, \mathbf{r}_2, \mathbf{r}_3) \nonumber \\ 
& \rightarrow \sum_{j_1,j_2,j_3} \sum_{\ell''_1} \sum_{L_1,L_2,L_3}   (4\pi)^{3} s_{\ell''_1}^{(\rm I)}\;g \nonumber \\ 
& \qquad \times  \mathcal{C}_{j_{1},j_{2},j_{3}}^{n_{1},n_{2},n_{3}}C_{L_1,L_2,L_3}\mathcal{G}_{L_1,L_2,L_{3}}\Upsilon_{L_1,L_2,L_{3}}  \nonumber \\ 
& \qquad  \times \mathcal{P}_{L_1,L_2,L_3}(\mathbf{\widehat{r}}_1,\mathbf{\widehat{r}}_2,\mathbf{\widehat{r}}_3) \xi_{L_2}^{\left[n'_2\right]}(r_2)i_{(\ell''_1),L_1,L_3}^{\left[n'_1\right],\left[n'_3\right]}(r_1,r_3),
\end{empheq}
with coefficients given in Table \ref{table:2} and the definition of $i_{(\ell''_1),L_1,L_3}$ given by Eq. (\ref{eq:radial_0_R1_T2}) and $\xi_L$ by Eq. (\ref{eq:1D-radial}).

\subsubsection{Second-Order Kernel Weighted by $\mu_{123}^{n_{123}}$} \label{sec:T4.1.7}
\qquad For term 7 in the list at the beginning of this section, we have $W^{(2)}\left(\mathbf{k}_{i}, \pm \mathbf{k}_{13}\right)$ with $i=\left\{1,2\right\}$. We remind the reader, if $i = 1$ we pick $\mathbf{-k}_{13}$ as an argument and if $i = 2$ we pick $\mathbf{k}_{13}$.

This implies that to evaluate this term in the most general possible way we will  use the same structure as Eq. (\ref{eq:W2W2}), which implies the result for the analysis of this term is given by Eqs.~(\ref{eq: R1_T2}), (\ref{eq: R1_n13=0_T2}), (\ref{eq: R1_n123=0_T2}) and (\ref{eq:R1_0_T2}): 

\begin{empheq}[box=\widefbox]{align}\label{eq: R7_T2}
&R_{7,(2);\left[n’_{1}\right], \left[n’_{2}\right], \left[n’_{3}\right], \left[n'_{13}\right],\left[n_{12}\right], \left[n_{13}\right], \left[n_{23}\right]}^{\left[n_{123}\right], \left[n_1\right], \left[n_2\right], \left[n_3\right]}(\mathbf{r}_1, \mathbf{r}_2, \mathbf{r}_3)  \nonumber \\ 
&  \rightarrow \sum_{j_1,j_2,j_3}\sum_{j_{12},j_{23},j_{13}}\sum_{\ell'_1,\ell'_2,\ell'_3}\sum_{\ell_1,\ell''_1} \sum_{L_1,L_2,L_3} (4\pi)^{8} \nonumber \\ 
&  \qquad  \times (256 \pi^{3})i^{n'_{13}+n_{123}} \;s_{\ell''_1}^{(\rm I)} \;w\; s_{\ell_1}^{(\rm I)}\;\mathcal{C}_{j_1,j_2,j_3}^{n_1,n_2,n_3}\nonumber\\ 
& \qquad \times c_{j_{12}}^{(n_{12})}c_{j_{13}}^{(n_{13})}c_{j_{23}}^{(n_{23})} \mathcal{C}_{\ell'_1,\ell'_2,\ell'_3} C_{L_1,L_2,L_3} \mathcal{G}_{L_1,L_2,L_3}\Upsilon_{L_1,L_2,L_3}\nonumber \\ 
& \qquad  \times  \left\{ \Gamma(n'_{13}-1)\Gamma(n_{123}-1)\right\}^{-1}\nonumber \\ 
& \qquad \times  \left\{ \left[ (-1)^{n'_{13} -1} - 1 \right]\left[ (-1)^{n_{123} -1} - 1 \right]\right\}^{-1} \nonumber \\ 
& \qquad  \times \mathcal{P}_{L_1,L_2,L_3}(\mathbf{\widehat{r}}_1,\mathbf{\widehat{r}}_2,\mathbf{\widehat{r}}_3) I_{(\ell_1,\ell'_1,\ell'_2,\ell'_3,\ell''_1),L_1,L_2,L_3}^{\left[n'_1\right],\left[n'_2\right],\left[n'_3\right]}(r_1,r_2,r_3).
\end{empheq}

\begin{empheq}[box=\widefbox]{align}\label{eq: R7_n13=0_T2}
&R_{7,(2);\left[n’_{1}\right], \left[n’_{2}\right], \left[n’_{3}\right], \left[0\right],\left[n_{12}\right], \left[n_{13}\right], \left[n_{23}\right]}^{\left[n_{123}\right], \left[n_1\right], \left[n_2\right], \left[n_3\right]}(\mathbf{r}_1, \mathbf{r}_2, \mathbf{r}_3) \nonumber \\ 
&  \rightarrow \sum_{j_1,j_2,j_3}\sum_{j_{12},j_{23},j_{13}}\sum_{\ell'_1,\ell'_2,\ell'_3}\sum_{\ell''_1} \sum_{L_1,L_2,L_3} (4\pi)^{15/2} (32\pi^2) \; i^{n_{123}}\nonumber \\ 
&  \qquad \times \frac{s_{\ell''_1}^{(\rm I)} \;w\; \;\mathcal{C}_{j_1,j_2,j_3}^{n_1,n_2,n_3}c_{j_{12}}^{(n_{12})}c_{j_{13}}^{(n_{13})}c_{j_{23}}^{(n_{23})} \mathcal{C}_{\ell'_1,\ell'_2,\ell'_3}}{\Gamma(n_{123}-1)\left[ (-1)^{n_{123} -1} - 1 \right]}\nonumber \\ 
& \qquad\times  C_{L_1,L_2,L_3} \mathcal{G}_{L_1,L_2,L_3}\Upsilon_{L_1,L_2,L_3}\nonumber \\ 
& \qquad  \times  \mathcal{P}_{L_1,L_2,L_3}(\mathbf{\widehat{r}}_1,\mathbf{\widehat{r}}_2,\mathbf{\widehat{r}}_3) \mathcal{I}_{(\ell'_1,\ell'_2,\ell'_3,\ell''_1),L_1,L_2,L_3}^{\left[n'_1\right],\left[n'_2\right],\left[n'_3\right]}(r_1,r_2,r_3).
\end{empheq}

\begin{empheq}[box=\widefbox]{align}\label{eq: R7_n123=0_T2}
&R_{7,(2);\left[n’_{1}\right], \left[n’_{2}\right], \left[n’_{3}\right], \left[n'_{13}\right],\left[n_{12}\right], \left[n_{13}\right], \left[n_{23}\right]}^{\left[0\right], \left[n_1\right], \left[n_2\right], \left[n_3\right]}(\mathbf{r}_1, \mathbf{r}_2, \mathbf{r}_3) \nonumber \\ 
&  \rightarrow \sum_{j_1,j_2,j_3}\sum_{j_{12},j_{23},j_{13}}\sum_{\ell_1,\ell''_1} \sum_{L_1,L_2,L_3} (4\pi)^{8} (8 \pi) \; i^{n_{123}}\nonumber \\ 
&  \qquad \times \frac{s_{\ell''_1}^{(\rm I)}s_{\ell_1}^{(\rm I)} \;w\; \;\mathcal{C}_{j_1,j_2,j_3}^{n_1,n_2,n_3}c_{j_{12}}^{(n_{12})}c_{j_{13}}^{(n_{13})}c_{j_{23}}^{(n_{23})}}{\Gamma(n_{13}-1)\left[ (-1)^{n_{13} -1} - 1 \right]}\nonumber \\ 
& \qquad  \times  C_{L_1,L_2,L_3} \mathcal{G}_{L_1,L_2,L_3}\Upsilon_{L_1,L_2,L_3}\nonumber \\ 
& \qquad \times\mathcal{P}_{L_1,L_2,L_3}(\mathbf{\widehat{r}}_1,\mathbf{\widehat{r}}_2,\mathbf{\widehat{r}}_3) \xi_{L_2}^{\left[n'_2\right]}(r_2) \mathbb{I}_{(\ell_1,\ell''_1),L_1,L_3}^{\left[n'_1\right],\left[n'_3\right]}(r_1,r_3).
\end{empheq}

\begin{empheq}[box=\widefbox]{align}\label{eq:R7_0_T2}
&R_{7,(2);\left[n’_{1}\right], \left[n’_{2}\right], \left[n’_{3}\right], \left[0\right],\left[n_{12}\right], \left[n_{13}\right], \left[n_{23}\right]}^{\left[0\right], \left[n_1\right], \left[n_2\right], \left[n_3\right]}(\mathbf{r}_1, \mathbf{r}_2, \mathbf{r}_3)  \nonumber \\ 
& \rightarrow \sum_{j_1,j_2,j_3}\sum_{j_{12},j_{23},j_{13}}\sum_{J_1,J_2,J_3} \sum_{l''_1} \sum_{L_1,L_2,L_3}   (4\pi)^{15/2} \nonumber \\ 
& \qquad  \times s_{\ell''_1}^{(\rm I)} \;w\;
\mathcal{C}_{j_1,j_2,j_3}^{n_1,n_2,n_3} c_{j_{12}}^{(n_{12})}c_{j_{13}}^{(n_{13})}c_{j_{23}}^{(n_{23})} C_{L_1,L_2,L_3} \mathcal{G}_{L_1,L_2,L_3} \Upsilon_{L_1,L_2,L_3}   \nonumber \\ 
& \qquad  \times \mathcal{P}_{L_1,L_2,L_3}(\mathbf{\widehat{r}}_1,\mathbf{\widehat{r}}_2,\mathbf{\widehat{r}}_3)   \xi_{L_2}^{\left[n'_2\right]}(r_2) i_{(\ell''_1),L_1,L_3}^{\left[n'_1\right],\left[n'_3\right]}(r_1,r_3).
\end{empheq}

\subsubsection{Tidal Tensor Kernel without $\mu_3$ Weight}\label{sec:T4.1.8}
\qquad For term 8 in the list at the beginning of this section, we have to evaluate $S^{(2)}\left(\mathbf{k}_{i}, \pm \mathbf{k}_{13}\right) \mu_{1}^{n_1}\mu_{2}^{n_2}$ following the same logic of the arguments as we did with term 7 in the previous subsection. Then, given the structure for $S^{(2)}\left(\mathbf{k}_{i}, \pm \mathbf{k}_{13}\right)$ by Eq. (\ref{eq:gen_S2}) and having only $\mu_{1}$ and $\mu_{2}$ our solution to this term is given by equations (\ref{eq:R5,2}) and (\ref{eq:R5_0,2}):

\begin{empheq}[box=\widefbox]{align}\label{eq:R8,2}
&R_{8,(2);\left[n’_{1}\right], \left[n’_{2}\right], \left[n’_{3}\right], \left[n'_{13}\right],\left[n_{12}\right], \left[n_{13}\right], \left[n_{23}\right]}^{ \left[n_1\right], \left[n_2\right]}(\mathbf{r}_1, \mathbf{r}_2, \mathbf{r}_3)  \nonumber \\ 
&   \rightarrow \sum_{j_1,j_2}\sum_{j_{12},j_{23},j_{13}}\sum_{\ell_1,\ell''_1} \sum_{L_1,L_2,L_3} (4\pi)^{8} (8 \pi)i^{n'_{13}} \nonumber \\ 
&  \qquad\times \frac{ \; s_{\ell_1}^{(\rm I)}\;s_{\ell''_1}^{(\rm I)}\; c_{j_1}^{(n_1),(n_2)} C_s\;}{\Gamma(n'_{13}-1)\left[ (-1)^{n'_{13} -1} - 1 \right]}\nonumber \\ 
& \qquad  \times  c_{j_{12}}^{(n_{12})}c_{j_{13}}^{(n_{13})}c_{j_{23}}^{(n_{23})} C_{L_1,L_2,L_3} \mathcal{G}_{L_1,L_2,L_{3}}\Upsilon_{L_1,L_2,L_3}\nonumber \\ 
& \qquad  \times \mathcal{P}_{L_1,L_2,L_3}(\mathbf{\widehat{r}}_1,\mathbf{\widehat{r}}_2,\mathbf{\widehat{r}}_3) \xi_{L_2}^{\left[n'_2\right]}(r_2) \mathbb{I}_{(\ell_1,\ell''_1),L_1,L_3}^{\left[n'_1\right],\left[n'_3\right]}(r_1,r_3).
\end{empheq}

\begin{empheq}[box=\widefbox]{align}\label{eq:R8_0,2}
&R_{8,(2);\left[n’_{1}\right], \left[n’_{2}\right], \left[n’_{3}\right], \left[0\right],\left[n_{12}\right], \left[n_{13}\right], \left[n_{23}\right]}^{ \left[n_1\right], \left[n_2\right];\left[0\right]}(\mathbf{r}_1, \mathbf{r}_2, \mathbf{r}_3) \nonumber \\ 
&  \rightarrow \sum_{j_1,j_2}\sum_{j_{12},j_{23},j_{13}} \sum_{\ell''_1} \sum_{L_1,L_2,L_3}   (4\pi)^{15/2} s_{\ell''_1}^{(\rm I)}\;C_s \nonumber \\ 
& \qquad \times  c_{j_1}^{(n_1),(n_2)} c_{j_{12}}^{(n_{12})}c_{j_{13}}^{(n_{13})}c_{j_{23}}^{(n_{23})} C_{L_1,L_2,L_3} \mathcal{G}_{L_1,L_2,L_{3}}\Upsilon_{L_1,L_2,L_{3}}  \nonumber \\ 
& \qquad  \times \mathcal{P}_{L_1,L_2,L_3}(\mathbf{\widehat{r}}_1,\mathbf{\widehat{r}}_2,\mathbf{\widehat{r}}_3) \xi_{L_2}^{\left[n'_2\right]}(r_2) i_{(\ell''_1),L_1,L_3}^{\left[n'_1\right],\left[n'_3\right]}(r_1,r_3).
\end{empheq}

\section{Discussion and Conclusion}\label{sec:4PCF_Main_Result}
\subsection{Master Equation for the 4PCF}
\qquad In $\S$\ref{Section T3} and $\S$\ref{Section T2}, we found 17 equations reproducing in full the 4PCF. Hence, our final result is:

\begin{empheq}[box=\widefbox]{align} \label{4PCF_Result}
&R(\mathbf{r}_1,\mathbf{r}_2,\mathbf{r}_3) = R_{3111}(\mathbf{r}_1,\mathbf{r}_2,\mathbf{r}_3) + R_{2211}(\mathbf{r}_1,\mathbf{r}_2,\mathbf{r}_3), 
\end{empheq}
where

\begin{empheq}[box=\widefbox]{align} \label{4PCF_Result_R3111}
&R_{3111}(\mathbf{r}_1,\mathbf{r}_2,\mathbf{r}_3) = 2 \times \sum_{\rm All} \left\{ K_{i; \; \Vec{i}_{\rm d}^{\;'}, \Vec{i}_{\rm d}}^{\Vec{e}_{\rm s}^{\;'}} \mathbb{P}_{i}^{\Vec{e}_{\rm s}^{\;'},\Vec{e}_{\rm s}} \left[R_{i,(3); \Vec{i}_{\rm d}^{\;'}, \Vec{i}_{\rm d}}^{\Vec{e}_{\rm s}^{\;'},\Vec{e}_{\rm s}}(\mathbf{r}_1; \mathbf{r}_2,\mathbf{r}_3 )  \right. \right. \\ \nonumber
& \quad \qquad \qquad \qquad \left. \left. + R_{i,(3); \Vec{i}_{\rm d}^{\;'}, \Vec{i}_{\rm d}}^{\Vec{e}_{\rm s}^{\;'}, \Vec{e}_{\rm s}}(\mathbf{r}_2; \mathbf{r}_1,\mathbf{r}_3) + R_{i,(3); \Vec{i}_{\rm d}^{\; '}, \Vec{i}_{\rm d}}^{\Vec{e}_{\rm s}^{\;'}, \Vec{e}_{\rm s}}(\mathbf{r}_3; \mathbf{r}_2,\mathbf{r}_1 ) \right. \right. \\ \nonumber
&  \quad \qquad \qquad \qquad \left. \left. + R_{i,(3); \Vec{i}_{\rm d}^{\; '}, \Vec{i}_{\rm d}}^{\Vec{e}_{\rm s}^{\;'}, \Vec{e}_{\rm s}}(\mathbf{r}_{13}; \mathbf{r}_{23},-\mathbf{r}_3) + R_{i,(3); \Vec{i}_{\rm d}^{\; '}, \Vec{i}_{\rm d}}^{\Vec{e}_{\rm s}^{\;'}, \Vec{e}_{\rm s}}(\mathbf{r}_{23}; \mathbf{r}_{13},-\mathbf{r}_3) \right. \right. \\ \nonumber
&  \quad\qquad \qquad \qquad \left. \left. + R_{i,(3); \Vec{i}_{\rm d}^{\; '}, \Vec{i}_{\rm d}}^{\Vec{e}_{\rm s}^{\;'}, \Vec{e}_{\rm s}}(-\mathbf{r}_{3}; \mathbf{r}_{23},\mathbf{r}_{13}) \right. \right. \\ \nonumber
&  \quad \qquad \qquad \qquad \left. \left. + R_{i,(3); \Vec{i}_{\rm d}^{\; '}, \Vec{i}_{\rm d}}^{\Vec{e}_{\rm s}^{\;'}, \Vec{e}_{\rm s}}(\mathbf{r}_{12}; -\mathbf{r}_{2},\mathbf{r}_{32}) + R_{i,(3); \Vec{i}_{\rm d}^{\; '}, \Vec{i}_{\rm d}}^{\Vec{e}_{\rm s}^{\;'}, \Vec{e}_{\rm s}}(-\mathbf{r}_{2}; \mathbf{r}_{12},\mathbf{r}_{32}) \right. \right. \\ \nonumber
& \quad \qquad \qquad \qquad \left. \left. + R_{i,(3); \Vec{i}_{\rm d}^{\; '}, \Vec{i}_{\rm d}}^{\Vec{e}_{\rm s}^{\;'}, \Vec{e}_{\rm s}}(\mathbf{r}_{32}; -\mathbf{r}_{2},\mathbf{r}_{12}) \right. \right. \\ \nonumber
&  \quad \qquad \qquad \qquad \left. \left. + R_{i,(3); \Vec{i}_{\rm d}^{\; '}, \Vec{i}_{\rm d}}^{\Vec{e}_{\rm s}^{\;'}, \Vec{e}_{\rm s}}( - \mathbf{r}_{1}; \mathbf{r}_{21},\mathbf{r}_{31}) + R_{i,(3); \Vec{i}_{\rm d}^{\; '}, \Vec{i}_{\rm d}}^{\Vec{e}_{\rm s}^{\;'}, \Vec{e}_{\rm s}}(\mathbf{r}_{21}; -\mathbf{r}_{1},\mathbf{r}_{31}) \right. \right. \\ \nonumber
& \quad \qquad \qquad \qquad \left. \left. + R_{i,(3); \Vec{i}_{\rm d}^{\; '}, \Vec{i}_{\rm d}}^{\Vec{e}_{\rm s}^{\;'},\Vec{e}_{\rm s}}(\mathbf{r}_{31}; \mathbf{r}_{21},-\mathbf{r}_{1})\right] \right\}, 
\end{empheq}
and 
\begin{empheq}[box=\widefbox]{align} \label{4PCF_Result_R2211}
&R_{2211}(\mathbf{r}_1,\mathbf{r}_2,\mathbf{r}_3) = 12 \times \sum_{\rm All}\left\{ \mathcal{K}_{j;\Vec{i}_{\rm d}^{\; '}}^{\Vec{e}_{\rm t}^{\; '},\Vec{e}_{\rm s}} \mathbb{P}_{j}^{\Vec{e}_{\rm t}^{\; '},\Vec{e}_{\rm s}}\left[ R_{j,(2);\Vec{i}_{\rm s}^{\; '}, \Vec{i}_{\rm d}^{\; '},  \Vec{i}_{\rm d}}^{\Vec{e}_{\rm t}^{\; '},\Vec{e}_{\rm s}}(\mathbf{r}_1, \mathbf{r}_2,  \mathbf{r}_3 )  \right. \right. \\ \nonumber
&  \quad \qquad \qquad \qquad \left. \left. + R_{j,(2);\Vec{i}_{\rm s}^{\; '}, \Vec{i}_{\rm d}^{\; '},  \Vec{i}_{\rm d}}^{\Vec{e}_{\rm t}^{\; '},\Vec{e}_{\rm s}}(\mathbf{r}_{13}, \mathbf{r}_{23},  -\mathbf{r}_3 ) + R_{j,(2);\Vec{i}_{\rm s}^{\; '}, \Vec{i}_{\rm d}^{\; '},  \Vec{i}_{\rm d}}^{\Vec{e}_{\rm t}^{\; '},\Vec{e}_{\rm s}}(\mathbf{r}_{12}, -\mathbf{r}_{2},  \mathbf{r}_{32} )  \right. \right. \\ \nonumber
&  \quad \qquad \qquad \qquad \left. \left. + R_{j,(2);\Vec{i}_{\rm s}^{\; '}, \Vec{i}_{\rm d}^{\; '},  \Vec{i}_{\rm d}}^{\Vec{e}_{\rm t}^{\; '},\Vec{e}_{\rm s}}(-\mathbf{r}_{1}, \mathbf{r}_{21}, \mathbf{r}_{31} )\right]\right\},
\end{empheq}
with $\mathbf{r}_{ij}\equiv \mathbf{r}_{i}-\mathbf{r}_j$. {Here, ``All'' denotes summing over all the sub- and super-indices\\ 
\noindent$\{i, j,  \vec{i}_{\rm s}, \vec{i}\;'_{\rm s},\vec{i}_{\rm d}, \vec{i}\;'_{\rm d}\}$ and $\{\vec{e}_{\rm s},  \vec{e}\;'_{\rm s}, \vec{e}_{\rm t}\}$. We have defined the set of {\it external} dependencies $\vec{e}$, which are in the lists of terms at the beginning of $\S$\ref{Section T3} and $\S$\ref{Section T2} as:

\begin{align}
&\Vec{e}_{\rm s}^{\;'} \equiv\left(\left[n_{1}^{'}\right],\left[n_{2}^{'}\right],\left[n_{3}^{'}\right]\right)\nonumber\\
&\Vec{e}_{\rm s} \equiv \left(\left[n_{1}\right], \left[n_{2}\right], \left[n_{3}\right] \right)\\
&\Vec{e}_{\rm t} \equiv \left(\left[n_{123}^{}\right]\right)\nonumber
\end{align}
Here we use ``${\rm s}$'' to denote the {\it single} subscripts associated with the exponents of wave number $k_{\{1,2,3\}}$ or the cosine of the angle between the line of sight and the wave vector $\mu_{\{1,2,3\}}\equiv \hat{{\bf k}}_{\{1,2,3\}}\cdot \hat{{\bf z}}$. We use ``${\rm t}$'' to denote the {\it triple} subscript associated with the exponent for the cosine of the angle between the line of sight and the sum of the three wave vectors $\mu_{123}\equiv \hat{{\bf k}}_{123}\cdot \hat{{\bf z}}$. In addition, for both the internal and external dependencies, we use a prime to distinguish the exponents for wave number ${k}$ and no prime for powers in the dot product between the two unit vectors, such as the $\mu_j\equiv \hat{\bf k}_j \cdot \hat{{\bf z}}$ or $\hat{\bf k}_i \cdot \hat{\bf k}_j$. In the same manner, we have defined the set of {\it internal} dependencies $\vec{i}$, which arose from our analysis of the structure of the trispectrum kernels, as: 

\begin{align}
&\Vec{i}_{\rm s}^{\;'} \equiv \left(\left[n_{1}^{'}\right], \left[n_{2}^{'}\right],\left[n_{3}^{'}\right]\right) \nonumber \\
&\Vec{i}_{\rm d}^{\;'} \equiv \left(\left[n_{12}^{'}\right],\left[n_{13}^{'}\right], \left[n_{23}^{'}\right]\right)\\ 
&\Vec{i}_{\rm d} \equiv \left(\left[n_{12}\right],\left[n_{13}\right], \left[n_{23}\right]\right), \nonumber
\end{align}
similarly, here we use ``${\rm d}$'' to denote the {\it double} subscripts.


We have introduced:
\begin{align}
\mathbb{P} \equiv  \left\{b_1, b_1f, b_2, b_2f, \cdots \right\}|^{\Vec{e}_{\rm s}^{\;'}, \Vec{e}_{\rm t}^{\;'}, \Vec{e}_{\rm s}}, \nonumber
\end{align}
as the set of all possible bias parameters that multiply the terms of the 4PCF. Depending on which 4PCF term we are evaluating, encoded by the external dependencies given by the vectors $\vec{e}$ indicated, we will choose a different element from the set given above. We have also introduced:
\begin{align}
K \equiv \begin{cases}
0 & \text{ if } i=\left\{ 1,\cdots,6\right\} \; {\rm and\;all\;of}\; n'_{1},\;n'_{2},\; n'_{3},\;n'_{23},\;n_{12},\;n_{13},\;n_{23} \neq 0  \\ 
0 & \text{ if } i=7 \; {\rm and\;all\;of}\;n'_{23},\;n_{12},\;n_{13},\;n_{23} \neq 0  \\ 
0 & \text{ if } i= \left\{8,9\right\}\; {\rm and\;all\;of}\; n'_{1},\;n'_{2},\;n'_{3} \neq 0\\ 
1 & \text{ otherwise },  
\end{cases}    \nonumber
\end{align}
and 

\begin{align}
\mathcal{K} \equiv \begin{cases}
0 & \text{ if } j=\left\{ 4,5,6,8\right\} \; {\rm and}\;  n_{123} \neq 0  \\ 
0 & \text{ if } j= \left\{4,8\right\}\; {\rm and\;all\;of}\; \;n_{3}, n_{123} \neq 0\\ 
1 & \text{ otherwise }, \nonumber
\end{cases}
\end{align}
which act as modified Kronecker deltas, ensuring only the expressions derived in $\S$\ref{Section T3} and $\S$\ref{Section T2} are non-zero. 

The expression $R_{1,(3)}$ is in Eq. (\ref{eq:R1,3}). The expression for $R_{2,(3)}$ is in Eq. (\ref{eq:R2,3}). The expression $R_{3,(3)}$ is in Eq. (\ref{eq:R5,3}). The expression $R_{4,(3)}$ is in Eq. (\ref{eq:R6,3}). The expression $R_{5,(3)}$ is in Eq. (\ref{eq:R7,3}). The expression $R_{6,(3)}$ is in Eq. (\ref{eq:R4,3}). The expression $R_{7,(3)}$ is in Eq. (\ref{eq:R3_T3111}). The expression $R_{8,(3)}$ is in Eqs. (\ref{eq:R8,3,1}) and (\ref{eq:R8,3,2}). The expression $R_{9,(3)}$ is in Eqs. (\ref{eq:R_9T3111}) and (\ref{eq:R_9,0,T3}). The expression $R_{1,(2)}$ is in Eqs. (\ref{eq: R1_T2}), (\ref{eq: R1_n13=0_T2}), (\ref{eq: R1_n123=0_T2}) and (\ref{eq:R1_0_T2}). The expression $R_{2,(2)}$ is in Eqs. (\ref{eq: R2_T2}), (\ref{eq: R2_n13=0_T2}), (\ref{eq: R2_n123=0_T2}) and (\ref{eq:R2_0_T2}). The expression $R_{3,(2)}$ is in Eqs. (\ref{eq: R3_T2}), (\ref{eq: R3_n13=0_T2}), (\ref{eq: R3_n123=0_T2}) and (\ref{eq:R3_0_T2}). The expression $R_{4,(2)}$ is in Eqs. (\ref{eq:R5,2}) and (\ref{eq:R5_0,2}). The expression $R_{5,(2)}$ is in Eqs. (\ref{eq: R5_n123=0_T2}) and (\ref{eq:R5_0_T2}). The expression $R_{6,(2)}$ is in Eqs. (\ref{eq:R6,2}) and (\ref{eq:R6_0,2}). The expression $R_{7,(2)}$ is in Eqs. (\ref{eq: R7_T2}), (\ref{eq: R7_n13=0_T2}), (\ref{eq: R7_n123=0_T2}) and (\ref{eq:R7_0_T2}). The expression $R_{8,(2)}$ is in Eqs. (\ref{eq:R8,2}), and (\ref{eq:R8_0,2}).

\subsection{Concluding Remarks on our 4PCF Model}

In this work, we have developed the first model of the redshift-space galaxy 4PCF. This model will enable future studies of the BAO features in the 4PCF, which this work is the first to identify. The 4PCF should contain these features as do the 3PCF \cite{LadoBAO, MorescoBAO, SlepianBAO, modeling3PCF} and 2PCF \cite{PeeblesBAO, Eisen2PCF, Cole05}. If detected at high significance, BAO features in the 4PCF can be used as a standard ruler exactly as is already done with the 2PCF \cite{Eisen2PCF, Cole05} and 3PCF \cite{DetectionBAO3pcf}. 

Given the large galaxy and quasar samples DESI and other upcoming spectroscopic datasets will offer, it will be desirable to exploit the 4PCF as an additional tool for constraining the cosmic expansion history via the BAO method. Our model will also enable measuring the matter density of the Universe, the logarithmic derivative of the linear growth rate, $f$, and galaxy bias parameters up to third order in Eulerian Standard Perturbation Theory (SPT). After measuring these cosmological parameters, our model will enable the self-calibration of the parity-odd 4PCF as explained in $\S$\ref{sec:intro}. In short, our redshift-space galaxy 4PCF model offers many avenues for probing the early and late-time Universe with the wealth of data that will soon become available.

\acknowledgments
\qquad We thank the Slepian research group for useful discussions and comments on this work. WOL is extremely thankful to Jessica Chellino, Farshad Kamalinejad, Matthew Reinhard and James Sunseri for their insights and help to produce the plots presented on this paper. We also thank Bob Cahn, David Schlegel, Ben Sherwin, and Victoria Williamson. This publication was made possible through the support of Grant 63041 from the John Templeton Foundation. The opinions expressed in this publication are those of the author(s) and do not necessarily reflect the views of the John Templeton Foundation. JH has received funding from the European Union’s Horizon 2020 research and innovation program under the Marie Sk\l{}odowska-Curie grant agreement No. 101025187.

\appendix

\section{Computing the Tree-Level Trispectrum \textit{Ab Initio}}\label{sec:Tree-Level_Trispectrum}
We compute the tree-level trispectrum in the same manner as in Appendix B of \cite{Gualdi3} and indicate where we find an incorrect result in the trispectrum computation of \cite{Gualdi1, Gualdi2, Gualdi3}. We start by expanding the trispectrum in a perturbative series up to $O((\tilde{\delta}^{(1)})^8)$:
\begin{align}
T(\mathbf{k}_1,\mathbf{k}_2,\mathbf{k}_3,\mathbf{k}_4) = T_{1111}(\mathbf{k}_1,\mathbf{k}_2,\mathbf{k}_3,\mathbf{k}_4) + T_{1122}(\mathbf{k}_1,\mathbf{k}_2,\mathbf{k}_3,\mathbf{k}_4) + T_{1113}(\mathbf{k}_1,\mathbf{k}_2,\mathbf{k}_3,\mathbf{k}_4).
\end{align}
The first term above corresponds to a product of power spectrum via Wick's theorem since all the density contrasts in it are Gaussian random fields. The connected tree-level trispectrum is therefore:
\begin{align}
T_{\rm C}(\mathbf{k}_1,\mathbf{k}_2,\mathbf{k}_3,\mathbf{k}_4) =  T_{1122}(\mathbf{k}_1,\mathbf{k}_2,\mathbf{k}_3,\mathbf{k}_4) + T_{1113}(\mathbf{k}_1,\mathbf{k}_2,\mathbf{k}_3,\mathbf{k}_4).
\end{align}

Comparing the above equation with our perturbative expansion in Eq. (\ref{eq:Trispectrum}) (and dropping the subscripts for brevity) we find:

\begin{align}\label{eq:T3_Appendix}
(2\pi)^{3}\delta_{\rm D}^{[3]}(\mathbf{k}_{1234}) T_{1113}(\mathbf{k}_1,\mathbf{k}_2,\mathbf{k}_3,\mathbf{k}_4) = \left<\widetilde{\delta}^{(1)}(\mathbf{k}_1)\widetilde{\delta}^{(1)}(\mathbf{k}_2)\widetilde{\delta}^{(1)}(\mathbf{k}_3)\widetilde{\delta}^{(3)}(\mathbf{k}_4) \right> + 3 \;{\rm perm.}
\end{align}

\begin{align}\label{eq:T2_Appendix}
(2\pi)^{3}\delta_{\rm D}^{[3]}(\mathbf{k}_{1234}) T_{1122}(\mathbf{k}_1,\mathbf{k}_2,\mathbf{k}_3,\mathbf{k}_4) = \left<\widetilde{\delta}^{(1)}(\mathbf{k}_1)\widetilde{\delta}^{(1)}(\mathbf{k}_2)\widetilde{\delta}^{(2)}(\mathbf{k}_3)\widetilde{\delta}^{(2)}(\mathbf{k}_4) \right> + 5\; {\rm perm.}
\end{align}. 

We proceed by computing the first term in the right-hand side of Eq. (\ref{eq:T3_Appendix}) using our model of the perturbative redshift-space galaxy density contrast Eq. (\ref{eq:delta_sg_z_kernel}):

\begin{align}\label{eq:delta_T3_Before_Wicks}
&\left<\widetilde{\delta}^{(1)}(\mathbf{k}_1)\widetilde{\delta}^{(1)}(\mathbf{k}_2)\widetilde{\delta}^{(1)}(\mathbf{k}_3)\widetilde{\delta}^{(3)}(\mathbf{k}_4) \right> \nonumber \\ 
& =  Z^{(1)}(\mathbf{k}_1)Z^{(1)}(\mathbf{k}_2)Z^{(1)}(\mathbf{k}_3)\frac{1}{(2\pi)^6}  \int d^3\mathbf{q}_1\;d^3\mathbf{q}_2\;d^3\mathbf{q}_3 \;\delta_{\rm D}^{[3]}(\mathbf{k}_4 - \mathbf{q}_1- \mathbf{q}_2- \mathbf{q}_3) Z^{(3)}_{\rm ns}(\mathbf{q}_1;\mathbf{q}_2,\mathbf{q}_3)  \nonumber \\ 
&  \times \left<\widetilde{\delta}^{(1)}(\mathbf{q}_1)\widetilde{\delta}^{(1)}(\mathbf{q}_2)\widetilde{\delta}^{(1)}(\mathbf{q}_3)\widetilde{\delta}^{(1)}(\mathbf{k}_1)\widetilde{\delta}^{(1)}(\mathbf{k}_2)\widetilde{\delta}^{(1)}(\mathbf{k}_3) \right>.
\end{align}
Before proceeding with the integrals we use Wick's theorem to evaluate the ensemble average of the product of our six Gaussian random fields:

\begin{align}\label{eq:delta_mistake_found}
&\left<\widetilde{\delta}^{(1)}(\mathbf{q}_1)\widetilde{\delta}^{(1)}(\mathbf{q}_2)\widetilde{\delta}^{(1)}(\mathbf{q}_3)\widetilde{\delta}^{(1)}(\mathbf{k}_1)\widetilde{\delta}^{(1)}(\mathbf{k}_2)\widetilde{\delta}^{(1)}(\mathbf{k}_3) \right>  \nonumber \\
& = \left< \widetilde{\delta}^{(1)}(\mathbf{q}_1) \widetilde{\delta}^{(1)}(\mathbf{k}_1)\right>\left< \widetilde{\delta}^{(1)}(\mathbf{q}_2) \widetilde{\delta}^{(1)}(\mathbf{k}_2)\right>\left< \widetilde{\delta}^{(1)}(\mathbf{q}_3) \widetilde{\delta}^{(1)}(\mathbf{k}_3)\right> + 5 \; {\rm perm.}\nonumber \\
& = (2\pi)^{9}\delta_{\rm D}^{[3]}(\mathbf{k}_1+\mathbf{q}_1)\delta_{\rm D}^{[3]}(\mathbf{k}_2+\mathbf{q}_2)\delta_{\rm D}^{[3]}(\mathbf{k}_3+\mathbf{q}_3) P(k_1)P(k_2)P(k_3) + 5 \; {\rm perm.}
\end{align}
We note there are a total of  15 possible ways for how to pair the wave vectors in the first equality, however only the six permutations shown above contribute to the tree-level term. The other permutations lead to a one-loop contribution as shown in Figure \ref{fig:Wick's_Theorem_diagram}; \textit{e.g.,}  $\left<\widetilde{\delta}^{(1)}(\mathbf{k}_1)\widetilde{\delta}^{(1)}(\mathbf{k}_2)\right> \left<\widetilde{\delta}^{(1)}(\mathbf{k}_3)\widetilde{\delta}^{(1)}(\mathbf{q}_3)\right> \left<\widetilde{\delta}^{(1)}(\mathbf{q}_1)\widetilde{\delta}^{(1)}(\mathbf{q}_2)\right> $. Also, since the $Z_{\rm ns}^{(3)}$ kernel breaks the symmetry between the wave-vector in the first argument and the wave-vectors in the second and third argument, any permutation that changes the position of the first wave-vector to be as either second or third argument will lead to a different interpretation of the $Z_{\rm ns}^{(3)}$---the configuration space results of $\S$\ref{Section T3} remain unchanged, except they will now be evaluated at different arguments as shown in Eq. (\ref{4PCF_Result_R3111})---which is where we find a discrepancy with \cite{Gualdi1, Gualdi2, Gualdi3}. The authors multiply Eq. (\ref{eq:delta_mistake_found}) by a factor of 6 to account for all permutations, but they do not consider the asymmetry of the $Z_{\rm ns}^{(3)}$ kernel. Since it is not symmetrized, we only can interchange the second and third wave-vectors, giving rise to a factor of two, not six. Therefore, we need to allow all three options for the first argument; each has two orderings of the second and third, resulting in six overall permutations. However, with the non-symmetric kernel, the expressions for each permutation are different depending on the first argument. Substituting the above result into Eq. (\ref{eq:delta_T3_Before_Wicks}) and computing the intermediate momentum variable integrals, we find the right-hand side of Eq. (\ref{eq:T3_Appendix}) to be:  

\begin{align}\label{eq:delta_T3_after_wicks}
&\left<\widetilde{\delta}^{(1)}(\mathbf{k}_1)\widetilde{\delta}^{(1)}(\mathbf{k}_2)\widetilde{\delta}^{(1)}(\mathbf{k}_3)\widetilde{\delta}^{(3)}(\mathbf{k}_4) \right> + 3 \;{\rm perm.} \nonumber \\ 
&= 2 Z^{(1)}(\mathbf{k}_1)Z^{(1)}(\mathbf{k}_2)Z^{(1)}(\mathbf{k}_3) Z_{\rm ns}^{(3)}(\mathbf{k}_1;\mathbf{k}_2,\mathbf{k}_3)P(k_1)P(k_2)P(k_{3})\;(2\pi)^3 \delta_{\rm D}^{[3]}(\mathbf{k}_{1234}) \nonumber \\
& \quad + 11 \;{\rm perm.},
\end{align}
which agrees with our result in Eq. (\ref{eq:T3}). 

\begin{figure}[h]
\centering
\includegraphics[scale=0.25]{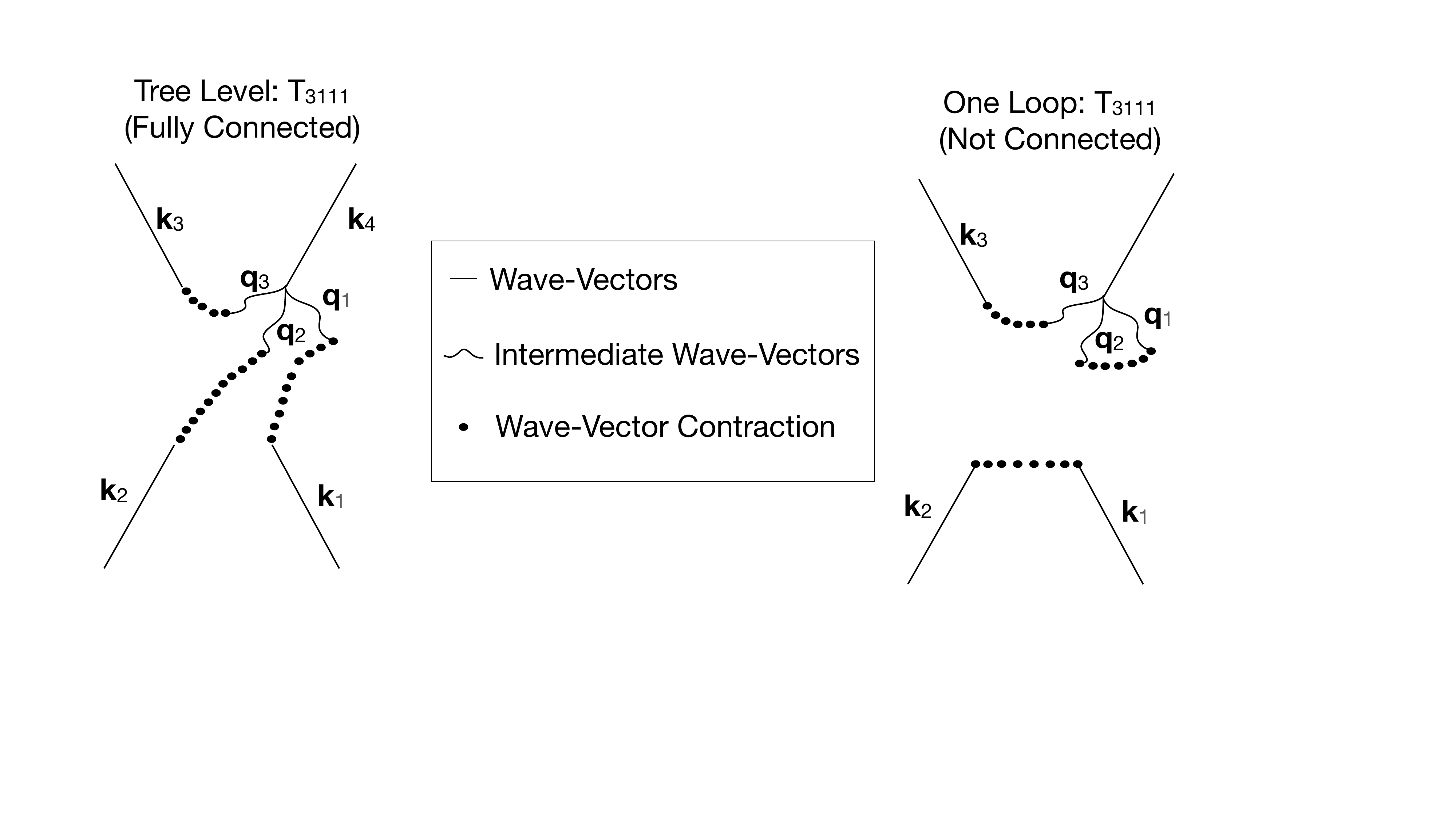}
\caption{Depiction of the Wick's theorem structure of our calculation via Feynman diagrams. The left diagram represents the connected tree-level contraction of $\left<\widetilde{\delta}^{(1)}(\mathbf{k}_1) \widetilde{\delta}^{(1)}(\mathbf{k}_2)\widetilde{\delta}^{(1)}(\mathbf{k}_3)\widetilde{\delta}^{(3)}(\mathbf{k}_4)\right>$ for $T_{3111}$. The right diagram represents a non-connected one-loop contraction of $\left<\widetilde{\delta}^{(1)}(\mathbf{k}_1) \widetilde{\delta}^{(1)}(\mathbf{k}_2)\widetilde{\delta}^{(1)}(\mathbf{k}_3)\widetilde{\delta}^{(3)}(\mathbf{k}_4)\right>$ for $T_{3111}$. The wording "Not Connected" in the right diagram matches the usage in field theory for diagrams in which the wave-vectors are not fully connected, and should not be confused with the cosmology meaning, where this phrase can refer to the purely Gaussian Random Field contributions to a correlation function or polyspectrum. The straight lines represent the wave-vectors, the wiggly lines the intermediate wave-vectors, and the dots the wave-vectors being contracted via Wick's theorem. For example, the dots join the wave-vector $\mathbf{k}_3$ with the intermediate wave-vector $\mathbf{q}_3$ in the left-hand  diagram, indicating that we have the product $\left<\widetilde{\delta}^{(1)}(\mathbf{k}_3)\widetilde{\delta}^{(1)}(\mathbf{q}_3)\right>$ via Wick's theorem. We do not show the diagrams for the contractions entering $T_{2211}$, but they can be developed analogously to those displayed here.}
\label{fig:Wick's_Theorem_diagram}
\end{figure}

We continue with the first term in the right-hand side of Eq. (\ref{eq:T2_Appendix}). Here we have two second-order density contrasts, which gives:

\begin{align}\label{eq:delta_T2_before_Wicks}
&\left<\widetilde{\delta}^{(1)}(\mathbf{k}_1)\widetilde{\delta}^{(1)}(\mathbf{k}_2)\widetilde{\delta}^{(2)}(\mathbf{k}_3)\widetilde{\delta}^{(2)}(\mathbf{k}_4) \right> \nonumber \\ 
& =  Z^{(1)}(\mathbf{k}_1)Z^{(1)}(\mathbf{k}_2)\frac{1}{(2\pi)^6}  \int d^3\mathbf{q}_1\;d^3\mathbf{q}_2\;\delta_{\rm D}^{[3]}(\mathbf{k}_3 - \mathbf{q}_1- \mathbf{q}_2)Z^{(2)}(\mathbf{q}_1,\mathbf{q}_2) \nonumber \\ 
&  \times \int d^3\mathbf{q}_3\;d^3\mathbf{q}_4\;\delta_{\rm D}^{[3]}(\mathbf{k}_4 - \mathbf{q}_3- \mathbf{q}_4) Z^{(2)}(\mathbf{q}_3,\mathbf{q}_4)  \nonumber \\ 
&  \times \left<\widetilde{\delta}^{(1)}(\mathbf{q}_1)\widetilde{\delta}^{(1)}(\mathbf{q}_2)\widetilde{\delta}^{(1)}(\mathbf{q}_3)\widetilde{\delta}^{(1)}(\mathbf{q}_4)\widetilde{\delta}^{(1)}(\mathbf{k}_1)\widetilde{\delta}^{(1)}(\mathbf{k}_2) \right>. 
\end{align}
As before we now evaluate the ensemble average product of six Gaussian random fields with Wick's Theorem: 

\begin{align}\label{eq:delta_product_T2211}
&\left<\widetilde{\delta}^{(1)}(\mathbf{q}_1)\widetilde{\delta}^{(1)}(\mathbf{q}_2)\widetilde{\delta}^{(1)}(\mathbf{q}_3)\widetilde{\delta}^{(1)}(\mathbf{q}_4)\widetilde{\delta}^{(1)}(\mathbf{k}_1)\widetilde{\delta}^{(1)}(\mathbf{k}_2) \right> \nonumber \\
& = 4\times \left\{ \left< \widetilde{\delta}^{(1)}(\mathbf{q}_1) \widetilde{\delta}^{(1)}(\mathbf{k}_1)\right>\left< \widetilde{\delta}^{(1)}(\mathbf{q}_4) \widetilde{\delta}^{(1)}(\mathbf{k}_2)\right>\left< \widetilde{\delta}^{(1)}(\mathbf{q}_2) \widetilde{\delta}^{(1)}(\mathbf{q}_3)\right> \right.\nonumber \\ 
& \left. \quad + \left< \widetilde{\delta}^{(1)}(\mathbf{q}_1) \widetilde{\delta}^{(1)}(\mathbf{k}_2)\right>\left< \widetilde{\delta}^{(1)}(\mathbf{q}_4) \widetilde{\delta}^{(1)}(\mathbf{k}_1)\right>\left< \widetilde{\delta}^{(1)}(\mathbf{q}_2) \widetilde{\delta}^{(1)}(\mathbf{q}_3)\right> \right\} \nonumber \\
& = 4\times (2\pi)^{9} P(k_1)P(k_2)P(q_3) \left\{\delta_{\rm D}^{[3]}(\mathbf{k}_1+\mathbf{q}_1)\delta_{\rm D}^{[3]}(\mathbf{k}_2+\mathbf{q}_4)\delta_{\rm D}^{[3]}(\mathbf{q}_2+\mathbf{q}_3) \right. \nonumber \\ 
& \left. \quad + \delta_{\rm D}^{[3]}(\mathbf{k}_2+\mathbf{q}_1)\delta_{\rm D}^{[3]}(\mathbf{k}_1+\mathbf{q}_4)\delta_{\rm D}^{[3]}(\mathbf{q}_2+\mathbf{q}_3)\right\}. 
\end{align}
As with the $T_{3111}$ term, we note there is another possible pairing of the wave vectors in the first equality, $\left<\widetilde{\delta}^{(1)}(\mathbf{k}_1)\widetilde{\delta}^{(1)}(\mathbf{k}_2)\right> \left<\widetilde{\delta}^{(1)}(\mathbf{q}_1)\widetilde{\delta}^{(1)}(\mathbf{q}_3)\right> \left<\widetilde{\delta}^{(1)}(\mathbf{q}_2)\widetilde{\delta}^{(1)}(\mathbf{q}_4)\right> $. However, as shown in Figure \ref{fig:Wick's_Theorem_diagram} for the $T_{3111}$ term, this contribution leads to a non-connected one-loop. Unlike the $Z_{\rm ns}^{(3)}$ term, $Z^{(2)}$ is symmetric, allowing us to perform all the internal permutations and include the factor of four as in \cite{Gualdi1, Gualdi2, Gualdi3}. Inserting Eq. (\ref{eq:delta_product_T2211}) into Eq. (\ref{eq:delta_T2_before_Wicks}), we find the right-hand side of Eq. (\ref{eq:T2_Appendix}) to be: 

\begin{align}
&\left<\widetilde{\delta}^{(1)}(\mathbf{k}_1)\widetilde{\delta}^{(1)}(\mathbf{k}_2)\widetilde{\delta}^{(2)}(\mathbf{k}_3)\widetilde{\delta}^{(2)}(\mathbf{k}_4) \right> + 5\; {\rm perm.}\nonumber \\ 
&= 4 \times (2\pi)^3 \delta_{\rm D}^{[3]}(\mathbf{k}_{123})Z^{(1)}(\mathbf{k}_1)Z^{(1)}(\mathbf{k}_2)P(k_1)P(k_2)\left\{Z^{(2)}(\mathbf{k}_1,-\mathbf{k}_{13}) Z^{(2)}(\mathbf{k}_2,\mathbf{k}_{13})P(k_{13})\right. \nonumber \\ 
&\left. \quad + Z^{(2)}(\mathbf{k}_1,-\mathbf{k}_{14}) Z^{(2)}(\mathbf{k}_2,\mathbf{k}_{14})P(k_{14}) \right\} + 5\; {\rm perm.},
\end{align}
which agrees with our result in Eq. (\ref{eq:T2}). 

\section{Conventions of the Paper}\label{sec:Conventions}
\subsection{Fourier Transform}
We use the convention for the inverse Fourier transform:

\begin{align}
    F(\mathbf{r}) = \int \frac{d^{3}\mathbf{k}}{(2\pi)^{3}} F(\mathbf{k})e^{-i\mathbf{k}\cdot\mathbf{r}},
\end{align}
which means the forward Fourier transform is: 

\begin{align}
    F(\mathbf{k}) = \int d^{3}\mathbf{r}\; F(\mathbf{r}) e^{i\mathbf{k}\cdot\mathbf{r}}. 
\end{align}

\subsection{4PCF Definitions}

\subsubsection{Definition of Arguments}
We present how the vectors $\mathbf{r}_1,\; \mathbf{r}_2$ and $\mathbf{r}_3$ are defined from the inverse Fourier transform of the trispectrum and how to cyclically sum all the terms of the 4PCF. To do this, we will use Eq. (\ref{eq:delta_T3_after_wicks}) as an example. The right-hand side of Eq. (\ref{eq:delta_T3_after_wicks}) is composed of two terms: products of RSD kernels and power spectra---with wave vectors $\mathbf{k}_1, \;\mathbf{k}_2$ and $\mathbf{k}_3$ as arguments---times a Dirac delta function of all the wave-vectors, and eleven permutations of this term. To simplify this example, let us define the product of RSD kernels and power spectra in the first term on the right-hand side (RHS) of Eq. (\ref{eq:delta_T3_after_wicks}) as:

\begin{align}
\tau(\mathbf{k}_1,\mathbf{k}_2, \mathbf{k}_3)_{\rm pre.} \equiv 2 Z^{(1)}(\mathbf{k}_1)Z^{(1)}(\mathbf{k}_2)Z^{(1)}(\mathbf{k}_3) Z_{\rm ns}^{(3)}(\mathbf{k}_1;\mathbf{k}_2,\mathbf{k}_3)P(k_1)P(k_2)P(k_{3}),
\end{align}
where the subscript pre. indicates $\tau$ does not account for any permutations. Taking the inverse Fourier transform of the first term in the RHS of Eq. (\ref{eq:delta_T3_after_wicks}), we find: 

\begin{align}
&{\rm FT}^{-1}\left({\rm R.H.S}\right)_{\rm pre.}\bigg{|}_{\rm Eq. (\ref{eq:delta_T3_after_wicks})} = \int\frac{d^{3}\mathbf{k}_{1}d^{3}\mathbf{k}_{2}d^{3}\mathbf{k}_{3}d^{3}\mathbf{k}_{4}}{(2\pi)^{12}} e^{-i\sum_{j=1}^{4} \mathbf{k}_{j}\cdot\mathbf{x}_{j}} \; \tau(\mathbf{k}_1,\mathbf{k}_2, \mathbf{k}_3)_{\rm pre.} \delta_{\rm D}^{[3]}(\mathbf{k}_{1234})\\ \nonumber
&  = \int \frac{d^{3}\mathbf{k}_{1}d^{3}\mathbf{k}_{2}d^{3}\mathbf{k}_{3}}{(2\pi)^{9}} e^{-i \mathbf{k}_{1}\cdot(\mathbf{x}_{1}-\mathbf{x}_4)} e^{-i \mathbf{k}_{2}\cdot(\mathbf{x}_{2}-\mathbf{x}_4)} e^{-i \mathbf{k}_{1}\cdot(\mathbf{x}_{3}-\mathbf{x}_4)} \; \tau(\mathbf{k}_1,\mathbf{k}_2, \mathbf{k}_3)_{\rm pre.} \\ \nonumber 
& \equiv \int \frac{d^{3}\mathbf{k}_{1}d^{3}\mathbf{k}_{2}d^{3}\mathbf{k}_{3}}{(2\pi)^{9}} e^{-i \mathbf{k}_{1}\cdot \mathbf{r}_{1}} e^{-i \mathbf{k}_{2}\cdot \mathbf{r}_{2}} e^{-i \mathbf{k}_{1}\cdot \mathbf{r}_{3}} \; \tau(\mathbf{k}_1,\mathbf{k}_2, \mathbf{k}_3)_{\rm pre.}. 
\end{align}
To obtain the second equality, we used the Dirac delta function to set $\mathbf{k}_4 = -(\mathbf{k}_1 + \mathbf{k}_2 + \mathbf{k}_3)$. In the third line, we then defined our vector arguments of the 4PCF as $\mathbf{r}_1 \equiv \mathbf{x}_{1}-\mathbf{x}_4 $, $\mathbf{r}_2 \equiv \mathbf{x}_{2}-\mathbf{x}_4 $, and $\mathbf{r}_3 \equiv \mathbf{x}_{3}-\mathbf{x}_4$. 

\subsubsection{Definition of Permutations}
To understand how to permute all the terms entering the trispectrum, we must first realize the permutation of $\tau(\mathbf{k}_1,\mathbf{k}_2, \mathbf{k}_3)$ implies a change of its arguments to other wave vectors such as $\tau_{\rm perm.} (\mathbf{k}_1,\mathbf{k}_2, \mathbf{k}_4)$ or simply the position of the arguments $\tau_{\rm perm.}(\mathbf{k}_1, \mathbf{k}_3, \mathbf{k}_2)$. If we continue using Eq. (\ref{eq:delta_T3_after_wicks}) as an example, $\tau_{\rm perm.}$ corresponds to any of the eleven permutations of the second term. We choose to analyze and evaluate $\tau_{\rm perm.} (\mathbf{k}_1,\mathbf{k}_2, \mathbf{k}_4)$: 

\begin{align}
&{\rm FT}^{-1}\left({\rm R.H.S}\right)_{\rm perm.}\bigg{|}_{\rm Eq. (\ref{eq:delta_T3_after_wicks})} = \int\frac{d^{3}\mathbf{k}_{1}d^{3}\mathbf{k}_{2}d^{3}\mathbf{k}_{3}d^{3}\mathbf{k}_{4}}{(2\pi)^{12}} e^{-i\sum_{j=1}^{4} \mathbf{k}_{j}\cdot\mathbf{x}_{j}} \; \tau_{\rm perm.} (\mathbf{k}_1,\mathbf{k}_2, \mathbf{k}_4) \delta_{\rm D}^{[3]}(\mathbf{k}_{1234})\\ \nonumber
&  = \int \frac{d^{3}\mathbf{k}_{1}d^{3}\mathbf{k}_{2}d^{3}\mathbf{k}_{4}}{(2\pi)^{9}} e^{-i \mathbf{k}_{1}\cdot(\mathbf{x}_{1}-\mathbf{x}_3)} e^{-i \mathbf{k}_{2}\cdot(\mathbf{x}_{2}-\mathbf{x}_3)} e^{-i \mathbf{k}_{4}\cdot(\mathbf{x}_{4}-\mathbf{x}_3)} \; \tau_{\rm perm.} (\mathbf{k}_1,\mathbf{k}_2, \mathbf{k}_4)\\ \nonumber
& = \int \frac{d^{3}\mathbf{k}_{1}d^{3}\mathbf{k}_{2}d^{3}\mathbf{k}_{4}}{(2\pi)^{9}} e^{-i \mathbf{k}_{1}\cdot (\mathbf{r}_{1}-\mathbf{r}_3)} e^{-i \mathbf{k}_{2}\cdot (\mathbf{r}_{2}-\mathbf{r}_3)} e^{i \mathbf{k}_{4}\cdot \mathbf{r}_{3}} \; \tau_{\rm perm.}(\mathbf{k}_1,\mathbf{k}_2, \mathbf{k}_4). 
\end{align}

The above result indicates that performing any permutation of the wave-vectors leads to a different combination of vectors. From the example above we have the following transformation: $R(\mathbf{r}_1, \mathbf{r}_2, \mathbf{r}_3)  \rightarrow R(\mathbf{r}_1 - \mathbf{r}_3, \mathbf{r}_2 - \mathbf{r}_3, - \mathbf{r}_3)$; where $R$ represents the result of taking the inverse Fourier transform of $\tau$. 

\section{Decoupling Denominators in Our Integrands}
\qquad To decouple the denominator we start by taking the Fourier transform of the most general $r^{n}$ with the power n, which can be any integer. Once we have this Fourier transform, we will express the resulting wave vector in terms of the Fourier transform, ultimately allowing us to decouple the denominator. To accomplish this task, we will use the same argument that classical electromagnetism uses to take the Fourier Transform of the Coulomb potential; the Fourier transform of the Yukawa potential in the limit $\lambda \rightarrow 0$ \cite{QFT, EM_Jackson}. 

First, we divide the work for $n< 0$ and $n> 0$. We do this because when $n< 0$, $y=r^{n}$ diverges as $r\rightarrow0$, but approaches zero as $r\rightarrow \infty$. If we examine the exponent on the Yukawa potential $e^{-\lambda r}$, we observe the exact same behavior (with the behavior shifted across the $y$ axis). Therefore, the product of $r^{n}\times e^{-\lambda r}$ only serves as a re-scaling for $r^{n}$ when $n<0$, simplifying the calculation of the Fourier transform. 

Hence, to find the Fourier transform of $r^{n}$ when $n>0$, we need to find the appropriate re-scaling. $y=r^n$ goes to $\infty$ as $r\rightarrow \infty$, the same as $e^{\lambda r}$. Therefore, when we analyze $r^{n}$ for $n>0$, we will use $e^{\lambda r}$ with a positive exponent instead of a negative one. The case $n=0$ we do not evaluate since this is simply a Dirac delta function. 

\subsection{Fourier Transform of $r^n$}
{\bf i)} $n<0$:
Here we take the Fourier transform of $r^n$ for $n<0$ with the appropriate re-scaling: 
\\
\begin{align}
&{\rm FT} \left\{ e^{-\lambda r} r^n\right\} (k) = \int d^{3}\mathbf{r}\;r^{n} e^{-\lambda r} e^{i\mathbf{k\cdot r}}  \nonumber \\ 
& \qquad \qquad \qquad = \int_{0}^{2\pi} d\phi  \int_{0}^{\pi}\sin \theta \; d\theta  \int_{0}^{\infty} dr \; r^{n+2} e^{-\lambda r} e^{i k r \cos \theta}  \;  \nonumber \\ 
& \qquad \qquad \qquad = \frac{2\pi}{ik} \int_{0}^{\infty} dr\; r^{n+1} e^{-\lambda r} \left[ e^{ikr} - e^{-ikr}\right] \nonumber \\ 
& \qquad \qquad \qquad =  \frac{2\pi}{ik} \left\{ \frac{(-1)^{n+2}\Gamma(n+2)}{(ik-\lambda)^{n+2}} - \frac{\Gamma(n+2)}{(ik + \lambda)^{n+2}} \right\}. 
\end{align}

For $\Gamma(n+2)$ to be finite, we must have $n > -2$, so our result only holds for $n=-1$. Taking the limit $\lambda \rightarrow 0$, we find: 

\begin{align}\label{eq:rn_gen}
{\rm FT}\left\{r^n\right\} (k) = \frac{2 \pi \Gamma (n+2)}{(ik)^{n+3}}\left[ (-1)^{n+2} - 1 \right]. 
\end{align}

Inserting $n=-1$ confirms that our result agrees with the Coulomb potential's Fourier transform \cite{EM_Jackson, QFT}:

\begin{align}
{\rm FT}\left\{r^{-1}\right\} (k) = \frac{4\pi}{k^2}. 
\end{align}

\noindent{\bf ii)} $n>0$:

For $n>0$, we take the Fourier transform of $r^n$ with the appropriate re-scaling:
\begin{align}
&\quad {\rm FT}\left\{ e^{\lambda r} r^n\right\} = \int r^{n} e^{\lambda r} e^{i\mathbf{k\cdot r}} d^{3}\mathbf{r} \nonumber \\ 
& \qquad \qquad \qquad = \int_{0}^{2\pi} d\phi \int_{0}^{\pi} \sin \theta \int_{0}^{\infty} dr \; r^{n+2} e^{\lambda r} e^{i\;kr\; \cos \theta} 
  \nonumber \\ 
& \qquad \qquad \qquad = \frac{2\pi}{ik} \int_{0}^{\infty} dr\; r^{n+1} e^{\lambda r} \left[ e^{ikr} - e^{-ikr}\right]  \nonumber \\ 
& \qquad \qquad \qquad =  \frac{2\pi}{ik} \left\{ \frac{(-1)^{n+2}\Gamma(n+2)}{(ik+\lambda)^{n+2}} - \frac{\Gamma(n+2)}{(ik - \lambda)^{n+2}} \right\}, 
\end{align}
which yields the same equation as Eq. (\ref{eq:rn_gen}) after taking the limit $\lambda \rightarrow 0 $. We note that for even $n$ (\textit{i.e.,} $n = 2,4,6,\cdots$), we obtain ${\rm FT}\left\{ r^n\right\} = 0 $ as expected. 

Given our results for $n<0$ and $n>0$, we re-write $n'=n+3$ and solve for $k^{-n^{'}}$ in terms of the Fourier transform: 

\begin{align}
\boxed{\frac{1}{k^{n'}} = \frac{i^{n'}}{2\pi \Gamma(n'-1) \left[ (-1)^{n'-1} - 1 \right]} \int d^{3}\mathbf{r}\;r^{n'-3} e^{i \mathbf{k \cdot r}} }.
\end{align}
After rewriting the dummy index, the result above only holds for $n'_{\rm even} \geq 2$. We note that, for the rest of this Appendix, we write $n'\rightarrow n$ for simplicity. 

\subsection{Decoupling Denominator Involving a Sum of Two Wave-Vectors}
With this result in hand, we now derive a general method for decoupling a denominator of the form $\mathbf{k = k_2 + k_3}$: 

\begin{align}
\frac{1}{k_{23}^{n_{23}}} = \frac{i^{n_{23}}}{2\pi \Gamma(n_{23}-1) \left[ (-1)^{n_{23}-1} - 1 \right]} \int d^{3}\mathbf{r}\;r^{n_{23}-3} e^{i \mathbf{k}_2 \cdot \mathbf{r}}e^{i \mathbf{k}_3 \cdot \mathbf{r}} 
\end{align}
with $k_{23}^{n_{23}} = \left | \mathbf{k}_2 + \mathbf{k}_3 \right |^{n_{23}}$. Using the plane-wave expansion on the exponents, we find:

\begin{align}\label{eq:k23}
&\frac{1}{k_{23}^{n_{23}}} = \frac{8\pi \; i^{n_{23}}}{ \Gamma(n_{23}-1) \left[ (-1)^{n_{23}-1} - 1 \right]} \sum_{\ell_2=0}^{\infty}\sum_{\ell_3=0}^{\infty}\sum_{m_2=-\ell_2}^{\ell_2}\sum_{m_3=\ell_3}^{\ell_3} (i)^{\ell_2 +\ell_3}  Y_{\ell_2 m_2}(\mathbf{\widehat{k}}_2)Y_{\ell_3 m_3}^{*}(\mathbf{\widehat{k}}_3) \nonumber \\
& \qquad \qquad \qquad \qquad \qquad \qquad \qquad \qquad \times \int  d^{3}\mathbf{r}\; j_{\ell_2}(k_2r)j_{\ell_3}(k_3r)Y_{\ell_2 m_2}^{*}(\mathbf{\widehat{r}})Y_{\ell_3 m_3}(\mathbf{\widehat{r}}) r^{n_{23}-3}  \nonumber \\ 
& \qquad  = \frac{8\pi \;i^{n_{23}}}{ \Gamma(n_{23}-1) \left[ (-1)^{n_{23}-1} - 1 \right]} \sum_{\ell_2=0}^{\infty}\sum_{m_2=-\ell_2}^{\ell_2} (-1)^{\ell_2} Y_{\ell_2 m_2}(\mathbf{\widehat{k}}_2)Y_{\ell_2 m_2}^{*}(\mathbf{\widehat{k}}_3) \nonumber \\
& \qquad \qquad \qquad \qquad \qquad \qquad \qquad \qquad \times \int_{0}^{\infty} dr\; j_{\ell_2}(k_2r)j_{\ell_2}(k_3r) r^{n_{23}-1}  \nonumber \\ 
& \qquad = \frac{8\pi \;i^{n_{23}}}{ \Gamma(n_{23}-1) \left[ (-1)^{n_{23}-1} - 1 \right]} \sum_{\ell_2} \sqrt{2\ell_2+1}\;\mathcal{P}_{\ell_2}(\mathbf{\widehat{k}}_2,\mathbf{\widehat{k}}_3) \nonumber \\
& \qquad \qquad \qquad \qquad \qquad \qquad \qquad \qquad \times \int_{0}^{\infty} dr \; j_{\ell_2}(k_2r)j_{\ell_2}(k_3r) r^{n_{23}-1}   \nonumber \\
& \qquad \equiv \frac{8\pi \;i^{n_{23}}}{ \Gamma(n_{23}-1) \left[ (-1)^{n_{23}-1} - 1 \right]} \sum_{\ell_2} s_{\ell_2}^{(\rm I)} \mathcal{P}_{\ell_2}(\mathbf{\widehat{k}}_2,\mathbf{\widehat{k}}_3) \int_{0}^{\infty}dr\; j_{\ell_2}(k_2r)j_{\ell_2}(k_3r) r^{n_{23}-1}. 
\end{align}
We have defined $s_{\ell_2}^{(\rm I)} $ in the last equality, and $\mathcal{P}_{\ell_2}(\mathbf{\widehat{k}}_2,\mathbf{\widehat{k}}_3) = \sqrt{4\pi}\;\mathcal{P}_{0,\ell_2,\ell_2}(\mathbf{\widehat{k}}_1,\mathbf{\widehat{k}}_2,\mathbf{\widehat{k}}_3)$ is defined by: 

\begin{align} \label{eq:iso_gen}
&\mathcal{P}_{j_1,j_2,j_3}(\mathbf{\widehat{k}}_1,\mathbf{\widehat{k}}_2,\mathbf{\widehat{k}}_3) = (-1)^{j_1+j_2+j_3}\sum_{m_{j_1} ,m_{j_2},m_{j_3}} \begin{pmatrix} j_1&j_2  &j_3  \\ m_{j_1}& m_{j_2} & m_{j_3} \\ \end{pmatrix} \nonumber \\
& \qquad \qquad \qquad\qquad \qquad \qquad \qquad \qquad \qquad \times Y_{j_1,m_{j_1}}(\mathbf{\widehat{k}}_1)Y_{j_2,m_{j_2}}(\mathbf{\widehat{k}}_2)Y_{j_3,m_{j_3}}(\mathbf{\widehat{k}}_3) 
 \end{align}
 where the $2\times 3$ matrix is a Wigner 3-$j$ symbol. The splitting approach here is motivated by a similar one used for the 3PCF in \cite{BaryonVelocity}.
 
\subsection{Decoupling Denominator Involving a Sum of Three Wave-Vectors}
\qquad Following the same approach, we can also derive an expression for a denominator involving the sum of three wave-vectors, $\mathbf{k}=\mathbf{k}_1+\mathbf{k}_2+\mathbf{k}_3$: 

\begin{align}\label{eq:k123}
&\frac{1}{k_{123}^{n_{123}}} = \frac{32 \pi^2 \; i^{n_{123}}}{ \Gamma(n_{123}-1) \left[ (-1)^{n_{123}-1} - 1 \right]} \sum_{\ell'_1,\ell'_2,\ell'_3} \mathcal{C}_{\ell'_1,\ell'_2,\ell'_3} \mathcal{P}_{\ell'_1,\ell'_2,\ell'_3}(\mathbf{\widehat{k}}_1,\mathbf{\widehat{k}}_2,\mathbf{\widehat{k}}_3) \nonumber \\ 
&\qquad \qquad \qquad \qquad \qquad \qquad  \times \int_{0}^{\infty}dr'\;r'^{n_{123}-1}\;j_{\ell'_1}(k_1r')j_{\ell'_2}(k_2r')j_{\ell'_3}(k_3r') 
\end{align}
where $\mathcal{C}_{\ell'_1,\ell'_2,\ell'_3}$ has the same definition as $\mathcal{C}_{j_1,j_2,j_3}$, which is given in Eq. (\ref{eq:C_cons}).

\section{Expanding a Dot Product into Isotropic Basis Functions}
Here we show how to convert a dot product of unit vectors, $\mathbf{\widehat{a}}$ and $\mathbf{\widehat{b}}$, raised to the power $n$, into a sum over the isotropic basis functions. We start by expressing it as a sum over Legendre polynomials:

\begin{align}\label{eq:dot_porduct_to_Legendre}
&(\mathbf{\widehat{a}}\cdot \mathbf{\widehat{b}})^{n} = \sum_{j=n,n-2,\cdots} \frac{n!(2j+1)}{2^{(n-j)/2}(\frac{1}{2}(n-j))!(j+n+1)!!}\mathcal{L}_{j}(\mathbf{\widehat{a}}\cdot \mathbf{\widehat{b}}) \nonumber \\
& \qquad \quad \; \equiv   \sum_{j} \overline{c}_{j}^{(n)} \mathcal{L}_{j}(\mathbf{\widehat{a}}\cdot \mathbf{\widehat{b}}),
\end{align}
where $\mathcal{L}_j$ denotes a Legendre polynomial of order $j$. Using the spherical harmonic addition theorem on Eq. (\ref{eq:dot_porduct_to_Legendre}), we find: 
\begin{align} \label{eq:exp_sph}
(\mathbf{\widehat{a}}\cdot \mathbf{\widehat{b}})^{n} = \sum_{j} \overline{c}_{j}^{(n)}  \frac{4\pi}{2j+1}\sum_{m_j=-j}^{j} Y_{j,m_j}(\mathbf{\widehat{a}}) Y_{j,m_j}^{*}(\mathbf{\widehat{b}}).  
\end{align}

Finally, using equation (3) of \cite{Iso_fun} to write the above result in terms of the 2-argument isotropic basis functions, we obtain: 

\begin{align}\label{eq:dot_to_iso_eq}
(\mathbf{\widehat{a}}\cdot \mathbf{\widehat{b}})^{n} \equiv 4\pi \sum_{j} (-1)^{j}\;\overline{c}_{j}^{(n)} \sqrt{2j+1}  \;\mathcal{P}_{j}(\mathbf{\widehat{a}},\mathbf{\widehat{b}}) \equiv 4 \pi \sum_{j}  \mathcal{P}_{j} (\mathbf{\widehat{a}},\mathbf{\widehat{b}}) c_{j}^{(n)}.  
\end{align}

\section{Reduction of Products of Isotropic Basis Functions}
We combine the products of two, three, and $N$ isotropic basis functions into a single 3-argument isotropic basis function. Let us mention that any 2-argument isotropic function can be turned into a 3-argument isotropic basis function since $\mathcal{P}_{J_2,J_3}(\mathbf{\widehat{k}}_2,\mathbf{\widehat{k}}_3) = \sqrt{4\pi} \;\mathcal{P}_{0,J_2,J_3}(\mathbf{\widehat{k}}_1,\mathbf{\widehat{k}}_2,\mathbf{\widehat{k}}_3)$, therefore the results presented on this appendix can be extrapolated to 2-argument isotropic functions. We will use the notation $\Lambda$ to refer to the set of angular momentum we are using on a given N-point isotropic basis; \textit{i.e.}, $\Lambda = \left\{ \ell_1,\ell_2,...,\ell_N \right\}$. Also, we use the notation $\mathbf{\widehat{K}}$ to refer to the set of unit wave-vectors; \textit{i.e.}, $\mathbf{\widehat{K}} = \left\{\mathbf{\widehat{k}}_1, \mathbf{\widehat{k}}_2,...,\mathbf{\widehat{k}}_N \right\}$. This is consistent with the notation and language as presented in \cite{Iso_fun}; however, in the work presented above, we show each angular momentum component explicitly; \textit{i.e.}, $\mathcal{P}_{\Lambda}(\mathbf{\widehat{K}})= \mathcal{P}_{J_1,J_2,J_3}(\mathbf{\widehat{k}}_1,\mathbf{\widehat{k}}_2,\mathbf{\widehat{k}}_3)$.
\\
\\
\subsection{Product of Two Isotropic Basis Functions}
We begin with the product of two 3-argument isotropic basis functions \cite{Iso_fun}:

\begin{align}\label{eq:Product_of_2_iso}
\mathcal{P}_{\Lambda^{(1)}}(\mathbf{\widehat{K}}) \mathcal{P}_{\Lambda^{(2)}}(\mathbf{\widehat{K}}) = \sum_{\Lambda^{(12)}} \varepsilon(\Lambda^{(12)}) \mathcal{G}_{\Lambda^{(1)} \Lambda^{(2)} \Lambda^{(12)}} \mathcal{P}_{\Lambda^{(12)}}(\mathbf{\widehat{K}}), 
\end{align}
$\varepsilon(\Lambda) \equiv (-1)^{\sum_i \Lambda_i}$, with $\Lambda_i$ referring to the different angular momenta involved. We proceed by multiplying both sides by $ \mathcal{P}_{\Lambda^{(12)}}^{*}(\mathbf{\widehat{K}})$ and integrating over their solid angle to find $\mathcal{G}_{\Lambda^{(1)} \Lambda^{(2)} \Lambda^{(12)}}$: 

\begin{align}
&\mathcal{G}_{\Lambda^{(1)} \Lambda^{(2)} \Lambda^{(12)}} = \int d\mathbf{\widehat{K}}\; \mathcal{P}_{\Lambda^{(1)}}(\mathbf{\widehat{K}}) \mathcal{P}_{\Lambda^{(2)}}(\mathbf{\widehat{K}})  \mathcal{P}_{\Lambda^{(12)}}(\mathbf{\widehat{K}}) \nonumber \\
&  =  \sum_{M^{(1)},M^{(2)},M^{(12)}} \mathcal{C}_{M^{(1)}}^{\Lambda^{(1)}} \mathcal{C}_{M^{(2)}}^{\Lambda^{(2)}} \mathcal{C}_{M^{(12)}}^{\Lambda^{(12)}} \nonumber \\ 
& \quad \times \int d\mathbf{\widehat{K}} \;\left[Y_{\Lambda_{1}^{(1)}M_{1}^{(1)}}(\mathbf{\widehat{k}}_1)Y_{\Lambda_{2}^{(1)}M_{2}^{(1)}}(\mathbf{\widehat{k}}_2)Y_{\Lambda_{3}^{(1)}M_{3}^{(1)}}(\mathbf{\widehat{k}}_3)\right] \nonumber \\ 
& \quad \times \left[Y_{\Lambda_{1}^{(2)}M_{1}^{(2)}}(\mathbf{\widehat{k}}_1)Y_{\Lambda_{2}^{(2)}M_{2}^{(2)}}(\mathbf{\widehat{k}}_2)Y_{\Lambda_{3}^{(2)}M_{3}^{(2)}}(\mathbf{\widehat{k}}_3)\right] \nonumber \\ 
&\quad \times  \left[Y_{\Lambda_{1}^{(12)}M_{1}^{(12)}}(\mathbf{\widehat{k}}_1)Y_{\Lambda_{2}^{(12)}M_{2}^{(12)}}(\mathbf{\widehat{k}}_2)Y_{\Lambda_{3}^{(12)}M_{3}^{(12)}}(\mathbf{\widehat{k}}_3)\right]. 
\end{align}
We have used $ \mathcal{P}_{\Lambda}^{*} = \varepsilon(\Lambda)\mathcal{P}_{\Lambda}$ and defined:

\begin{align}
\mathcal{C}_{M}^{\Lambda} \equiv (-1)^{\Lambda_1 + \Lambda_2 + \Lambda_3} \begin{pmatrix}
\Lambda_1 & \Lambda_2 & \Lambda_3 \\
M_1 & M_2 & M_3 \\
\end{pmatrix}.
\end{align}

To continue, we will integrate only over one of the angular components---since all components will yield the same result---and later will put the full result together: 

\begin{align}
\int d\mathbf{\widehat{k}}_1\; Y_{\Lambda_{1}^{(1)}M_{1}^{(1)}}(\mathbf{\widehat{k}}_1) Y_{\Lambda_{1}^{(2)}M_{1}^{(2)}}(\mathbf{\widehat{k}}_1) Y_{\Lambda_{1}^{(12)}M_{1}^{(12)}}(\mathbf{\widehat{k}}_1) = G_{M_{1}^{(1)} M_{1}^{(2)} M_{1}^{(12)}}^{\Lambda_{1}^{(1)} \Lambda_{1}^{(2)} \Lambda_{1}^{(12)}},
\end{align}
with $G$ being the Gaunt integral, defined as: 

\begin{align}
&G_{M_{1}^{(1)} M_{1}^{(2)} M_{1}^{(12)}}^{\Lambda_{1}^{(1)} \Lambda_{1}^{(2)} \Lambda_{1}^{(12)}} \equiv\\
&\qquad\qquad\sqrt{\frac{(2\Lambda_{1}^{(1)}+1)(2\Lambda_{1}^{(2)}+1)(2\Lambda_{1}^{(12)}+1)}{4\pi}}  \begin{pmatrix}
\Lambda_1^{(1)} & \Lambda_1^{(2)} & \Lambda_1^{(12)} \\
0 & 0 & 0\\
\end{pmatrix} \begin{pmatrix}
\Lambda_1^{(1)} & \Lambda_1^{(2)} & \Lambda_1^{(12)} \\
M_1^{(1)} & M_1^{(2)} & M_1^{(12)} \\
\end{pmatrix}.\nonumber
\end{align}
Therefore, we find that:

\begin{align}
\mathcal{G}_{\Lambda^{(1)}\Lambda^{(2)}\Lambda^{(12)}} = \sum_{M^{(1)},M^{(2)},M^{(12)}} \mathcal{C}_{M^{(1)}}^{\Lambda^{(1)}} \mathcal{C}_{M^{(2)}}^{\Lambda^{(2)}} \mathcal{C}_{M^{(12)}}^{\Lambda'} \prod_{j=1}^{3} G_{M_{j}^{(1)} M_{j}^{(2)} M_{j}^{(12)}}^{\Lambda_{j}^{(1)} \Lambda_{j}^{(2)} \Lambda_{j}^{(12)}}. 
\end{align}

\subsection{Product of Three Isotropic Basis Functions}
We will show how the product of three 3-argument isotropic basis functions can be combined into a single 3-argument isotropic basis function: 

\begin{align}\label{eq:Product_of_3_iso}
\mathcal{P}_{\Lambda^{(1)}}(\mathbf{\widehat{K}}) \mathcal{P}_{\Lambda^{(2)}}(\mathbf{\widehat{K}}) \mathcal{P}_{\Lambda^{(3)}}(\mathbf{\widehat{K}}) = \sum_{\Lambda^{(123)}} \mathcal{G}_{\Lambda^{(1)} \Lambda^{(2)}\Lambda^{(3)}  \Lambda^{(12)} \Lambda^{(123)}} \mathcal{P}_{\Lambda^{(123)}}(\mathbf{\widehat{K}}). 
\end{align}
\\
Applying the result from Eq. (\ref{eq:Product_of_2_iso}) on the first two isotropic basis functions, we find:

\begin{align}
\mathcal{P}_{\Lambda^{(1)}}(\mathbf{\widehat{K}}) \mathcal{P}_{\Lambda^{(2)}}(\mathbf{\widehat{K}}) \mathcal{P}_{\Lambda^{(3)}}(\mathbf{\widehat{K}}) = \sum_{\Lambda^{(12)}} \varepsilon(\Lambda^{(12)}) \mathcal{G}_{\Lambda^{(1)} \Lambda^{(2)}\Lambda^{(12)}} \mathcal{P}_{\Lambda^{(12)}}(\mathbf{\widehat{K}}) \mathcal{P}_{\Lambda^{(3)}}(\mathbf{\widehat{K}}).  
\end{align}
Applying Eq. (\ref{eq:Product_of_2_iso}) once again to the right-hand side, we find: 

\begin{align}
&\sum_{\Lambda^{(12)}} \varepsilon(\Lambda^{(12)}) \mathcal{G}_{\Lambda^{(1)} \Lambda^{(2)}\Lambda^{(12)}}^{(2)} \mathcal{P}_{\Lambda^{(12)}}(\mathbf{\widehat{K}}) \mathcal{P}_{\Lambda^{(3)}}(\mathbf{\widehat{K}}) \nonumber \\ &\qquad = \sum_{\Lambda^{(12)} \Lambda^{(123)}} \varepsilon(\Lambda^{(12)}) \varepsilon(\Lambda^{(123)}) \mathcal{G}_{\Lambda^{(1)} \Lambda^{(2)}\Lambda^{(12)}} \mathcal{G}_{\Lambda^{(12)} \Lambda^{(3)}\Lambda^{(123)}} \mathcal{P}_{\Lambda^{(123)}}(\mathbf{\widehat{K}}).  
\end{align}
Therefore, by defining:

\begin{align}
\mathcal{G}_{\Lambda^{(1)} \Lambda^{(2)}\Lambda^{(3)}  \Lambda^{(12)} \Lambda^{(123)}} \equiv \sum_{\Lambda^{(12)}} \varepsilon(\Lambda^{(12)}) \varepsilon(\Lambda^{(123)}) \mathcal{G}_{\Lambda^{(1)} \Lambda^{(2)}\Lambda^{(12)}} \mathcal{G}_{\Lambda^{(12)} \Lambda^{(3)}\Lambda^{(123)}},
\end{align}
 we obtain our desired result, Eq. (\ref{eq:Product_of_3_iso}). 

\subsection{Product of $N$ Isotropic Basis Functions}
With the above results for the product of three 3-argument isotropic basis functions, we see that a pattern emerges when solving for $\mathcal{G}$. From this pattern we obtain a simple equation for $\mathcal{G}$ that allows us to compute it for a product of $n$ 3-argument isotropic basis functions: 

\begin{align}\label{eq:Product_of_n_iso}
\mathcal{P}_{\Lambda^{(1)}}(\mathbf{\widehat{K}}) \times \cdots \times \mathcal{P}_{\Lambda^{(n)}}(\mathbf{\widehat{K}}) = \sum_{\Lambda^{(12\cdots n)}} \mathcal{G}_{\Lambda^{(1)} \cdots \Lambda^{(n)} \Lambda^{(12)}\cdots\Lambda^{(12\cdots n)}} \mathcal{P}_{\Lambda^{(12\cdots n)}}(\mathbf{\widehat{K}}). 
\end{align}
with $\mathcal{G}_{\Lambda^{(1)} \cdots \Lambda^{(n)} \Lambda^{(12)}\cdots\Lambda^{(12\cdots n)}}$ given by: 

\begin{align}
&\mathcal{G}_{\Lambda^{(1)} \cdots \Lambda^{(n)} \Lambda^{(12)}\cdots \Lambda^{(12\cdots n)}} = \varepsilon(\Lambda^{(12)})\times \cdots \times\varepsilon(\Lambda^{(12\cdots n)})  \\ 
& \qquad\qquad \qquad \qquad \qquad \times \sum_{\Lambda^{(12)},\cdots,\Lambda^{(12\cdots n-1)}} \mathcal{G}_{\Lambda^{(1)}\Lambda^{(2)}\Lambda^{(12)}}\times \cdots \times \mathcal{G}_{\Lambda^{(12 \cdots n-1)}\Lambda^{(n)} \Lambda^{(12  \cdots n)}}.\nonumber
\end{align}

\section{Splitting Products of Mixed-Space Isotropic Basis Functions}
From Eq. (\ref{eq:exp_expansion_iso}) we see that we can express the exponential term from the inverse Fourier transform as a product of three 2-argument isotropic basis functions. These isotropic basis functions have unit vectors and unit wave-vectors as arguments. Therefore, this section will demonstrate how to split the unit vector and unit wave-vector parts of these isotropic basis functions as:

\begin{align}\label{eq:product_iso_kr_into_k_r}
&\mathcal{P}_{\ell'_1}(\mathbf{\widehat{k}}_1,\mathbf{\widehat{r}}_1) \mathcal{P}_{\ell'_2}(\mathbf{\widehat{k}}_2,\mathbf{\widehat{r}}_2) \mathcal{P}_{\ell'_3}(\mathbf{\widehat{k}}_3,\mathbf{\widehat{r}}_3) = \nonumber \\ 
& \qquad \qquad \qquad \qquad \qquad \sum_{L_1,L_2,L_3}\Upsilon_{L_1,L_2,L_3} \mathcal{P}_{L_1,L_2,L_3}(\mathbf{\widehat{k}}_1,\mathbf{\widehat{k}}_2,\mathbf{\widehat{k}}_3)  \mathcal{P}_{L_1,L_2,L_3}(\mathbf{\widehat{r}}_1,\mathbf{\widehat{r}}_2,\mathbf{\widehat{r}}_3),
\end{align}
where

\begin{align}\label{eq:Upsilon_splitting}
&\Upsilon_{L_1,L_2,L_3} = \int d\Omega_{k_1}d\Omega_{r_1}d\Omega_{k_2}d\Omega_{r_2}d\Omega_{k_3}d\Omega_{r_3 }\;\mathcal{P}_{\ell'_1}(\mathbf{\widehat{k}}_1,\mathbf{\widehat{r}}_1) \mathcal{P}_{\ell'_2}(\mathbf{\widehat{k}}_2,\mathbf{\widehat{r}}_2) \mathcal{P}_{\ell'_3}(\mathbf{\widehat{k}}_3,\mathbf{\widehat{r}}_3) \nonumber \\ 
& \qquad \qquad \qquad \times \mathcal{P}_{L_1,L_2,L_3}^{*}(\mathbf{\widehat{k}}_1,\mathbf{\widehat{k}}_2,\mathbf{\widehat{k}}_3)  \mathcal{P}_{L_1,L_2,L_3}^{*}(\mathbf{\widehat{r}}_1,\mathbf{\widehat{r}}_2,\mathbf{\widehat{r}}_3). 
\end{align}
We then find: 

\begin{align}
& \Upsilon_{L_1,L_2,L_3} = \sum_{M_1,M_2,M_3}\frac{(-1)^{L_1+L_2+L_3}}{\sqrt{(2L_1+1)(2L_2+1)(2L_3+1)}}\begin{pmatrix} L_1 & L_2 &L_3  \\ M_1 &  M_2& M_3 \\ \end{pmatrix}^{2} \Kd_{L_1,\ell'_1} \Kd_{L_2,\ell'_2} \Kd_{L_3,\ell'_3}\nonumber \\ 
& \qquad \qquad = \quad \frac{(-1)^{L_1+L_2+L_3}}{\sqrt{(2L_1+1)(2L_2+1)(2L_3+1)}}, 
\end{align}
where $\Kd$ is the Kronecker delta, and the last equality holds by NIST DLMF 34.3.18.

\section{Averaging Over the Line of Sight}
Throughout this paper we have averaged over the line of sight given the isotropy of our Universe. Here we show how to write the result of this integral in terms of the 3-argument isotropic basis functions. We have:

\begin{align}\label{eq:avg_z}
&\left< \mu_{1}^{n_1}\mu_{2}^{n_2}\mu_{3}^{n_3} \right>_{\rm l.o.s} = \int d\Omega_z \; \mu_{1}^{n_1}\mu_{2}^{n_2}\mu_{3}^{n_3} \nonumber \\
& \qquad \qquad \qquad \quad  = (4 \pi)^{3} \sum_{j_1,j_2,j_3 =0}^{n_1,n_2,n_3} \sum_{m_{j_1} = -j_1,m_{j_2}= -j_2,m_{j_3}= -j_3}^{j_1,j_2,j_3} \frac{c_{j_1}^{(n_1)}c_{j_2}^{(n_2)}c_{j_3}^{(n_3)}}{(2j_1+1)(2j_2+1)(2j_3+1)} \nonumber \\
&\qquad \qquad \qquad \qquad \quad \times Y_{j_1,m_{j_1}}(\mathbf{\widehat{k}}_1)Y_{j_2,m_{j_2}}(\mathbf{\widehat{k}}_2)Y_{j_3,m_{j_3}}(\mathbf{\widehat{k}}_3)\int d\Omega_z \; Y_{j_1,m_{j_1}}^{*}(\mathbf{\widehat{z}})Y_{j_2,m_{j_2}}^{*}(\mathbf{\widehat{z}})Y_{j_3,m_{j_3}}^{*}(\mathbf{\widehat{z}})  \nonumber \\
& \qquad \qquad \qquad \quad = (4 \pi)^{3} \sum_{j_1,j_2,j_3 =0}^{n_1,n_2,n_3} \; \; \sum_{m_{j_1} = -j_1,m_{j_2}= -j_2,m_{j_3}= -j_3}^{j_1,j_2,j_3} \frac{c_{j_1}^{(n_1)}c_{j_2}^{(n_2)}c_{j_3}^{(n_3)}}{(2j_1+1)(2j_2+1)(2j_3+1)} \nonumber \\
& \qquad \qquad \qquad \qquad \quad \times Y_{j_1,m_{j_1}}(\mathbf{\widehat{k}}_1)Y_{j_2,m_{j_2}}(\mathbf{\widehat{k}}_2)Y_{j_3,m_{j_3}}(\mathbf{\widehat{k}}_3) G_{j_1 j_2 j_3}^{m_{j_1} m_{j_2} m_{j_3}}.    
\end{align}

In the second equality we have used the result Eq. (\ref{eq:exp_sph}) to express all the $\mu_{i}$ in terms of spherical harmonics. Next, we decomposed the Gaunt integral into 3-$j$ symbols to combine the spherical harmonics into an isotropic basis function:

\begin{align}
\left< \mu_{1}^{n_1}\mu_{2}^{n_2}\mu_{3}^{n_3} \right>_{\rm l.o.s} = (4 \pi)^{3} \sum_{j_1,j_2,j_3 }\mathcal{C}_{j_1,j_2,j_3}^{n_1,n_2,n_3} \mathcal{P}_{j_1,j_2,j_3}(\mathbf{\widehat{k}}_1,\mathbf{\widehat{k}}_2,\mathbf{\widehat{k}}_3),
\end{align}
where 
\begin{align}\label{eq:C_cons}
&\mathcal{C}_{j_1,j_2,j_3}^{n_1,n_2,n_3} \equiv \frac{c_{j_1}^{(n_1)}c_{j_2}^{(n_2)}c_{j_3}^{(n_3)}}{(2j_1+1)(2j_2+1)(2j_3+1)} \mathcal{C}_{j_1,j_2,j_3} \begin{pmatrix} j_1&j_2  &j_3  \\ 0& 0 &  0\\ \end{pmatrix} \nonumber \\ 
&\qquad \; \; \; \; \; \;= \frac{c_{j_1}^{(n_1)}c_{j_2}^{(n_2)}c_{j_3}^{(n_3)}}{\sqrt{4\pi (2j_1+1)(2j_2+1)(2j_3+1)}} \begin{pmatrix} j_1&j_2  &j_3  \\ 0& 0 &  0\\ \end{pmatrix}.
\end{align}

\section{Decoupling the Power Spectrum}
\qquad In this section, we demonstrate the decoupling of a power spectrum with the magnitude of the sum of two wave-vectors as an argument, $P(k_{13})$:

\begin{align}\label{eq:Decoupled_P(k13)}
&P(k_{13}) = \int_{0}^{\infty} ds \; s^{2} \; j_{0}(k_{13}s) \xi_0(s)\nonumber \\ 
& \qquad \quad  = \sum_{\ell''_1,m''_1} (-1)^{\ell''_1} \int_{0}^{\infty} ds\; s^2 \; \xi_0(s) j_{\ell''_1}(k_1s)j_{\ell''_1}(k_3s) Y_{\ell''_1,m''_1}(\mathbf{\widehat{k}}_1)Y_{\ell''_1,m''_1}^{*}(\mathbf{\widehat{k}}_3) \nonumber \\ 
& \qquad \quad = \sum_{\ell''_1} s_{\ell''_1}^{(\rm I)} \mathcal{P}_{\ell''_1}(\mathbf{\widehat{k}}_1,\mathbf{\widehat{k}}_3) \int_{0}^{\infty} ds\; s^2 \; \xi_0(s) j_{\ell''_1}(k_1s)j_{\ell''_1}(k_3s). 
\end{align}
Here, $s_{\ell''_1}^{(\rm I)}$ has the same definition as in the last line of Eq. (\ref{eq:k23}). In the first equality we used the inverse Fourier transform of the power spectrum, which is the 2PCF, $\xi_0$, and in the second line, we used the result equation (B1) in \cite{modeling3PCF}, duplicated below: 

\begin{align}
&j_{L}(k\left| \mathbf{r}_1 - \mathbf{r}_2\right|)Y_{LM}(\widehat{\mathbf{r}_1-\mathbf{r}_2}) = 4\pi \sum_{L_1,M_1}\sum_{L_2,M_2} i^{L_2-L_1 + L} j_{L_1}(kr_1)j_{L_2}(kr_2) \mathcal{C}_{L_1L_2L} \begin{pmatrix} L_1 & L_2 & L  \\ 0 & 0 & 0 \\ \end{pmatrix}\nonumber \\ 
&\qquad \qquad \qquad \qquad \qquad \qquad \qquad \times \begin{pmatrix} L_1 & L_2 & L  \\ M_1 & M_2 & M \\ \end{pmatrix} Y_{L_1M_1}^{*}(\widehat{\mathbf{r}}_1) Y_{L_2M_2}^{*}(\widehat{\mathbf{r}}_2). 
\end{align}
Choosing $L=0$ and $M=0$ renders the result used on Eq. (\ref{eq:Decoupled_P(k13)}). $\mathcal{C}_{L_1L_2L}$ is defined in Eq. (\ref{eq:C_cons}).

\section{Approach to Numerical Integrations}
We explain our method for efficient numerical computation of the integrals  (\ref{eq:1D-radial}), (\ref{eq:2D-radial}), (\ref{eq:2_Sph.Bess_int}), (\ref{eq:R_int_T3111}), (\ref{eq:3_Sph.Bess_int}), (\ref{eq:fancy_R_int}), (\ref{eq:R1_radial_n13_n_123}), (\ref{eq:4_Sph.Bessel_int}), (\ref{eq:prime_4_sph.Bessel_int}), (\ref{eq:radial_n13_R1_T2}), (\ref{eq:3_prime_Sph.Bess_int}), (\ref{eq:radial_n13_R1_T2}), (\ref{eq:radial_0_R1_T2}) and (\ref{eq:2_Sph.Bess_int_prime}). We begin by computing different powers of $k$ by taking successive multiplications, \textit{i.e.}, we first form $k$, then $k^2 = k \times k$, $k^3 = k^2 \times k$, \textit{etc}. This saves taking the log and then exponentiating as would be done numerically if we used the $**$ operator or \textsc{pow} function in \textsc{python}. We  compute in the same way any powers of $r$ we will require. 

Next, we compute the sBF, $j_{\ell}$, on a 2D grid in $k$ and $r$. We do this efficiently by explicitly looping over values of $r$, but at each value of $r$, using a vector of $k$.\footnote{Here we do not mean a 3D spatial vector, but rather just a list of values.} This approach is possible because \texttt{scipy} has a vectorized sBF, which can take a vector $kr$ at each fixed $r$ step in the loop. The sBFs cannot take a tensor such as we would have if we tried to insert an outer-product of our $k$ vector and a vector of the $r$ values into the sBF, so the approach above is the maximum vectorization achievable.

We may then use tensor operations to compute our integral efficiently with the \texttt{numpy} function \texttt{einsum}. This function uses the Einstein summation convention to compute different operations on any multi-dimensional array. Let us illustrate how this works with the most challenging integral of this paper, Eq. (\ref{eq:4_Sph.Bessel_int}). 

In Eq. (\ref{eq:4_Sph.Bessel_int}), we have a product of four sBFs and $k^{n+2}P(k)$.  As mentioned above, each sBF is a set of values on a 2D grid of $r$ and $k$. Hence, each sBF can be viewed as an $n \times m$ matrix, with the rows representing the corresponding $r$ values and the columns representing each value of $k$. Therefore, each entry in the matrix is a value of the sBF for a specific $r\times k$. At the time of integration, we must be careful as the number of columns, $m$, must be the same for all terms; however the number of rows, $n$, need not be the same. On the other hand, the term $k^{n+2}P(k)$ is a 1D numerical vector, of length $m$. To symbolize that $k$ enters \texttt{einsum} as a 1D numerical vector, we will denote it as $\underline{\mathbf{k}}$. 

With this in hand, if we define $\mathcal{J}_{(\ell),ij} \equiv j_{\ell}(\underline{\mathbf{k}}_{i}r_{j})$ to represent the matrix of the sBF, where $i$ and $j$ indicate the elements of the vectors $\underline{\mathbf{k}}$ and $r$, respectively, and $\hat{\mathcal{E}}$ to denote \texttt{einsum}, we have:

\begin{align}
&\hat{\mathcal{E}}\left[j_{\ell}(\underline{\mathbf{k}}r_{n}),j_{\ell'}(\underline{\mathbf{k}}r'_{f}), j_{\ell''}(\underline{\mathbf{k}}s_{h}), j_{L}(\underline{\mathbf{k}} r_{i,g}) \underline{\mathbf{k}}^{n+2}P(\underline{\mathbf{k}})\right]_{n,f,h,g} \nonumber \\
& \qquad\qquad= \sum_m \mathcal{J}_{(\ell),mn}\; \mathcal{J}_{(\ell'),mf}\;
\mathcal{J}_{(\ell''),mh} \; \mathcal{J}_{(L),mg} \; k_{m}^{n+2}P(k_m)|_{n,f,g,h},
\end{align}
where the subscript $\left\{n,f,g,h\right\}$ tells us we are only showing one term out of the 4D array that \texttt{einsum} would return. As can also be seen pictorially in Figure \ref{fig:Numerical_integral}, \texttt{einsum} obtains the integral over $k$ as a sum over the values in the array of $k$, for each specific value of $r,r',s,r_i$.

\subsection{Convergence Tests}
To ensure that our integrals are converging to the correct answer, we performed the integrals with different numbers of $k$ points---from 200 to 800 points---and summed the resulting array corresponding to the $r$ variable while keeping $r_i$ constant. As can be seen in Figures \ref{fig:f_convergence_test} - \ref{fig:h_convergence_test}, convergence of the integration is achieved once we see the result of the sum does not change as we increase the number of $k$ points. Our result shows that using 200 points for $k$ is an ideal number, since the integrals have already converged the evaluation will be faster the lower the number of $k$ points. We note that the $f'$ integral is the only one for which we find a more stringent requirement; there, we need at least $1,250$ $k$ points to obtain convergence, as shown in Figure \ref{fig:f_prime_convergence_test}.

We have evaluated the $k$ integral from 0.0001 to 3 $\left[{\rm Mpc}/h\right]^{-1}$, and the $r'$, $r$, $r_i$ and $s$ variables from 0.0001 to 200 $\left[{\rm Mpc}/h\right]$. For the radial variables, we kept the number of points constant, using 200 points for the $r$ and $s$ variables and 80 points for the $r_i$ variable. We use 200 points for $r$ and $s$ because they are intermediate variables that get integrated out numerically and want to maximize their convergence in the same fashion as we do with $k$; $r_i$ is the resulting variable and the number of points we chose is a preference for the resolution of our plots. Since we used the same sampling for the $r'$ and $r$ variables, and we wanted to save the computational cost of the 3sBF integral, we chose $r'= 100\;\left[{\rm Mpc}/h\right]$ for Eq. (\ref{eq:3_Sph.Bess_int}); likewise we chose $r'= s = 100\;\left[{\rm Mpc}/h\right]$ for Eq. (\ref{eq:4_Sph.Bessel_int}) during the 4sBF integration.

\begin{figure}[h]
\centering
\includegraphics[scale=0.7]{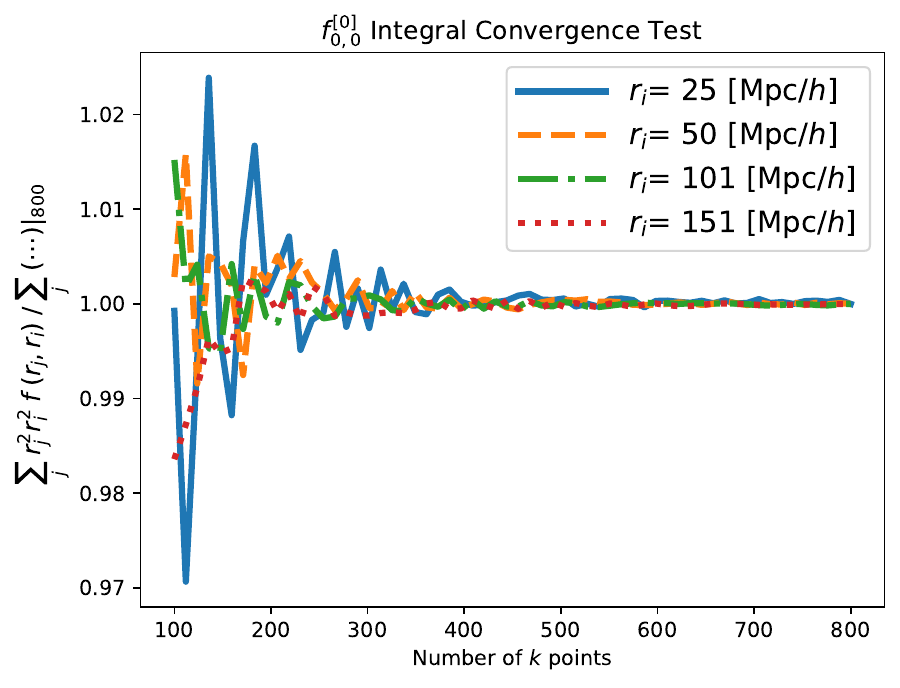}
\caption{Here, we show the convergence of Eq. (\ref{eq:2_Sph.Bess_int}) as a function of the number of $k$ points. The $y$ axis shows the normalized result, obtained by summing all the values of our integral array corresponding to the $r_j$ variable, appropriately weighted while keeping $r_i$ constant. Each line in the plot is normalized by its value when integrating Eq. (\ref{eq:2_Sph.Bess_int}) with 800 sample points in $k$. The $x$ axis shows the number of sample points used in $k$. The legend indicates the values of $r_i$ for which results are displayed. The plot shows that performing the integral with 200 points for $k$ is an ideal choice, as the integral has already converged within better than 1\%.}
\label{fig:f_convergence_test}
\end{figure}

\begin{figure}[h]
\centering
\includegraphics[scale=0.7]{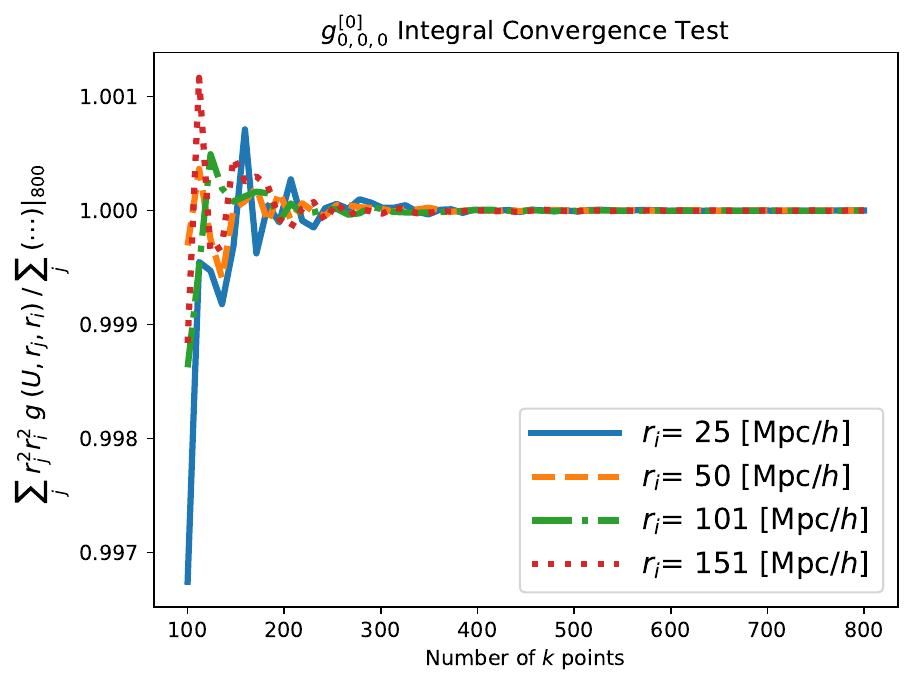}
\caption{Same as previous Figure but for Eq. (\ref{eq:3_Sph.Bess_int}).}
\label{fig:g_convergence_test}
\end{figure}

\begin{figure}[h]
\centering
\includegraphics[scale=0.7]{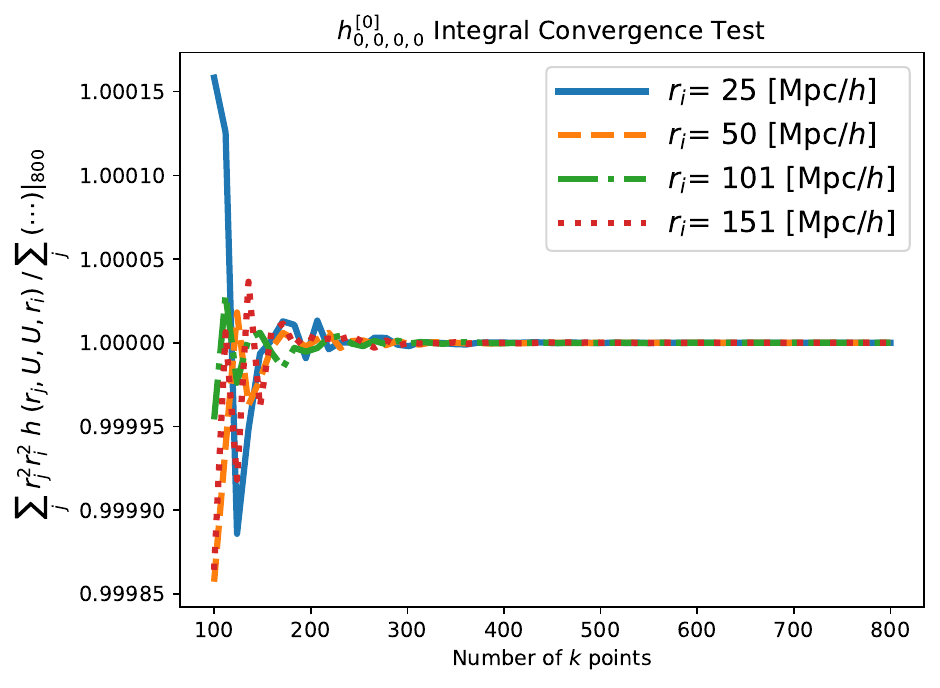}
\caption{Same as previous Figure but for Eq. (\ref{eq:4_Sph.Bessel_int}).}
\label{fig:h_convergence_test}
\end{figure}

\begin{figure}[h]
\centering
\includegraphics[scale=0.7]{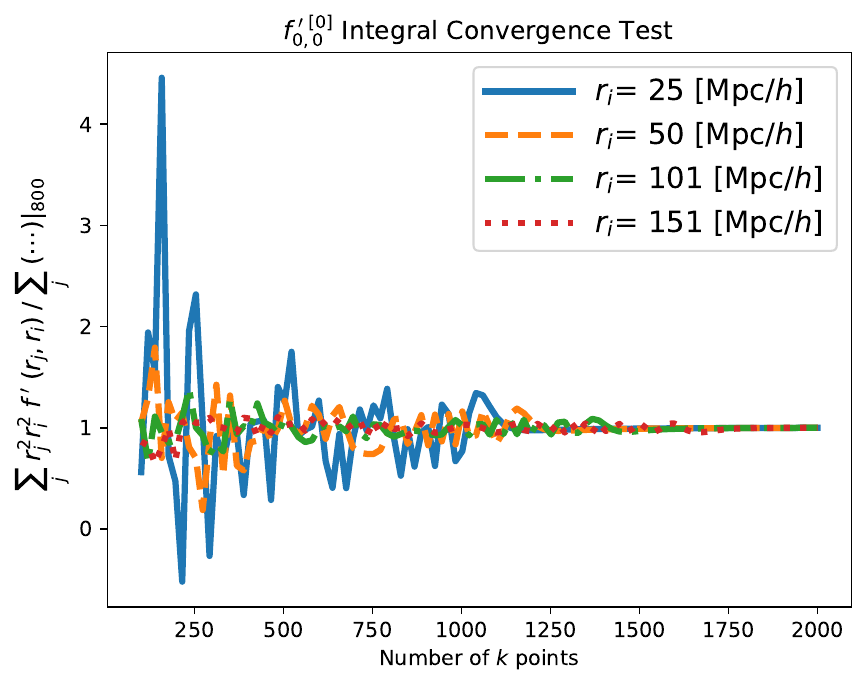}
\caption{Same as previous Figure but for (\ref{eq:2_Sph.Bess_int_prime}); here our maximum number of sample points is $2,000$. We see that convergence of this integral requires evaluating it with at least 1,250 sample points in $k$. This behavior is distinct from that of all the other integrals of this work, which converge with far fewer sample points in $k$.}
\label{fig:f_prime_convergence_test}
\end{figure}

\begin{figure}[h]
\centering
\includegraphics[scale=0.7]{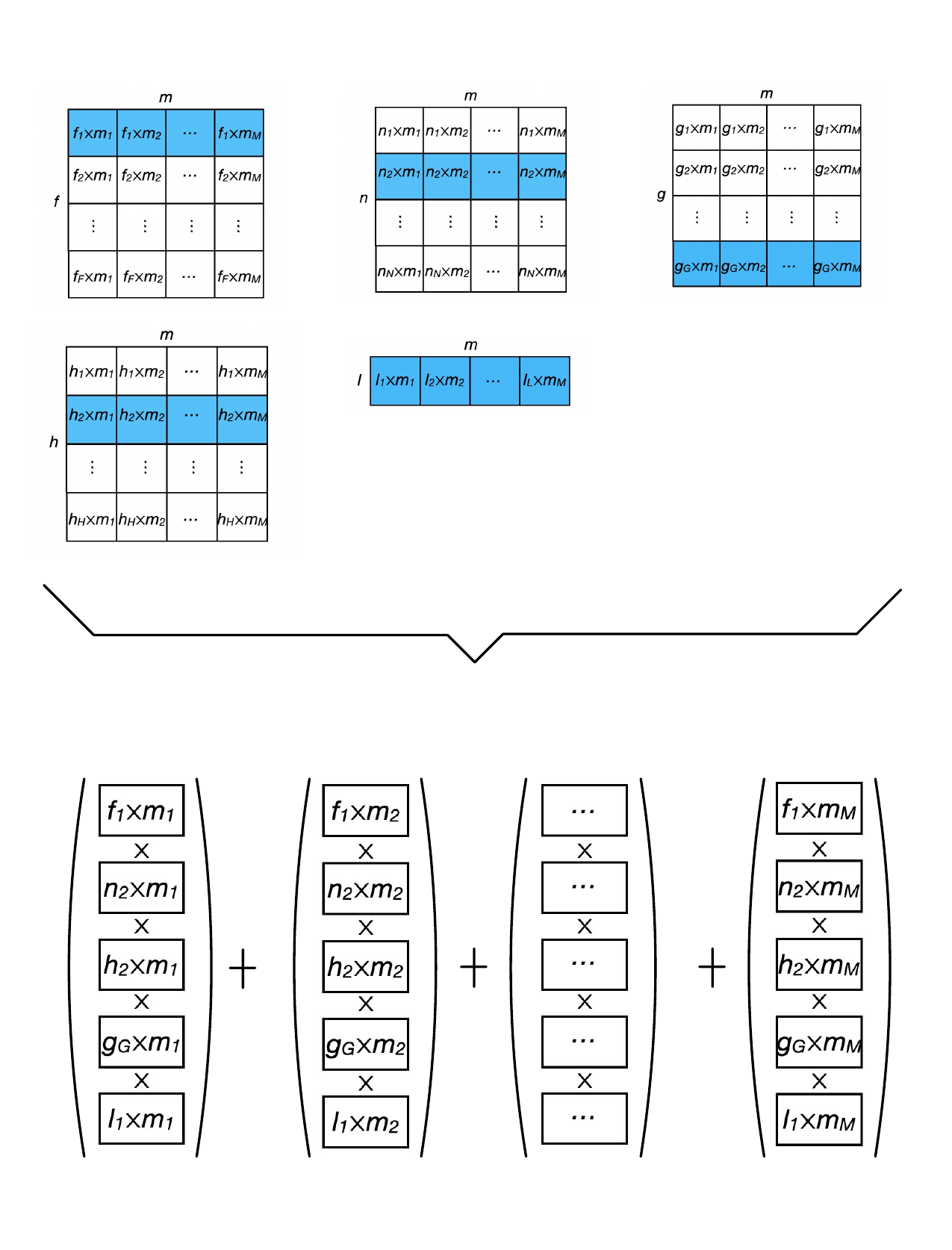}
\caption{Depiction of the process by which our \texttt{einsum}-based integration scheme proceeds. At the top, we have four 2D matrices, representing four sBFs, and a 1D vector representing the term $k^{n+2} P(k)$. Each matrix (and the vector) has a blue highlighted row, which indicates that it has been selected by \texttt{einsum} to be evaluated. The square braces represent \texttt{einsum}'s selecting and grouping all the terms such that at the bottom of the depiction, we find the previously highlighted rows divided by column entry. Row entries with the same column index are multiplied by each other and then added to the other columns; \textit{i.e.,} all the values in the array associated with the same column index $i = \{1,2,\cdots,M\}$ are multiplied with each other and later added to the remaining product of numbers.}
\label{fig:Numerical_integral}
\end{figure}

\clearpage

\section{Analytic Approach to the Radial Integrals } \label{Sec: Radial Integrals}
We evaluate analytically the radial integrals required by this work to explain the behavior we see in their plots in the main text; we then plot our analytic expressions using \texttt{Mathematica} where relevant. This offers a good check on our numerical integrations as well, as \texttt{Mathematica} uses its own integration algorithms. 

\subsection{One, Two, Three and Four sBF Integrals with or without Power Spectrum}\label{Sec J1}
We start with the 1D transform integral Eq. (\ref{eq:1D-radial}): 

\begin{align}
    \xi_{j}^{\left[n'\right]}(r_i) = \int_{0}^{\infty} \frac{dk_i}{2\pi^2} \; k_{i}^{n'+2} j_{j}(k_ir_i)P(k_i), \nonumber  
\end{align}
and look at its asymptotic behavior as $r$ grows large. In the integrand, only the spherical Bessel depends on $r$. Therefore, we need look only at the asymptotic behavior of the spherical Bessel function:

\begin{align}
    j_{l}(x) \rightarrow \frac{\sin\left(x-(l\pi/2)\right)}{x}, \nonumber
\end{align}
where $x\equiv k_ir_i$. This implies that $\xi_{j}^{\left[n'\right]}(r_i) \rightarrow 1/r_i^{n'+2}$ at large scales. To remove this overall fall-off, we weight the integrals by $r_i^{n'+2}$ in Figure \ref{fig:1-d_int}. 

Next, let us represent the sBF as an integral of the complex exponential against a Legendre polynomial. To do so, we begin with the plane-wave expansion: 

\begin{align}
    e^{i\mathbf{k}\cdot\mathbf{r}_i} = \sum_{\ell} i^{\ell} (2\ell +1) j_{\ell} (kr_i)\mathcal{L}_{\ell}(\mu),
\end{align}
with $\mu \equiv \mathbf{\widehat{k}} \cdot \mathbf{\widehat{r}}_i $. We multiply both sides by another Legendre polynomial of a different order, and then integrate over $\mu$ (using the orthogonality of the Legendre polynomials) to find:

\begin{align}\label{eq:Spherical_Bessel_C_Exponential}
    j_{\ell}(kr_i) = \frac{i^{-\ell}}{2\ell+1}\int_{-1}^{1}\mathcal{L}_{\ell}(\mu)\;e^{i\mathbf{k}\cdot\mathbf{r}_i} \;d\mu.   
\end{align}
With this in hand, we take the partial derivative with respect to $r_i$ of Eq. (\ref{eq:1D-radial}): 

\begin{align}
    \partial_{r_i}\xi_{j}^{\left[n'\right]}(r_i) = \int_{0}^{\infty} \frac{dk_i}{2\pi^2} \; k_{i}^{n'+2}P(k_i) \partial_{r_i} j_{j}(k_ir_i),
\end{align}
where the partial derivative only acts on the sBF. Using Eq. (\ref{eq:Spherical_Bessel_C_Exponential}) for the sBF to take the derivative, we find:

\begin{align}
    \partial_{r_i} j_{j}(k_ir_i) = \sum_{\ell} \frac{(2\ell+1)i^{1-j+\ell}}{2j+1} k \; j_{\ell}(kr_i) \int_{-1}^{1} \mu \; \mathcal{L}_{j} (\mu) \; \mathcal{L}_{\ell} (\mu)\; d\mu, 
\end{align}
where we have used the plane-wave expansion once again to write the integrand only in terms of Legendre polynomials times $\mu$. This result implies $\partial_{r_i}\xi_{0}^{\left[n'\right]}(r_i) = -\xi_{1}^{\left[n'+1 \right]}(r_i) $. Therefore we see that the derivatives of the curves in the top rectangular panel of Figure \ref{fig:1-d_int}, for $n'=-1$, $n'=0$ and $j=0$, correspond to the curves for $n'= 0$, $n'=1$ and $j=1$, respectively, in the middle rectangular panel of that Figure. 

We continue with the double-sBF integrals, Eq. (\ref{eq:2_Sph.Bess_int}) and Eq. (\ref{eq:2_Sph.Bess_int_prime}):

\begin{align}
&f_{\ell,L}^{\left[n'\right]}(r,r_i) \equiv \int_{0}^{\infty}\frac{dk_i}{2 \pi^2} \; k_{i}^{n'+2} j_{\ell}(k_ir)j_{L}(k_i r_i)P(k_i) \nonumber \\
&f_{\ell,L}^{'\left[n'\right]}(r,r_i) \equiv \int_{0}^{\infty}\frac{dk_i}{2 \pi^2} \; k_{i}^{n'+2} j_{\ell}(k_ir)j_{L}(k_i r_i). \nonumber
\end{align}

We take a power-law power spectrum, $P(k)\sim 1/k$, for Eq. (\ref{eq:2_Sph.Bess_int}), which means the $k$ in the integrand will differ by a power of $1$ compared to Eq. (\ref{eq:2_Sph.Bess_int_prime}). Besides this difference, given Figure \ref{fig:fint2} and Figure \ref{fig:fprime_int}, we assume the analytic results shown below apply to both equations, \textit{i.e.}, $f \sim f'$. 

We start the analysis with Eq. (\ref{eq:2_Sph.Bess_int_prime}) for the case $\ell = 0$, $L=0$ and $n'=0$, and use the orthogonality relation of the sBFs to find:

\begin{align}\label{eq:2_sph_Bessel_delta_relation}
    f_{0,0}^{'\left[0\right]}(r,r_i) = \int_{0}^{\infty}\frac{dk_i}{2 \pi^2} \; k_{i}^{2} j_{0}(k_ir)j_{0}(k_i r_i) = \frac{1}{4\pi r_i^2}\delta^{[1]}_{ \rm D}(r-r_i).
\end{align}
This result explains why the integral is non-vanishing mainly along the diagonal in the upper panels of Figure \ref{fig:fint2} and Figure \ref{fig:fprime_int} (for Eq. (\ref{eq:2_Sph.Bess_int}) and Eq. (\ref{eq:2_Sph.Bess_int_prime}), respectively). 

Next, we evaluate  Eq.(\ref{eq:2_Sph.Bess_int}) (with $P(k_i) \sim 1/k_i$) for $\ell = 0$, $L=1$ and $n'=0$. We use the result in Eq. (D1) of \cite{SlepianDecoupling}: 
\begin{align}
    \int_{0}^{\infty} x\; j_{\ell+1}(ax) j_{\ell}(bx)\; dx = \frac{b^{\ell}}{a^{\ell+2}}, \quad b<a; \quad 0, \quad a<b,
\end{align}
finding:

\begin{align}\label{eq:f1_analitical}
f_{0,1}^{\left[0\right]}(r,r_i) \sim 1/r_i^2, \quad r<r_i; \quad 0, \quad r_i<r.
\end{align}
The above result explains why, above the diagonal ($r \geq r_i$), the integral is zero or close to zero, and why most of the non-vanishing integral is below the diagonal, for the lower 2D plots in Figures \ref{fig:fint2} and \ref{fig:fprime_int} (for Eq. (\ref{eq:2_Sph.Bess_int}) and Eq. (\ref{eq:2_Sph.Bess_int_prime}), respectively).  

We now evaluate the triple-sBF integrals, Eq. (\ref{eq:3_Sph.Bess_int}) and Eq. (\ref{eq:3_prime_Sph.Bess_int}):

\begin{align}
&g_{\ell',\ell,L}^{\left[n'\right]}(r',r,r_i) \equiv \int_{0}^{\infty} \frac{dk_i}{2\pi^2} \; k_i^{n'+2} j_{\ell'}(k_ir')j_{\ell}(k_ir)j_{L}(k_ir_i)P(k_i)\nonumber\\
&g_{\ell',\ell,L}^{'\left[n'\right]}(r',r,r_i) \equiv \int_{0}^{\infty} \frac{dk_i}{2\pi^2} \; k_i^{n'+2} j_{\ell'}(k_ir')j_{\ell}(k_ir)j_{L}(k_ir_i). \nonumber 
\end{align}

We again use a power-law power spectrum, $P(k)\sim 1/k$ for Eq. (\ref{eq:3_Sph.Bess_int}), which means the $k$ in the integrand will differ by a power of $1$ compared to Eq. (\ref{eq:3_prime_Sph.Bess_int}). Besides this difference, given Figure \ref{fig:gint2} and Figure \ref{fig:gprime_int}, we assume the analytic results shown below apply to both equations, \textit{i.e.}, $g \sim g'$. 

~\cite{Whelan1993} gives a result for the $g'$ integral for  $n'=0$, $\ell =0$, $\ell'=L$, which can be extended to our case $\ell=L=0$. We find:

\begin{align}\label{eq:g000_result}
    g_{0,0,0}^{'\left[0\right]}(r',r,r_i) = \theta(r',r,r_i)\frac{\pi}{4\;r'\;r\;r_i}
\end{align}
with 

\begin{align}
\theta(r',r,r_i) = \left\{\begin{matrix}
1, \quad \left|r-r'\right|<r_i<\left|r+r'\right| \\
\qquad \; 0, \quad \left|r-r'\right|> r_i \; {\rm or} \; \left|r+r'\right|< r_i \\
\quad \; \; 1/2, \quad \left|r-r'\right| =  r_i \; {\rm or} \; \left|r+r'\right|=  r_i. 
\end{matrix}\right.
\end{align}

The result above explains the behavior seen in the upper panels of Figures \ref{fig:gint2} and \ref{fig:gprime_int} (for Eq. (\ref{eq:3_Sph.Bess_int}) and Eq. (\ref{eq:3_prime_Sph.Bess_int}), respectively). The conditions imposed by $\theta$ explain the rectangular boundary in these figures. Specifically, we have substituted $r'=100 \;\left[{\rm Mpc}/h\right]$ into the above conditions, which results in the rectangular behavior with boundary $r = -r_i+100$. 

\begin{figure}[h]
\centering
\includegraphics[scale=0.85]{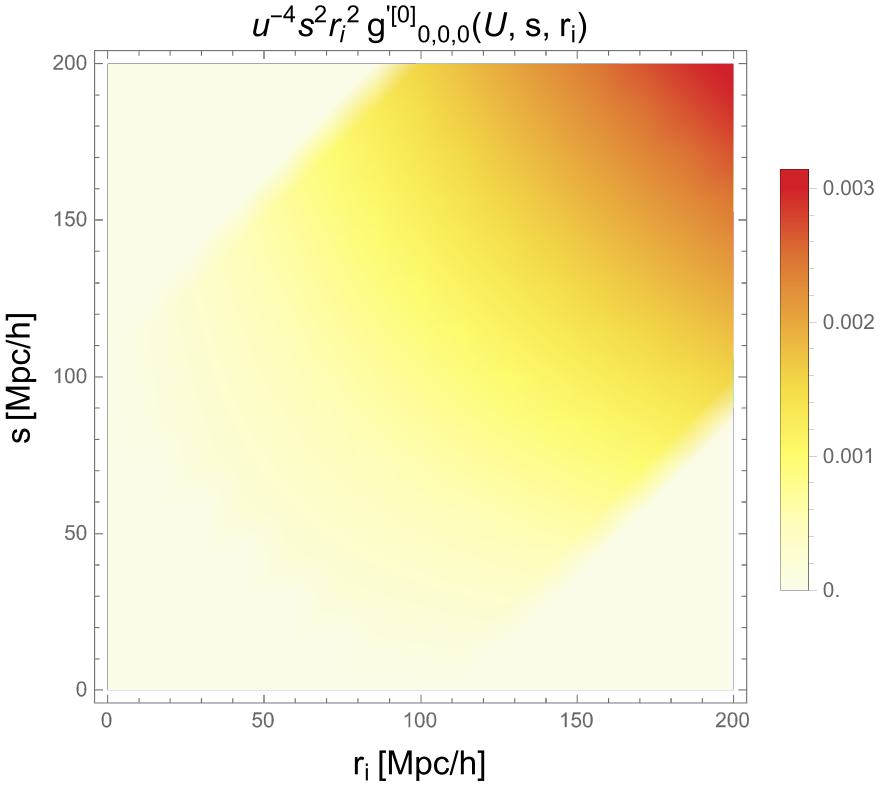}
\caption{We show the analytic expression of Eq. (\ref{eq:g000_result}) for $r^2r_i^2/u^4 g_{0,0,0}^{'\left[0\right]}(U,r,r_i)$, with $u \equiv 10\;\left[{\rm Mpc}/h\right]$ and $U \equiv 100 \;\left[{\rm Mpc}/h\right]$. This figure shows that our analytic result reproduces the \textit{upper panel} in Figure \ref{fig:gprime_int}, both as regards the rectangular and the monotonically increasing behavior.}
\label{fig:g000 analytical}
\end{figure}

We continue by evaluating Eq. (\ref{eq:3_prime_Sph.Bess_int}) for the case $\ell'=0$, $\ell=0$, $L=1$ and $n'=0$, yielding:

\begin{align}\label{eq:g_prime_analytical0}
    g_{0,0,1}^{'\left[0\right]}(r',r,r_i) = \frac{2}{\pi}\left\{ H(u-r_i)I_{u>}(u,r_i) + H(r_i-u)I_{u<}(u,r_i)  \right\} \big|_{u_{-}}^{u_{+}},
\end{align}
as demonstrated in Eq. (47) of \cite{Sph_Bessel_Integral_kiersten}. $u$ is an intermediate variable to be evaluated at $u_+ = r'+r$ and $u_-=\left| r' - r \right|$. $H$ is the Heaviside step function: 
\begin{align}
 H(x) \equiv \left\{\begin{matrix}
1, \quad x>0 \\ 0,\quad x<0,
\end{matrix}\right.
\end{align}

and 

\begin{align}\label{eq:I_function_from_g001}
&I_{u>}(u,r_i) = -\frac{1}{2r_i}{\rm ln}\left[u^2 - r_i^2\right] + \frac{1}{2r_i^2}\left[ u \;\tanh^{-1}\left(\frac{r_i}{u}\right)+\frac{r_i}{2}{\rm ln}(u^2-r_i^2)\right] \nonumber \\
&I_{u<}(u,r_i) =  -\frac{2u}{3r_i} + \frac{1}{3}{\rm ln}\left( \frac{r_i+u}{r_i-u} \right) \nonumber \\
& \qquad \qquad \qquad \quad + \frac{1}{6r_i^2}\left[  2u^2 \;\tanh^{-1}(r_i/u)+r_i \left( 2u + r_i {\rm ln}\left( \frac{r_i-u}{r_i+u}\right) \right) \right].  
\end{align}
We note that we have re-derived the expressions shown above to verify the (novel) results of \cite{Sph_Bessel_Integral_kiersten}, and they can be combined as:
\begin{align}
    g_{0,0,1}^{'\left[0\right]}(r',r,r_i) = \frac{1}{2rr'r_i}\left\{ H(u-r_i)\mathcal{I}_{u>}(u,r_i) + H(r_i-u)\mathcal{I}_{u<}(u,r_i)  \right\} \big|_{u_{+}}^{u_{-}},
\end{align}
with
\begin{align}
&\mathcal{I}_{u>}(u,r_i) = 1 - \frac{u}{r_i} \tanh^{-1}{\left(\frac{r_i}{u}\right)} \\
&\mathcal{I}_{u<}(u,r_i) = 1 - \frac{u}{r_i} \tanh^{-1}{\left(\frac{u}{r_i}\right)}.
\end{align}

Eq. (\ref{eq:g_prime_analytical0}) shows that we still expect the square behavior (given by the Heaviside function) in our plots but with different behavior across the diagonal (given by the $I$ functions,) as can be seen in the lower panels of Figures \ref{fig:gint2} and \ref{fig:gprime_int} (for Eq. (\ref{eq:3_Sph.Bess_int}) and Eq. (\ref{eq:3_prime_Sph.Bess_int}), respectively). For the rest of this section, we will refer to the evaluation of the four terms  in Eq.(\ref{eq:g_prime_analytical0}) in a simplified way: 

\begin{align}\label{eq:g_prime_analytical1}
    g_{0,0,1}^{\left[0\right]}(r',r,r_i) = \frac{2}{\pi} \mathcal{H}(u_\pm,r_i)I(u_\pm,r_i). 
\end{align}

\begin{figure}[h]
\centering
\includegraphics[scale=0.85]{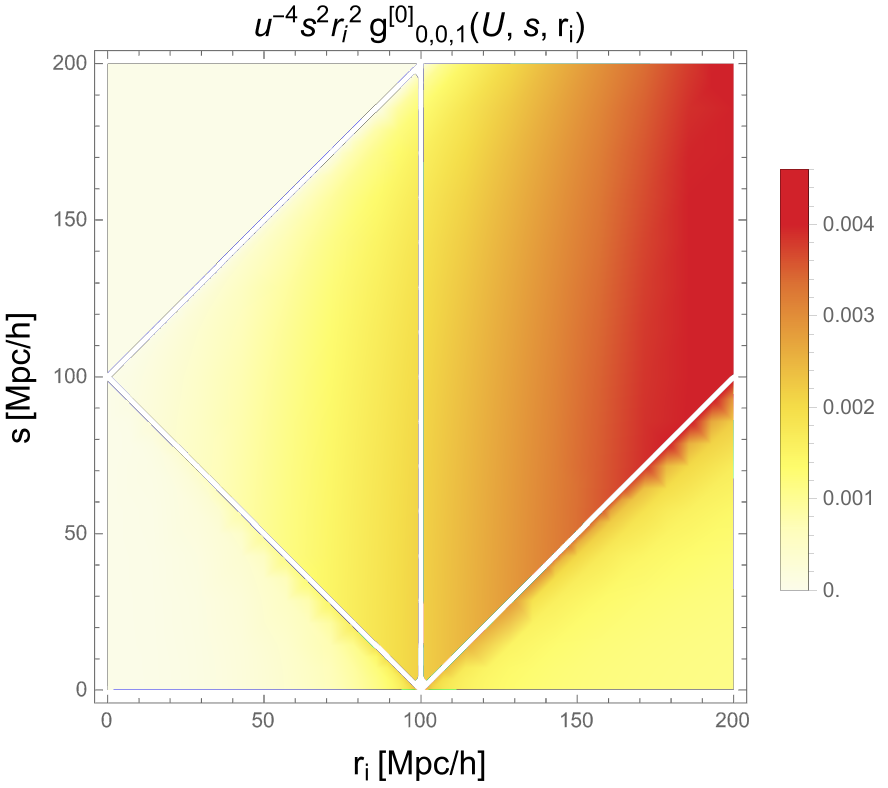}
\caption{Here we show Eq. (\ref{eq:g_prime_analytical0}) for $r^2r_i^2/u^4 g_{0,0,1}^{'\left[0\right]}(U,r,r_i)$, with $u \equiv 10\;\left[{\rm Mpc}/h\right]$ and $U \equiv 100 \;\left[{\rm Mpc}/h\right]$. We see that our analytic result reproduces the \textit{lower panel} in Figure \ref{fig:gprime_int}, both as regards the rectangular behavior, and the asymmetry across the diagonal $r=r_i$. The white lines on the boundary and at $r_1 =100$ are places where \texttt{Mathematica} could not obtain a numerical result. We believe this is caused by the inequalities defining the Heaviside function and theta function, which are distributed into the integral's bounds and preclude evaluation along those boundaries.}
\label{fig:g001 analytical}
\end{figure}

Finally, we must evaluate the  quadruple-sBF integrals, Eqs. (\ref{eq:4_Sph.Bessel_int}) and (\ref{eq:prime_4_sph.Bessel_int}):

\begin{align}
&h_{\ell,\ell',\ell'',L}^{\left[n'\right]} (r,r',s,r_i) = \int_{0}^{\infty}\frac{dk_i}{2\pi^2} \; k_i^{n'+2} j_{\ell}(k_i r) j_{\ell'}(k_i r') j_{\ell''}(k_i s) j_{L}(k_i r_i)P(k_i),\nonumber \\ 
&h_{\ell,\ell',\ell'',L}^{'\left[n'\right]} (r,r',s,r_i) = \int_{0}^{\infty}\frac{dk_i}{2\pi^2} \; k_i^{n'+2} j_{\ell}(k_i r) j_{\ell'}(k_i r') j_{\ell''}(k_i s) j_{L}(k_i r_i). \nonumber 
\end{align}

We again use a power-law power spectrum, $P(k)\sim 1/k$, in Eq. (\ref{eq:4_Sph.Bessel_int}), which means that the $k$ in the integrand will differ by a power of $1$ compared with Eq. (\ref{eq:prime_4_sph.Bessel_int}). Besides this difference, given Figure \ref{fig:h_int} and Figure \ref{fig:hprime_int}, we assume the analytic results shown below apply to both equations, \textit{i.e.}, $h \sim h'$. 

Starting with the case $\ell,\ell',\ell'',L = 0$, $n'=0$, and using Eq. (\ref{eq:2_sph_Bessel_delta_relation}) and Eq. (2.2) in \cite{4SBF_integral} to write the integral as:

\begin{align}
&h_{0,0,0,0}^{'\left[0\right]} (r,r',s,r_i) = \frac{1}{\pi^3}\int_{0}^{\infty}dR\; R^2\left( \int_{0}^{\infty}dk_i\;k_i^2  j_{0}(k_i r) j_{0}(k_i r') j_{0}(k_i R) \right) \nonumber \\
& \qquad \qquad \qquad \qquad \qquad \qquad \qquad \times \left( \int_{0}^{\infty}dk^{'}_i\;k_i^{'2} j_{0}(k^{'}_i s) j_{0}(k^{'}_i r_i) j_{0}(k^{'}_i R)\right).
\end{align}
 Using Eq. (\ref{eq:g000_result}) we find:

\begin{align}\label{eq:h_analitycal0}
    h_{0,0,0,0}^{'\left[0\right]} (r,r',s,r_i) = \frac{1}{16\pi r\;r'\;s\;r_i}\int_{0}^{\infty}dR \; \theta(r,r',R)\theta(s,r_i,R). 
\end{align}
Figure \ref{fig:h0000 analytical} shows this integral with $r'=s'=100 \; [{\rm Mpc}/h]$). It is notable that the resultant function of $r$ and $r_i$ increases monotonically starting at the cut-off $100 \; [{\rm Mpc}/h]$, and reproduces the upper panels of Figures \ref{fig:h_int} and \ref{fig:hprime_int} (for Eq. (\ref{eq:4_Sph.Bessel_int}) and Eq.(\ref{eq:prime_4_sph.Bessel_int}), respectively). 

\begin{figure}[h]
\centering
\includegraphics[scale=0.85]{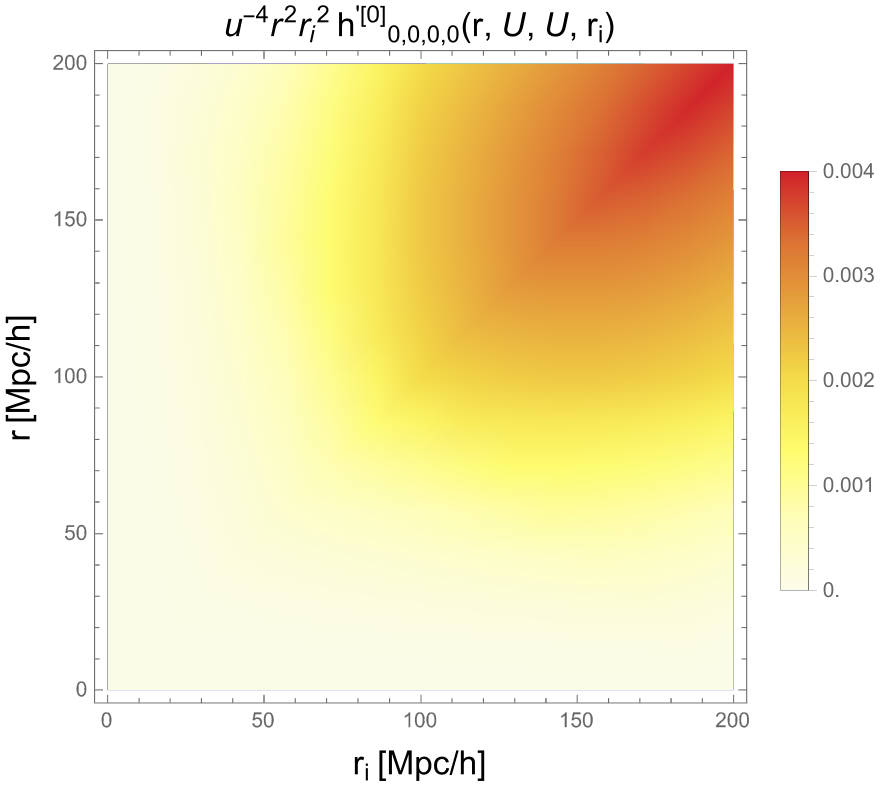}
\caption{Here we show Eq. (\ref{eq:h_analitycal0}) for $r^2r_i^2/u^4 \; h_{0,0,0,0}^{'\left[0\right]}(r,U,U,r_i)$ with $u \equiv 10\;\left[{\rm Mpc}/h\right]$ and $U \equiv 100 \;\left[{\rm Mpc}/h\right]$.  We see monotonically increasing behavior with an oval-shaped concentration near the upper right-hand corner. The oval features left-hand and lower boundaries are at roughly $r=r_i\approx 100 \;\left[{\rm Mpc}/h\right]$. We see that our analytic result reproduces  the \textit{upper panel} in Figure \ref{fig:hprime_int}, both as regards the monotonically increasing, oval-shaped behavior, and the symmetry.}
\label{fig:h0000 analytical}
\end{figure}

Continuing with the case $\ell,\ell',\ell'' = 0$, $L=1$ $n'=0$, the integral becomes:

\begin{align}
&h_{0,0,0,1}^{'\left[0\right]} (r,r',s,r_i) = \frac{1}{\pi^3}\int_{0}^{\infty}dR\; R^2\left( \int_{0}^{\infty}dk_i\;k_i^2  j_{0}(k_i r) j_{0}(k_i r') j_{0}(k_i R) \right) \nonumber \\
& \qquad \qquad \qquad \qquad \qquad \qquad \qquad \times \left( \int_{0}^{\infty}dk^{'}_i\;k_i^{'2} j_{0}(k^{'}_i s) j_{1}(k^{'}_i r_i) j_{0}(k^{'}_i R)\right). 
\end{align}
The results for the two integrals in parentheses are given by Eqs. (\ref{eq:g000_result}) and \ref{eq:g_prime_analytical1}), allowing us to obtain:
\begin{align}\label{eq:hprime_1_analitical}
&h_{0,0,0,1}^{'\left[0\right]} (r,r',s,r_i) = \frac{1}{2\pi^3\;r\;r'}\int_{0}^{\infty}dR\; R\; \theta(r,r',R)\mathcal{H}(u_{\pm,h},R)I(u_{\pm,h},R),  
\end{align}
with $u_{+,h} = s+r_i$ and $u_{-,h} = \left|s-r_i\right|$. We now redistribute the term $\mathcal{H}(u_{\pm,h},R)I(u_{\pm,h},R)$ such that the integral becomes:

\begin{align}
&h_{0,0,0,1}^{'\left[0\right]} (r,r',s,r_i) = \frac{1}{2\pi^3\;r\;r'}\int_{0}^{\infty}dR\; R\;\theta(r,r',R)\left[H(u_{+,h}-R)I_{u>}(u_{+},R) \right. \nonumber \\ 
& \qquad \qquad \qquad \qquad \qquad \qquad \quad \left. - H(R-u_{+,h})I_{u<}(u_{+,h},R) - H(u_{-,h}-R)I_{u>}(u_{-,h},R) \right. \nonumber \\ 
& \qquad \qquad \qquad \qquad \qquad \quad \qquad \left. + H(R-u_{-,h})I_{u<}(u_{-,h},R)\right].  \end{align}
We then use the definitions of $\theta$ and $H$ to simplify the integral bounds, finding:
\begin{align}
&h_{0,0,0,1}^{'\left[0\right]} (r,r',s,r_i) = \frac{1}{2\pi^3\;r\;r'}\left\{\int_{\left|r-r'\right|}^{r+r'< r_i+s }dR\;R\; I_{u>}(u_{+,h},R) \right. \nonumber \\ 
& \qquad \qquad \qquad \qquad \qquad \qquad \quad \left.+ \int_{\left|r-r'\right|>r_i+s}^{r+r'}dR\;R\; I_{u<}(u_{+,h},R)\right.\nonumber \\ 
& \qquad \qquad \qquad \qquad \qquad \qquad \quad \left. - \int_{\left|r-r'\right|}^{(r+r')<\left|r_i-s\right|}dR\;R\; I_{u>}(u_{-,h},R) \right.\nonumber \\ 
& \qquad \qquad \qquad \qquad \qquad \qquad \quad \left.- \int_{\left|r-r'\right|>\left|r_i-s\right|}^{r+r'}dR\; R\;I_{u<}(u_{-,h},R)\right\}. 
\end{align}

The bounds of integration above can be understood as follows: $r+r'< \left|r_i\pm s\right|$ means that at any value of $r_i$, $s$, $r$, and $r'$ we will evaluate the result of the integral at $r+r'$ as long as it is smaller than the combination $|r_i\pm s|$. Likewise, for $\left|r-r'\right|>\left|r_i\pm s\right|$, we will evaluate the result of the integral at $\left|r-r'\right|$ as long as it is greater than the combination $\left|r_i\pm s\right|$. These bounds of integration are obtained from analyzing the $\theta$ and Heaviside function. From $\theta$ we have a non-zero result when $\left|r-r'\right|\leq R \leq r+r'$. The Heaviside can give us either an upper bound or lower bound depending on its arguments. For example, if $H(u-R)$, then the integral is non-zero only when $u>R$ creates an upper bound; if $H(R-u)$, then we have a lower bound given by $R>u$. We note that, because of the Heaviside function, $R$ can never be equal to $u$; this is why we cannot evaluate the above integrals at $u$. We have computed the above integrals with \texttt{Mathematica} to produce the 3D plot of Figure \ref{fig:h0001_analytical}, which reproduces our numerical result in the lower panels of Figures \ref{fig:h_int} and \ref{fig:hprime_int}.

\begin{figure}[h]
\centering
\includegraphics[scale=0.85]{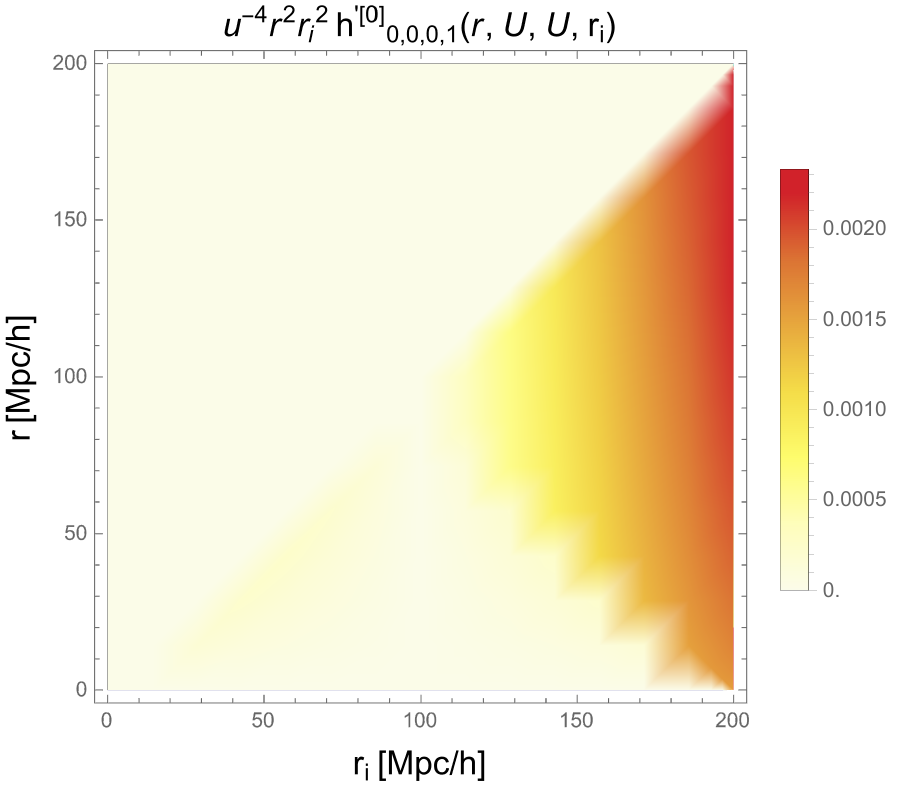}
\caption{Here we show $r^2 \;r_i^2/u^4\; h_{0,0,0,1}^{'\left[0\right]} (r,U,U,r_i)$, where $u = 10\;\left[{\rm Mpc}/h\right]$ and $U = 100\;\left[{\rm Mpc}/h\right]$. We see monotonic increase in the region $r_i>r$, with the behavior turning on approximately across the line $r=-r_i$. We see that our analytic result reproduces the \textit{lower panel} in Figure \ref{fig:hprime_int}, both overall and as regards the asymmetry across the diagonal $r=r_i$.}
\label{fig:h0001_analytical}
\end{figure}

\subsection{Integrals Over Products of sBF Integrals}
\subsubsection{Radial Integrals from $\S$\ref{Section T3}}
We now evaluate the radial integrals that combine the integrals of Eqs. (\ref{eq:2_Sph.Bess_int}), (\ref{eq:2_Sph.Bess_int_prime}), (\ref{eq:3_Sph.Bess_int}), (\ref{eq:3_prime_Sph.Bess_int}), (\ref{eq:4_Sph.Bessel_int}) and (\ref{eq:prime_4_sph.Bessel_int})----\textit{i.e.} the integrals already explored in $\S$\ref{Sec J1}---to integrate out some of their variables, yielding functions of just $r_1$, $r_2$ and $r_3$. We start with the radial integrals in $\S$\ref{Section T3}, specifically with Eq. (\ref{eq:2D-radial}):

\begin{align}
F_{(\ell_2),L_2,L_3}^{\left[n'_2\right],\left[n'_3\right]}(r_2,r_3) = \int_{0}^{\infty} dr \; r\; f_{\ell_2,L_2}^{\left[n'_2\right]}(r,r_2)f_{\ell_2,L_3}^{\left[n'_3\right]}(r,r_3),\nonumber
\end{align}
for the case $\ell_2=L_2=L_3=n'_2=n'_3=0$. Writing all the terms out, including the $f$ integrals, we find:

\begin{align}\label{eq:F0_int_numerically}
&F_{(0),0,0}^{\left[0\right],\left[0\right]}(r_2,r_3) = \frac{1}{4\pi^4}\int_{0}^{\infty} dk_2 \;k_2^2 \; j_0(k_2r_2) P(k_2)\int_{0}^{\infty} dk_3\; k_3^2\; j_0(k_3r_3)P(k_3)\nonumber \\ 
& \qquad \qquad \qquad \qquad \times \int_{0}^{\infty} dr \;r\;j_0(k_2r)j_0(k_3r),
\end{align}
and if we approximate the $r$ integral as a $\delta_{ \rm D}(k_2-k_3)/k_2^2$ (based on our assumption above $f\sim f'$), and take the integral over $k_2$ (or $k_3$), we obtain: 

\begin{align}
F_{(0),0,0}^{\left[0\right],\left[0\right]}(r_2,r_3) \sim \int_{0}^{\infty} dk_3 \;k_3^2 \; j_0(k_3r_2) j_0(k_3r_3)P(k_3)^2. 
\end{align}
If we further approximate $P(k)\sim 1/k$, the integral can be evaluated to:

\begin{align}
F_{(0),0,0}^{\left[0\right],\left[0\right]}(r_2,r_3) \sim \int_{0}^{\infty} dk_3 \; \; j_0(k_3r_2) j_0(k_3r_3) = \frac{\pi}{2r_>}, 
\end{align}
where $r_>$ is the larger of $r_2$ and $r_3$. The above result explains the symmetry across the diagonal in the upper 2D plot in Figure \ref{fig:2D_int}. Noting that we multiply the above result by $r_2^2$ and $r_3^2$ gives us the monotonically growing behavior we observe in the upper 2D plot of Figure \ref{fig:2D_int}.

Next, let us look at this integral for the case $\ell_2=L_3=n'_2=n'_3=0$, $L_2=1$. Following the same logic as the previous case, we find: 

\begin{align}
F_{(0),1,0}^{\left[0\right],\left[0\right]}(r_2,r_3) \sim \int_{0}^{\infty} dk_3 \;k_3^2 \; j_1(k_3r_2) j_0(k_3r_3)P^2(k_3),    
\end{align}
and if we approximate $P(k)\sim1/k$, we find:

\begin{align}\label{eq:F1_analytical}
F_{(0),1,0}^{\left[0\right],\left[0\right]}(r_2,r_3) \sim \int_{0}^{\infty} dk_3 \; j_1(k_3r_2) j_0(k_3r_3) = 0,\quad r_2=r_3. 
\end{align}
The above result explains why the integral vanishes along the diagonal in the lower 2D plot of Figure \ref{fig:2D_int}. We must realize the above integral is almost identical to the integral in Eq.(\ref{eq:f1_analitical})---except for a power of k---and the reason why we can take $F_{(0),1,0}^{\left[0\right],\left[0\right]}\sim f_{0,1}^{\left[0\right]}$; demonstrating the reason why the lower 2D plot of Figure \ref{fig:fint2} and Figure \ref{fig:2D_int} look almost identical. 

Next, we evaluate Eq. (\ref{eq:R_int_T3111}) for the case: 

\begin{align}\label{eq: R_000_analytical_inital}
&R_{(0,0,0,0),0,0,0}^{\left[0\right],\left[0\right],\left[0\right]}(r_1,r_2,r_3) \equiv \int_{0}^{\infty} dr' \; r' \;f_{0,0}^{\left[0\right]}(r',r_1) \int_{0}^{\infty} dr \;r \nonumber \\ 
& \qquad \qquad \qquad \qquad \qquad \qquad \qquad \qquad \times g_{0,0,0}^{\left[0\right]}(r',r,r_2) g_{0,0,0}^{\left[0\right]}(r',r,r_3). 
\end{align}
As we have shown in the evaluation of the spherical Bessel integrals above and as we can see in the 2D plots, the behavior of $f \sim f'$ and $g \sim g'$, at least for the shown cases. To evaluate the above integral we will make use of this similarity. Eqs. (\ref{eq:2_sph_Bessel_delta_relation}) and (\ref{eq:g000_result}) give us the expression for $f_{0,0}^{\left[0\right]}$ and $g_{0,0,0}^{\left[0\right]}$, respectively. So, if we introduce these expressions into Eq.(\ref{eq: R_000_analytical_inital}) and integrate them over the $r'$ integral, we obtain: 

\begin{align}\label{eq:R0_analytical}
&R_{(0,0,0,0),0,0,0}^{\left[0\right],\left[0\right],\left[0\right]}(r_1,r_2,r_3) \sim \frac{1}{r_1^3 r_2 r_3}\int_0^{\infty} dr\; \frac{\theta(r_1,r,r_2)\theta(r_1,r,r_3)}{r}, 
\end{align}
and if we compute the above integral with \texttt{Mathematica} we reproduce Figure \ref{fig:R0_analytical} and Figure \ref{fig:3d_R0_analytical}. One can observe that \texttt{Mathematica} produces a divergence (white area in the 2D plot and grey area in the 3D plot) which might be due to the $1/r$ in the integrand which ultimately is a reflection of the assumptions we have made to arrive at the above expression. However, as can be seen, these two figures show the same behavior as the lower 2D plot of Figure \ref{fig:R_int}, demonstrating that the plot of Figure \ref{fig:R_int} seems to be correct. Further investigation will be needed to understand why our assumptions above produce a divergence in this analytic result. 

\begin{figure}[h]
\centering
\includegraphics[scale=0.85]{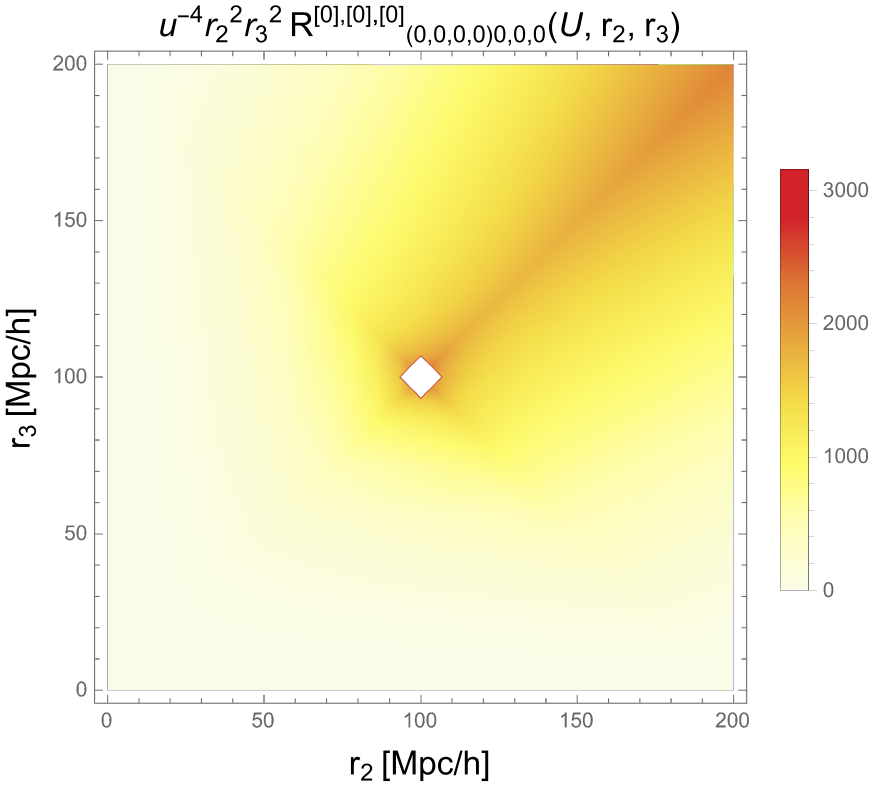}
\caption{We show our analytic result for $r_2^2 r_3^2/u^4\times R_{(0,0,0,0)0,0,0}^{\left[0\right],\left[0\right],\left[0\right]} (U,r_2,r_3)$, using Eq. (\ref{eq:R0_analytical}) with $u = 10\;\left[{\rm Mpc}/h\right]$ and $U = 100\;\left[{\rm Mpc}/h\right]$. We see monotonically increasing behavior with a divergence centered at $r_2 = r_3 =100 \left[{\rm Mpc}/h\right]$ and an oval-shaped structure on the right-hand side of the plot. This figure is nearly identical---save for the divergence---to the \textit{upper panel} of Figure \ref{fig:R_int}, demonstrating that our analytic result is a good approximation to the true integral. We ascribe the divergence here to the $1/r$ in our analytic integrand.}
\label{fig:R0_analytical}
\end{figure}

\begin{figure}[h]
\centering
\includegraphics[scale=0.65]{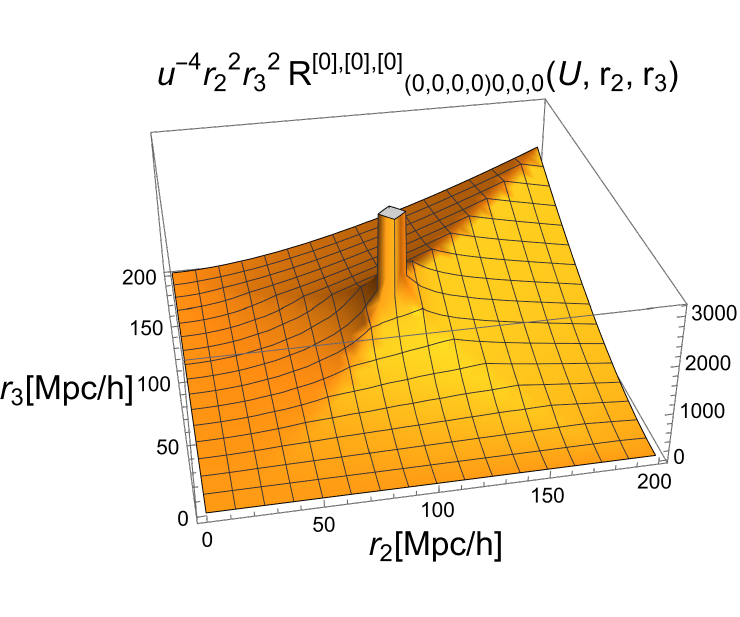}
\caption{3D plot of Figure \ref{fig:R0_analytical} to show more clearly structure where \texttt{Mathematica} finds a divergence in our analytic approximation of $r_2^2 r_3^2/u^4\times R_{(0,0,0,0)0,0,0}^{\left[0\right],\left[0\right],\left[0\right]} (U,r_2,r_3)$ through Eq. (\ref{eq:R0_analytical}). }
\label{fig:3d_R0_analytical}
\end{figure}

Looking next at the $L_2=1$ case and assuming once again $f \sim f'$ and $g \sim g'$, we find: 

\begin{align}
&R_{(0,0,0,0),0,1,0}^{\left[0\right],\left[0\right],\left[0\right]}(r_1,r_2,r_3) \sim \frac{1}{r_1^2 r_3}\int_0^{\infty} dr\; I(u_{\pm,R},r)\mathcal{H}(u_{\pm,R},r)\theta(r_1,r_3,r), 
\end{align}
where we have used the result of Eq. (\ref{eq:g_prime_analytical1}) for the term $g_{0,1,0}^{\left[0\right]}(r',r,r_2)$, $u_{+,R} = r_1+r_2$ and $u_{-,R} = \left|r_1-r_2\right|$. The above integral is similar to that of Eq. (\ref{eq:hprime_1_analitical}), so if we re-distribute the $ I(u_{\pm,R},r)\mathcal{H}(u_{\pm,R},r)$ term and the integral bounds, we find

\begin{align}\label{eq:R001_analytical}
&R_{(0,0,0,0),0,1,0}^{\left[0\right],\left[0\right],\left[0\right]}(r_1,r_2,r_3) \sim \frac{1}{r_1^2 r_3}\left\{\int_{\left|r_1-r_3\right|}^{r_1+r_3< r_1+r_2 }dr\; I_{u>}(u_{+,R},r) \right. \nonumber \\ 
& \qquad \qquad \qquad \qquad \qquad \qquad \quad \left.+ \int_{\left|r_1-r_3\right|>r_1+r_2}^{r_1+r_3}dr\; I_{u<}(u_{+,R},r)\right.\nonumber \\ 
& \qquad \qquad \qquad \qquad \qquad \qquad \quad \left. - \int_{\left|r_1-r_3\right|}^{r_1+r_3<\left|r_1-r_2\right|}dr\; I_{u>}(u_{-,R},r) \right.\nonumber \\ 
& \qquad \qquad \qquad \qquad \qquad \qquad \quad \left.- \int_{\left|r_1-r_3\right|>\left|r_1-r_2\right|}^{r_1+r_3}dr\;I_{u<}(u_{-,R},r)\right\}.
\end{align}

We have evaluated the above integrals in \texttt{Mathematica} and produced the 2D plot of Figure \ref{fig:R1_analytical} and 3D plot of Figure \ref{fig:Analytcal_R010}. \texttt{Mathematica} produces a divergence---white area in the 2D plot and grey area in the 3D plot. The divergence is caused because the integral and the result of the integral depend on $\tanh^{-1}$, which diverges when its argument approaches 1. Therefore, the divergence is the region where the ratio $r/u_{+,R}$, $u_{+,R}/r$, $r/u_{-,R}$, and/or $u_{-,R}/r$ approach 1. We have evaluated numerically the bounds of the integral from $0$ to $200\;\left[{\rm Mpc}/h\right]$, therefore by choosing $r_1 = 100\;\left[{\rm Mpc}/h\right]$ (\textit{i.e.}, $u_{\pm,R} = \left|100 \pm r_2\right|$) we observe the divergence only occurs for $r_2 \geq 100\;\left[{\rm Mpc}/h\right]$. However, as can be seen, the two figures produced in \texttt{Mathematica} show behavior similar to the lower 2D plot of Figure \ref{fig:R_int}.  

\begin{figure}[h]
\centering
\includegraphics[scale=0.85]{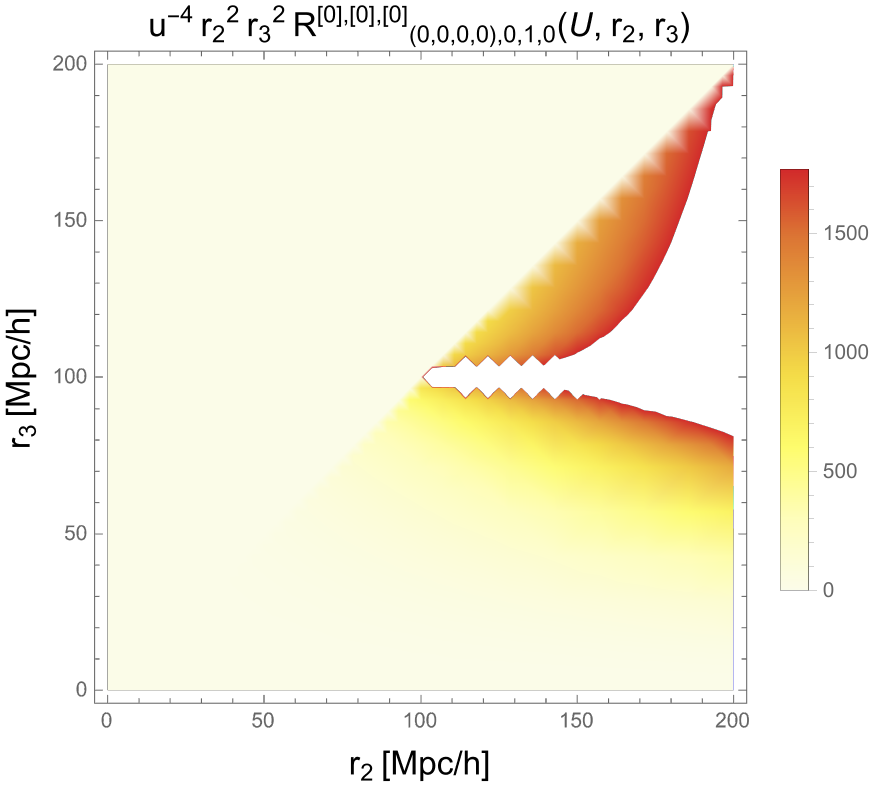}
\caption{Here we show $r_2^2r_3^2/u^4\times R_{(0,0,0,0)0,1,0}^{\left[0\right],\left[0\right],\left[0\right]} (U,r_2,r_3)$, using Eq. (\ref{eq:R001_analytical}) with $u \equiv 10\;\left[{\rm Mpc}/h\right]$ and $U \equiv 100\;\left[{\rm Mpc}/h\right]$. We see monotonic increasing as we go to larger $r_2$ and $r_3$ with oval-like behavior on the right-hand side. Around this oval-like behavior, \texttt{Mathematica} produces a divergence. To understand this better, we display a 3D plot in Figure \ref{fig:Analytcal_R010}. We see that the behavior of the figure above is a good approximation---except for the divergence---to that in the \textit{lower panel} of Figure \ref{fig:R_int}. The divergence is due to dependence on $\tanh^{-1}$, which diverges when its argument approaches unity. Therefore, the divergence covers the region where the ratio $r/u_{+,R}$, $u_{+,R}/r$, $r/u_{-,R}$, and/or $u_{-,R}/r$ approach unity. We have evaluated numerically the bounds of the integral from $0$ to $200\;\left[{\rm Mpc}/h\right]$, therefore by choosing $r_1 = 100\;\left[{\rm Mpc}/h\right]$ (\textit{i.e.}, $u_{\pm,R} = \left|100 \pm r_2\right|$) we observe the divergence only occurs for $r_2 \geq 100\;\left[{\rm Mpc}/h\right]$.}
\label{fig:R1_analytical}
\end{figure}

\begin{figure}[h]
\centering
\includegraphics[scale=0.65]{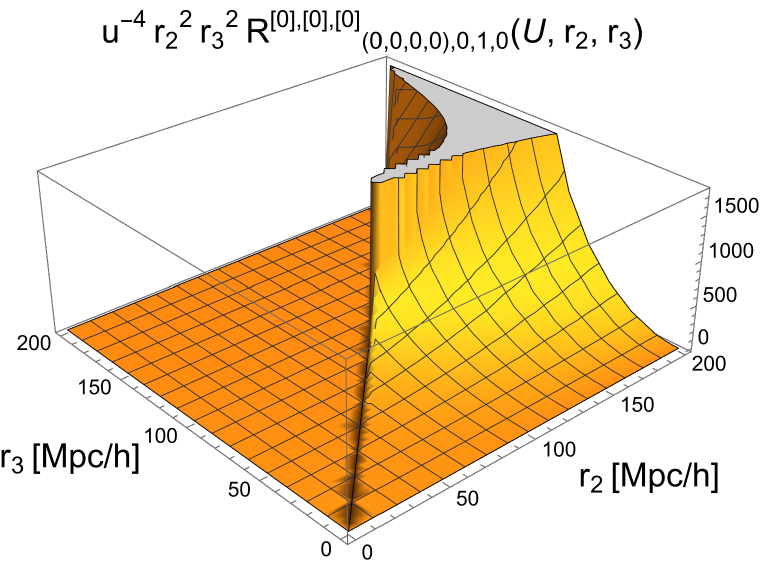}
\caption{3D plot of Figure \ref{fig:R1_analytical} to show more clearly structure where \texttt{Mathematica} finds a divergence in our analytic approximation of $r_2^2r_3^2/u^4\times R_{(0,0,0,0)0,1,0}^{\left[0\right],\left[0\right],\left[0\right]} (U,r_2,r_3)$ as in Eq. (\ref{eq:R001_analytical}).}
\label{fig:Analytcal_R010}
\end{figure}

We continue our analysis with Eq. (\ref{eq:fancy_R_int}) for the case: 

\begin{align}
&\mathcal{R}_{(0,0,0),0,0,0}^{\left[0\right],\left[0\right],\left[0\right]}(r_1,r_2,r_3) \equiv \int_{0}^{\infty} dr' \; r' \;f_{0,0}^{\left[0\right]}(r',r_1)f_{0,0}^{\left[0\right]}(r',r_2)f_{0,0}^{\left[0\right]}(r',r_3), 
\end{align}
and assuming once again $f \sim f'$, the integral becomes:

\begin{align}\label{eq:Fancy_R0_analytical}
&\mathcal{R}_{(0,0,0),0,0,0}^{\left[0\right],\left[0\right],\left[0\right]}(r_1,r_2,r_3) \sim \frac{1}{r_1^2r_2^2r_3^2} \int_{0}^{\infty} dr' \; r'\; \delta_{ \rm D}^{\left[1\right]}(r'-r_1) \delta_{ \rm D}^{\left[1\right]}(r'-r_2) \delta_{ \rm D}^{\left[1\right]}(r'-r_3)  \nonumber \\ 
& \qquad \qquad \qquad \qquad \quad \;  = \frac{\delta_{ \rm D}^{\left[1\right]}(r_1-r_2) \delta_{ \rm D}^{\left[1\right]}(r_1-r_3)}{r_1r_2^2r_3^2}.
\end{align}
If $r_1 = 100 \;\left[{\rm Mpc}/h\right]$, then we expect to see non-zero behavior when $r_2 =100\;\left[{\rm Mpc}/h\right]$ and $r_3=100\;\left[{\rm Mpc}/h\right]$. The upper 2D plot of Figure \ref{fig:fancy_R_int} shows our expectations to be true with behavior only around $r_2=r_3=100\;\left[{\rm Mpc}/h\right]$ and symmetry across the diagonal. 

\noindent Looking at the next case for Eq. (\ref{eq:fancy_R_int}) with $L_2=1$, we have: 
\begin{align}
&\mathcal{R}_{(0,0,0),0,1,0}^{\left[0\right],\left[0\right],\left[0\right]}(r_1,r_2,r_3) \equiv \int_{0}^{\infty} dr' \; r' \;f_{0,0}^{\left[0\right]}(r',r_1)f_{0,1}^{\left[0\right]}(r',r_2)f_{0,0}^{\left[0\right]}(r',r_3), 
\end{align}
and using Eq. (\ref{eq:2_sph_Bessel_delta_relation}) for the $f$ integrals (assuming $f\sim f'$), we obtain:

\begin{align}\label{eq:Fancy_R1_analytical}
&\mathcal{R}_{(0,0,0),0,1,0}^{\left[0\right],\left[0\right],\left[0\right]}(r_1,r_2,r_3) \sim \frac{1}{r_1 r_3^2}f_{0,1}^{\left[0\right]}(r_1,r_2)\delta_{ \rm D}^{\left[1\right]}(r_1-r_3). 
\end{align}
If we now apply the result of Eq. (\ref{eq:f1_analitical}) for $f_{0,1}^{\left[0\right]}$, we find:

\begin{align}\label{eq:R1_analitical}
&\mathcal{R}_{(0,0,0),0,1,0}^{\left[0\right],\left[0\right],\left[0\right]}(r_1,r_2,r_3) \sim \frac{1}{r_1 r_2^2 r_3^2}\delta_{ \rm D}^{\left[1\right]}(r_1-r_3), \quad r_1<r_2; \quad \sim 0, \quad r_2<r_1. 
\end{align} 
Comparing our result above with the lower panel Figure \ref{fig:fancy_R_int}, we find agreement. The Dirac delta function tells us to expect non-zero behavior around the line $r_3=r_1$. Indeed, observing the lower panel of Figure \ref{fig:fancy_R_int} we only see behavior along $r_3=r_1$ (with $r_1 = 100\;\left[{\rm Mpc}/h\right]$) when $r_2>r_1$, which is the exact condition we must satisfy for the integral not to vanish. 

\subsubsection{Radial Integrals from $\S$\ref{Section T2}}

We now evaluate the radial integrals from $\S$\ref{Section T2} and start by evaluating Eq. (\ref{eq:R1_radial_n13_n_123}) for the case:

\begin{align}
&I_{(0,0,0,0),0,0,0}^{\left[0\right],\left[0\right],\left[0\right]}(r_1,r_2,r_3) =\int_{0}^{\infty} dr' \;r' f_{0,0}^{\left[0\right]}(r',r_2) \int_{0}^{\infty}dr \;r\int_{0}^{\infty}ds\; s^2 \;\xi_0(s) \nonumber \\ 
& \quad \qquad \qquad \qquad\qquad \qquad \qquad \qquad   \times  h_{0,0,0,0}^{\left[0\right]} (r,r',s,r_1) h_{0,0,0,0}^{'\left[0\right]}(r,r',s,r_3). 
\end{align}

We will assume that $f\sim f'$, $h \sim h'$ and $\xi_{0}(r)\sim 1/r^2$ such that we can use Eq. (\ref{eq:2_sph_Bessel_delta_relation}) and Eq. (\ref{eq:h_analitycal0}) to arrive at: 

\begin{align}
&I_{(0,0,0,0),0,0,0}^{\left[0\right],\left[0\right],\left[0\right]} (r_1,r_2,r_3)\sim \frac{1}{r_1 r_2^2r_3}\int_{0}^{\infty}\frac{dr}{r}\int_{0}^{\infty} \frac{ds}{s^2}\int_{0}^{\infty} dR\; \theta(r,r_2,R)\theta(s,r_1,R)\nonumber \\
& \qquad \qquad \qquad \qquad  \times \int_{0}^{\infty}dR' \;\theta(r,r_2,R')\theta(s,r_3,R').
\end{align}

We will evaluate the $s$ integral first and approximate it as:

\begin{align}\label{eq:thetaXtheta_delta}
\int_{0}^{\infty}\frac{ds}{s^2}\; \theta(s,r_1,R)\theta(s,r_3,R') \sim \delta_{\rm D}^{\left[1\right]}(r_1-R) \delta_{\rm D}^{\left[1\right]} (r_3-R'),
\end{align}
based on Figure \ref{fig:theta_theta_50} and Figure \ref{fig:theta_theta_100}. The two figures show a divergence near the location at which we choose $R$ and $R^{'}$, justifying our assumption to approximate the above integral as the product of two Dirac delta functions.

\begin{figure}[h]
\centering
\includegraphics[scale=0.85]{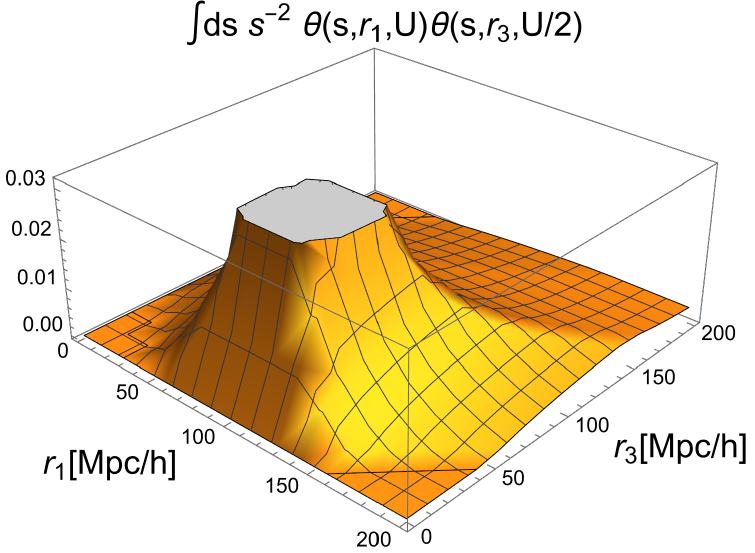}
\caption{3D plot of the integral Eq. (\ref{eq:thetaXtheta_delta}) with $R= U \equiv 100 \;\left[{\rm Mpc}/h\right]$ and $R^{'} = U/2 = 50 \;\left[{\rm Mpc}/h\right]$. The plot shows a square divergence centered at $r_1=R=U$ and $r_3=R^{'} = U/2$. This plot is the first of two and serves as an illustration that we can approximate Eq. (\ref{eq:thetaXtheta_delta}) as a product of two Dirac delta functions.}
\label{fig:theta_theta_50}
\end{figure}

\begin{figure}[h]
\centering
\includegraphics[scale=0.85]{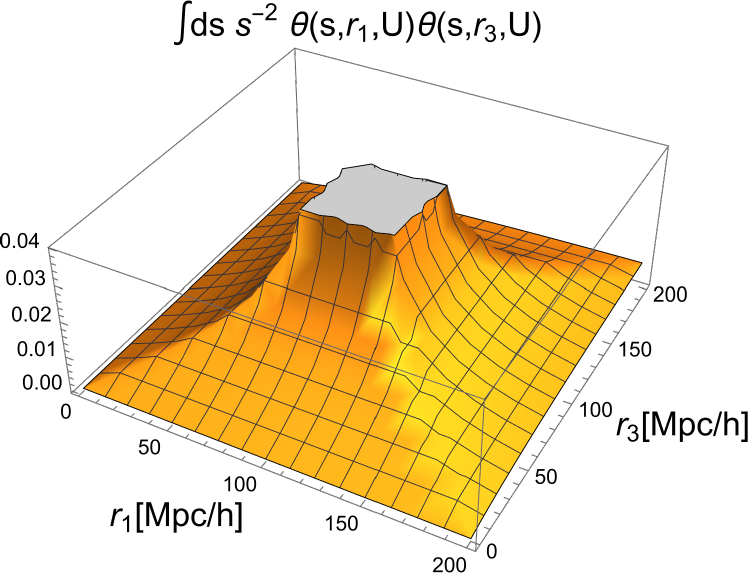}
\caption{3D plot of the integral Eq. (\ref{eq:thetaXtheta_delta}) with $R= U \equiv 100 \;\left[{\rm Mpc}/h\right]$ and $R^{'}  = U$. The plot shows a square divergence centered at $r_1=R=U$ and $r_3=R^{'} = U$. This plot is the second of two and serves as an illustration that we can approximate Eq. (\ref{eq:thetaXtheta_delta}) as a product of two Dirac delta functions.}
\label{fig:theta_theta_100}
\end{figure}

Using the above result to integrate over the $R$ and $R'$ integral, we find: 

\begin{align}
I_{(0,0,0,0),0,0,0}^{\left[0\right],\left[0\right],\left[0\right]}(r_1,r_2,r_3) \sim \frac{1}{r_1 r_2^2r_3}\int_{0}^{\infty}\frac{dr}{r}\; \theta(r,r_2,r_1)\theta(r,r_2,r_3). 
\end{align}

The resulting integral is the same as that which we found in Eq. (\ref{eq:R0_analytical}), except that in the above equation, $r_2$ takes the place of $r_1$, and the denominator has $r_2^2$ instead of $r_2^3$. Therefore, we expect to observe the same behavior as for the integral $R_{(0,0,0),0,0,0}^{\left[0\right],\left[0\right],\left[0\right]}$. Comparing the upper panel of Figure \ref{fig:R_int} with the upper panel of Figure \ref{fig:I_int}, we find the same behavior as expected. 

We continue with the same integral, but analyzing the case $L_3=1$:

\begin{align}
&I_{(0,0,0,0),0,0,1}^{\left[0\right],\left[0\right],\left[0\right]}(r_1,r_2,r_3) = \int_{0}^{\infty} dr' \;r' f_{0,0}^{0}(r',r_2) \int_{0}^{\infty}dr \;r\int_{0}^{\infty}ds\; s^2 \;\xi_0(s) \nonumber \\ 
& \quad \qquad \qquad \qquad\qquad \qquad \qquad \qquad   \times  h_{0,0,0,0}^{\left[0\right]} (r,r',s,r_1) h_{0,0,0,1}^{'\left[0\right]}(r,r',s,r_3). 
\end{align}

We make the same assumptions as before and use Eq.(\ref{eq:hprime_1_analitical}) to rewrite the last term, finding: 

\begin{align}
&I_{(0,0,0,0),0,0,1}^{\left[0\right],\left[0\right],\left[0\right]} (r_1,r_2,r_3) \sim \frac{1}{r_1 r_2^2}\int_{0}^{\infty}\frac{dr}{r}\int_{0}^{\infty} \frac{ds}{s}\int_{0}^{\infty} dR\; \theta(r,r_2,R)\theta(s,r_1,R)\nonumber \\
& \qquad \qquad \qquad \qquad  \times \int_{0}^{\infty}dR' \;\theta(r,r_2,R')\mathcal{H}(u_{\pm},R')I(u_{\pm},R'),
\end{align}
with $u_{\pm,I} \equiv \left|s\pm r_3 \right|$. Performing the $r$ integral first, we obtain:

\begin{align}
\int_{0}^{\infty}\frac{dr}{r}\; \theta(r,r_2,R)\theta(r,r_2,R') \sim \delta_{\rm D}^{\left[1\right]}(r_2-R) \delta_{\rm D}^{\left[1\right]}(r_2-R'), 
\end{align}
which allows us to evaluate the $R$ and $R'$ integrals to find: 

\begin{align}\label{eq:I_001_analytical}
I_{(0,0,0,0),0,0,1}^{\left[0\right],\left[0\right],\left[0\right]}(r_1,r_2,r_3)  \sim \frac{1}{r_1 r_2^2}\int_{0}^{\infty}\frac{ds}{s} \; \theta(s,r_1,r_2)\mathcal{H}(u_{\pm,I},s)I(u_{\pm,I},s),
\end{align}
where $u_{\pm,I} \equiv \left|r_2 \pm r_3 \right|$. The above integral is similar to the integral we found in Eq. (\ref{eq:hprime_1_analitical}) but integrates over the variable $s$, and weighted by $1/s$. Therefore, we have evaluated it in the same manner and produced the 2D plot of Figure \ref{fig:I001_analytical} and 3D plot of Figure \ref{fig:I_001 analitycal}. The divergence is caused because the integral and the result of the integral depend on $\tanh^{-1}$, which diverges when its argument approaches unity. Therefore, the divergence is the region where the ratio $s/u_{+,I}$, $u_{+,I}/s$, $s/u_{-,I}$, and $u_{-,I}/s$ approach unity. We have evaluated numerically the bounds of the integral from $0$ to $200\;\left[{\rm Mpc}/h\right]$, therefore by choosing $r_2 = 100\;\left[{\rm Mpc}/h\right]$ (\textit{i.e.}, $u_{\pm,I} = \left|100 \pm r_3\right|$) we observe the divergence only occurs after $r_3 = 100\;\left[{\rm Mpc}/h\right]$. However, as can be seen, these two figures show the same behavior as the lower 2D plot of Figure \ref{fig:I_int}, demonstrating that the latter seems to be correct.

\begin{figure}[h]
\centering
\includegraphics[scale=0.7]{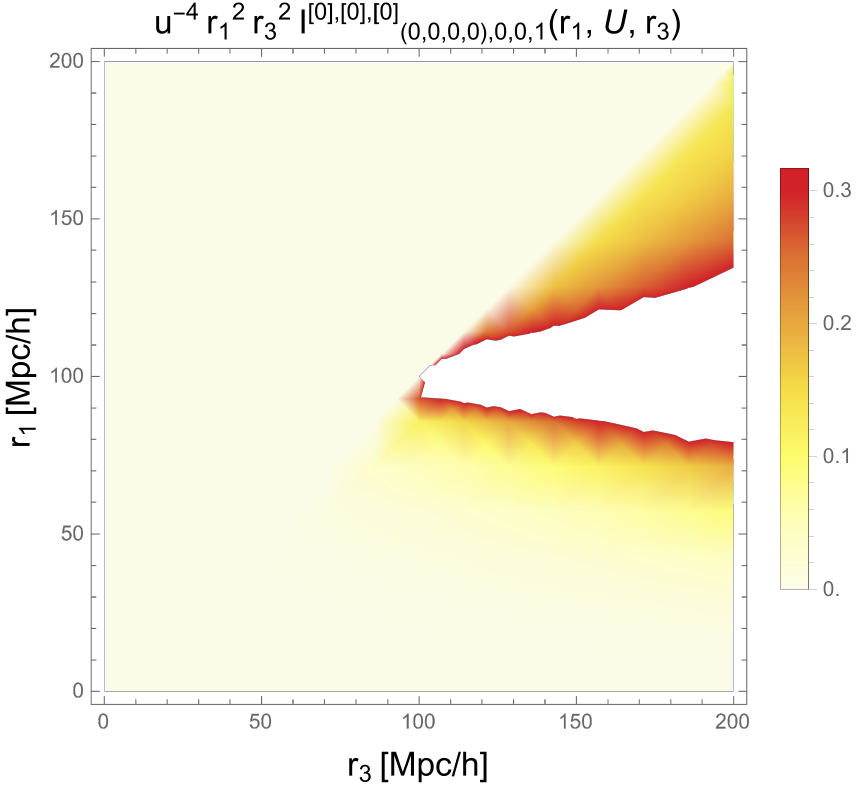}
\caption{Here we show $r_1^2r_3^2/u^4\times I_{(0,0,0,0)0,0,1}^{\left[0\right],\left[0\right],\left[0\right]} (r_1,U,r_3)$, for Eq. (\ref{eq:I_001_analytical}) with $u \equiv 10\;\left[{\rm Mpc}/h\right]$ and $U \equiv 100\;\left[{\rm Mpc}/h\right]$. We see the integral increasing around the line $r_3=100\left[{\rm Mpc}/h\right]$, but only in the lower-diagonal elements, $r_1<r_3$. Around the line $r_3=100\left[{\rm Mpc}/h\right]$, \texttt{Mathematica} produces a divergence. To understand this better we show a 3D plot in Figure \ref{fig:I_001 analitycal}. As can be seen above, the behavior of the above figure is a good approximation---except for the divergence---to the \textit{lower panel} of Figure \ref{fig:I_int}. The divergence due to dependence on $\tanh^{-1}$, which diverges when its argument approaches unity. Therefore, the divergence is over the region where the ratio $s/u_{+,I}$, $u_{+,I}/s$, $s/u_{-,I}$, and/or $u_{-,I}/s$ approach unity. We have evaluated numerically the bounds of the integral from $0$ to $200\;\left[{\rm Mpc}/h\right]$, therefore by choosing $r_2 = 100\;\left[{\rm Mpc}/h\right]$ (\textit{i.e.}, $u_{\pm,I} = \left|100 \pm r_3\right|$) we observe the divergence only occurs after $r_3 = 100\;\left[{\rm Mpc}/h\right]$.}
\label{fig:I001_analytical}
\end{figure}

\begin{figure}[h]
\centering
\includegraphics[scale=0.85]{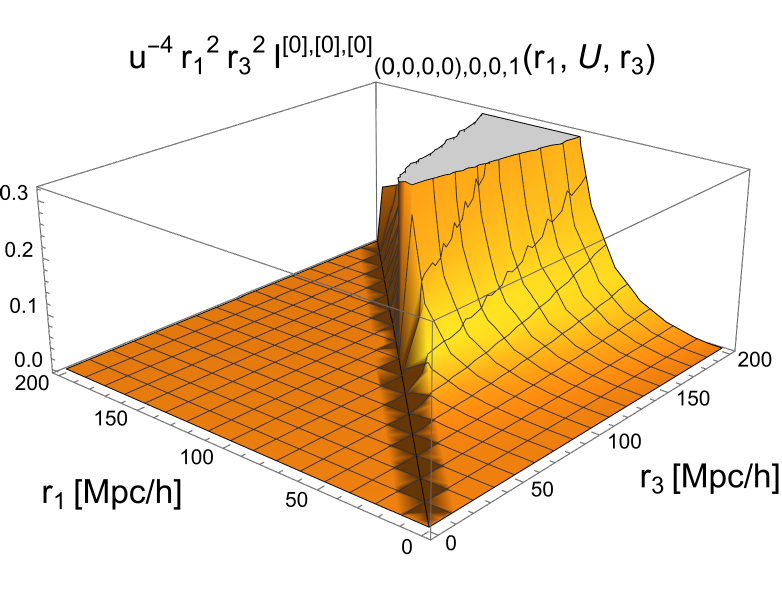}
\caption{3D plot of Figure \ref{fig:I001_analytical} to show more clearly structure where \texttt{Mathemtica} finds a divergence in our analytic approximation of $r_1^2r_3^2/u^4\times I_{(0,0,0,0)0,0,1}^{\left[0\right],\left[0\right],\left[0\right]} (r_1,U,r_3)$ through Eq. (\ref{eq:I_001_analytical}).}
\label{fig:I_001 analitycal}
\end{figure}

Next, we evaluate Eq. (\ref{eq:radial_n123_R1_T2}) for the case:
\begin{align}
&\mathcal{I}_{(0,0,0,0)0,0,0}^{\left[0\right],\left[0\right],\left[0\right]}(r_1,r_2,r_3) = \int_{0}^{\infty} dr'\;r'\; f_{0,0}^{\left[0\right]}(r',r_2) \int_{0}^{\infty} ds\;s^2\;\xi_{0}(s)\nonumber \\
&\qquad \qquad \qquad \qquad \qquad \quad  \times   g_{0,0,0}^{\left[0\right]}(r',s,r_1)g_{0,0,0}^{'\left[0\right]}(r',s,r_3).
\end{align}

Making the assumption that $f\sim f'$, $g \sim g'$ and $\xi_{0}(r)\sim 1/r^2$ such that we can use Eq. (\ref{eq:2_sph_Bessel_delta_relation}) and Eq. (\ref{eq:g000_result}), we find: 

\begin{align}
\mathcal{I}_{(0,0,0,0)0,0,0}^{\left[0\right],\left[0\right],\left[0\right]} (r_1,r_2,r_3)\sim \frac{1}{r_1 r_2^3r_3} \int_{0}^{\infty} \; \frac{ds}{s^2}\; \theta(r_2,s,r_1)\theta(r_2,s,r_3)
\end{align}
after integrating over the $r'$ integral. The above integral we have evaluated before as an approximation of two Dirac delta functions. We find:

\begin{align}\label{eq:I0_sim_R0}
\mathcal{I}_{(0,0,0,0)0,0,0}^{\left[0\right],\left[0\right],\left[0\right]}(r_1,r_2,r_3) \sim \frac{1}{r_1 r_2^3r_3}\delta_{\rm D}^{\left[1\right]}(r_2-r_1) \delta_{\rm D}^{\left[1\right]}(r_2-r_3).
\end{align}
The result above matches exactly that which we found in Eq. (\ref{eq:Fancy_R0_analytical}) except for the power of $r_3$. Therefore, the above results explains why the upper 2D plot of Figure \ref{fig:fancy_I_int} and Figure \ref{fig:fancy_R_int} are almost identical. 

We continue with Eq. (\ref{eq:radial_n123_R1_T2}) for the case:
\begin{align}
&\mathcal{I}_{(0,0,0,0)0,0,1}^{\left[0\right],\left[0\right],\left[0\right]}(r_1,r_2,r_3) = \int_{0}^{\infty} dr'\;r'\; f_{0,0}^{\left[0\right]}(r',r_2) \int_{0}^{\infty} ds\;s^2\;\xi_{0}(s)\nonumber \\
&\qquad \qquad \qquad \qquad \qquad \quad  \times    g_{0,0,0}^{\left[0\right]}(r',s,r_1)g_{0,0,1}^{'\left[0\right]}(r',s,r_3).
\end{align}
We will demonstrate below why the lower 2D plots of Figure \ref{fig:fancy_R_int} and Figure \ref{fig:fancy_I_int} look the same. To get this done, we  start by making the assumption $f\sim f'$, which allows us to use the result in Eq. (\ref{eq:2_sph_Bessel_delta_relation}), $g \sim g'$ so that we can distribute the sBF's without having to carry the power spectrum and $\xi_0(s)\sim 1/s^2$. Hence, the above integral becomes:
 
\begin{align}
&\mathcal{I}_{(0,0,0,0)0,0,1}^{\left[0\right],\left[0\right],\left[0\right]}(r_1,r_2,r_3) \sim \frac{1}{r_2}\int_{0}^{\infty} ds \int_{0}^{\infty} dk_1 \;k_1^2\; j_0(k_1r_2)j_0(k_1s)j_0(k_1r_1)\nonumber \\ 
&\qquad \qquad \qquad \qquad \qquad \qquad \quad \times \int_{0}^{\infty} dk_3\;k_3^2\;j_0(k_3r_2)j_0(k_3s)j_1(k_3r_3).
\end{align}
If we perform the $s$ integral, we find: 

\begin{align}
\int_{0}^{\infty} ds \;j_0(k_1s)j_0(k_3s) = \frac{\pi}{2\;k_>},
\end{align}
where $k_>$ is the greater of $k_1$ and $k_3$. Therefore, we obtain:

\begin{align}\label{eq:I_001_Analytial_analysis}
&\mathcal{I}_{(0,0,0,0)0,0,1}^{\left[0\right],\left[0\right],\left[0\right]}(r_1,r_2,r_3) \sim \frac{1}{r_2} \int_{0}^{\infty} dk_1 \;\frac{k_1^2}{k_>}\; j_0(k_1r_2)j_0(k_1r_1)\int_{0}^{\infty} dk_3\;k_3^2\;j_0(k_3r_2)j_1(k_3r_3).
\end{align}

Let us evaluate the above integral when $k_>=k_1$: The $k_1$ integral can be approximated as $\delta^{\left[1\right]}(r_2-r_1)/r_1^2$ as we have done with $f_{0,0}^{\left[0\right]}$ and the $k_3$ integral can also be approximated as  $f_{0,1}^{\left[0\right]}(r_2,r_3)$. Therefore, we can conclude $\mathcal{I}_{(0,0,0,0)0,0,1}^{\left[0\right],\left[0\right],\left[0\right]}(r_1,r_2,r_3) \sim \mathcal{R}_{(0,0,0)0,0,1}(r_1,r_2,r_3)$. To see that these two equations are almost identical we have moved the spherical Bessel order from $j_2=1 \rightarrow j_2 = 0$ and $j_3=0\rightarrow j_3=1$, which implies also making a change of variable in the right-hand side of Eq. (\ref{eq:Fancy_R1_analytical}) of $r_1\rightarrow r_2$, $r_2 \rightarrow r_3$ and $r_3 \rightarrow r_1$. Then, we evaluate Eq.(\ref{eq:I_001_Analytial_analysis}) when $k_> = k_3$: The $k_1$ integral becomes exactly $\delta^{\left[1\right]}(r_2-r_1)/r_1^2$, and the $k_3$ integral can be approximated once again as $f_{0,1}^{\left[0\right]}(r_2,r_3)$, implying that $\mathcal{I}_{(0,0,0,0)0,0,1}^{\left[0\right],\left[0\right],\left[0\right]}(r_1,r_2,r_3) \sim \mathcal{R}_{(0,0,0)0,0,1}(r_1,r_2,r_3)$. Hence, we can conclude $\mathcal{I}_{(0,0,0,0)0,0,1}^{\left[0\right],\left[0\right],\left[0\right]}(r_1,r_2,r_3) \sim \mathcal{R}_{(0,0,0)0,0,1}(r_1,r_2,r_3)$, which explains why the lower 2D plots of Figure \ref{fig:fancy_R_int}
and Figure \ref{fig:fancy_I_int} are almost identical. 

Next, we evaluate Eq. (\ref{eq:radial_n13_R1_T2}) for the case:
\begin{align}
\mathbb{I}_{(0,0),0,0}^{\left[0\right],\left[0\right]}(r_1,r_3) = \int_{0}^{\infty} dr'\;r'\; \int_{0}^{\infty} ds\;s^2\;\xi_{0}(s) g_{0,0,0}^{\left[0\right]}(r',s,r_1)g_{0,0,0}^{'\left[0\right]}(r',s,r_3). 
\end{align}
We assume once again $\xi_0(s)\sim1/s^2$ and $g \sim g'$, so that we can use Eq. (\ref{eq:g000_result}) to find:

\begin{align}
&\mathbb{I}_{(0,0),0,0}^{\left[0\right],\left[0\right]}(r_1,r_3) \sim \frac{1}{r_1r_3} \int_{0}^{\infty} dr'\; \frac{1}{r'}\int_{0}^{\infty} ds\; \frac{\theta(r',s,r_1)\theta(r',s,r_3)}{s^2} \nonumber \\ 
& \qquad \qquad  \qquad \; \sim \frac{1}{r_1r_3} \int_{0}^{\infty} dr'\; \frac{\delta_{\rm D}^{\left[1\right]}(r'-r_1) \delta_{\rm D}^{\left[1\right]}(r'-r_3)}{r'}, 
\end{align}
where we have used Eq. (\ref{eq:thetaXtheta_delta}) to go from the first line to the second one. Finally, evaluating the $r'$ integral we find:
\begin{align}
&\mathbb{I}_{(0,0),0,0}^{\left[0\right],\left[0\right]}(r_1,r_3) \sim \frac{\delta_{\rm D}^{\left[1\right]} (r_1-r_3)}{r_1^2r_3}. 
\end{align}
The above result explains why we see non-zero values only along the diagonal $r_1 = r_3$ for the upper 2D plot of Figure \ref{fig:II_int}. 

Let us continue to the case with $L_3=1$:

\begin{align}
\mathbb{I}_{(0,0),0,1}^{\left[0\right],\left[0\right]}(r_1,r_3) = \int_{0}^{\infty} dr'\;r'\; \int_{0}^{\infty} ds\;s^2\;\xi_{0}(s) g_{0,0,0}^{\left[0\right]}(r',s,r_1)g_{0,0,1}^{'\left[0\right]}(r',s,r_3). 
\end{align}

We will express the above equation in terms of the sBFs and use again the assumptions $\xi_{0}(s)\sim 1/s^2$ and $P(k)\sim 1/k$. Hence, the above integral becomes: 

\begin{align}
&\mathbb{I}_{(0,0),0,1}^{\left[0\right],\left[0\right]}(r_1,r_3) \sim \int_{0}^{\infty} dr'\; r' \int_{0}^{\infty} ds \int_{0}^{\infty} dk_1\; k_1\; j_0(k_1r')j_0(k_1s) j_0(k_1r_1)\nonumber \\ 
& \qquad \qquad \qquad \qquad \qquad \qquad \quad  \times \int_{0}^{\infty} dk_3\; k_3^2\; j_0(k_3r')j_0(k_3s) j_1(k_3r_1).   
\end{align}
The $r'$ integral can be approximated by $\delta_{\rm D
}^{\left[1\right]}(k_1 - k_3)/k_1$ as we showed in Eq. (\ref{eq:2_sph_Bessel_delta_relation}) by assuming $f \sim f'$. So, we obtain: 

\begin{align}
&\mathbb{I}_{(0,0),0,1}^{\left[0\right],\left[0\right]}(r_1,r_3) \sim \int_{0}^{\infty} dk_1\; k_1\; j_0(k_1r_1) j_1(k_1r_3) \int_{0}^{\infty} dx \; j_0^2(x) \nonumber \\ 
& \qquad \qquad \qquad =\frac{\pi}{2} \int_{0}^{\infty} dk_1\; k_1\; j_0(k_1r_1) j_1(k_1r_3), 
\end{align}
where $x\equiv k_1 s$. The above integral is the same as those we see in Eq. (\ref{eq:f1_analitical}) and Eq. (\ref{eq:F1_analytical}). Thus, the above expression explains why we see that the lower 2D plots of Figure \ref{fig:fint2}, Figure \ref{fig:2D_int} and Figure \ref{fig:II_int} are almost identical to each other. 

Lastly, we evaluatie Eq. (\ref{eq:radial_0_R1_T2}) for the case $\ell''_1 = L_1 = L_3 = n'_1 = n'_3 = 0$:

\begin{align}
i_{(0),0,0}^{\left[0\right],\left[0\right]}(r_1,r_3) = \int_{0}^{\infty}ds \; s^2\;\xi_0(s) f_{0,0}^{\left[0\right]}(s,r_1) f_{0,0}^{'\left[0\right]}(s,r_3). 
\end{align}
The $f'$ integral we can re-express in terms of the Dirac delta function using Eq. (\ref{eq:2_sph_Bessel_delta_relation}), and if we perform the $s$ integral we arrive at: 

\begin{align}\label{eq:anaytical_i}
i_{(0),0,0}^{\left[0\right],\left[0\right]}(r_1,r_3) = \frac{1}{4\pi}\xi_{0}(r_3)f_{0,0}^{\left[0\right]}(r_3,r_1). 
\end{align}

As shown in Figure \ref{fig:analytical_evaluation_of_i}, the above expression gives the same 2D plot as we can see in the \textit{upper panel} of Figure \ref{fig:i_int}. The above expression also tells us that even though we started with what looked like a symmetric function, the fact that $f'$ is not dependent on the power spectrum creates an asymmetry across the diagonal $r_1=r_3$. The result above is dependent on $f$, which we showed in Eq. (\ref{eq:2_sph_Bessel_delta_relation}) can be approximated as a Dirac delta function, explaining why mainly the non-zero values for Figure \ref{fig:i_int} are seen along the diagonal. The $f$ function is also dependent on the full power spectrum and therefore sensible to BAO features, showing along the diagonal a bump at $r_1 = r_3 = 100\;\left[{\rm Mpc}/h\right]$. If we fix $r_1 = 100\;\left[{\rm Mpc}/h\right]$ and let $r_3$ vary, $\xi_{0}(r_3)$ reproduces the 2PCF behavior along that line with a first peak at $r_3 \approx 0 \;\left[{\rm Mpc}/h\right]$ and the BAO peak at $r_3 = 100\;\left[{\rm Mpc}/h\right]$. 

Finally, we evaluate the case $\ell''_1 = L_3 = n'_1 = n'_3 = 0$ and $L_1=1$. Following the same procedure as above we arrive at:

\begin{align}
i_{(0),1,0}^{\left[0\right],\left[0\right]}(r_1,r_3) = \frac{1}{4\pi}\xi_{0}(r_3)f_{0,1}^{\left[0\right]}(r_3,r_1) \sim  \left\{\begin{matrix}
\xi_0(r_3)/r_1^2, \quad r_3<r_1\\ \;0, \qquad \qquad \; \; r_1\leq r_3,
\end{matrix}\right.
\end{align}
with the last equality holding true via Eq. (\ref{eq:f1_analitical}) if we let $P(k)\sim 1/k$. The above expression explains why the integral values vanish above the diagonal $r_3>r_1$ and why we only observe non-zero values below the diagonal for the lower 2D plot of Figure \ref{fig:i_int}. 

\begin{figure}[h]
\centering
\includegraphics[scale=0.85]{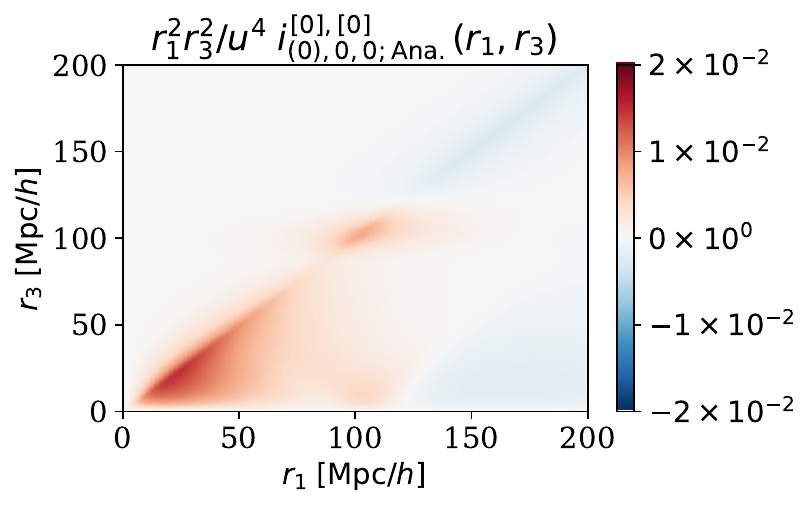}
\caption{Analytic evaluation of Eq.(\ref{eq:radial_0_R1_T2}) for $\ell''_1=L_1=L_3=n'_1=n'_3=0$ using Eq.(\ref{eq:anaytical_i}). The above plot shows the same diagonal behavior and bump around $r_1 = r_3 = 100 \;\left[{\rm Mpc}/h\right]$ as we see in the \textit{upper panel} of Figure \ref{fig:i_int}, adding confidence that the figure computed with the full power spectrum is indeed correct. We described in the paragraph below Eq. (\ref{eq:anaytical_i}) how the dependence of $i$ on the function $f$ explains the behavior along the diagonal, because $f$ can be approximated as a Dirac delta function. $f$ also depends on the full power spectrum and shows BAO features at $r_1 = r_3 = 100\;\left[{\rm Mpc}/h\right]$. If we fix $r_1 = 100\;\left[{\rm Mpc}/h\right]$ and let $r_3$ vary, $\xi_{0}(r_3)$ reproduces the 2PCF behavior along that line with a first peak at $r_3 \approx 0 \;\left[{\rm Mpc}/h\right]$ and the BAO peak at $r_3 = 100\;\left[{\rm Mpc}/h\right]$. We have defined $u \equiv 10\;\left[{\rm Mpc}/h\right]$.}
\label{fig:analytical_evaluation_of_i}
\end{figure}


\clearpage


\begin{thebibliography}{99}

\bibitem{Aviles}
A. Aviles, A. Banerjee, N. Gustavo and Z. Slepian, “Clustering in massive neutrino cosmologies via Eulerian Perturbation Theory,” JCAP, Volume 2021, Issue 11, id.028, 43 pp., Nov. 2021. 

\bibitem{AvilesMG}
A. Aviles and G. Niz,  "Galaxy three-point correlation function in modified gravity," Physical Review D, Volume 107, Issue 6, article id.063525, March 2023. 

\bibitem{deMattia2021eboss}
A. de Mattia et al., "The completed SDSS-IV extended Baryon Oscillation Spectroscopic Survey: measurement of the BAO and growth rate of structure of the emission line galaxy sample from the anisotropic power spectrum between redshift 0.6 and 1.1", MNRAS Volume 501, Issue 4, p.5616-5645, Mar. 2021.

\bibitem{Eggemeier_Bias}
A. Eggemeier, et al., "Testing one-loop galaxy bias: Joint analysis of power spectrum and bispectrum," Physical Review D, Volume 103, Issue 12, article id.123550, June 2021. 

\bibitem{Sanchez}
A.G. Sanchez, et al., "The clustering of galaxies in the completed SDSS-III Baryon Oscillation Spectroscopic Survey: Cosmological implications of the configuration-space clustering wedges," MNRAS, Volume 464, Issue 2, p.1640-1658, Jan. 2017.

\bibitem{Tamone2020eboss}
A. Tamone, et al., "The completed SDSS-IV extended baryon oscillation spectroscopic survey: growth rate of structure measurement from anisotropic clustering analysis in configuration space between redshift 0.6 and 1.1 for the emission-line galaxy sample," MNRAS, Volume 499, Issue 4, pp.5527-5546, Dec. 2020.

\bibitem{Jain}
B. Jain and E. Bertschinger, "Second-order power spectrum and nonlinear evolution at hight redshift," ApJ, Volume 431, p. 495, Aug. 1994.

\bibitem{Sabiu_4PCF_measurment}
C.G. Sabiu et al., "Graph Database Solution for Higher-order Spatial Statistics in the Era of Big Data," ApJ Journal Supplement Series, Volume 242, Issue 2, article id. 29, 8 pp. (2019), June 2019. 

\bibitem{3PCFbiasConstraint}
C. K. McBride et al., "Three-point Correlation Functions of SDSS Galaxies: Constraining Galaxy-mass Bias," ApJ, Volume 739, Issue 2, article id. 85, 21 pp. (2011). Oct. 2011


\bibitem{Whelan1993}
C. T. Whelan, "On the evaluation of integrals over three spherical Bessel functions" J. Phys. B: At. Mol. Opt. Phys. 26 L823, Sept., 1993.  

\bibitem{Bertolini_Trispectrum}
D. Bertolini et al., "The trispectrum in the Effective Field Theory of Large Scale Structure," JCAP, Issue 06, article id. 052, June 2016. 


\bibitem{Eisen2PCF}
D. Eisenstein, H. Seo, and M. White, “On the Robustness of the Acoustic Scale in the Low-Redshift Clustering of Matter,” ApJ, Volume 664, Issue 2, pp. 660-674., Aug. 2007. 

\bibitem{EisenBAO}
D. Eisenstein and W. Hu, “Baryonic Features in the Matter Transfer Function,” ApJ, Volume 496, Issue 2, pp. 605-614., March 1998. 

\bibitem{DES}
DES Collaboration, "The Dark Energy Survey: more than dark energy - an overview," MNRAS Volume 460, Issue 2, p.1270-1299, Aug, 2016. 

\bibitem{itrispectrum_measurment}
D. Gualdi and L. Verde, "Integrated trispectrum detection from BOSS DR12 NGC CMASS," JCAP, Volume 2022, Issue 09, id.050, 36 pp., May 2017.  

\bibitem{Gualdi1}
D. Gualdi, H. Gil-Marín, and L. Verde, “Joint analysis of anisotropic power spectrum, bispectrum and trispectrum: application to N-body simulations,” JCAP, Volume 2021, Issue 07, id.008, 44 pp., July 2021. 

\bibitem{Gualdi3}
D. Gualdi, M. Marena, B. Joachimi and O. Lahav, "Maximal compression of the redshift-space galaxy power spectrum and bispectrum," MNRAS, Volume 476, Issue 3, pp. 4045–4070, May 2018.

\bibitem{Gualdi2}
D. Gauldi, S. Novell, H. Gil-Marín and L. Verde, “Matter trispectrum: theoretical modelling and comparison to N-body simulations,” JCAP, Issue 01, article id. 015 (2021), Jan. 2021. 

\bibitem{SDSS}
D.G. York, et al., "The Sloan Digital Sky Survey: Technical Summary," AJ, , Volume 120, Issue 3, pp. 1579-1587, Sept. 2002.

\bibitem{Jeong}
D. Jeong, “Cosmology with high (z>1) redshift galaxy surveys,” University of Texas at Austin, Aug. 2010

\bibitem{LadoBAO}
D.W. Pearson and L. Samushia, "A Detection of the Baryon Acoustic Oscillation features in the SDSS BOSS DR12 Galaxy Bispectrum," MNRAS, Volume 478, Issue 4, p.4500-4512, Aug. 2018. 

\bibitem{eboss}
eBoss collaboration et al., “Completed SDSS-IV extended Baryon Oscillation Spectroscopic Survey: Cosmological implications from two decades of spectroscopic surveys at the Apache Point Observatory,” Physical Review D, Volume 103, Issue 8, article id.083533, April 2021.

\bibitem{Bernardeau1995}
F. Bernardeau, L. Kofman, "Properties of the Cosmological Density Distribution Function," APJ, v.443, p.479, April 1995.

\bibitem{Bernardeau_review}
F. Bernardeau, S. Colombi, E. Gaztanaga and R. Scoccimarro, “Large-scale structure of the Universe and cosmological perturbation theory,” Physics Reports, Volume 367, Issue 1-3, p. 1-248, Sept. 2002.

\bibitem{Beutler}
F. Beutler, et al., "The clustering of galaxies in the completed SDSS-III Baryon Oscillation Spectroscopic Survey: anisotropic galaxy clustering in Fourier space," MNRAS, Volume 466, Issue 2, p.2242-2260, April 2017.

\bibitem{Farshad_neutrino}
F. Kamalinejad and Z. Slepian, “A Non-Degenerate Neutrino Mass Signature in the Galaxy Bispectrum,” eprint arXiv:2011.00899, Nov. 2020. 

\bibitem{Bouchet}
F.R. Bouchet, et al., "Perturbative Lagrangian approach to gravitational instability," Astronomy and Astrophysics, v.296, p.575, Aoril 1995. 

\bibitem{ZwickyCatalogue}
F. Zwicky, "Catalogue of galaxies and of clusters of galaxies, Vol. I," CIT, Vol I to Vol. VI, 1961-1968. 

\bibitem{D'Amico_Bispectrum}
G. D'Amico, et. al., "The BOSS bispectrum analysis at one loop from the Effective Field Theory of Large-Scale Structure," eprint arXiv:2206.08327, June 2022. 

\bibitem{studyLSS3}
H. Gil-Marín et al., “The clustering of galaxies in the SDSS-III Baryon Oscillation Spectroscopic Survey: RSD measurement from the power spectrum and bispectrum of the DR12 BOSS galaxies,” MNRAS, Volume 465, Issue 2, p.1757-1788, Feb. 2017. 

\bibitem{LAMOST}
H. Yan, et al., "Overview of the LAMOST survey in the first decade," The Innovation, Volume 3, article id. 100224., March 2022. 

\bibitem{DeLaBella3sbf}
L. Fonseca De La Bella et al., "The matter power spectrum in redshift space using effective field theory," JCAP, Issue 11, article id. 039 (2017), Nov. 2017. 

\bibitem{Bautista2021eboss}
J. Bautista, et al., "The completed SDSS-IV extended Baryon Oscillation Spectroscopic Survey: measurement of the BAO and growth rate of structure of the luminous red galaxy sample from the anisotropic correlation function between redshifts 0.6 and 1," MNRAS, Volume 500, Issue 1, pp.736-762, Jan. 2021.

\bibitem{J.Chellino}
J. Chellino and Z. Slepian, "Triple-Spherical Bessel Function Integrals with Exponential and Gaussian Damping: Towards An Analytic N-Point Correlation Function Covariance Model," PRSA, Volume 479, Issue 2276, article id.20230138, Aug. 2023.

\bibitem{FOG}
J.C. Jackson, "Fingers of God a critique of Ree's theory of primordial gravitational radiation," MNRAS, Volume 156, p. 1P, 1972. 

\bibitem{EM_Jackson}
J.D. Jackson, "Classical Electrodynamics," New York :Wiley, 1999. 


\bibitem{Hamilton}
J. Hamilton, “Linear Redshift Distortions: A Review,” ASSL Series vol no: 231, ISBN: 079235074X, p.185, 1998. 


\bibitem{Hou2018eboss}
J. Hou, et al. "The clustering of the SDSS-IV extended Baryon Oscillation Spectroscopic Survey DR14 quasar sample: anisotropic clustering analysis in configuration space," MNRAS, Volume 480, Issue 2, p.2521-2534, Oct. 2018. 

\bibitem{Hou2021eboss}
J. Hou, et al. "The completed SDSS-IV extended Baryon Oscillation Spectroscopic Survey: BAO and RSD measurements from anisotropic clustering analysis of the quasar sample in configuration space between redshift 0.8 and 2.2," MNRAS, Volume 500, Issue 1, pp.1201-1221, Jan. 2021.

\bibitem{Hou2023:MGreview}
J. Hou et al., "Cosmological Probes of Structure Growth and Tests of Gravity", Universe Volume 9, number 7, p. 302, June 2023.

\bibitem{hou-analytic-covar}
J. Hou, R. Cahn, O.H.E. Philcox and Z. Slepian, "Analytic Gaussian covariance matrices for galaxy N -point correlation functions," Physical Review D, Volume 106, Issue 4, article id.043515, Aug. 2022.

\bibitem{Parity-odd}
J. Hou, R. Cahn and Z. Slepian, “Measurement of Parity-Odd Modes in the Large-Scale 4-Point Correlation Function of SDSS BOSS DR12 CMASS and LOWZ Galaxies,” MNRAS, Volume 522, Issue 4, pp.5701-5739, May 2023.

\bibitem{Indefinite_Integrals_SBF}
J. K. Bloomfield, S. H. P. Face and Z. Moss, "Indefinite Integrals of Spherical Bessel Functions," eprint arXiv:1703.06428, Mar. 2017. 

\bibitem{Fry4PCF_BBGKY1}
J.N. Fry, "The four-point function in the BBGKY hierarchy," APJ Part 1, vol. 262, Nov. 15, 1982, p. 424-431, Nov. 1982

\bibitem{Fry4PCF_BBGKY2}
J.N. Fry, "Galaxy N-point correlation functions - Theoretical amplitudes for arbitrary N," APJ Part 2, vol. 277 p. L5-L8, Feb. 1984. 

\bibitem{Fry4PCFdef}
J.N. Fry and J. Peebles, "Statistical analysis of catalogs of extragalactic objects. IX. The four-point galaxy correlation function," APJ Part 1, vol. 221, Apr. 1, 1978, p. 19-33. April 1978

\bibitem{Grieb}
J.N. Grieb, et al., "The clustering of galaxies in the completed SDSS-III Baryon Oscillation Spectroscopic Survey: Cosmological implications of the Fourier space wedges of the final sample," MNRAS, Volume 467, Issue 2, p.2085-2112, May 2017.

\bibitem{PeeblesBAO}
J. Peebles and J. Yu, “Primeval Adiabatic Perturbation in an Expanding Universe,” ApJ, vol. 162, p.815, Jan. 1970. 

\bibitem{BondBAO1}
J.R. Bond and G. Efstathiou, "Cosmic background radiation anisotropies in Universes dominated by nonbaryonic dark matter," Astrophysical Journal, Part 2, vol. 285, Oct. 15, 1984, p. L45-L48, Oct. 1984. 

\bibitem{BondBAO2}
J.R. Bond and G. Efstathiou, "The statistics of cosmic background radiation fluctuations," MNRAS, vol. 226, June 1, 1987, p. 655-687, June 1987.


\bibitem{KiDS}
J.T.A. de Jong et al., "The Kilo-Degree Survey", The Messenger, vol. 154, p. 44-46, Dec. 2013.


\bibitem{Bias1}
K. Chan, R. Scoccimarro and R. Sheth, “Gravity and large-scale non-local bias,” Physical Review D, vol. 85, Issue 8, id. 083509, Apr. 2012.

\bibitem{HETDEX}
K. Gebhardt et al. "The Hobby-Eberly Telescope Dark Energy Experiment (HETDEX) Survey Design, Reductions, and Detections"  The ApJ Volume 923, Issue 2, id.217, 39 pp., Dec. 2021. 
/
\bibitem{Sph_Bessel_Integral_kiersten}
K. Meigs and Z. Slepian, "On a General Method for Resolving Integrals of Multiple Spherical Bessel Functions Against Power Laws into Distributions," eprint arXiv:2112.07809, Dec. 2021. 

\bibitem{BOSS}
K.S. Dawson, et al., "The Baryon Oscillation Spectroscopic Survey of SDSS-III," AJ,  Volume 145, Issue 1, article id. 10, 41 pp. (2013), Jan. 2013.

\bibitem{eBOSS}
K.S. Dawson, et al., "The SDSS-IV Extended Baryon Oscillation Spectroscopic Survey: Overview and Early Data," AJ, Volume 151, Issue 2, article id. 44, 34 pp. (2016), Feb. 2016. 

\bibitem{Higher_order_info}
L.Samushia, Z. Slepian, and F. Villaescusa-Navarro, "Information of Higher Order Galaxy Correlation Functions," MNARS Volume 505, Issue 1, pp.628-641, July 2021. 

\bibitem{Bias2}
M. Abidi, T. Baldauf, “Cubic halo bias in Eulerian and Lagrangian space,” JCAP, Issue 07, article id. 029 (2018), July 2018. 

\bibitem{LickCatalogue}
M.E. Brown and E. J. Groth, "The Shane Wirtanen Counts: Observability of the galaxy correlation function," ApJ Part 1, vol. 338, p. 605-617, March 1989. 


\bibitem{Ivanov_Bispectrum}
M. M. Ivanov, et al., "Cosmology with the galaxy bispectrum multipoles: Optimal estimation and application to BOSS data," Physical Review D, Volume 107, Issue 8, article id.083515, April 2023.

\bibitem{MorescoBAO}
M. Moresco et al., "C3: Cluster Clustering Cosmology. II. First Detection of the Baryon Acoustic Oscillations Peak in the Three-point Correlation Function of Galaxy Clusters," ApJ, Volume 919, Issue 2, id.144, 13 pp., Oct. 2021. 

\bibitem{DESTUDY}
M. Moresco et al., "Disentangling interacting dark energy cosmologies with the three-point correlation function," MNRAS Volume 443, Issue 4, p.2874-2886 Oct. 2014.

\bibitem{QFT}
M. Peskin and D. Schroeder, "An Introduction to Quantum Field Theory," Westview Press, Chicago, 1995.

\bibitem{TsedrikDE}
M. Tsedrik, et. al., "Interacting dark energy from the joint analysis of the power spectrum and bispectrum multipoles with the EFTofLSS," MNRAS, Volume 520, Issue 2, pp.2611-2632, April 2023. 

\bibitem{studyLSS1}
N. Agarwal, V. Desjacques, D. Jeong and F. Schmidt, “Information content in the redshift-space galaxy power spectrum and bispectrum”, JCAP, Volume 2021, March 2021. 

\bibitem{modified_gr1}
N. Bartolo, E. Bellini, D. Bertacca, S. Matarrese, “Matter bispectrum in cubic Galileon cosmologies,” JCAP, Issue 03, article id. 034, (2013).

\bibitem{Kaiser}
N. Kaiser, "Clustering in real space and in redshift space," MNRAS, Volume 225, pp. 1-21, Jul. 1987. 

\bibitem{studyLSS2}
N. Sugiyama, S. Shun, F. Beutler, H. Seo, “Towards a self-consistent analysis of the anisotropic galaxy two- and three-point correlation functions on large scales: application to mock galaxy catalogues,” MNRAS, Volume 501, Issue 2, pp.2862-2896, Feb. 2021. 

\bibitem{PFS}
N. Tamura, et al., "Prime Focus Spectrograph (PFS) for the Subaru telescope: overview, recent progress, and future perspectives," Proceedings of the SPIE, Volume 9908, id. 99081M 17 pp. (2016), Aug. 2016. 

\bibitem{Philcox-Parity}
O.H.E. Philcox, "Probing parity violation with the four-point correlation function of BOSS galaxies," Physical Review D, Sep. 2022. 

\bibitem{Philcox_4PCF_measurment}
O.H.E. Philcox, J. Hou and Z. Slepian, "A First Detection of the Connected 4-Point Correlation Function of Galaxies Using the BOSS CMASS Sample," eprint arXiv:2108.01670, Aug. 2021. 

\bibitem{LadoRSDBispectrum}
P. Gagrani and L. Samushia, "Information Content of the Angular Multipoles of Redshift-Space Galaxy Bispectrum, MNRAS, Volume 467, Issue 1, p.928-935, May 2017. 
 
\bibitem{DESI}
P. Martini et al., "Overview of the Dark Energy Spectroscopic Instrument," Proceedings of the SPIE, Volume 10702, id. 107021F 11 pp. (2018). July 2018. 

\bibitem{Bias3}
P. McDonald and A. Roy, “Clustering of dark matter tracers: generalizing bias for the coming era of precision LSS,” JCAP, Issue 08, id. 020 (2009), Aug. 2009.

\bibitem{Planckdata}
Planck Collaboration, "Planck 2015 results. I. Overview of products and scientific results," Astronomy \& Astrophysics, Volume 594, id.A1, 38 pp., Sept. 2016.

\bibitem{SunyaevBAO}
R.A. Sunyaev and Ya. B. Zel'dovich, "The Observations of Relic Radiation as a Test of the Nature of X-Ray Radiation from the Clusters of Galaxies," Astrophysics and Space Physics, Vol. 4, p.173, Nov. 1972. 


\bibitem{Iso_fun}
R. Cahn and Z. Slepian, “Isotropic N-Point Basis Functions and Their Properties,” Journal of Physics A: Mathematical and Theoretical, Volume 56, Issue 32, id.325204, 38 pp. Aug. 2023.


\bibitem{hou_bao}
J. Hou, Z. Slepian, and D. Jamieson, ``Can Baryon Acoustic Oscillations Illuminate the Parity-Violating Galaxy 4PCF?'', eprint arXiv:2410.05230, Oct. 2024.

\bibitem{cahn_prl}
R. Cahn, Z. Slepian, and J. Hou, "A Test for Cosmological Parity Violation Using the 3D Distribution of Galaxies," eprint arXiv:2110.12004, Oct. 2021. 


\bibitem{Euclid}
R. Laurejis et al., "Euclid Definition Study Report," eprint arXiv:1110.3193, Oct. 2011. 

\bibitem{4SBF_integral}
R. Mehrem, "Analytical evaluation of an infinite integral over four spherical Bessel functions",
Applied Mathematics and Computation,Volume 219, Issue 21, May 2013.

\bibitem{Sph_Bessel_int_Rami}
R. Mehrem, "The Plane Wave Expansion, Infinite Integrals and Identities involving Spherical Bessel Functions," Appl. Math. Comput. 217 (2011) 5360, Feb. 2011. 

\bibitem{Neveux2020eboss}
R. Neveux, et al., "The completed SDSS-IV extended Baryon Oscillation Spectroscopic Survey: BAO and RSD measurements from the anisotropic power spectrum of the quasar sample between redshift 0.8 and 2.2" MNRAS, Volume 499, Issue 1, pp.210-229, Nov. 2020.

\bibitem{Scoccimarro1998}
R. Scoccimarro, H. Couchman and J. Frieman, “The Bispectrum as a Signature of Gravitational Instability in Redshift Space,” ApJ, Volume 517, Issue 2, pp. 531-540, June 1999. 

\bibitem{4MOST}
R.S. De Jong, et al., "4MOST: Project overview and information for the First Call for Proposals," The Messenger, vol. 175, p. 3-11, March 2019. 

\bibitem{modified_gr2}
S. Alam, et al., “Towards testing the theory of gravity with DESI: summary statistics, model predictions and future simulation requirements,” JCAP, Volume 2021, Issue 11, id.050, 105 pp., Nov. 2021. 

\bibitem{Cole05}
S. Cole, et al., "The 2dF Galaxy Redshift Survey: power-spectrum analysis of the final data set and cosmological implications," MNRAS, Volume 362, Issue 2, pp. 505-534, Sept. 2005

\bibitem{Satpathy}
S. Satpathy, et al., "The clustering of galaxies in the completed SDSS-III Baryon Oscillation Spectroscopic Survey: on the measurement of growth rate using galaxy correlation functions," MNRAS, Volume 469, Issue 2, p.1369-1382, Aug. 2017. 


\bibitem{Fabrikant}
V. Fabrikant, “Elementary exact evaluation of infinite integrals of the product of several spherical bessel
functions, power and exponential,” Quarterly of Applied Mathematics, Jan. 2013

\bibitem{Roman}
V. P. Bailey, et al., "Nancy Grace Roman Space Telescope Coronagraph Instrument Overview and Status," Proceedings of the SPIE, Volume 12680, id. 126800T 11 pp. Oct. 2023. 

\bibitem{HuBAO}
W. Hu and N. Sugiyama, "Small-Scale Cosmological Perturbations: an Analytic Approach," ApJ, v.471, p.542, Nov. 1996.

\bibitem{Kobayashi}
Y. Kobayashi, "Fast computation of the non-Gaussian covariance of redshift-space galaxy power spectrum multipoles," PRD, Volume 108, Issue 10, article id.103512, Nov. 2023. 

\bibitem{LSST}
Z. Ivezić et al., "LSST: From Science Drivers to Reference Design and Anticipated Data Products,"  ApJ Volume 873, Issue 2, article id. 111, 44 pp. (2019). March 2019. 

\bibitem{modeling3PCF}
Z. Slepian and D. Eisenstein, “Modelling the large-scale redshift-space 3-point correlation function of galaxies,” MNRAS, Volume 469, Issue 2, p.2059-2076, Aug. 2017. 

\bibitem{BaryonVelocity}
Z. Slepian and D. Eisenstein, "On the signature of the baryon-dark matter relative velocity in the two- and three-point galaxy correlation functions," MNRAS, Volume 448, Issue 1, p.9-26, March 2015.

\bibitem{DetectionBAO3pcf}
Z. Slepian et al., "Detection of Baryon Acoustic Oscillation Features in the Large-Scale 3-Point Correlation Function of SDSS BOSS DR12 CMASS Galaxies," Volume 469, Issue 2, p.1738-1751, Aug. 2014.

\bibitem{SlepianBAO}
Z. Slepian et al., "The large-scale three-point correlation function of the SDSS BOSS DR12 CMASS galaxies", MNRAS Volume 468, Issue 1, p.1070-1083, June 2017. 


\bibitem{SlepianDecoupling}
Z. Slepian, "On Decoupling the Integrals of Cosmological Perturbation Theory," MNRAS, Volume 507, Issue 1,p. 1337-1360, Oct. 2021. 








 






\end{thebibliography}
\end{document}